\documentclass[11pt]{article}

\usepackage[utf8]{inputenc}
\usepackage{graphicx}
\usepackage{tikz}
\usepackage{pgfplots}
\usepackage{pgfopts}
\usepackage{rotating}
\usepackage{float}
\usepackage{xcolor}
\usepackage{colortbl}
\usepackage{longtable}
\usepackage{fancyvrb}
\usepackage{mdframed}
\usepackage{blindtext}
\usepackage{amsmath}
\usepackage{tkz-tab}
\usepackage{adjustbox}
\usepackage{amssymb}
\usepackage{amsbsy}
\usepackage{booktabs}
\usepackage{caption}
\usepackage{parskip}
\usepackage[algoruled, linesnumbered]{algorithm2e}
\usepackage{listings}
\usepackage{enumitem}
\usepackage{dirtytalk}
\usepackage{tcolorbox}
\usepackage{multirow}
\usepackage[hyphens]{url}
\usepackage{hyperref}
\hypersetup{
    hidelinks,
    breaklinks
}

\usetikzlibrary{shapes, arrows, positioning, calc, fit, backgrounds}
\usetikzlibrary{shapes.geometric}
\usepackage{tikz}
\usetikzlibrary{shapes.geometric, arrows, positioning}

\tikzset{
  decision/.style = {diamond, draw, fill=blue!10,
                     text badly centered, inner sep=4pt},
  action/.style   = {rectangle, draw, rounded corners, fill=gray!10,
                     text centered, inner sep=4pt},
  arrow/.style    = {->, thick}
}

\tcbuselibrary{skins}
\tcbuselibrary{breakable}
\tcbset{
  width=\textwidth,
  center,
  colback=blue!5!white
}

 \newcommand{\ourapproach}{DBTuneSuite\xspace}
 \newcommand{\authorcontributions}[1]{%
}

\usepackage{enumitem}
\setlist{noitemsep, topsep=2pt}

\newcolumntype{C}[1]{>{\centering\arraybackslash}m{#1}}

\title{\textbf{\ourapproach: An Extendible Experimental Suite to Test the Time Performance of Multi-layer Tuning Options on Database Management Systems}}

\author{
Amani Agrawal\textsuperscript{1},
Tianxin Wang\textsuperscript{1},
Dennis Shasha\textsuperscript{1}\thanks{Corresponding author: \texttt{shasha@cs.nyu.edu}} \\
\textsuperscript{1}Department of Computer Science, Courant Institute of Mathematical Sciences,\\
New York University, New York, NY 10012, USA \\
\texttt{aa10733@nyu.edu, tw3090@nyu.edu, shasha@cs.nyu.edu}
}

\date{January 2026}

\begin{document}

\maketitle

\begin{abstract}
\ourapproach is a suite of experiments on four widely deployed free database systems to test their performance under various query/upsert loads and under various tuning options. The suite provides: (i) scripts to generate data and to install and run tests, making it expandable to other tests and systems; (ii) suggestions of which systems work best for which query types; and (iii) quantitative evidence that tuning options widely used in practice can behave very differently across systems. This paper is most useful for database system engineers, advanced database users and troubleshooters, and students.
\end{abstract}


\vspace{0.5em}

\section{Introduction}

Tuning benchmarks help decide two questions: 

(i) Which tuning choices should one make to increase  the performance of a given workload on a given database system?

(ii) After the above is determined and an application's query profile is established, which database management system should one use  for that application? 

Our contributions apply to both of the above points and give tools to readers for their own experiments:
\begin{itemize}
\item 
We present a  variety of tuning options at all levels of a database application, from data loading to index selection to query writing to programming language-database interaction to table design. These choices have all proven useful to the last author in helping big companies in finance, telecommunications, and pharmaceuticals, make their databases run faster. For each option, we present experimental results on four popular free database systems: MySQL, MariaDB, DuckDB, and PostgreSQL.
\item 
We provide scripts and data for readers to adapt to their own applications and systems. The code and data used to run the experiments is available at: 
\href{https://github.com/pequeniajugar/DBTuneSuite.git}
{\texttt{GitHub Repository: Database Tuning}}.

\end{itemize}

{\em Note on paper Organization: }
After presenting related work and the common data layouts, we devote one section to each tuning option. We present the setup and the qualitative conclusions in the body and present the detailed quantitative results in the appendices, one appendix per section. A reader who wants to understand the effect of each choice may read just the body. The reader who wants to understand the detailed results may read the corresponding appendices as well.

\section{Related Work}

Because \ourapproach compares tuning strategies on different database management systems, our related work is divided into (i) benchmarks to compare different database management systems and (ii) experimental works on database tuning.


\subsection{Benchmarking Different Database Systems}

Comparative benchmarks across various database systems highlight performance trade‑offs dependent on workload complexity and infrastructure configurations. Andjelić \textit{et~al.} \cite{andjelic2008} initially demonstrated significant performance advantages for MySQL, particularly using the MyISAM storage engine, over PostgreSQL in basic SQL operations such as \texttt{SELECT}, \texttt{INSERT}, \texttt{DELETE}, and \texttt{ORDER BY} on datasets of about 40,000 rows. Their findings recommended MySQL for applications prioritizing raw query speed, whereas PostgreSQL provided better consistency and robustness.

In contrast, Salunke \& Ouda (2024) \cite{salunke2024} presented a different scenario, using a custom benchmarking framework designed specifically for continuous-authentication applications. Their experiments on one-million-row tables indicated PostgreSQL significantly outperformed MySQL in \texttt{SELECT} queries, exhibiting latencies roughly 13 times lower in basic scenarios and about nine times faster with \texttt{WHERE} clauses. Additionally, PostgreSQL maintained low latency under mixed \texttt{SELECT} and \texttt{INSERT} workloads, while MySQL experienced notable latency increases. Thus, PostgreSQL exhibited superior concurrency performance and stability. The authors used InnoDB as the storage engine for MySQL, which tended to suffer under concurrent read-write workloads in their setup. This may partly explain the observed performance gap, because PostgreSQL’s MVCC (Multi-Version Concurrency Control) model handles such scenarios more efficiently.

Tongkaw \& Tongkaw (2016)\cite{tongkaw2016} performed OLTP benchmarks on Xen virtual machines, comparing MariaDB 10.0.21 with MySQL 5.6. The divergence from Salunke’s results stems from (i) the markedly smaller working set, which fits comfortably in the MySQL buffer pool, and (ii) the presence of  virtualization that enabled MySQL to achieve  roughly twice MariaDB’s throughput at the highest concurrency level tested (1000 threads).

Comparisons between MySQL and MariaDB also reveal nuanced performance differences. Min (2018) \cite{min2018} showed that both systems can match the performance of commercial databases such as MS SQL Server when indexing is used and data sizes remain moderate, suggesting that they are viable open‑source alternatives in typical read‑heavy scenarios on small databases.

Further comparative benchmarks between PostgreSQL and MariaDB by Han \& Choi (2024)\cite{han2024} investigated performance under different storage bandwidth conditions. The authors populated a 60 GB Sysbench OLTP database (ten 25M‑row tables) for write‑heavy tests and a 5 GB TPC‑H scale‑factor‑5 database for read‑heavy analytics. Pure \texttt{SELECT} workloads on TPC‑H showed PostgreSQL processing up to three times faster once bandwidth exceeded 1GB/s, whereas mixed \texttt{SELECT + INSERT} OLTP loads favored MariaDB below 500 MB/s but flipped in PostgreSQL’s favor at higher bandwidths.

DuckDB, a relatively new entrant introduced by Raasveldt \& Muhleisen (2019)\cite{raasveldt2019}, was compared against SQLite, MonetDBLite, and HyPer using analytical TPC‑H queries at scale factors 1 (approx   1 GB) and 10 (approx 10 GB), with demonstrations scaling beyond SF 20 on commodity laptops. At the larger scales, DuckDB significantly outperformed the other embedded engines, benefiting from its vectorized columnar execution model that handles substantial analytical workloads efficiently on modest hardware.

Krocz \textit{et al.} (2020)\cite{krocz2020} conducted comprehensive comparisons involving MySQL, MariaDB, PostgreSQL, and the in‑memory database H2. Their study indicated that PostgreSQL consistently outperformed MySQL and MariaDB in complex queries involving \texttt{JOIN} and \texttt{UPDATE} operations, while MariaDB remained competitive with PostgreSQL on simpler operations such as \texttt{SELECT} and \texttt{DELETE}. MySQL showed marked degradation once table sizes grew beyond about 100k rows—dropping from sub‑second response times at 10k rows to over 8s at 500k rows—whereas PostgreSQL maintained sub‑2s latencies across the same range.

These benchmark studies reveal distinct and sometimes contradictory results when comparing database systems. This implies that benchmarking on an application's specific queries and available hardware is a useful exercise in choosing a system. The benchmarks that we offer on MySQL, PostgreSQL, MariaDB, and DuckDB using the InnoDB storage engine can easily be modified to use different storage engines, table and index structures, and different queries.



\subsection{Reviews of Tuning Methods -- non-experimental}

Database tuning involves a variety of techniques across multiple layers of the system, from table formulation to indexes and data page layout. Previous work has focused primarily on physical layer tuning strategies. 

Mbaiossoum  et al.~\cite{mbaiossoumdatabase} provide a review of many important physical design options such as indexing, materialized views, horizontal fragmentation, parallel execution, object grouping, and query rewriting. Their study evaluates the applicability of these techniques across different types of database systems, suggesting that although originally developed for relational databases, many remain relevant in modern or hybrid architectures.

Sanders and Shin~\cite{926306} study denormalization as a tuning method, who argue that selectively introducing redundancy into database schema can reduce join complexity and improve query performance. Their evaluation highlights the trade-off between storage redundancy and analytical performance gain, particularly in data warehouse contexts.

Shasha and Bonnet ~\cite{10.1145/1024694.1024720} categorize database tuning into configuration, schema, index, query, concurrency control, and application interface tuning. Their work focuses on relational database systems, presenting detailed tuning actions with empirical evidence. This structured classification highlights how tuning is not confined to low-level system parameters but includes logical and workload-aware strategies. \ourapproach takes this wide spectrum view of database tuning and proposes experiments for modern free database systems.

These works present detailed and practical approaches for improving database system performance. They provide strong theoretical evidence for the significance of applying database tuning knobs and document the history of how DBMS users have managed heavy workloads. Our work builds on these established methods and incorporates additional experiments to evaluate whether some of these approaches, such as indexing and query rewriting, remain effective in modern relational DBMSs.

\subsection{Reviews of Tuning Methods -- experimental}

Some authors have benchmarked various tuning knobs, again mostly at the physical level, across different database systems to evaluate their effectiveness under a variety workloads.

Raman et al.~\cite{10.1145/3662165.3662764} propose a  benchmarking framework to assess the behavior of indexing strategies when varying the data sortedness in an ingestion workload (two learned structures (ALEX, LIPP), a traditional B+-tree, and the RocksDB LSM-tree). The results show that data sortedness  has a large influence on indexing performance.  For example, learned index ALEX performs 60\% better on ordered data than on disordered data. 
The B+ Tree remains stable regardless of order, while RocksDB gains 1.7× throughput on ordered streams but incurs up to 2.7× more compactions under disorder. \ourapproach extends that work by comparing the performance of clustered and non-clustered indexes as well as hash structures and on a variety of query types. The results surprised us.

Martins et al.~\cite{martins2021comparing}  test the performance gains from indexing through experiments on Oracle and PostgreSQL using the TPC-H benchmark. 
\\ \\
 Sobri et al.~\cite{sobri2022study} present a benchmark to test the impact of database connection pool size in microservice architectures using PostgreSQL. They simulate workloads of 500–2000 requests per second with the Vegeta tool and compare single-instance and multi-instance deployments. Their results show that small pool sizes cause severe latency under load, while a pool size of just  five offers good  performance in multi-instance setups, underscoring the importance of connection tuning. \ourapproach vastly increases the ranges of pool sizes and explores the impact on different systems.
\\ \\
Kanellis et al.~\cite{254280} evaluate configuration tuning through large-scale experiments on PostgreSQL and Cassandra using YCSB-A (write-heavy) and YCSB-B (read-heavy) workloads. Configuration tuning can involve 100s of parameters (involving commit log size, cache size, size triggers for forcing data to disk, concurrent threads, memtable size), but the most important, according to the authors have to do with the memtable managment and number of threads. The authors assert that automatically tuning parameters can yield performance improvements by a factor of two or more. 
Because \ourapproach avoids benchmarking different configuration parameter settings (such settings  differ from system to system), we consider the Kanellis et al. work to be complementary to ours. 

These works present benchmark experiments on various database tuning knobs and offer recommendations on how to improve them in specific situations and DBMSs to maximize system performance. Building upon these prior studies, our work complements them by introducing an experimental suite that extends the range of evaluated tuning knobs and broadens the scope of tested DBMSs. In particular, we not only revisit the knobs examined in previous benchmarks but also incorporate additional parameters and modern DBMS features that have emerged in recent years. 

\subsection{Automatic Configuration Tuning}

Several automatic tuning frameworks try to find the best settings for configuration parameters like those studied by Kanellis et al. \\ ~\cite{254280}. For example, Van Aken et al. \cite{van2017automatic} introduced OtterTune, which uses learned models to predict good configurations  for a given workload. Wang et al.\cite{wang2021udo} propose UDO, a framework that performs offline tuning of configuration knobs but also index tuning and picking among different pre-supplied transaction templates. 

\subsection{Contributions of \ourapproach}

As we have seen, the experimental literature does an excellent job in  the study of configuration parameter tuning  (having to do with physical resources like memory management and threads) and some index tuning of various kinds. The difference in our strategy is to benchmark tuning knobs at several levels,   from schema design to query design to indexes to  data loading. These knobs are ones that the third author has used extensively in industrial consulting, where he found that query level tuning  often has the greatest payoff. 

Running our benchmark on a given database system will help
\begin{itemize}
\item 
the implementers of those database systems to improve performance;
\item 
 advanced application developers to decide on schema design, query writing methods, index selection, and data loading;
\item 
developers and students to design experiments that are relevant to their applications -- regarding schema design, data loading procedures, queries, and repeated tests to measure variance; 
\item 
technical leads to choose a database management system appropriate for their applications as we summarize in Section \ref{conclusions}.
\item 
advanced application developers to choose tuning options, the effectiveness of which can vary from system to system and are sometimes quite surprising   as discussed  in Section \ref{surprises}.

\end{itemize}


\section{Materials and Methods}

This paper is an experimental paper, so the methods are divided among many sections:
\begin{itemize}
\item 
Data setup in Section \ref{shared-settings}.
\item 
Individual experiments having to do with tuning choices starting in Section \ref{dataload} and until Section \ref{connectionpooling}.
\end{itemize}

\section{Shared Schemas, Data, and Computational Environment} \label{shared-settings}

{\bf DBMS Versions Used in Our Experiments }
\begin{itemize}
    \item MariaDB 11.4
    \item DuckDB 1.1
    \item PostgreSQL 13.20
    \item MySQL 9.1.0 
\end{itemize}

All of our experiments share some common storage engine and hardware platforms.


{\bf Default Storage Engines }
MySQL and MariaDB use the InnoDB storage engine by default. MySQL, MariaDB, and PostgreSQL all use B+ Tree structures for indexing, while DuckDB adopts Adaptive Radix Tree (ART). In the experiment of Section \ref{Hash-vs-btree}, we consider Hashing as well.

{\bf Default Hardware Environments}
All benchmarks were run on the NYU CIMS compute server \texttt{crunchy5.cims.nyu.edu}, which is equipped with four 16-core AMD Opteron 6272 processors (64 physical cores in total, 2.1 GHz), 256 GB of RAM, and Red Hat Enterprise Linux 9 as the host OS. Each query was executed ten times and averaged. 

{\bf Default Schemas}
Because several of our tests use either (i) a particular Employee schema or (ii) the TPCH (Transaction Processing Council benchmark H) schema,  we describe each of those database structures in turn.
\subsection{Employee}
 \label{subsec:benchmark employee}

 We designed a synthetic benchmark dataset named \texttt{employee} to evaluate index performance under controlled conditions. The dataset is uniformly distributed and supports recommended input size: $n=10^i$.  

\begin{lstlisting}[language=SQL, 
                   label={lst:employee schema}, 
                   caption={Schema and Index Definitions for the \texttt{employee} Table} ]
-- Table schema
employee(ssnum, name, lat, longitude, hundreds1, hundreds2);

-- Index definitions
NONCLUSTERED INDEX c ON employee(hundreds1) WITH FILLFACTOR = 100;
NONCLUSTERED INDEX nc ON employee(hundreds2);
INDEX nc3 ON employee (ssnum, name, hundreds2);
INDEX nc4 ON employee (lat, ssnum, name);
\end{lstlisting}

Since DuckDB does not support configuring the fillfactor parameter, we excluded this setting from our DuckDB experiments. 

\paragraph{Column Descriptions:}
\begin{itemize}
  \item \textbf{ssnum:} Primary key, integer values in the range $[1, n]$.
  \item \textbf{name:} Strings in the format \texttt{Employee1} to \texttt{Employee(n/100)}.
  \item \textbf{lat:} Integer values in the range $[100, n/100]$.
  \item \textbf{long:} Integer values in the range $[100, n/100]$.
  \item \textbf{hundreds1:} Integer values in the range $[100, n/100]$.
  \item \textbf{hundreds2:} Integer values in the range $[100, n/100]$.
\end{itemize}

\paragraph{Data Properties::}
\begin{itemize}
  \item Column \texttt{lat}, \texttt{long}, \texttt{hundreds1}, \texttt{hundreds2} contains $n/100$ consistent unique values and every unique value is held in 100 rows.
  \item \texttt{name} contains $n/100$ unique values and every unique value is held in 100 rows.
  \item \texttt{ssnum} and \texttt{hundreds1} are sorted in ascending order. 
  \item The values in columns \texttt{lat}, \texttt{long}, \texttt{hundreds2} are  permuted copies of the \texttt{hundreds1} column values, with each of the three columns containing a different permutation.

\end{itemize}



\subsection{TPC-H}
\label{subsec:benchmark tpch}

We use the industry-standard \href{https://www.tpc.org/TPC\_Documents\_Current\_Versions/pdf/TPC-H\_v3.0.1.pdf}{ TPC-H Benchmark Specification} (accessed February 2025)  benchmark to evaluate engine performance on realistic workloads. Developed by the Transaction Processing Performance Council (TPC), the TPC-H benchmark models a product order and supply chain environment.

The benchmark supports multiple dataset sizes, known as \emph{scale factors}, and we focus on the $0.01667$ and $1.667$ scale factors, which result in approximately $10^5$ and $10^7$ rows respectively in the largest table (\texttt{lineitem}).

\begin{lstlisting}[language=SQL, 
                   label={lst:tpch-schema}, 
                   caption={Schema for the TPC-H Benchmark}]
-- lineitem table
lineitem(
  l_orderkey      INTEGER REFERENCES orders(o_orderkey),
  l_partkey       INTEGER REFERENCES part(p_partkey),
  l_suppkey       INTEGER REFERENCES supplier(s_suppkey),
  l_linenumber    INTEGER,
  l_quantity      DECIMAL(15,2),
  l_extendedprice DECIMAL(15,2),
  l_discount      DECIMAL(15,2),
  l_tax           DECIMAL(15,2),
  l_returnflag    CHAR(1),
  l_linestatus    CHAR(1),
  l_shipdate      DATE,
  l_commitdate    DATE,
  l_receiptdate   DATE,
  l_shipinstruct  CHAR(25),
  l_shipmode      CHAR(10),
  l_comment       VARCHAR(44),
  PRIMARY KEY (l_orderkey, l_linenumber)
);

-- supplier table
supplier(
  s_suppkey   INTEGER PRIMARY KEY,
  s_name      CHAR(25),
  s_address   VARCHAR(40),
  s_nationkey INTEGER REFERENCES nation(n_nationkey),
  s_phone     CHAR(15),
  s_acctbal   DECIMAL(15,2),
  s_comment   VARCHAR(101)
);

-- nation table
nation(
  n_nationkey INTEGER PRIMARY KEY,
  n_name      CHAR(25),
  n_regionkey INTEGER REFERENCES region(r_regionkey),
  n_comment   VARCHAR(152)
);

-- region table
region(
  r_regionkey INTEGER PRIMARY KEY,
  r_name      CHAR(25),
  r_comment   VARCHAR(152)
);
\end{lstlisting}


\subsection{Forcing the Use of Indexes}\label{Forcing the Use of Indexes}

In several of our experiments, we force the use of indexes (either hash or B+ Tree, or Adaptive Radix Tree). We do this, for example, to see whether one kind of indexing is better than another or better than scanning. Here we show how to force an index on each of the studied systems:

\begin{itemize}
    \item DuckDB:
    \begin{lstlisting}[language=SQL,caption={DuckDB enforces the use of indexes by adjusting specific configuration parameters. The index\_scan\_percentage and index\_scan\_max\_count settings define thresholds for enabling index scans. Specifically, an index scan is triggered instead of a full table scan if the number of matching rows is less than MAX(index\_scan\_max\_count, index\_scan\_percentage × total\_row\_count).},label={lst:duckdb force index}] 
SET index_scan_percentage = 1;
SET index_scan_max_count = 1000000;
\end{lstlisting} 

    \item MariaDB:
    \begin{lstlisting}[language=SQL,caption={On MariaDB, indexes are enforced by explicitly specifying index names in the query using the FORCE INDEX clause.},label={lst:mariadb force index}] 
SELECT cols FROM table FORCE INDEX (idx_col) WHERE col = {value};
\end{lstlisting} 

    \item MySQL:
    \begin{lstlisting}[language=SQL,caption={On MySQL, indexes are enforced by explicitly specifying index names in the query using the FORCE INDEX clause.},label={lst:mysql force index}] 
SELECT cols FROM table FORCE INDEX (idx_col) WHERE col = {value};
\end{lstlisting} 

    \item PostgreSQL:
    \begin{lstlisting}[language=SQL,caption={On PostgreSQL, we enforce index usage by modifying planner configuration parameters to disable sequential scans and encourage index-based access.},label={lst:duckdb force index}] 
SET enable_seqscan TO off;
\end{lstlisting} 

\end{itemize}

\section{Data Loading}\label{dataload}
This experiment evaluates the performance of different methods of inserting data into a database:
\begin{enumerate}
\item 
The first method is to insert data  row by row, committing after each row.
\item 
The second is batch loading in which inserts are done in various-sized batches (original file is divided into files having [100 / 1000 / 10000 / 50000 / 100000] rows, each of which is loaded as a separate transaction ).
\item 
The third method is direct path loading, which bypasses the SQL engine’s conventional processing and writes data directly to storage structures.
\end{enumerate} 
This experiment compares these strategies on medium-to-large datasets to assess their impact on load time and resource utilization.

\begin{enumerate}

\item
\textbf{Experiment Goal:} \\[3pt]
The goal of the experiment is to evaluate the performance impact of the different data loading strategies. 
 \\

\item 
\textbf{Experiment Setup:}
\\ The experiment adopts the \texttt{lineitem} table, which follows the TPC-H schema described in Section~\ref{subsec:benchmark tpch} but with no indexes, no primary keys, and no foreign keys. We test two sizes of the table:  $\mathbf{10^{5}}$ and $\mathbf{10^{7}}$ rows. 
The queries shown in Listings \ref{lst:insert}, \ref{lst:batching} and \ref{lst:direct-path} upload data to the \texttt{lineitem} table.
\\[0.75\baselineskip] 
\begin{minipage}
{\linewidth} 
\begin{lstlisting}[language=SQL,caption={Query to insert data row-by-row in the lineitem table.},label={lst:insert}] 
INSERT INTO lineitem (
    l_orderkey, l_partkey, l_suppkey, l_linenumber,
    l_quantity, l_extendedprice, l_discount, l_tax,
    l_returnflag, l_linestatus, l_shipdate, l_commitdate,
    l_receiptdate, l_shipinstruct, l_shipmode, l_comment
) VALUES (%s, %s, %s, %s, %s, %s, %s, %s, %s, %s, %s, %s, %s, %s, %s, %s)
\end{lstlisting} 
\begin{lstlisting}[language=SQL,caption={Query to insert data by batching the data in the lineitem table.},label={lst:batching}] 
LOAD DATA LOCAL INFILE '{batch_file}' INTO TABLE lineitem
FIELDS TERMINATED BY '|' LINES TERMINATED BY '\n';
\end{lstlisting} 

\begin{lstlisting}[language=SQL,caption={Query to insert data through direct path in the lineitem table.},label={lst:direct-path}] 
LOAD DATA INFILE '{file_path}'
INTO TABLE lineitem
FIELDS TERMINATED BY '|'
LINES TERMINATED BY '\n';
\end{lstlisting} 
\end{minipage}

\item
Please refer to \textbf{Appendix \ref{dataloading-appendix}} for the Quantitative Experimental Results and to \href{https://github.com/pequeniajugar/DBTuneSuite/tree/main/scripts/bulk_loading_data}{\texttt{Bulk Loading Experiment Scripts}} for the code to run those experiments.

\item
\textbf{Qualitative Conclusions}
\\
The data loading strategy has a significant impact on performance in our experiment. Row-by-row insertion, while straightforward, introduces high overhead from frequent commits and logging, making it unsuitable for large datasets.

Batching inserts proved to be a highly effective approach, offering substantial speedups across all engines. For large tables, it  strikes a good balance between efficiency and portability—especially in engines where direct path loading is unstable.

Direct path loading, when well-supported, can offer dramatic performance gains. On both datasets, it was often the fastest method, outperforming batching by wide margins in some engines, especially on DuckDB and MariaDB.

The results are consistent with the findings of Shasha and Bonnet \\ ~\cite{10.1145/1024694.1024720}, who concluded that direct path loading is often the most efficient approach for bulk data loading.

\end{enumerate}
\section{Data Loading with Different Batch Sizes}
Testing different batch sizes for data loading simply means inserting data in groups of varying sizes (e.g., 100 rows vs. 1 000 rows) to see which group size loads the whole dataset fastest and most efficiently.

\begin{enumerate}

\item
\textbf{Experiment Goal:} \\[3pt]
The goal of this experiment is to quantify how varying batch sizes — e.g., single-row, medium, and large batch inserts — affect data-load performance.

\item 
\textbf{Experiment Setup:}
\\ The experiment adopts the \texttt{lineitem} table, which follows the TPC-H schema described in Section~\ref{subsec:benchmark tpch} but with no indexes, no primary keys, and no foreign keys at two scales: $\mathbf{10^{5}}$ and $\mathbf{10^{7}}$ rows. 
The experiments begin with an empty table. The original file is divided into sub-files of varying sizes — [1 / 100 / 1,000 / 10,000 / 50,000 / 100,000] for a $\mathbf{10^{5}}$ row file, and [20,000 / 40,000 / 60,000 / 80,000 / 100,000] for a $\mathbf{10^{7}}$ row file. Each sub-file is loaded sequentially into the table as an individual batch.

The queries shown in Listing \ref{lst:batch} is used to load the data set in varying batch sizes. 
\\[0.75\baselineskip] 
\begin{minipage}
{\linewidth} 
\begin{lstlisting}[language=SQL,caption={Query to load batches of data into a table.},label={lst:batch}] 
LOAD DATA LOCAL INFILE '{batch_file}' INTO TABLE lineitem
FIELDS TERMINATED BY '|' LINES TERMINATED BY '\n';
\end{lstlisting} 
\end{minipage}

\item
Please refer to \textbf{Appendix \ref{batchloading-appendix}} for the Quantitative Experimental Results and to \href{https://github.com/pequeniajugar/DBTuneSuite/tree/main/scripts/bulk_loading_data}{\texttt{Batch Loading Experiment Scripts}} for the code to run those experiments.
\\ 

\item
\textbf{Qualitative Conclusions}
\\
Across the 10 million-row test, bigger batches always won on DuckDB: moving from 20,000 to 100,000 rows roughly halved load time. For MariaDB larger batches never hurt,  though the difference was minimal compared to the other engines. 
For PostgreSQL, batching initially improved performance but helped little beyond a batch size of 10,000 for the $10^7$-row dataset, with larger batch sizes sometimes causing slight performance degradation.
MySQL exhibited a similar trend, with performance stabilizing after the batch size reached 20,000. Batch loading is a good idea except for a few cases in $10^5$ rows, which shows worsening of time complexity as batch sizes increases from 1 to 100 on MariaDB and MySQL and after 10,000 for PostgreSQL. The results are generally consistent with the findings of Shasha and Bonnet~\cite{10.1145/1024694.1024720}, who concluded that throughput increases with larger batch sizes until it stabilizes.


\end{enumerate}

\section{Hash vs B+tree}\label{Hash-vs-btree}
B+ Trees and Hash structures are two commonly used data structures  in databases.
A B+ Tree  is a balanced tree structure optimized for efficient range queries, but capable of point, multipoint,  extremal queries (e.g. min/max), and prefix queries (e.g. name = "Sm*"). 
 In a B+ tree, all data values reside in the leaf nodes, while internal nodes serve to navigate from higher level nodes to lower level ones. The leaf nodes are often linked in order, from left to right, enabling fast sequential access.

In contrast, a Hash structure uses a hash function to map keys directly to specific locations, offering near-constant-time performance for exact-match queries (e.g., WHERE id = 123). 

\begin{enumerate}
    \item 
\textbf{Experiment Setup:}

We conduct experiments at two data scales: $10^5$ and $10^7$ rows. The \texttt{employee(ssnum, name, lat, longitude, hundreds1, hundreds2)} table (described in Section~\ref{subsec:benchmark employee}) is used throughout this experiment.

To isolate the effects of the data structure, all default indexes are removed. For each sub-experiment, a different data structure is applied to evaluate its performance impact.

For MySQL and MariaDB, the \texttt{MEMORY} storage engine is used, as it supports both B+ Trees and Hash structures. In contrast, DuckDB  supports only an Adaptive Radix Tree (ART), which functions similarly to B+ Trees; thus, we treat ART as a B+ Tree-equivalent and exclude Hash structure experiments for DuckDB. So the DuckDB measurements serve only to compare different systems on B+ Tree-like structures.

\begin{lstlisting}[language=SQL, label={lst:Hash-index}, caption={Hash index creation on three columns of \texttt{employee}}]
CREATE INDEX idx_ssnum ON employee (ssnum) USING Hash;
CREATE INDEX idx_hundreds1 ON employee (hundreds1) USING Hash;
CREATE INDEX idx_longitude ON employee (longitude) USING Hash;
\end{lstlisting}

\begin{lstlisting}[language=SQL, label={lst:btree-index}, caption={B+ Tree  creation on three columns of \texttt{employee}}]
CREATE INDEX idx_ssnum ON employee (ssnum) USING BTREE;
CREATE INDEX idx_hundreds1 ON employee (hundreds1) USING BTREE;
CREATE INDEX idx_longitude ON employee (longitude) USING BTREE;
\end{lstlisting}

On all database management systems, our tests include directives within the queries to force the database management system to use  indexes when present.  Section~\ref{Forcing the Use of Indexes} describes how to force the use of indexes.

We include two kinds of queries: multipoint  and point. We exclude range queries, because MySQL, MariaDB and PostgreSQL do not use hash indexes for range queries. 

\end{enumerate}
\subsection{Multipoint Queries}

\begin{enumerate}
\item
\textbf{Experiment Goal:} \\[3pt]
This experiment aims to compare the effectiveness of B+ Trees and Hash structures on multipoint queries across different database systems.
\\
\item 
\textbf{Experiment Setup:} \\[3pt]
This experiment compares two cases: either use a B+ Tree  or a Hash structure. Each case is tested on two dataset sizes, $10^5$ and $10^7$ rows. For evaluation, we issue a multipoint query returning 100 rows.

\begin{lstlisting}[language=SQL, label={lst:range-query}, caption={Multipoint query on \texttt{hundreds1}}]
SELECT * FROM employee WHERE hundreds1 = {value};
\end{lstlisting}

\item
Please refer to \textbf{Appendix \ref{hashvsbtree_multipoint-appendix}} for the Quantitative Experimental Results and to \href{https://github.com/pequeniajugar/DBTuneSuite/tree/main/scripts/hash_vs_btree_multipoint}{\texttt{Hash vs B+ Tree Multipoint Experiment Scripts}} for the code to run those experiments.

\item
\textbf{Qualitative Conclusions}

Across both dataset scales, MySQL consistently exhibited the fastest response times overall, profited greatly from indexes, but showed no preference for B+ trees vs. Hash structure. PostgreSQL showed a similar trend, with a marginal edge for Hash Index on the $10^5$-row dataset.  DuckDB benefited from its B+ tree-like index only on the $10^7$-row index. MariaDB enjoyed near-identical performance between Hash and B+ Tree structures for the two data sizes and that performance was better than using no index. 

\end{enumerate}

\subsection{Point Query} 

\begin{enumerate}
\item
\textbf{Experiment Goal:} \\[3pt]
This experiment aims to compare the effectiveness of B+ Tree and Hash indexes for point queries across different database systems.
\\
\item 
\textbf{Experiment Setup:} \\[3pt]
This experiment compares two data structures:  B+ Trees and Hash structures. Each data structure is tested on the usual two dataset sizes of $10^5$ and $10^7$. For evaluation, we issue a point query returning exactly one row. 

\begin{lstlisting}[language=SQL, label={lst:hash point-query}, caption={Point query using attribute \texttt{ssnum}}]
SELECT * FROM employee WHERE ssnum = 150;
\end{lstlisting}

\item
Please refer to \textbf{ Appendix \ref{hashvsbtree_point-appendix}} for the Quantitative Experimental Results and to \href{https://github.com/pequeniajugar/DBTuneSuite/tree/main/scripts/hash_vs_btree_point}{\texttt{Hash vs B+ Tree Point Experiment Scripts}} for the code to run those experiments.

\item
\textbf{Qualitative Conclusions}
\\
Across both dataset scales, for PostgreSQL, MySQL, and MariaDB,  B+ trees and Hash Structures performed equally well on point queries. MySQL and PostgreSQL benefited the most from either index compared to no index. MySQL consistently demonstrated the fastest response times with  PostgreSQL a close second.  DuckDB, which only supports B+ Tree-like structures, barely benefited from its index. 

The overall qualitative conclusion from these experiments for these particular database management systems is: use the B+-tree. It supports more types of queries (e.g. range queries and prefix queries) and is equally performant even on point queries. This result surprised us, but it seems to hold for these systems.

\end{enumerate}
\section{Clustered Index, Non-Clustered Index, and No Index} \label{clustered}

A {\em clustered index} is a type of index and table organization where the rows in  table $T$ are physically stored in an order determined by the index (e.g. rows are sorted in the case of a B+ Tree). A {\em non-clustered} index does not change the order of the rows in table $T$, but the leaves of the data structure consist of pointers to rows in the table $T$. 

In the InnoDB storage engine, only the primary key is clustered. Therefore this experiment adds two  new key columns called \say{ssnumpermuted1} and \say{ssnumpermuted2} to the Employee table. 

\begin{itemize}
    \item \textbf{Experiment Setup:}

We use the table \texttt{employee\_ssnum(ssnum, ssnumpermuted1, ssnumpermuted2, name, lat, longitude, hundreds1, hundreds2)}, which is an extension of \texttt{employee(ssnum, name, lat, longitude, hundreds1, hundreds2)} (see Section~\ref{subsec:benchmark employee} for a detailed definition). That is, in addition to the original attributes, \texttt{employee\_ssnum} includes two extra columns: \texttt{ssnumpermuted1} and \texttt{ssnumpermuted2}, which are different permutations of the \texttt{ssnum} attribute. 

To support three different indexing choices (clustered, non-clustered, no index), we define a primary key on \texttt{ssnum}, a non-clustered index named \texttt{nc} on \texttt{ssnumpermuted1}, and no index on \texttt{ssnumpermuted2}. In this set of experiments, we compare query performance across three scenarios: using \texttt{ssnum} with a clustered B+ tree index, \texttt{ssnumpermuted1} with a non-clustered B+ tree index, and \texttt{ssnumpermuted2} with no index.\label{employee_ssnum}

 On all database management systems, our tests include directives within the queries to force the database management system to use non-clustered indexes when present.  Section~\ref{Forcing the Use of Indexes} describes how to force the use of indexes.

\end{itemize}

\subsection{Point Query}
\begin{enumerate}
\item 
\textbf{Experiment Goal:}

The goal of this experiment is to examine whether the clustered index will improve the performance of point queries (i.e., equality queries that return at most one row) compared to a non-clustered index and no index. 

\item
\textbf{Experiment Setup:}

Experiments are conducted on two data sizes: $10^5$ and $10^7$ rows. 






The experiment compares three distinct cases: using a clustered index, using a non-clustered index, and using no index. Since DuckDB does not support clustered indexes, we skipped the clustered index cases of DuckDB. The queries all return 1 row and are defined as follows:

\begin{lstlisting}[language=SQL, label={lst:clustered}, caption={Query using a clustered index on \texttt{ssnum}}]
select * from employee_ssnum where ssnum = $VALUE; 
\end{lstlisting}

\begin{lstlisting}[language=SQL, label={lst:nonclustered}, caption={Query using a non-clustered index on \texttt{ssnumpermuted1}}]
select * from employee_ssnum where ssnumpermuted1 = $VALUE;
\end{lstlisting}

\begin{lstlisting}[language=SQL, label={lst:noindex}, caption={Query without any index usage on \texttt{ssnumpermuted2}}]
select * from employee_ssnum where ssnumpermuted2 = $VALUE;
\end{lstlisting}

\item
Please refer to \textbf{ Appendix \ref{clustered_point-appendix}} for the Experimental Results and to \href{https://github.com/pequeniajugar/DBTuneSuite/tree/main/scripts/clustered_nonclustered_noIndex/point}{\texttt{Clustered, Non-Clustered and No Index Point Query Experiment Scripts}} for the code to run those experiments.




\item
\textbf{Qualitative Conclusions }

On MariaDB, PostgreSQL and MySQL, clustered and non-clustered indexes enjoy very similar performance on point queries, while the two index types outperform  no index.  DuckDB shows minimal improvement on the smaller dataset but gains meaningful benefits from non-clustered indexing for the larger one.

Indexing also improves performance stability. Clustered indexes generally yield the lowest response time variance on MariaDB and MySQL, especially on the larger dataset. DuckDB and PostgreSQL also exhibits reduced variance under indexing at scale, though the effect is less pronounced.

\end{enumerate}


\subsection{Range Queries}
\begin{enumerate}
\item 
\textbf{Experiment Goal:}
\\
The goal of this experiment is to test whether the clustered index will improve the performance of range queries compared to a non-clustered index and no index. 
\\
\item
\textbf{Experiment Setup:}

Experiments are conducted on two data sizes: $10^5$ and $10^7$ rows. 

The experiment compares three distinct cases: using a clustered index, using a non-clustered index, and using no index. Because DuckDB does not support clustered indexes, we omitted  clustered index tests for DuckDB. The queries all return 1000 rows and are defined as follows:

\begin{lstlisting}[language=SQL, label={lst:clustered}, caption={Query using a clustered index on \texttt{ssnum}}]
select * from employee_ssnum where ssnum BETWEEN $VALUE1 AND $VALUE2; 
\end{lstlisting}

\begin{lstlisting}[language=SQL, label={lst:nonclustered}, caption={Query using a non-clustered index on \texttt{ssnumpermuted1}}]
select * from employee_ssnum where ssnumpermuted1 BETWEEN $VALUE1 AND $VALUE2;
\end{lstlisting}

\begin{lstlisting}[language=SQL, label={lst:noindex}, caption={Query without any index usage (on \texttt{ssnumpermuted2})}]
select * from employee_ssnum where ssnumpermuted2 BETWEEN $VALUE1 AND $VALUE2;
\end{lstlisting}

\item
Please refer to \textbf{ Appendix \ref{clustered_range-appendix}} for the Quantitative Experimental Results and to \href{https://github.com/pequeniajugar/DBTuneSuite/tree/main/scripts/clustered_nonclustered_noIndex/range}{\texttt{Clustered, Non-Clustered and No Index Range Experiment Scripts}} for the code to run those experiments.




\item
\textbf{Qualitative Conclusions }

Indexing sometimes improves range query performance with benefits and costs varying by DBMS and dataset size.

MySQL, PostgreSQL and MariaDB benefit significantly, especially on larger datasets, where clustered indexes deliver the best performance with over 30× speedups. On smaller datasets, both index types perform similarly. DuckDB sees degraded performance with its (non-clustered) indexes at both scales. 

In terms of stability, PostgreSQL, MariaDB and MySQL generally achieve the most consistent performance with clustered indexes. DuckDB shows little variance across configurations.

The overall conclusion is to use clustered indexes for range queries whenever possible. Again, this is not surprising, but the magnitude of the effect is.
\end{enumerate}

These results are consistent with the overall trends reported in Section 6.5 of the study by Martins et al.~\cite{martins2021comparing}, which also found that enabling indexes generally improves query performance over full-table scans. Our experiment extends this observation by differentiating clustered and non-clustered indexes, showing that clustered indexes tend to provide greater benefits for range-based queries.

As noted in the work of Mbaiossoum et al. \\ ~\cite{mbaiossoumdatabase}, indexing is an important technique for database tuning, particularly for OLAP databases.
\section{Covering Indexes } \label{coveringindexes}
A covering index is an index that contains all the columns referenced in a query’s \texttt{SELECT} and \texttt{WHERE} clauses. A covering index is {\em ordered} if the order of the attributes in that index consist of the attributes in the \texttt{where} clause followed by the attributes in the \texttt{select} clause. For example, for a query of the form \say{select X from R where Y = ...}, an ordered covering index would order the attributes (Y, X). Intuitively this ordering is useful because a query could find the sought-after Y value in, say, a B+ tree, and then search for the corresponding X values. An unordered covering index in this case would be an index that orders the attributes as (X, Y).

\begin{enumerate}
\item 
\textbf{Experiment Goal:}

The goal of this experiment is to examine whether ordered covering indexes have performance benefits compared to unordered covering indexes and to non-clustered non-covering indexes, particularly in read-intensive scenarios. 

\item 
\textbf{Experiment Setup:}

 Experiments are conducted on two data sizes: $10^5$ and $10^7$ rows. We use the  \texttt{employee(ssnum, name, lat, longitude, hundreds1, hundreds2)} table (detailed definition in section \ref{subsec:benchmark employee}) for the experiment. In addition to using the index nc4(lat, ssnum, name), we also drop it and create a new index nc5(ssnum, name, lat) in order to implement different index configurations during the experiment.

The experiment compares three contrasting cases: query with an ordered covering index, query with an unordered covering index, and  query with a non-clustered non-covering index. For both the ordered and unordered covering indexes, we execute the same query but apply different index structures. The queries are defined as follows:

\begin{lstlisting}[language=SQL, label={lst:ordered-covering}, caption={Query using an ordered covering index nc4 on \texttt{(lat, ssnum, name)}}]
SELECT ssnum, name FROM employee WHERE lat = $value1;
\end{lstlisting}

\begin{lstlisting}[language=SQL, label={lst:unordered-covering}, caption={Query using an unordered covering index nc5 on \texttt{(ssnum, name, lat)}}]
DROP INDEX nc4 ON employee; 
CREATE INDEX nc5 ON employee(ssnum, name, lat); 
SELECT ssnum, name FROM employee WHERE lat = $value1;
\end{lstlisting}

\begin{lstlisting}[language=SQL, label={lst:clustered-noncovering}, caption={Query using a non-clustered non-covering index c on \texttt{(hundreds1)}}]
select * from employee where hundreds1 = $value1; 
\end{lstlisting}

On all database management systems, our tests include directives within the queries to force the database management system to use non-clustered indexes when present.  Section~\ref{Forcing the Use of Indexes} describes how to force the use of indexes.

\item
Please refer to \textbf{ Appendix \ref{covering-appendix}} for the Quantitative Experimental Results and to \href{https://github.com/pequeniajugar/DBTuneSuite/tree/main/scripts/covering_index}{\texttt{Covering Index Experiment Scripts}} for the code to run those experiments.

\item
\textbf{Qualitative Conclusions}

In conclusion, for MySQL, PostgreSQL, and MariaDB, ordered covering indexes generally deliver the best performance and stability across both the $10^5$ and $10^7$ row datasets, clearly outperforming non-ordered covering indexes in these systems.


However, the performance of ordered covering indexes and non-clustered non-covering indexes is almost identical. In PostgreSQL, the ordered covering index performs slightly better, whereas in MariaDB the non-clustered non-covering index has only a slight advantage. 

We were surprised by these results, because we expected covering indexes to perform better than non-clustered indexes, because coveringg indexes eliminate the need to access the $100$ rows where a given $lat$ value is present. That should be a savings   because the data is not clustered based on $lat$, so those rows would probably require accesses to nearly $100$ different pages, especially on the $10^7$ row table. (Note that the values of $lat$ are shuffled when the tables are generated.) These results contrast with those of  Shasha and Bonnet~\cite{10.1145/1024694.1024720}, who found that ordered covering indexes had the best performance. The reason may have to do with the  slower disks of the time of those experiments, which rendered the penalty of accessing many pages  high. 


\end{enumerate}
\section{For Which Access Fraction Is Scanning Better Than a Non-Clustered Index}\label{scan-vs-index}
A table scan refers to the sequential traversal of all rows in a table, typically performed when no applicable index exists or when the query optimizer determines a scan to be the most efficient access path. In contrast, a non-clustered index provides an auxiliary data structure that maps key values to physical row pointers, allowing the database engine to efficiently locate and retrieve individual target rows without scanning the entire table.

 \subsection{Range Query}
\begin{enumerate}
\item
\textbf{Experiment Goal:} \\[3pt]
The goal of this experiment is to determine at what point a full table scan becomes more efficient than using a non-clustered, non-covering index in range queries.
\\
\item 
\textbf{Experiment Setup:}
\\The experiments use the schema \texttt{employee(ssnum, name, lat, longitude, hundreds1, hundreds2)} described in section \ref{subsec:benchmark employee} at two data sizes of $10^5$ and $10^7$ rows.  

The experiment compares two distinct cases: using a non-clustered index and using no index (i.e. scanning). For each case, we evaluate five range queries which access the following fractions of data: 1\%, 5\%, 10\%, 20\%, and 40\% (for $10^5$ rows, these queries return $10^3$ rows, $5\times10^3$ rows, $10^5$ rows, $2\times10^4$ rows and $4\times10^4$ rows; for $10^7$ rows, these queries return $10^5$ rows, $5\times10^5$ rows, $10^6$ rows, $2\times10^6$ rows and $4\times10^6$ rows).

\begin{itemize}
    \item 
This query utilizes the \texttt{nc} index, which is a non-clustered index defined on the \texttt{hundreds2} column. The where clause \texttt{\$VALUE} is selected to control the access fraction. 
Since each hundreds2 value is present in  0.1\% of the rows and the hundreds2 value begins from 100, for the dataset with $10^5$ rows, \texttt{\$VALUE} is set to 110 for 1\%, 150  for 5\%, 200 for 10\%, 300 for 20\%, and 500 for 40\%. 

For the $10^7$ row dataset, \texttt{\$VALUE} is set  to 1100 (for 1\% of the data), then respectively to 5100 (for 5\%), 10100  (10\%), 20100 (20\%), and 40100 (40\%). 
\begin{lstlisting}[language=SQL, label={lst:range-query-indexed}, caption={Range query using a non-clustered index on \texttt{hundreds2}}]
SELECT * FROM employee WHERE hundreds2 < $VALUE;
\end{lstlisting}

    \item 
This query accesses the \texttt{longitude} column, which has no index defined. The same threshold values \texttt{X} are used as in the indexed query to achieve the same access fractions across both indexed and non-indexed cases.
\begin{lstlisting}[language=SQL, label={lst:range-query-noindex}, caption={Range query without index on \texttt{longitude}}]
SELECT * FROM employee WHERE longitude < $VALUE;
\end{lstlisting}

On all database management systems, our tests include directives within the queries to force the database management system to use non-clustered indexes when present.  Section~\ref{Forcing the Use of Indexes} describes how to force the use of indexes.

\end{itemize}

\item
Please refer to \textbf{ Appendix \ref{accessfraction_range-appendix}} for the Quantitative Experimental Results and to \href{https://github.com/pequeniajugar/DBTuneSuite/tree/main/scripts/fraction_scan_win/range}{\texttt{Scan Can Win Range Query Experiment Scripts}} for the code to run those experiments.

\item 
\textbf{Qualitative  Conclusions}

Overall,  behaviors vary greatly across database management system, dataset sizes, and access fractions for range queries. DuckDB consistently favors full table scan on both datasets.  MySQL shows a clear and consistent preference for non-clustered indexing on  datasets at low access fractions; however, indexing loses its advantage at 40\%, where scan becomes faster. PostgreSQL  benefits at least marginally from indexing across almost all settings except at access fraction 40\% on the larger dataset. MariaDB exhibits a crossover behavior: indexing is superior up to 20\% access on the smaller dataset, and up to 10\%–20\% on the larger dataset, after which scan becomes more efficient. 
DuckDB is consistently the best performer for range queries.

In terms of response time variability, indexing generally provides more stable performance than full table scan across most systems and access fractions. MySQL and PostgreSQL consistently show lower standard deviation under indexing, especially at smaller access fractions where scan exhibits high variance. MariaDB benefits from indexing stability at low access levels, but this advantage diminishes or reverses at higher access fractions. DuckDB, in contrast, shows better stability with scan, reflecting its columnar design optimized for sequential access.


    
\end{enumerate}
\subsection{Multipoint Query}

\begin{enumerate}
\item 
\textbf{Experiment Goal:}
    
The goal of this experiment is to determine at what point a full table scan becomes more efficient than using non-clustered indexes on multipoint queries.

\item \textbf{Experiment Setup:}

Experiments are conducted on two data sizes: $10^5$ and $10^7$ rows. We use the \texttt{scanwin\_multipoint(onepercent1, onepercent2, fivepercent1, fivepercent2, tenpercent1, tenpercent2, twentypercent1, twentypercent2)} table (detailed definition as following Schema~\ref{lst:scanwin_multipoint schema}) for the experiments.

\begin{lstlisting}[language=SQL, 
                   label={lst:scanwin_multipoint schema}, 
                   caption={Schema and Index Definitions for the \texttt{scanwin\_multipoint} Table}]
-- Table schema
scanwin_multipoint(onepercent1, onepercent2, fivepercent1, fivepercent2, tenpercent1, tenpercent2, twentypercent1, twentypercent2);

-- Index definitions
CREATE INDEX idx_onepercent1 ON scanwin_multipoint(onepercent1);
CREATE INDEX idx_fivepercent1 ON scanwin_multipoint(fivepercent1);
CREATE INDEX idx_tenpercent1 ON scanwin_multipoint(tenpercent1);
CREATE INDEX idx_twentypercent1 ON scanwin_multipoint(twentypercent1);
\end{lstlisting}

\textbf{{Column Descriptions:}}

Assume the table contains n rows.
\begin{itemize}
  \item \textbf{onepercent1, onepercent2:} Each column contains integers covering the full range from 1 to 100, with each value repeated exactly $0.01 \times n$ times.  
  This simulates 1\% selectivity in a dataset, because there are $0.01 \times n$ distinct values and all values appear with equal frequency (e.g. each value has $10^3$ rows in a dataset of $10^5$ rows).

  \item \textbf{fivepercent1, fivepercent2:} Values fully span [1, 20], with each value appearing $0.05 \times n$ times. This represents 5\% selectivity with uniform duplication.

  \item \textbf{tenpercent1, tenpercent2:} Values cover [1, 10], with each value repeated $0.1 \times n$ times, yielding 10\% selectivity.

  \item \textbf{twentypercent1, twentypercent2:} Each column contains all integers from 1 to 5, and each value is repeated $0.2 \times n$ times. This creates 20\% selectivity with evenly distributed repetitions.
\end{itemize}

The experiment compares two contrasting cases: with a nonclustered index and without any index. The queries are defined as follows:

\begin{itemize}
    \item The following queries are used for nonclustered index cases, each of the multipoint queries was executed on a column with a dedicated non-clustered index.
    \begin{lstlisting}[language=SQL, 
                       label={lst:scanwin_multipoint_index},
                       caption={Multipoint Queries on Indexed Columns with Varying Access Fractions}]
-- access fraction: 1%
select * from scanwin_multipoint where onepercent1 = $VALUE; 
-- access fraction: 5%
select * from scanwin_multipoint where fivepercent1 = $VALUE;
-- access fraction: 10%
select * from scanwin_multipoint where tenpercent1 = $VALUE; 
-- access fraction: 20%
select * from scanwin_multipoint where twentypercent1 = $VALUE; 
\end{lstlisting}

    \item The following queries are used for scan cases, each of the multipoint queries was executed on a column without any index. 
    \begin{lstlisting}[language=SQL, 
                       label={lst:scanwin_multipoint_scan},
                       caption={Multipoint Queries on Non-Indexed Columns with Varying Access Fractions}]
-- access fraction: 1%
select * from scanwin_multipoint where onepercent2 = $VALUE; 
-- access fraction: 5%
select * from scanwin_multipoint where fivepercent2 = $VALUE;
-- access fraction: 10%
select * from scanwin_multipoint where tenpercent2 = $VALUE; 
-- access fraction: 20%
select * from scanwin_multipoint where twentypercent2 = $VALUE; 
\end{lstlisting}

On all database management systems, our tests include directives within the queries to force the database management system to use non-clustered indexes when present.  Section~\ref{Forcing the Use of Indexes} describes how to force the use of indexes.

\end{itemize}

\item Please refer to \textbf{ Appendix \ref{accessfraction_multipoint-appendix}} for the Quantitative Experimental Results and to \href{https://github.com/pequeniajugar/DBTuneSuite/tree/main/scripts/fraction_scan_win/multipoint}{\texttt{Scan Can Win Range Multipoint Query Experiment Scripts}} for the code to run those experiments.

\item \textbf{Qualitative Conclusions:}

Across both the $10^5$ and $10^7$ row datasets, the effectiveness of non-clustered indexing for multipoint queries varies notably by system and access fraction. MySQL consistently benefits from indexing, achieving substantial and stable speedups across all settings. In contrast, MariaDB favors indexing only at low access fractions on smaller datasets; as data scale or access fraction increases, full scans become increasingly advantageous. PostgreSQL generally performs better with indexing on the smaller dataset, but favors scanning on the larger one, though the performance gap narrows at higher access levels. DuckDB consistently favors scanning across most settings, with indexing offering a clear advantage only at 1\% access on the $10^5$-row dataset. DuckDB, when scanning, is also the fastest system for multipoint queries across all access fractions for the $10^7$ row table, though not for the $10^5$ row table.

On both datasets, MySQL consistently exhibits lower response time variability with indexing, indicating stable performance benefits. DuckDB generally shows marginally more stable behavior with indexing on smaller dataset, while with scanning on larger dataset. MariaDB experiences increased variance under indexing, particularly at high access fractions on the smaller dataset and low access fractions on the larger dataset. PostgreSQL presents mixed results—indexing initially increases variability but becomes more stable or comparable to scanning as the access fraction grows.
\end{enumerate}

\section{Index on Small Table}



\begin{itemize}
    \item 
\textbf{Experiment Schema and Data Setup:} \\[3pt]
These experiments are conducted on two data sizes: $10^3$ and $10^4$ rows, which represent the small tables. We use the benchmark \texttt{'employee\_ssnum(ssnum, ssnumpermuted1, ssnumpermuted2, name, lat, longitude)'} , which is similar to the benchmark \texttt{'employee\_ssnum(ssnum, ssnumpermuted1, ssnumpermuted2, name, lat, longitude, hundreds1, hundreds2)'} (detailed definition in section \ref{employee_ssnum}) for the experiments after removing the columns \texttt{'hundreds1' and 'hundreds2'}. The idea is to create a keyed table of modest width and number of rows that is frequently queried and updated. Tables with such characteristics  might appear (though with different  fields) might appear in online ecommerce applications.

On all database management systems, our tests include directives within the queries to force the database management system to use non-clustered indexes when present.  Section~\ref{Forcing the Use of Indexes} describes how to force the use of indexes.
\end{itemize}

\subsection{Update}
\begin{enumerate}
    \item \textbf{Experiment Goal:}
    
    The goal of this experiment is to evaluate the impact of clustered indexes on update performance in small tables.

    \item  \textbf{Experiment Setup:}

\textbf{The experiment compares three distinct cases:} one using a clustered index, one using a non-clustered index and one with no index. Each case is tested on two dataset sizes. For comparison, the average response time over 10 runs is recorded. 

To evaluate a concurrent update workload on a small table, for the $10^3$-row dataset, this experiment launches two concurrent processes, each executing 100 \texttt{UPDATE} statements in parallel. In total, 200 rows are randomly selected for modification. For the $10^4$-row dataset, the experiment launches 10 concurrent processes, each executing 100 \texttt{UPDATE} statements in parallel. In total, 1000 rows are randomly selected for modification. Since DuckDB does not support multiprocessing, we omitted DuckDB from this experiment. 

The queries included in the processes are defined as follows: 

\begin{itemize}
    \item 
This query uses the index automatically created by the primary key, which is a clustered index on \texttt{'ssnum'}.
\begin{lstlisting}[language=SQL, 
                   label={lst:update-clustered},
                   caption={Update Query Using a Clustered Index on \texttt{hundreds1}}]
UPDATE employee SET name = '{new_name}', WHERE ssnum = {value};
\end{lstlisting}

    \item 
This query uses index \texttt{nc}, which is a non-clustered index on \texttt{'ssnumpermuted1'}.
\begin{lstlisting}[language=SQL, 
                   label={lst:update-nonclustered},
                   caption={Update Query Using a Non-Clustered Index on \texttt{ssnumpermuted1}}]
UPDATE employee SET name = '{new_name}', WHERE ssnumpermuted1 = {value};
\end{lstlisting}

    \item 
This query does not use any indexes.
\begin{lstlisting}[language=SQL, 
                   label={lst:update-noindex},
                   caption={Update Query Without Any Index (on \texttt{ssnumpermuted2})}]
UPDATE employee SET name = '{new_name}', WHERE ssnumpermuted2 = {value};
\end{lstlisting}

\end{itemize}

\item
Please refer to \textbf{ Appendix \ref{smalltable-appendix}} for the Quantitative Experimental Results and to \href{https://github.com/pequeniajugar/DBTuneSuite/tree/main/scripts/index_on_small_table/update}{\texttt{Index on Small Table Update Query Experiment Scripts}} for the code to run those experiments.
\item 
\textbf{Qualitative Conclusions:}
Across both small datasets ($10^3$ and $10^4$ rows), indexing consistently enhanced concurrent \texttt{UPDATE} performance and stability. Clustered indexes were generally the most effective, with MariaDB and MySQL achieving up to 61.6× and 42.3× speedups, respectively.

Non-clustered indexes also delivered strong performance in several cases—nearly matching or even slightly outperforming clustered indexes on MariaDB for the $10^4$ dataset, and performing competitively on MySQL and PostgreSQL on the $10^3$ dataset. Overall, their performance remained close to that of clustered indexes, being slightly better in some systems and slightly worse in others.

The results are consistent with the findings of Shasha and Bonnet \\ ~\cite{10.1145/1024694.1024720}, who concluded that applying a clustered index on small tables enables row-level locking and improves concurrent update throughput, making it an effective tuning strategy for high-concurrency workloads.

For stability, both index types reduced response time variability compared to the no-index case. Non-clustered indexes yielded the most consistent performance on MariaDB and MySQL, while PostgreSQL saw slightly better stability with clustered indexing. These findings highlight that optimal indexing strategies vary by DBMS, but indexing in general is critical for improving performance under concurrent updates on small tables.

\end{enumerate}
\subsection{Search}
\begin{enumerate}
    \item \textbf{Experiment Goal:}
    
    The goal of this experiment is to evaluate the impact of clustered indexes and nonclustered indexes and non-clustered indexes on point searching performance in small tables.
\\
    \item  \textbf{Experiment Setup:}

\textbf{The experiment compares three distinct cases:} one using a clustered index, one using a non-clustered index and one with no index 
Each case is tested on two dataset sizes. For comparison, the average response time over 10 runs is recorded. 

To simulate a concurrent search  workload on a small table, for the $10^3$-row dataset, the experiment launches two concurrent processes, each executing 100 point \texttt{SEARCH}  queries. In total, 200 rows are randomly selected, applying pressure on the database system. For the $10^4$-row dataset, the experiment launches ten concurrent processes, each executing 100 point \texttt{SEARCH} queries. In total, 1000 rows are randomly selected, applying pressure on the database system. 

The queries included in the processes are defined as follows: 

\begin{itemize}
    \item 
This query uses the index automatically created by the primary key, which is a clustered index on \texttt{'ssnum'}.
\begin{lstlisting}[language=SQL, 
                   label={lst:select-clustered}, 
                   caption={Point Query Using a Clustered Index on \texttt{ssnum}}]
SELECT count(*) FROM employee WHERE ssnum = {value};
\end{lstlisting}

    \item 
This query uses index \texttt{nc}, which is a non-clustered index on \texttt{'ssnumpermuted1'}.
\begin{lstlisting}[language=SQL, 
                   label={lst:select-nonclustered}, 
                   caption={Point Query Using a Non-Clustered Index on \texttt{ssnumpermuted1}}]
SELECT count(*) FROM employee WHERE ssnumpermuted1 = {value};
\end{lstlisting}

    \item 
This query does not use any indexes.
\begin{lstlisting}[language=SQL, 
                   label={lst:select-noindex}, 
                   caption={Point Query Without Index on \texttt{ssnumpermuted2}}]
SELECT count(*) FROM employee WHERE ssnumpermuted2 = {value};
\end{lstlisting}

\end{itemize}

\item
Please refer to \textbf{ \ref{smalltable-appendix}} for the Quantitative Experimental Results and to \href{https://github.com/pequeniajugar/DBTuneSuite/tree/main/scripts/index_on_small_table/search}{\texttt{Index on Small Table Search Query Experiment Scripts}} for the code to run those experiments.

\item 
\textbf{Qualitative Conclusions:}

Across both small datasets, indexing significantly improved concurrent point query performance and stability. Clustered indexes generally offered the highest speedups on MariaDB (up to 6.2×) and MySQL (4.9×). In contrast, PostgreSQL consistently benefited more from non-clustered indexes, which outperformed clustered indexes on both datasets.

Non-clustered indexes performed comparably to clustered ones on MySQL and MariaDB, though the gains were slightly smaller. PostgreSQL stood out as the only system where non-clustered indexes provided the best performance overall.

In terms of stability, non-clustered indexes achieved the lowest response time variance on MariaDB and PostgreSQL, indicating more consistent performance. On MySQL, clustered indexing was slightly more stable under concurrent load.

While Section 6.5 of the study by Martins et al. \\ ~\cite{martins2021comparing} focused on large-scale datasets, our results on smaller tables exhibit a similar pattern where indexing accelerates search, albeit with smaller performance gains.

As noted in the work of Mbaiossoum et al. \\ ~\cite{mbaiossoumdatabase}, indexing is an important technique for database tuning, particularly for OLAP databases.

\end{enumerate}

\section{Looping in Stored Procedures}
Looping replaces a single set-based range query with a stored-procedure loop that issues one parameterized point query per iteration. While this approach can fit naturally into a procedural control flow, it entails  additional overhead, adding statement-execution overhead. 

\begin{enumerate}

\item
\textbf{Experiment Goal:} \\[3pt]
The aim is to quantify the performance impact of replacing a single range‐scan query  with a stored-procedure loop that issues one parameterized point lookup for each \texttt{l\_partkey} value from 1 to 199. 
\\
\item 
\textbf{Experiment Setup:}
\\ The experiment uses the TPC-H schema (described in Section \ref{subsec:benchmark tpch}) at two scales - $\mathbf{10^{5}}$ and $\mathbf{10^{7}}$ rows.
The experiments contain two cases: using stored procedures for looping and not using stored procedures for looping. The queries are listed in Listings \ref{lst:loop-range} and \ref{lst:loop-proc}. Both return the rows with \texttt{l\_partkey} $<$ 200.
\\[0.75\baselineskip] 
\begin{minipage}
{\linewidth} 
\begin{lstlisting}[language=SQL,caption={Range query (no loop).},label={lst:loop-range}] 
SELECT * FROM lineitem WHERE l_partkey < 200; 
\end{lstlisting} 
\begin{lstlisting}[language=SQL,caption={Loop issuing 199 point lookups.},label={lst:loop-proc}] 
DELIMITER // 
CREATE PROCEDURE get_lineitems() 
BEGIN 
    DECLARE i INT DEFAULT 1; 
    PREPARE stmt FROM 'SELECT * FROM lineitem WHERE l_partkey = ?'; 
    
    WHILE i < 200 DO 
        SET @param = i; 
        EXECUTE stmt USING @param; 
        SET i = i + 1; 
    END WHILE; 
    DEALLOCATE PREPARE stmt; 
END // 
DELIMITER ; 

CALL get_lineitems(); 
\end{lstlisting} 

\end{minipage}
\item
Please refer to \textbf{ Appendix \ref{looping-appendix}} for the  Quantitative Experimental Results and to \href{https://github.com/pequeniajugar/DBTuneSuite/tree/main/scripts/looping_in_stored_procedures}{\texttt{Looping in Stored Procedures Experiment Scripts}} for the code to run those experiments.

\item
\textbf{Qualitative Conclusions}
\\
Retrieving data in loops appears never to be beneficial and often greatly hurts performance, which is consistent with the findings of Shasha and Bonnet~\cite{10.1145/1024694.1024720}.
Issuing hundreds of point look-ups from inside a stored procedure (loop) replaces one optimized set scan with hundreds of statement entries, multiplying the statement execution overhead.

At 10\textsuperscript{5} rows looping slowed MySQL by 19\% and PostgreSQL by a factor of 5. Surprisingly, MariaDB was not much affected, indicating an  efficient stored procedure implementation. 

At 10\textsuperscript{7} rows, looping pushed PostgreSQL latency up by two orders of magnitude while giving MariaDB  a modest   3\% penalty. For larger sizes, PostgreSQL was the fastest 
database management system when not using loops. 

\end{enumerate}
\section{Looping with Cursors}
A cursor is a way to iterate through the rows of a result set, one row at a time. It's like a pointer that allows a program to move through the data retrieved by a query. Thus, it is a mechanism to process data row by row, rather than operating on the entire result set at once.

Because the qualitative conclusions are similar to those having to do with looping (i.e. cursors are bad), we place all the description of these experiments in Appendix \ref{cursor-appendix} and to \href{https://github.com/pequeniajugar/DBTuneSuite/tree/main/scripts/loop_with_cursor}{\texttt{Looping with Cursors Experiment Scripts}} for the code to run those experiments.

\section{Retrieve Needed Columns}
Retrieving only the needed columns refers to selecting specific fields in a query rather than fetching all columns regardless of necessity. This technique reduces the amount of data read from disk or memory, leading to lower I/O and improved query performance. Sometimes, all the needed columns may be in an index, eliminating the need to fetch data from the data table.

\begin{enumerate}

\item
\textbf{Experiment Goal:} \\[3pt]
The goal of the experiment is to quantify the performance impact of retrieving only the necessary columns in a query instead of retrieving all the columns. While selecting all columns offers convenience and flexibility, it can lead to unnecessary data retrieval, increased I/O, and slower query execution, particularly on wide tables. 
\item 
\textbf{Experiment Setup:}
\\ The experiment uses the TPC-H schema (described in \ref{subsec:benchmark tpch}) at two scales - $\mathbf{10^{5}}$ and $\mathbf{10^{7}}$ rows.
The queries shown in Listings \ref{lst:all-col} and \ref{lst:no-all-col} both retrieve all rows from the \texttt{lineitem} table, but differ in the number of columns being retrieved. 
\\[0.75\baselineskip] 
\begin{minipage}
{\linewidth} 
\begin{lstlisting}[language=SQL,caption={Query to fetch all columns of the lineitem table.},label={lst:all-col}] 
SELECT * FROM lineitem; 
\end{lstlisting} 
\begin{lstlisting}[language=SQL,caption={Query to fetch selected columns of the lineitem table.},label={lst:no-all-col}] 
SELECT l_orderkey, l_partkey, l_suppkey, l_shipdate, l_commitdate FROM lineitem;
\end{lstlisting} 
\end{minipage}

\item
Please refer to \textbf{ Appendix \ref{retrieve-appendix}} for the Quantitative Experimental Results and to \href{https://github.com/pequeniajugar/DBTuneSuite/tree/main/scripts/retrieve_needed_columns}{\texttt{Retrieve Needed Columns Experiment Scripts}} for the code to run those experiments.
\\

\item
\textbf{Qualitative Conclusions}
\\
Retrieving needed columns is consistently beneficial to performance in our experiments. While selecting all columns via 'SELECT *' may offer convenience, it  introduces unnecessary I/O and memory overhead - especially on wide tables. The finding is in line with the results reported by Shasha and Bonnet~\cite{10.1145/1024694.1024720}.

By retrieving only the necessary fields, engines reduce data movement and improve cache efficiency, yielding both faster and more stable query execution.
At $10^5$ rows, the improvement was already significant across all engines.
At $10^7$ rows, the benefits became dramatic, with some engines running over three times faster.

The particularly strong performance of DuckDB illustrates the strong advantage of retrieving only needed columns on columnar execution engines.  
\end{enumerate}
\section{Eliminate Unneeded DISTINCT clause}
This experiment evaluates the impact of eliminating the  `DISTINCT` keyword when unnecessary (i.e. when the result of a query would not include duplicates even without that keyword) in SQL queries to measure performance gains across different database systems. The baseline query applies `DISTINCT` to a field that is already unique due to a  join condition, while the optimized version removes the keyword.

\begin{enumerate}

\item
\textbf{Experiment Goal:} \\[3pt]
The goal of this experiment is to quantify the performance impact of removing unnecessary `DISTINCT` clauses in SQL queries. When a query includes `DISTINCT`, it can introduce extra computation, including sorting or hashing for deduplication. By eliminating such operations,  query execution time may decrease, especially on large datasets, without affecting the correctness of the result. This experiment tests this hypothesis across multiple systems and across uniform and fractal data distributions. \\

\item 
\textbf{Experiment Setup:}
The experiment uses three tables: \texttt{employee} (Listing~\ref{lst:employee}), \texttt{student} (Listing~\ref{lst:student}), and \texttt{techdept} (Listing~\ref{lst:techdept}) at two scales – $\mathbf{10^{5}}$ and $\mathbf{10^{7}}$ rows. For each scale, we conducted the experiments on both a uniformly distributed dataset and a fractally distributed dataset.

\begin{table}[H]
  \centering
  \caption{Schemas for \texttt{employee}, \texttt{student}, and \texttt{techdept} tables}
  \label{tab:employee-student-techdept-schema}

  \vspace{0.75\baselineskip}
  \begin{minipage}{\linewidth}
  
  \begin{lstlisting}[language=SQL, caption={Schema for the \texttt{employee} table.}, label={lst:employee}]
CREATE TABLE employee (
    ssnum INT PRIMARY KEY,
    name VARCHAR(25),
    dept VARCHAR(25),
    salary DECIMAL(10,2),
    numfriends INT
);
  \end{lstlisting}

  \begin{lstlisting}[language=SQL, caption={Schema for the \texttt{student} table.}, label={lst:student}]
CREATE TABLE student (
    ssnum INT PRIMARY KEY,
    name VARCHAR(50),
    course VARCHAR(50),
    grade INT
);
  \end{lstlisting}

  \begin{lstlisting}[language=SQL, caption={Schema for the \texttt{techdept} table.}, label={lst:techdept}]
CREATE TABLE techdept (
    dept VARCHAR(25) PRIMARY KEY,
    manager VARCHAR(25),
    location VARCHAR(50)
);
  \end{lstlisting}

  \end{minipage}
\end{table}

 The \texttt{employee} table uses \texttt{ssnum} as the primary key and assigns the \texttt{dept} column either uniformly or with skew to model real-world data imbalance. The \texttt{student} table partially overlaps with \texttt{employee} via \texttt{ssnum}, while \texttt{techdept} provides metadata for each department and joins on \texttt{dept}. In the fractal setting, 20\% of departments account for 80\% of employee assignments, and the size of \texttt{techdept} is scaled accordingly.
\\

The queries shown in Listings \ref{lst:not-eliminated} and \ref{lst:eliminated} retrieve data from the \texttt{employee} and \texttt{techdept} tables using a join condition on the \texttt{dept} attribute on uniformly distributed data and fractally distributed data.
\\[0.75\baselineskip] 
\begin{minipage}
{\linewidth} 
\begin{lstlisting}[language=SQL,caption={Query with   DISTINCT (though unneeded).},label={lst:not-eliminated}] 
SELECT DISTINCT ssnum
FROM employee, techdept
WHERE employee.dept = techdept.dept;
\end{lstlisting} 
\begin{lstlisting}[language=SQL,caption={Query without DISTINCT},label={lst:eliminated}] 
SELECT ssnum
FROM employee, techdept
WHERE employee.dept = techdept.dept;
\end{lstlisting} 
\end{minipage}

\item
Please refer to \textbf{ Appendix \ref{distinct-appendix}} for the Quantitative Experimental Results and to \href{https://github.com/pequeniajugar/DBTuneSuite/tree/main/scripts/unneeded_distinct_eliminate}{\texttt{Eliminate Unneeded DISTINCT clause Experiment Scripts}} for the code to run those experiments.
\\

\item
\textbf{Qualitative Conclusions}
\\
Eliminating unnecessary \texttt{DISTINCT} clauses improves performance in our experiments, which is consistent with the results reported by Shasha and Bonnet~\cite{10.1145/1024694.1024720}. 
 On large datasets, the difference became even more pronounced, revealing how scalability is hindered by needless result materialization and memory usage.

Notably, \emph{MySQL} was unable to execute the query with \texttt{DISTINCT} on a large dataset in a reasonable amount of time, underscoring the risk of relying on this clause without considering its necessity. These results suggest that query authors should avoid using \texttt{DISTINCT} unless it is functionally required.

\end{enumerate}

\section{Correlated Subqueries vs. Temporary Tables} \label{correlated-vs-temporary}
A \emph{correlated subquery}, is a nested \texttt{SELECT} query that refers to a table in the outer query. We compare a correlated subquery formulation with two others.

A \emph{temporary table formulation}  builds a temporary table (corresponding roughly to what is done in the nested query) and then performs a join with another table (corresponding to the table in the outer query).  

A \emph{WITH clause formulation} builds a temporary using a WITH clause. 

\begin{enumerate}

\item
\textbf{Experiment Goal:} \\[3pt]
The goal of this experiment is to quantify how three SQL formulations -- \emph{correlated subquery}, \emph{temporary table}, and \emph{WITH-clause} -- perform relative to one another on uniform and fractally distributed data. 

\item 
\textbf{Experiment Setup:}

The experiment uses three tables: \texttt{employee} (Listing~\ref{lst:employee}), \texttt{student} (Listing~\ref{lst:student}), and \texttt{techdept} (Listing~\ref{lst:techdept}) - evaluated at two data scales: $\mathbf{10^5}$ and $\mathbf{10^7}$ rows, with both uniform and fractal data distributions.

The queries shown in Listings \ref{lst:subquery}, \ref{lst:join} and \ref{lst:with} are used to access the rows of the group.
\\[0.75\baselineskip] 
\begin{minipage}
{\linewidth} 
\begin{lstlisting}[language=SQL,caption={Query to find the employee (or employees) per department who earns the maximum salary in that department, using a correlated subquery.},label={lst:subquery}] 
SELECT ssnum 
FROM employee e1 
WHERE salary =
    (SELECT max(salary)
     FROM employee e2
     WHERE e2.dept = e1.dept);
\end{lstlisting} 
\begin{lstlisting}[language=SQL,caption={Query to find the employee (or employees) per department who earns the maximum salary in that department, using a temporary table},label={lst:join}] 
SELECT e1.ssnum
FROM employee e1
JOIN (
    SELECT dept, MAX(salary) AS bigsalary
    FROM employee
    GROUP BY dept
) e2 
ON e1.dept = e2.dept AND e1.salary = e2.bigsalary;
\end{lstlisting} 
\begin{lstlisting}[language=SQL,caption={Query to find the employee (or employees) per department who earns the maximum salary in that department, using a WITH Clause},label={lst:with}] 
WITH max_salary_per_dept AS (
    SELECT dept, MAX(salary) AS bigsalary
    FROM employee
    GROUP BY dept
)
SELECT e1.ssnum
FROM employee e1
JOIN max_salary_per_dept m
ON e1.dept = m.dept AND e1.salary = m.bigsalary;
\end{lstlisting} 
\end{minipage}

\item
Please refer to \textbf{ Appendix \ref{correlated-appendix}} for the Quantitative Experimental Results and to \href{https://github.com/pequeniajugar/DBTuneSuite/tree/main/scripts/correlated_subquery}{\texttt{Correlated Subqueries Experiment Scripts}} for the code to run those experiments.
\\ 

\item
\textbf{Qualitative Conclusions}
\\
In our tests, it is always better to use either the Temporary Table or WITH-clause formulation than the correlated subquery formulation. The only system which executes correlated subqueries well is DuckDB. Surprisingly, no formulation of the correlated  query or any of its variants performed well for any system except DuckDB at the $10^7$ row level. 


Overall, the correlated style adds unpredictability without delivering any consistent benefit, while the join (or an equivalent in-lined \texttt{WITH}-clause) remains the safest, most reliable way to express this lookup.

\end{enumerate}
\section{Aggregate Maintenance -- triggers }
Aggregate maintenance refers to the automatic updating of precomputed summary values (e.g., totals, averages) in a separate table when related base data changes.

Triggers are database mechanisms that automatically execute specified logic in response to events like INSERT, UPDATE, or DELETE.

When used together, triggers serve as the mechanism to implement aggregate maintenance. For example, when a new order is inserted into a transaction table, a trigger can be fired to immediately adjust the corresponding total in an aggregate table to ensure that summary data remains up-to-date.

\begin{itemize}
    \item 
\textbf{Experiment Schema and Data Setup:} \\[3pt]
We designed a synthetic benchmark dataset named \texttt{store} to evaluate Aggregate Maintenance performance under controlled conditions. The dataset is uniformly distributed as described below. 

\begin{lstlisting}[language=SQL, 
                   label={lst:employee schema}, 
                   caption={Schema and Index Definitions for the \texttt{store} Table}]
-- Table schema
orders(ordernum, itemnum, quantity, storeid, vendorid);
store(storeid, name);
item(itemnum, price); 
vendorOutstanding(vendorid, amount);
storeOutstanding(storeid, amount);

-- Index definitions
create clustered index i_item on item(itemnum);
\end{lstlisting}

\textbf{Column Descriptions:}

\begin{itemize}
  \item \textbf{orders (ordernum, itemnum, quantity, storeid, vendorid):}
    \begin{itemize}
      \item \textbf{ordernum:} Primary key, integers from $1$ to $n$ without repetition.
      \item \textbf{itemnum:} Permuted assignment of integers from $1$ to $n/100$, with each value repeated 100 times.
      \item \textbf{quantity:} Permuted values derived from $[1, n/100]$, repeated 100 times and then mapped to $[1, 100]$ using modulo.
      \item \textbf{storeid:} Permuted assignment of integers from $1$ to $n/100$, repeated 100 times.
      \item \textbf{vendorid:} Permuted assignment of integers from $1$ to $n/100$, repeated 100 times.
    \end{itemize}

  \item \textbf{store (storeid, name):}
    \begin{itemize}
      \item \textbf{storeid:} Primary key, integers from $1$ to $n/100$.
      \item \textbf{name:} Store names in the format \texttt{Store1}, \texttt{Store2}, $\dots$, \texttt{Store}$n/100$.
    \end{itemize}

  \item \textbf{item (itemnum, price):}
    \begin{itemize}
      \item \textbf{itemnum:} Primary key, integers from $1$ to $n/100$.
      \item \textbf{price:} Integer values in $[1, 100]$, derived from a permuted version of item numbers modulo 100.
    \end{itemize}

  \item \textbf{vendorOutstanding (vendorid, amount):}
    \begin{itemize}
      \item \textbf{vendorid:} Primary key, integers from $1$ to $n/100$.
      \item \textbf{amount:} Permuted assignment of integers from $1$ to $n/100$.
    \end{itemize}

  \item \textbf{storeOutstanding (storeid, amount):}
    \begin{itemize}
      \item \textbf{storeid:} Primary key, integers from $1$ to $n/100$.
      \item \textbf{amount:} Permuted assignment of integers from $1$ to $n/100$.
    \end{itemize}
\end{itemize}


\textbf{Table Descriptions:}

  To reflect trigger-based summarization and performance benchmarking, two additional tables are included: \texttt{storeOutstanding} and \texttt{vendorOutstanding}. These tables store pre-aggregated total amounts  per store and per vendor, respectively. Instead of recomputing the sums from the \texttt{orders} table during each query, these trigger-maintained summary tables can be queried directly.


    \item 
\textbf{Experiment Setup:}

These two SQL trigger definitions are used to automatically maintain aggregate values in \texttt{vendorOutstanding} and \texttt{storeOutstanding} tables after a new order is inserted into the \texttt{orders} table.

\begin{enumerate}
  \item updateVendorOutstanding Trigger: \\
  Event: Fires after an \texttt{INSERT} operation on the \texttt{orders} table. \\
  Logic: It calculates the total value of the new order (by multiplying \texttt{NEW.quantity} with the corresponding item's price from the \texttt{item} table), and adds it to the \texttt{amount} field of the row in the \texttt{vendorOutstanding} table matching the \texttt{vendorid}.
  
  \item updateStoreOutstanding Trigger: \\
  This trigger performs a similar update on the \texttt{storeOutstanding} table by adding the calculated order value to the corresponding \texttt{storeid} row after a new order is inserted.
\end{enumerate}

Together, these triggers ensure that whenever a new order is recorded, the corresponding vendor and store’s outstanding balances are updated automatically. This is a classic example of aggregate maintenance using \texttt{AFTER INSERT} triggers.

\begin{lstlisting}[language=SQL, 
                   label={lst:employee schema}]
DELIMITER $$

CREATE TRIGGER updateVendorOutstanding
AFTER INSERT ON orders
FOR EACH ROW
BEGIN
    UPDATE vendorOutstanding
    SET amount = amount + (NEW.quantity * (
        SELECT price
        FROM item
        WHERE item.itemnum = NEW.itemnum
    ))
    WHERE vendorid = NEW.vendorid;
END $$

DELIMITER;


DELIMITER $$

CREATE TRIGGER updateStoreOutstanding
AFTER INSERT ON orders
FOR EACH ROW
BEGIN
    UPDATE storeOutstanding
    SET amount = amount + (NEW.quantity * (
        SELECT price
        FROM item
        WHERE item.itemnum = NEW.itemnum
    ))
    WHERE storeid = NEW.storeid;
END $$

DELIMITER;
\end{lstlisting}

\end{itemize}

\subsection{Data Insertion With and Without Triggers}

\begin{enumerate}
\item
\textbf{Experiment Goal:} \\[3pt]
The goal of this experiment is to evaluate the performance impact of using aggregate maintenance triggers during data insertion and assess the trade-off between data freshness and insertion efficiency.

\item 
\textbf{Experiment Setup:} \\[3pt]
\textbf{This experiment compares two configurations:} one with triggers enabled, where each insert automatically updates the corresponding tables through \texttt{AFTER INSERT} triggers, and the other with triggers disabled, where no automatic aggregate maintenance is performed during insertion. Each configuration is tested on two dataset sizes: $10^5$ and $10^7$ rows. For evaluation, 1000 rows are inserted one by one into the \texttt{orders} table. This insertion process is repeated 10 times for each case. After each round, only the inserted rows are removed to restore the database to its original state before insertion, ensuring consistency across runs. The average insertion response time is then reported.

Since DuckDB does not support triggers, we omit DuckDB from this comparison.  

\item
Please refer to \textbf{ Appendix \ref{triggers_insert-appendix}} for the Quantitative Experimental Results and to \href{https://github.com/pequeniajugar/DBTuneSuite/tree/main/scripts/Aggregate_Maintenance_triggers%20}{\texttt{Aggregate Maintainance Triggers Experiment Scripts}} for the code to run those experiments.

\item
\textbf{Qualitative Conclusions}

Across both the $10^5$ and $10^7$ datasets, enabling aggregate maintenance using \texttt{AFTER INSERT} triggers consistently led to performance degradation for MySQL, MariaDB, and PostgreSQL. The impact was particularly severe for MySQL on the larger dataset, where insertion latency increased by over 120×. In contrast, MariaDB and PostgreSQL experienced only modest slowdowns on both datasets, with insertion times increasing by approximately 1–1.2×.

In terms of consistency, the effects of triggers varied across systems. On the $10^7$ dataset, MySQL exhibited a substantial increase in variability, while PostgreSQL also became more variable, though to a lesser extent. MariaDB, on the other hand, showed improved consistency with triggers.
\end{enumerate}
\subsection{Data Retrieval With and Without Triggers}

\begin{enumerate}
\item
\textbf{Experiment Goal:} \\[3pt]
The goal of this experiment is to evaluate the performance impact of using redundant tables for precomputed aggregate values versus computing aggregates dynamically at query time through multi-table joins and group-by queries.
\\
\item 
\textbf{Experiment Setup:} \\[3pt]
\textbf{This experiment includes two subexperiments:} one focuses on aggregate maintenance for \texttt{vendors}, and the other for \texttt{stores}. Both subexperiments evaluate the performance impact of maintaining precomputed aggregate values using redundant tables and \texttt{AFTER INSERT} triggers.

\textbf{All the subexperiments compare two cases:} one with triggers enabled, where each insert automatically updates the corresponding tables through \texttt{AFTER INSERT} triggers, and the other with triggers disabled, where no automatic aggregate maintenance is performed during insertion. 

The queries are defined as follows:
    
    \begin{lstlisting}[language=SQL, 
                       label={lst:vendor-query-no-trigger},
                       caption={Query for vendor outstanding amount without trigger }]
select sum(orders.quantity*item.price)
		from orders,item
		where orders.itemnum = item.itemnum
		and orders.vendorid = {value}; 
\end{lstlisting}

    \begin{lstlisting}[language=SQL, 
                       label={lst:vendor-query-with-trigger},
                       caption={Query for vendor outstanding amount with trigger (retrieved from preaggregated table)}]
select amount from vendorOutstanding where vendorid = {value};
\end{lstlisting}

    \begin{lstlisting}[language=SQL, 
                       label={lst:store-query-no-trigger},
                       caption={Query for store outstanding amount without trigger }]
select sum(orders.quantity*item.price)
                from orders,item, store
                where orders.itemnum = item.itemnum
                  and orders.storeid = store.storeid
                  and store.storeid = {value};
\end{lstlisting}

    \begin{lstlisting}[language=SQL, 
                       label={lst:store-query-with-trigger},
                       caption={Query for store outstanding amount with trigger (retrieved from preaggregated table)}]
select amount from storeOutstanding where storeid = {value}; 
\end{lstlisting}

Each case is tested on two dataset sizes: $10^5$ and $10^7$ rows. For evaluation, all the rows are inserted one by one into the \texttt{orders} table. This insertion process is repeated 10 times for each case. The average insertion response time is then reported.

Since DuckDB does not support triggers, we skip the experiments of DuckDB with triggers.
\\ 
\item
Please refer to \textbf{ Appendix \ref{triggers_retrieve-appendix}} for the Quantitative Experimental Results and to \href{https://github.com/pequeniajugar/DBTuneSuite/tree/main/scripts/Aggregate_Maintenance_triggers%20}{\texttt{Aggregate Maintainance Triggers Experiment Scripts}} for the code to run those experiments.

\item
\textbf{Qualitative Conclusions}

Overall, using \texttt{AFTER INSERT} triggers for aggregate maintenance significantly enhances both query performance and execution stability across database systems. MySQL and MariaDB benefit the most, particularly for larger tables, where triggers enable efficient query execution that would otherwise be prohibitively slow. PostgreSQL also gains in consistency, though performance improvements  vary. These findings highlight the effectiveness of trigger-based aggregate maintenance in optimizing analytical query workloads. 

Our results are consistent with the findings of Mbaiossoum et al. \\ ~\cite{mbaiossoumdatabase}, who identified materialized view maintenance as a key technique for database tuning, particularly for managing aggregate data. 

\item
\textbf{Qualitative Conclusions Regarding Triggers: when is the retrieval benefit worth the insertion cost}

Triggers will hurt insert performance but help query performance. The question is when is the tradeoff worth it. If queries for the total sales per vendor or per store are fairly frequent relative to the number of inserts, then the benefit is worth the price of increased trigger overhead. If such queries are infrequent, then we should not set up a trigger. 

For each system, we give the tradeoff point of how often a total per store or per vendor query must occur relative to the number of inserts for the trigger to be worthwhile: 

\begin{itemize}
    \item Insert Overhead:
The additional cost incurred per insertion due to trigger-based maintenance is calculated as:
\[
\text{Insert Overhead} = T_{\text{insert, trigger}} - T_{\text{insert, no trigger}}
\]

\item Query Benefit:
The time saved per query by using pre-aggregated results maintained by triggers:
\[
\text{Query Benefit} = T_{\text{query, no trigger}} - T_{\text{query, trigger}}
\]

\item Threshold $N_s$:
The maximum number of insertions between queries for triggers to be worthwhile. Intuitively, if this number is high for a system, then triggers tend to be more beneficial.
\[
N_s = \frac{\text{Query Benefit}}{\text{Insert Overhead}}
\]

\end{itemize}

\begin{table}[H]
  \centering
  \caption{Threshold $N_s$ (Inserts per Query) below which point  Triggers are worthwhile on $10^5$ Dataset. For example if there are fewer than roughly 3250 inserts between queries, then the trigger is worthwhile on MariaDB.}
  \label{tab:trigger_threshold_10e5}
  \begin{tabular}{llccc}
    \toprule
    \textbf{Query Type} & \textbf{System} & \textbf{Insert Overhead (s)} & \textbf{Query Benefit (s)} & \textbf{Threshold $N_s$} \\
    \midrule
    \multirow{4}{*}{Vendor}
        & MariaDB     & 0.000016   & 0.052     & 3250.00 \\
        & MySQL       & 0.00361    & 0.182     & 50.42   \\
        & PostgreSQL  & 0.00020    & 0.038     & 190.00  \\
    \midrule
    \multirow{4}{*}{Store}
        & MariaDB     & 0.000016   & 0.052     & 3250.00 \\
        & MySQL       & 0.00361    & 0.167     & 46.26   \\
        & PostgreSQL  & 0.00020    & 0.024     & 120.00  \\
    \bottomrule
  \end{tabular}
\end{table}


\begin{table}[H]
  \centering
  \caption{Threshold $N_s$ (Inserts per query) below which poiont  Triggers are worthwhile on $10^7$ Dataset. For example, if there are fewer than 62 inserts between queries, then it is worthwhile to use a trigger on MySQL. By contrast, if there are fewer than roughly 590,000 inserts between queries,  triggers are worthile on MariaDB.}
  \label{tab:trigger_threshold_10e7}
  \begin{tabular}{llccc}
    \toprule
    \textbf{Query Type} & \textbf{System} & \textbf{Insert Overhead (s)} & \textbf{Query Benefit (s)} & \textbf{Threshold $N_s$} \\
    \midrule
    \multirow{4}{*}{Vendor}
        & MariaDB     & 0.000011   & 6.50     & 590909.09 \\
        & MySQL       & 0.31731    & 19.74    & 62.22     \\
        & PostgreSQL  & 0.000278   & 1.054    & 3791.37   \\
    \midrule
    \multirow{4}{*}{Store}
        & MariaDB     & 0.000011   & 6.57     & 597272.73 \\
        & MySQL       & 0.31731    & 20.03    & 63.13     \\
        & PostgreSQL  & 0.000278   & 0.873    & 3139.57   \\
    \bottomrule
  \end{tabular}
\end{table}

\end{enumerate}


For each system, we give the tradeoff point of how often a total per store or per vendor query must occur relative to the number of inserts for the trigger to be worthwhile (see Table \ref{tab:trigger_threshold_10e7}). The larger the number in the last column, the more favorable it is to use  a trigger. For example, if there are fewer than 3,000 inserts between queries that can make use of the aggregated tables, then triggers are worthwhile on PostgreSQL.

\section{Vertical Partitioning}
This experiment evaluates alternative physical layouts and projection patterns for table scans to gauge their relative efficiency on a uniformly distributed dataset. In the \emph{no-vertical-partitioning} layout, an original table \texttt{account} is unchanged; scans  fetch the needed columns based on the query. In the \emph{vertical-partitioning} layout,  table \texttt{account} is split into two narrower tables \texttt{account1} and \texttt{account2} sharing the primary key: one storing frequently accessed columns and another holding the remaining columns. By comparing the four combinations -- all columns versus frequent column scans on a vertically partitioned and non-vertically partitioned layout --  this experiment measures how queries and physical partitioning interact to influence  scan performance.

\begin{itemize}

    \item \textbf{Experiment Setup:}
\\ The experiment uses three tables: \texttt{account} (\ref{lst:account}), \texttt{account1} (\ref{lst:account1}), \texttt{account2} (\ref{lst:account2}) at two scales - $\mathbf{10^{5}}$ and $\mathbf{10^{7}}$ rows. 

\begin{table}[H]
  \centering
  \caption{Sizes of column fields in \texttt{account} table (approximate)}
  \label{tab:size-in-bytes}

  \begin{tabular}{lccc}
    \toprule
    \textbf{System} & \textbf{\texttt{id INT}} & \textbf{\texttt{balance TEXT}} & \textbf{\texttt{homeaddress TEXT}} \\
    \midrule
    MySQL        & 4 bytes & 9 bytes & 2 kilobytes \\
    PostgreSQL   & 4 bytes & 11 bytes & 2 kilobytes \\
    MariaDB      & 4 bytes & 9 bytes & 2 kilobytes \\
    DuckDB       & 4 bytes & 9 bytes & 2 kilobytes \\
    \bottomrule
  \end{tabular}
\end{table}

\vspace{0.75\baselineskip}
\begin{lstlisting}[language=SQL,caption={Schema for the \texttt{account} table.},label={lst:account}] 
TABLE account (
    id INT primary key, 
    balance text, 
    homeaddress text
);
\end{lstlisting} 
\begin{lstlisting}[language=SQL,caption={Schema for the \texttt{account1} table.},label={lst:account1}] 
TABLE account1 (
    id INT primary key, 
    balance text
);
\end{lstlisting} 
\begin{lstlisting}[language=SQL,caption={Schema for the \texttt{account2} table.},label={lst:account2}] 
TABLE account2 (
    id INT primary key, 
    homeaddress text
);
\end{lstlisting} 
    
\end{itemize}

\subsection{Scan Query}
\begin{enumerate}

\item
\textbf{Experiment Goal:} \\[3pt]
The goal of this experiment is to quantify the performance impact of vertical partitioning under two common projection patterns: retrieving all columns or retrieving only a frequently accessed subset of columns. The un-partitioned layout retains all fields, thus avoiding join overhead when all columns are required, but incurs extra overhead for row-oriented layouts when queries touch only frequently accessed  attributes. Partitioning the table \texttt{account} vertically into \texttt{account1} and \texttt{account2} where the two fragments share the same key but \texttt{account1} has the frequently accessed attributes and \texttt{account2} has the others, reduces the data fetched for queries on the frequently accessed attributes but   obliges the database engine to execute a join when all columns are required. The hypothesis is that vertical partitioning will yield clear speed-ups for partial-column scans, while offering having a negative effect when all columns must be fetched. \\

\item 
\textbf{Experiment Setup:}

The queries shown in Listings \ref{lst:all-field-partition}, \ref{lst:all-field-no-partition}, \ref{lst:part-field-partition} and \ref{lst:part-field-no-partition} are used to access all/part columns in different vertical partitioning layouts.
\\[0.75\baselineskip] 
\begin{minipage}
{\linewidth} 
\begin{lstlisting}[language=SQL,caption={Query to access all fields in account with vertical partitioning requires a join.},label={lst:all-field-partition}] 
SELECT account1.id, balance, homeaddress from account1, account2 WHERE account1.id = account2.id;
\end{lstlisting} 
\begin{lstlisting}[language=SQL,caption={Query to access all fields in \texttt{account} without vertical partitioning.},label={lst:all-field-no-partition}] 
SELECT * FROM account;
\end{lstlisting} 
\begin{lstlisting}[language=SQL,caption={Query to access only id and balance fields with vertical partitioning.},label={lst:part-field-partition}] 
SELECT id, balance FROM account1;
\end{lstlisting} 
\begin{lstlisting}[language=SQL,caption={Query to access only id and balance fields without vertical partitioning.},label={lst:part-field-no-partition}] 
SELECT id, balance FROM account;
\end{lstlisting} 
\end{minipage}

\item
Please refer to \textbf{ Appendix \ref{vertical_partition-appendix}} for the Quantitative Experimental Results and to \href{https://github.com/pequeniajugar/DBTuneSuite/tree/main/scripts/vertical_partition_scan}{\texttt{Vertical Partitioning Scan Query Experiment Scripts}} for the code to run those experiments.
\\ 

\item
\textbf{Qualitative Conclusions}
\\
Across both the $10^5$ and $10^7$ row datasets, vertical partitioning showed mixed effects depending on query patterns and database engine architecture.

Vertical partitioning seems to be very helpful for partial-field access on MariaDB and MySQL. 
It provides  no advantage on DuckDB (which is already column-oriented) and little advantage on PostgreSQL.
\end{enumerate}
\subsection{Point Query}
This experiment evaluates the performance effects of vertical partitioning under point query workloads. In the no-vertical-partitioning layout, all attributes are stored in a single table. In the vertical-partitioning layout, the data is split into two narrower tables that share a primary key: one holding frequently accessed columns, and the other storing the remaining attributes.

To simulate skewed access patterns, the workload consists of point queries that all filter on the primary key but vary in the attributes retrieved. A fraction of the queries retrieve only the frequently accessed attributes, while the rest retrieve the full set of attributes. 

\begin{enumerate}

\item
\textbf{Experiment Goal:} \\[3pt]
The goal of this experiment is to quantify how vertical partitioning interacts with a subset of attributes versus all attribute point queries. 

\item 
\textbf{Experiment Setup:}
\\ The experiment uses three tables: \texttt{account} (Listing \ref{lst:account}), \texttt{account1} (Listing \ref{lst:account1}), \texttt{account2} (Listing \ref{lst:account2}) at two scales - $\mathbf{10^{5}}$ and $\mathbf{10^{7}}$ rows and uses the database engines explain in Section \ref{shared-settings}.

The queries shown in Listings \ref{lst:with-vp-all}, \ref{lst:with-vp-partial}, \ref{lst:without-vp-all} and \ref{lst:without-vp-partial} are used to access all/partial columns with/without vertical partitioning layouts.

\vspace{0.75\baselineskip}
\begin{minipage}
{\linewidth} 
\begin{lstlisting}[language=SQL,caption={Query to access all columns with vertical partitioning.},label={lst:with-vp-all}] 
Select account1.id, balance, homeaddress from account1, account2 where account1.id = account2.id and account1.id = 32099;
\end{lstlisting} 
\begin{lstlisting}[language=SQL,caption={Query to access partial columns with vertical partitioning.},label={lst:with-vp-partial}] 
SELECT id, balance FROM account1 where id = 32099;
\end{lstlisting} 
\begin{lstlisting}[language=SQL,caption={Query to access all columns without vertical partitioning.},label={lst:without-vp-all}] 
SELECT * FROM account where id = 32099;
\end{lstlisting} 
\begin{lstlisting}[language=SQL,caption={Query to access partial columns without vertical partitioning.},label={lst:without-vp-partial}] 
SELECT id, balance FROM account where id = 32099;
\end{lstlisting} 
\end{minipage}

In the experiments of Appendix \ref{vertical_partition-appendix}, we will mix queries that require all attributes in account (or a join between account1 and account2) with queries that require only the columns in account1. Fraction $f$ will be the portion of the queries that requires only the columns of account1. For example when $f = 0.9$ then 9 out of 10 queries require only the columns of account1. 

In our experiments, we evaluate six different values of $f$: 0, 0.2, 0.4, 0.6, 0.8, and 1, under both vertical partitioning and non-partitioned settings, in order to examine the conditions under which the benefits of vertical partitioning for point queries diminish. For each value of $f$, we issue 100 queries.


\item
Please refer to \textbf{Appendix \ref{vertical_partition-appendix}} for the Quantitative Experimental Results and to \href{https://github.com/pequeniajugar/DBTuneSuite/tree/main/scripts/vertical_partition_point}{\texttt{Vertical Partitioning Point Query Experiment Scripts}} for the code to run those experiments.
\\ 
\item
\textbf{Qualitative Conclusions}
\\
The strategy of vertical partitioning consistently hurt point query performance for both moderate ($10^5$ rows) and large ($10^7$ rows) dataset sizes. Dividing tables into vertical segments introduced significant overhead, especially on MySQL, where response times rose significantly at lower query fractions. Even when queries targeted only commonly accessed columns, vertical partitioning yielded no performance advantages. Across all evaluated systems -- DuckDB, MariaDB, MySQL, and PostgreSQL -- and various query fractions, vertical partitioning never helped appreciably. Instead, it typically reduced performance or had minimal effect, indicating that vertical partitioning is not beneficial for point queries. Our findings  contrast with the  results of Shasha and Bonnet~\cite{10.1145/1024694.1024720}, who reported that vertical partitioning improves performance when most queries access only a subset of attributes. This is of course especially true for columnar databases like DuckDB.

\end{enumerate}

\section{Denormalization}\label{denormalization-section}
Normalization is the process of structuring data into tables, so each relationship is represented just once. This eliminates redundancy while preserving logical integrity.  By contrast, denormalization deliberately combines  tables to create  wider tables in order to eliminate joins.   This intentional redundancy may speed up read-heavy workloads.

\begin{enumerate}

\item
\textbf{Experiment Goal:} \\[3pt]
The aim is to quantify the performance impact of denormalizing the \texttt{lineitem} tble by incorporating information from three small dimension tables (\texttt{region}, \texttt{nation}, \texttt{supplier}).  Denormalization eliminates a three-way join and may therefore reduce CPU time on reads that filter by region. 
\\
\item 
\textbf{Experiment Setup:}
\\ The experiment uses the TPC-H schema (described in section \ref{subsec:benchmark tpch}) at two scales - $\mathbf{10^{5}}$ and $\mathbf{10^{7}}$ rows.

Normalized and denormalized queries are shown in the listings \ref{lst:denorm-norm} and \ref{lst:denorm-flat}. Both return all \texttt{lineitem} rows whose supplier's region is \emph{Europe}.
\\ \\
\begin{minipage}{\linewidth}
\begin{lstlisting}[language=SQL,caption={Query on normalized data requires a three-way join on the supplier, nation and region tables.},label={lst:denorm-norm}]
SELECT  L_ORDERKEY, L_PARTKEY, L_SUPPKEY, L_LINENUMBER,
        L_QUANTITY, L_EXTENDEDPRICE, L_DISCOUNT, L_TAX,
        L_RETURNFLAG, L_LINESTATUS, L_SHIPDATE, L_COMMITDATE,
        L_RECEIPTDATE, L_SHIPINSTRUCT, L_SHIPMODE, L_COMMENT,
        R_NAME
FROM    lineitem  AS L,
        supplier  AS S,
        nation    AS N,
        region    AS R
WHERE   L.L_SUPPKEY   = S.S_SUPPKEY
  AND   S.S_NATIONKEY = N.N_NATIONKEY
  AND   N.N_REGIONKEY = R.R_REGIONKEY
  AND   R.R_NAME      = 'EUROPE';
\end{lstlisting}

\begin{lstlisting}[language=SQL,caption={Query on denomralized data requires no joins. },label={lst:denorm-flat}]
SELECT  L_ORDERKEY, L_PARTKEY, L_SUPPKEY, L_LINENUMBER,
        L_QUANTITY, L_EXTENDEDPRICE, L_DISCOUNT, L_TAX,
        L_RETURNFLAG, L_LINESTATUS, L_SHIPDATE, L_COMMITDATE,
        L_RECEIPTDATE, L_SHIPINSTRUCT, L_SHIPMODE, L_COMMENT,
        R_REGION
FROM    lineitemdenormalized
WHERE   R_REGION = 'EUROPE';
\end{lstlisting}
\end{minipage}

\item
Please refer to \textbf{Appendix \ref{denormalization-appendix}} for the Quantitative Experimental Results and to \href{https://github.com/pequeniajugar/DBTuneSuite/tree/main/scripts/denormalization}{\texttt{Denormalization Experiment Scripts}} for the code to run those experiments.

\item
\textbf{Qualitative Conclusions}
\\
Denormalization is neither universally beneficial nor universally harmful. 
Its effectiveness depends on the interplay between dataset size and each engine’s join and scan strategies.  
For modest data volumes, denormalization can cut latency dramatically (up to 44\% on MariaDB), but on larger datasets, disk access costs may outweigh the join overhead, rendering denormalization ineffective or even harmful.

Our results generally align with Sanders and Shin’s findings \\ ~\cite{926306} which showed 
 that denormalization can improve performance by reducing join overhead on smaller datasets.  However, on modern systems with optimized join algorithms, denormalization's benefits diminish or even reverse on larger datasets, suggesting that application designers should evaluate denormalization options based on data scale, query patterns, and the underlying engine.

\end{enumerate}
\section{Connection Pooling}\label{connectionpooling}

Connection pooling is a technique used to manage database connections efficiently by reusing a set of established connections rather than creating and closing a new one for every user request. When a request needs a database connection, it borrows one from the pool; after the request is completed, the connection is returned to the pool and can be reused.

In contrast, a simple connection (non-pooled connection) creates a brand-new connection to the database for each request, and closes it immediately after the request completes.

\begin{enumerate}

\item
\textbf{Experiment Goal:} \\
The goal of this experiment is to evaluate the impact of connection pooling on database performance under concurrent load, particularly in scenarios where the number of client threads exceeds the maximum number of allowable connections on the server.
\\
\item 
\textbf{Experiment Setup:}
Experiments are conducted on two data sizes: $10^5$ and $10^7$ rows. We use the \texttt{employee(ssnum, name, lat, longitude, hundreds1, hundreds2)} table (detailed definition in section \ref{subsec:benchmark employee}) for the experiments. 

The experiment compares two distinct connection strategies: using connection pooling or using simple connections.

In each test, the number of client threads is set to 10, 100, and 500. Each thread establishes a connection (or borrows one from the pool) and performs 5 insert statements. For connection pooling, the pool size is set to match the configured maximum number of database connections. Thus, in different experimental cases, we test three distinct maximum connection limits with and without pooling: 25, 50, and 100.

The insertion query is defined as following:

\begin{lstlisting}[language=SQL, label={lst:schema}]
INSERT INTO employee (ssnum, name, lat, longitude, hundreds1, hundreds2)
VALUES ({ssnum}, {name}, {lat}, {longitude}, {hundreds1}, {hundreds2});
\end{lstlisting}

For PostgreSQL, the parameter max\_connections is a postmaster-level setting, which means it can only take effect after the server is restarted. The configuration must be updated (e.g., via ALTER SYSTEM SET or by editing postgresql.conf) and then applied through a full service restart. Consequently, our PostgreSQL benchmarking scripts explicitly restart the server for each tested max\_connections value. In contrast, for MySQL, and MariaDB, the max\_connections setting (or its equivalent) can be modified at runtime without restarting the service, so no restart step is required in their experiments.

For DuckDB, since the system does not expose the maximum connection parameter, the simple connection case uses the default settings. In the connection pooling case, only the pool size is varied.
\\
\item
Please refer to \textbf{Appendix \ref{connection_pooling-appendix}} for the Quantitative Experimental Results and to \href{https://github.com/pequeniajugar/DBTuneSuite/tree/main/scripts/connection_pooling}{\texttt{Connection Pooling Experiment Scripts}} for the code to run those experiments.

\item
\textbf{Qualitative Conclusions}

Our evaluation on both $10^5$ and $10^7$ row datasets shows that connection pooling can be an effective strategy for improving performance and stability, especially under medium to high concurrency for MariaDB, MySQL, and PostgreSQL. For DuckDB, simple connections tend to perform slightly better than connection pooling in most cases, but the difference in performance is negligible.

On both data sizes, as max\_connection and pool size increase, pooling continues to improve performance in MySQL, PostgreSQL, and MariaDB, although the improvements tend to diminish, especially when max\_connection and pool size reach 100.

For DuckDB on the smaller dataset, connection pooling with a moderate pool size yields benefits under moderate thread counts (e.g., pool size = 50 with 10 threads, and pool size = 25 or 50 with 100 threads). However, as the thread count increases further, the advantage of pooling diminishes.

Pooling also enhances execution stability for traditional DBMSs on the smaller dataset. On the larger dataset, simple connections exhibit more stable performance. For DuckDB, simple connections consistently deliver more stable results than connection pooling.

In sum, connection pooling can be useful for scaling client-server databases like MySQL, PostgreSQL and MariaDB, delivering significant speedups and better predictability. By contrast, DuckDB tends to benefit more from its default simple connections under light workloads, as pooling offers little improvement and can sometimes even degrade performance.

Our findings regarding connection pooling on MySQL, MariaDB, and PostgreSQL are largely in line with prior works by Sobri et al. \\ ~\cite{sobri2022study} and Shasha and Bonnet~\cite{10.1145/1024694.1024720}, 
which similarly observed that pooling enhances performance and stability. While the Sobri et al. study identified an optimal configuration of around five connections per instance, our experiments found that moderate pool sizes (around 25–50 connections per instance) achieve the best balance between throughput and stability under higher concurrency levels. 

Note however that this observation is not universal. For example, for DuckDB, the effect of pooling appears limited, possibly due to its in-process execution model, for which simple connections already perform  well.


\end{enumerate}

\section{Results: Some Surprises from the Experiments} \label{surprises}

The biggest surprise to us was how large the effect of some tuning knobs were in some systems and how insignificant they were for others. For example, the systems vary widely in the degree to which clustered indexes are superior over non-clustered indexes. Here were some other surprises, though they are again system-dependent:

\begin{itemize}
\item
B+-trees are better than Hash structures in almost all cases, including multipoint queries (equality queries on non-keys). Hash structures perform  either equally well or negligibly better than B+ trees  for point queries. Please see section \ref{Hash-vs-btree}. We had expected Hash Structures to dominate for point and possibly even multipoint queries.
\item 
Non-clustered non-covering indexes perform as well as covering indexes.  Please see section \ref{coveringindexes}. We had expected covering indexes to be better especially for tables having $10^7$ rows.
\item 
Non-clustered indexing performed surprisingly well even  for range queries accessing up to 20\% of the rows on several systems. Please see section \ref{scan-vs-index}. We had expected scans to dominate before this point.
\item 
Most systems handle correlated subqueries very badly compared to using temporary tables. Please see Section \ref{correlated-vs-temporary}. We had expected correlated subqueries to be well optimized after all the decades of work on query processing.
\item 
Denormalization is  unhelpful on most modern systems even for query-only workloads.  Please see Section \ref{denormalization-section}. We had expected the elimination of joins (conferred by denormalization) to be helpful, but  most systems  support efficient foreign key indices.
\end{itemize}

\section{Limitations of \ourapproach}

Computational experiments, by their very nature, take place on particular versions of database management systems on particular data on particular operating systems and on particular hardware. A natural question therefore is to ask how generalizable these results are. The frank answer is that we don't know for sure. Some new hardware/database system version might appear that completely upends these results. 
That is why the conclusions regarding each tuning knob and  the comparative speeds of the different systems should be understood as being a snapshot rather than a long-term truth. 

Dramatic results that hold across many systems, e.g. the benefit of large batches when loading or retrieving only needed columns,  are likely to  generalize to other systems and generations of hardware. On the other hand, the benefit of query rewriting strategies having to do with correlated subqueries may not be generalizable. These remarks, however, are conjectures.
 
That said, our contribution goes beyond the snapshots and conjectured generalizability. We provide the scripts for our experiments, so readers can take the scripts and apply them to different  database system versions, hardware platforms and so on. Readers can also adapt the code to apply the experiments to their application workloads while testing the same or slight variants of the tuning knobs.

\section{Conclusions}\label{conclusions}

Database tuning involves changes at all layers of the application hierarchy from hardware to configuration parameters to indexes to query formulation to schema design. Some tuning tasks have been fully automated (notably configuration parameters) and others partly automated (notably, index choice) in the state of the art. 

By contrast, \ourapproach shows that some of the largest tuning gains result from the highest layers of the application, e.g. in query formulation or connection pooling. \ourapproach has also shown that some tuning strategies have large effects on some systems and minor effects on others. 

\ourapproach also shows some basic observations about the systems tested: MySQL performs well for modest size databases (under 1 million rows) especially with indexes.  PostgreSQL, MySQL, MariaDB use indexes much better than DuckDB, but DuckDB excels at scans, especially for large tables.

\begin{table}[H]
\caption{Best system(s) per operation in our experiments on $10^7$ row tables. See also the decision tree of Figure \ref{fig:decision-tree-3-children-2x}. (Full disclosure: the authors have no financial, familial, or emotional interest in any of these products.)}
\begin{tabular}{|l|c|c|c|c|}
\hline
Operation & MySQL & MariaDB & Postgres & DuckDB \\ \hline
Data Loading & & & & $\checkmark$ \\ \hline
Point query clustered & $\checkmark$ & &$\checkmark$  & \\ \hline
Point Query, non-clustered &$\checkmark$ & & $\checkmark$&  \\ \hline
Multipoint query & $\checkmark$ & & $\checkmark$ &  \\ \hline
Range query, clustered & $\checkmark$ & & $\checkmark$ & \\ \hline
Correlated subquery & & & & $\checkmark$ \\ \hline
\end{tabular}
\end{table}

Finally, \ourapproach provides a framework upon which the community can build application-specific benchmarks across data systems, schemas, and data distributions.

\begin{figure}
\centering
\begin{tikzpicture}[node distance=2cm, auto]

\node (dec0) [decision, align=left, xshift=2cm, yshift=-1cm]
{Query\\Type?};

\node (dec1) [action, below of=dec0, align=left,
              xshift=-2.0cm, yshift=-1.0cm]
{Data loading:\\DuckDB is\\fastest};

\node (dec2) [action, below of=dec0, align=left,
              xshift=1cm, yshift=-1.5cm]
{Point\\query};

\node (dec3) [action, below of=dec2, align=left,
              xshift=-2.0cm, yshift=-1.0cm]
{Point query,\\clustered index:\\MySQL and\\Postgres are\\fastest};

\node (dec4) [action, below of=dec2, align=left,
              xshift=3cm, yshift=-2.0cm]
{Point query,\\non-clustered\\index:\\MySQL and\\Postgres are\\fastest};

\node (dec5) [action, right of=dec0, align=left,
              xshift=2.0cm, yshift=2.0cm]
{Multipoint\\query:\\MySQL and\\Postgres are\\fastest};

\node (dec6) [action, right of=dec0, align=left,
              xshift=5.5cm]
{Range query,\\clustered index:\\MySQL and\\Postgres are\\fastest};

\node (dec7) [action, below of=dec0, align=left,
              xshift=7.0cm, yshift=-3.0cm]
{Correlated\\subquery:\\DuckDB is\\fastest};

\draw [arrow] (dec0) -- node {Loading} (dec1);
\draw [arrow] (dec0) -- node {Point} (dec2);
\draw [arrow] (dec2) -- node {Clustered} (dec3);
\draw [arrow] (dec2) -- node {Non-clustered} (dec4);
\draw [arrow] (dec0) -- node {Multipoint} (dec5);
\draw [arrow] (dec0) -- node {Range} (dec6);
\draw [arrow] (dec0) -- node {Correlated} (dec7);

\end{tikzpicture}
\caption{\textbf{Decision tree for engine choice based on workload.}
Indexes are particularly effective for MySQL  and Postgres. DuckDB is very fast for scans.}
\label{fig:decision-tree-3-children-2x}
\end{figure}
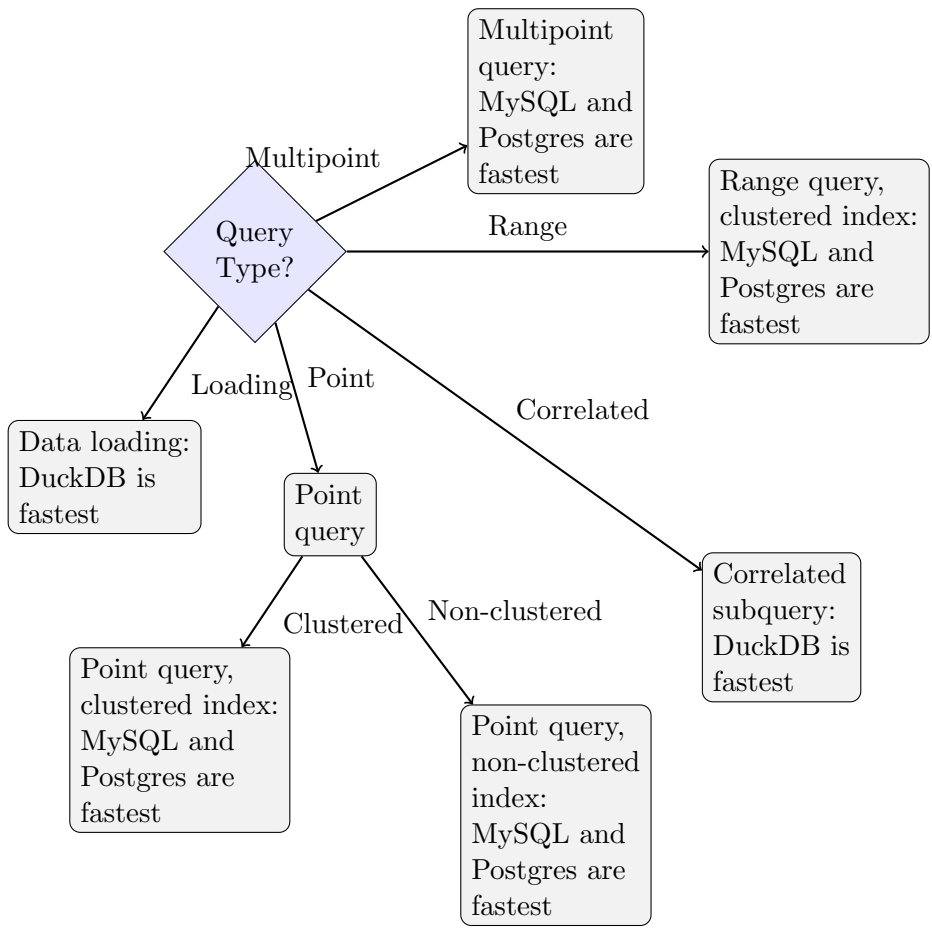

\clearpage

Data Availability: The software and  data are available at \url{https://github.com/pequeniajugar/dbtunning_experiements/tree/main}

Acknowledgment:
We would thank Philippe Bonnet for his helpful suggestions of systems to test.

Partial Funding: NYU Wireless

Conflict of Interest: The authors declare no conflict of interest. The funder had no role in the design or implementation or analysis of these experiments.

Corresponding Authors: all of us.

\appendix

\section[\appendixname~\thesection]{Data Loading: Quantitative Experimental Results}\label{dataloading-appendix}
\begin{figure}[H]
  \centering
  \includegraphics[width=0.8\textwidth]{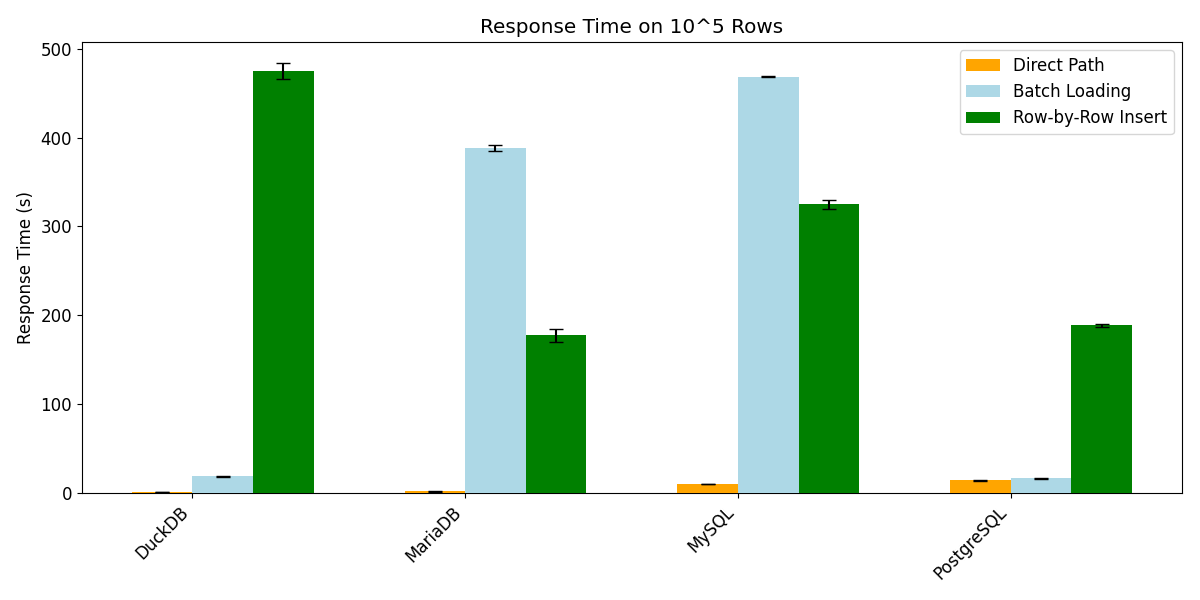}
  \caption{Direct path significantly improved the performance across all the engines.}
  \label{fig: Direct path loading results for 10^5 dataset}
\end{figure}

\textbf{On the $10^5$ row dataset} (Figure \ref{fig: Direct path loading results for 10^5 dataset}), direct path loading dramatically outperformed both batch loading and row-by-row insertion across all database engines that support it.
DuckDB achieved the fastest load time with direct path, running approximately $18.9\times$ faster than batch loading and $472.5\times$ faster than row-by-row insertion.
MariaDB saw a $186.7\times$ speedup over batch loading and $85.4\times$ over row-wise insertion, while MySQL achieved $45.7\times$ and $31.7\times$ improvements, respectively. PostgreSQL had a 2 second speed up with direct path as compared to batch loading and the row-wise insert proved to be the worst method significantly.
The performance difference between batching and row-wise insertion was substantial across engines.
These results demonstrate that batching inserts significantly improves load time in DuckDB, but notably, in MySQL and MariaDB, row-by-row insertion was faster than conventional batching. 
\\ \\
Nonetheless, direct path loading consistently offered the most efficient mechanism for batch data ingestion across all supported systems.
See Table \ref{tab:direct-10e5} for the full set of average response times and standard deviations.

\begin{table}[H]
  \centering
  \caption{Average Response Times along with their Standard Deviations for Direct Path Experiment on a $10^5$ Row Dataset (in seconds)}
  \label{tab:direct-10e5}

  \label{tab:direct-response-10e5}
  \begin{tabular}{lccc}
    \toprule
    \textbf{System} & \textbf{Direct Path} & \textbf{Batch loading (100 rows)} & \textbf{Row-by-row insert} \\
    \midrule
    DuckDB     & 1.004 (0.11) & 19.09 (0.69) & 474.44 (8.99) \\
    MariaDB     & 2.08 (0.22) & 388.41 (3.04) & 177.71 (7.55) \\
    MySQL       & 10.24 (0.39) & 468.42 (0.80) & 324.76 (5.48) \\
    PostgreSQL  & 14.44 (0.32) & 16.84 (0.28) & 188.77 (1.63)\\
    \bottomrule
  \end{tabular}

\end{table}

\begin{figure}[H]
  \centering
  \includegraphics[width=0.8\textwidth]{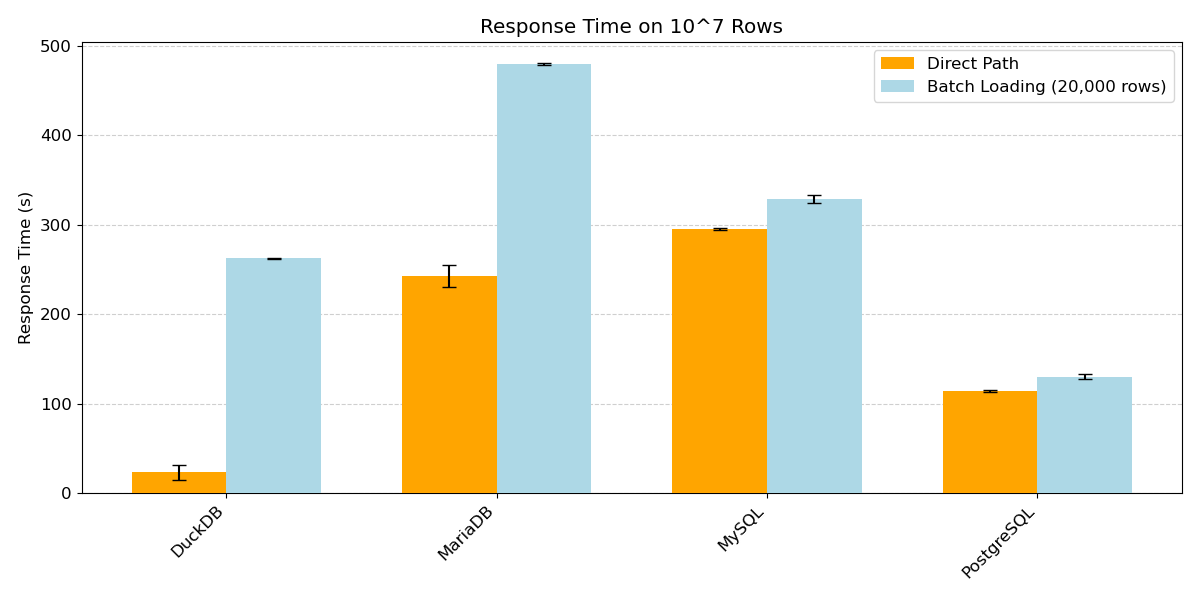}
  \caption{Direct path significantly improved the performance across all the engines. }
  \label{fig: Direct Path results for 10^7 dataset}
\end{figure}

\textbf{On the $10^7$ row dataset} (Figure \ref{fig: Direct Path results for 10^7 dataset}), all database systems benefit significantly from direct path loading. Among them, DuckDB achieves the best performance with this method. DuckDB and MariaDB show the largest improvements, by factors of 11.2 and 2.0, respectively, compared to conventional bulk loading with a batch size of 20,000. In contrast, PostgreSQL and MySQL exhibit more moderate gains. 



Row-by-row insertion was excluded from this experiment as it proved impractically slow on all engines during earlier runs — taking several hours or more — making it unsuitable for datasets of this scale.
These results reaffirm that direct path loading and batch loading, when well-implemented, offer significant performance benefits for large-scale ingestion.

See Table \ref{tab:direct-10e7} for the full set of average response times and standard deviations.

\begin{table}[H]
  \centering
  \caption{Average Response Times along with their Standard Deviations for Direct Path Experiment on a $10^7$ Row Dataset (in seconds)}
  \label{tab:direct-10e7}

  \label{tab:direct-response-10e7}
  \begin{tabular}{lccc}
    \toprule
    \textbf{System} & \textbf{Direct Path} & \textbf{Batch loading (20,000 rows)}\\
    \midrule
    DuckDB     & 23.43 (8.17) & 262.55 (0.42)\\
    MariaDB     & 243.03 (12.55) & 479.24 (1.17)\\
    MySQL       & 295.09 (1.27) & 328.91 (4.12)\\
    PostgreSQL  & 114.64 (1.03) &  130.20 (3.08)\\
    \bottomrule
  \end{tabular}
\end{table}

\section[\appendixname~\thesection]{Batch Loading: Quantitative Experimental Results}\label{batchloading-appendix}
\begin{figure}[H]
  \centering
  \includegraphics[width=0.8\textwidth]{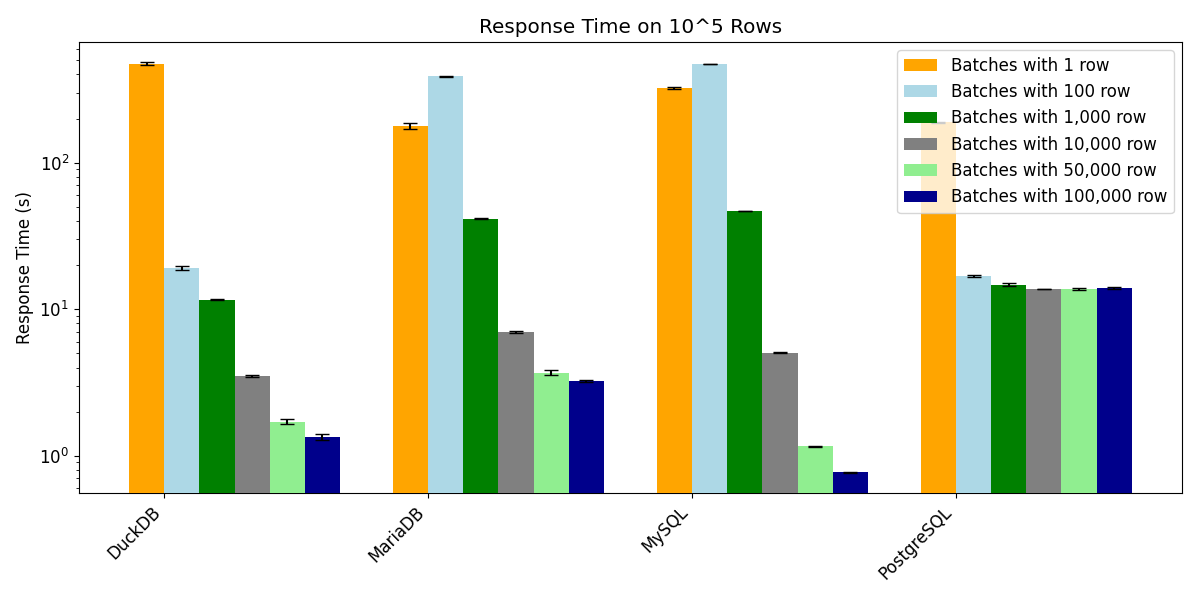}
  \caption{There is a clear improvement in the performance of data loading across all engines as batch size increases on the $10^5$ row dataset except on PostgreSQL after 10,000 batch.}
  \label{fig: Batching results for 10^5 dataset}
\end{figure}

\textbf{On the $10^{5}$ row dataset} (Table \ref{fig: Batching results for 10^5 dataset}), every engine accelerated sharply as batch size increased, often by one to two orders of magnitude.

\begin{itemize}
\item
DuckDB sped up by roughly \textbf{$136\times$} when moving from single-row inserts to 10,000 row batches, and about \textbf{$350\times$} at 100,000 rows, all while keeping standard deviation below 5\%.
\item
For MariaDB,   batches of 10,000 rows improved load time by about \textbf{$26\times$}, and 100,000 rows lifted the gain to roughly \textbf{$55\times$} compared to single-row inserts.
\item
For MySQL,  10,000 row batches cut runtime by around \textbf{$64\times$}; 100,000 row batches slashed it by an impressive \textbf{$421\times$}, compared to single-row inserts.
\item
In PostgreSQL, switching from row-wise inserts to batched inserts improved performance, but increasing the batch size beyond 1,000 produced no significant gains; at 50,000 and 100,000 performance even slowed slightly.
\end{itemize}

In short, batching is indispensable: groups of 10,000 rows or more reduce ingest time by one to two orders of magnitude across engines except PostgreSQL, while very small batches can  underperform even row-by-row inserts.

See Table \ref{tab:batch-10e5} for the full set of average response times and standard deviations.

\begin{table}[H]
  \centering
  \caption{Average Response Times along with their Standard Deviations for data loading on different batch sizes on a $10^5$ row dataset (in seconds)}
  \label{tab:batch-10e5}

  \label{tab:batch-response-10e5}
    \resizebox{\textwidth}{!}{%
  \begin{tabular}{@{}lccccccc@{}}
    \toprule
    \textbf{System} & \textbf{1 row} & \textbf{100 rows} &
    \textbf{1,000 rows} & \textbf{10,000 rows} & \textbf{50,000 rows} &
    \textbf{100,000 rows} \\
    \midrule
    DuckDB      & 474.4 (8.99)  & 19.09 (0.69)  & 11.62 (0.15) & 3.498 (0.079) & 1.710 (0.066) & 1.348 (0.060) \\
    MariaDB     & 177.7 (7.55)  & 388.4 (3.04)  & 41.43 (0.19) & 6.942 (0.105) & 3.680 (0.150) & 3.222 (0.060) \\
    MySQL       & 324.8 (5.48)  & 468.4 (0.80)  & 46.61 (0.18) & 5.048 (0.019) & 1.157 (0.009) & 0.772 (0.007) \\
    PostgreSQL  & 188.77 (1.63) & 16.84 (0.28)  & 14.69 (0.31) & 13.72 (0.69) & 13.82 (0.24) & 13.88 (0.20) \\

    \bottomrule
  \end{tabular}}
\end{table}

\begin{figure}[H]
  \centering
  \includegraphics[width=0.8\textwidth]{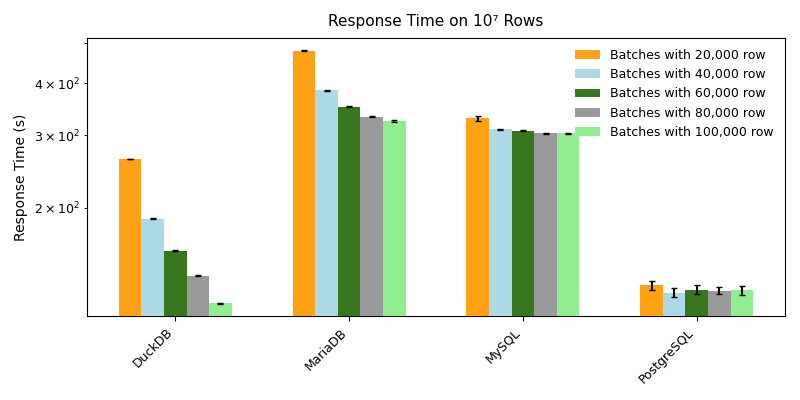}
  \caption{There is a noticeable improvement in data loading performance for MariaDB and DuckDB as the batch size increases for the $10^7$-row dataset, while PostgreSQL and MySQL show generally stable performance across different batch sizes.}
  \label{fig: Batching results for 10^7 dataset}
\end{figure}

\textbf{On the $10^{7}$ row dataset} (Table \ref{fig: Batching results for 10^7 dataset}), increasing batch size beyond 20,000 rows continued to improve performance, though the improvements were less  dramatic than for the smaller dataset.

\begin{itemize}
\item
DuckDB. Raising the batch from 20,000 to 100,000 rows cut load time by roughly \textbf{$2.2\times$}, with a  sub-$1\%$ standard deviation throughout.
\item
MariaDB showed an improvement as we move from 20,000 to 100,000 rows with a speedup of 1.3$\times$.
\item
The performance of MySQL improves by a factor of 1.06 as the batch size increases from 20,000 to 40,000, after which it remains generally stable despite further increases in batch size.
\item
The performance of PostgreSQL remains stable across all batch sizes. The largest improvement occurs between batch sizes of 20,000 and 40,000, with a speedup of 1.04×. Beyond this point, the performance remains generally consistent and occasionally shows slight degradation.
\end{itemize}

\begin{table}[H]
  \centering
  \caption{Average Response Times along with their Standard Deviations for data loading on different batch sizes on a $10^7$ row dataset (in seconds)}
  \label{tab:batch-10e7}

  \label{tab:batch-response-10e7}
  \resizebox{\textwidth}{!}{%
  \begin{tabular}{@{}lccccccc@{}}
    \toprule
    \textbf{System} & \textbf{20,000 rows} &
    \textbf{40,000 rows} & \textbf{60,000 rows} & \textbf{80,000 rows} &
    \textbf{100,000 rows} \\
    \midrule
    DuckDB      & 262.56 (0.43) & 188.39 (0.54) & 157.789 (0.35) & 137.18 (0.30) & 117.95 (0.32) \\
    MariaDB     & 479.24 (1.17) & 384.67 (1.27) & 351.28 (0.95) & 332.47 (1.35) & 324.24 (1.18) \\
    MySQL       & 328.91 (4.12) & 309.58 (0.67) &  307.41 (0.72) &  302.98 (0.87) &  303.03 (0.62) \\
    PostgreSQL  & 130.2 (3.08) & 124.95 (3.01) &  127.04 (3.02) &  126.20 (2.40) &  126.53 (3.08) \\
    \bottomrule
  \end{tabular}}
\end{table}

\section[\appendixname~\thesection]{Hash vs B+ Tree for Multipoint Query: Quantitative Experimental Results}\label{hashvsbtree_multipoint-appendix}
\begin{figure}[H]
  \centering
  \includegraphics[width=1\textwidth]{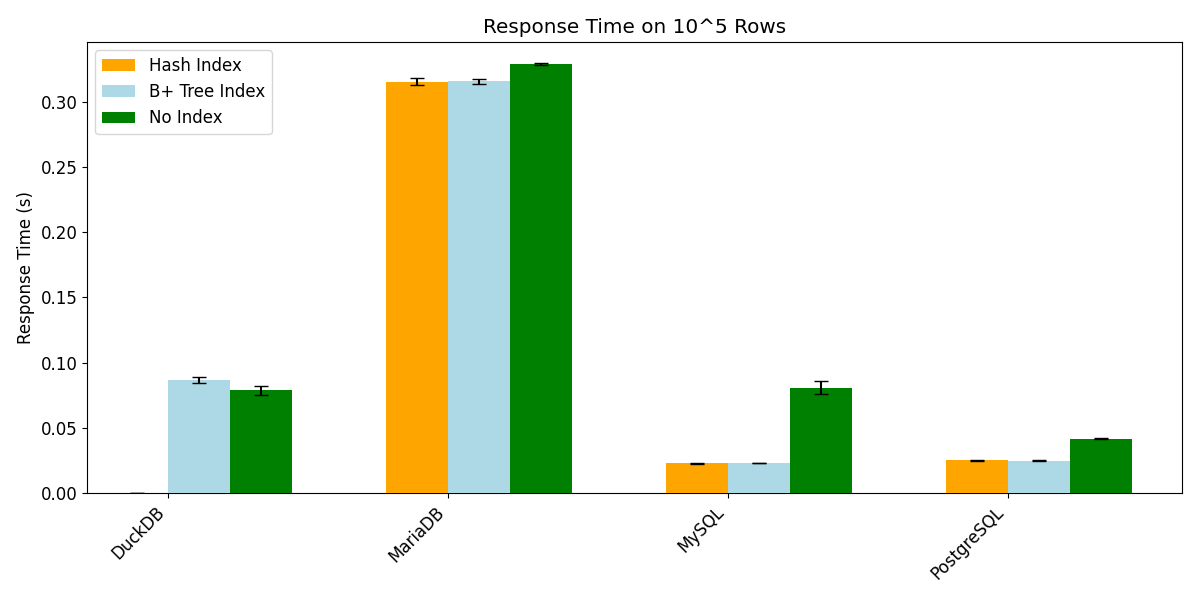}
  \caption{Impact of Hash Structure vs. B+ Tree  on Multipoint Query Performance on $10^5$ rows. DuckDB supports only a B+ tree-like structure. The performance of the two kinds of data structures on MariaDB, MySQL and PostgreSQL are similar. While no index outperforms the indexes in DuckDB, indexes help for MariaDB, MySQL and PostgreSQL. }
  \label{fig: Hashvsbtree multipoint results for 10^5 dataset}
\end{figure}

\textbf{On the $10^5$ row dataset}, (shown in Figure~\ref{fig: Hashvsbtree multipoint results for 10^5 dataset}) DuckDB, with its B+ Tree-like structure, achieves a response time of 0.087 seconds with minimal variance and recording 0.078 seconds for without index.
MariaDB, MySQL and PostgreSQL shows no performance difference between the two indexing strategies. Indexing is significantly better than no indexing for MySQL and slightly better for MariaDB and PostgreSQL.

Refer to Table~\ref{tab:Hashvsbtree_multipoint5} for the full set of average response times and their standard deviations. 

\begin{table}[H]
  \centering
  \caption{Average Response Time (Standard Deviation) for Hash and B+ Tree  Experiments in Multipoint Query on a $10^5$-Row Dataset (in seconds).}
  \label{tab:Hashvsbtree_multipoint5}
    \begin{tabular}{lccc}
        \toprule
        \textbf{System} & \textbf{Hash Index} & \textbf{B+ Tree Index} & \textbf{No Index}\\
        \midrule
        DuckDB      & N.A & 0.087 (0.0020) & 0.078 (0.003) \\
        MariaDB     & 0.32 (0.0022) & 0.32 (0.0018) & 0.33 (0.0006) \\
        MySQL       & 0.023 (0.0003) & 0.023 (0) & 0.08 (0.005)\\
        PostgreSQL  & 0.025 (0.0004) & 0.024 (0.0003) & 0.041 (0.0004) \\
        \bottomrule
    \end{tabular}
\end{table}

\begin{figure}[H]
  \centering
  \includegraphics[width=0.8\textwidth]{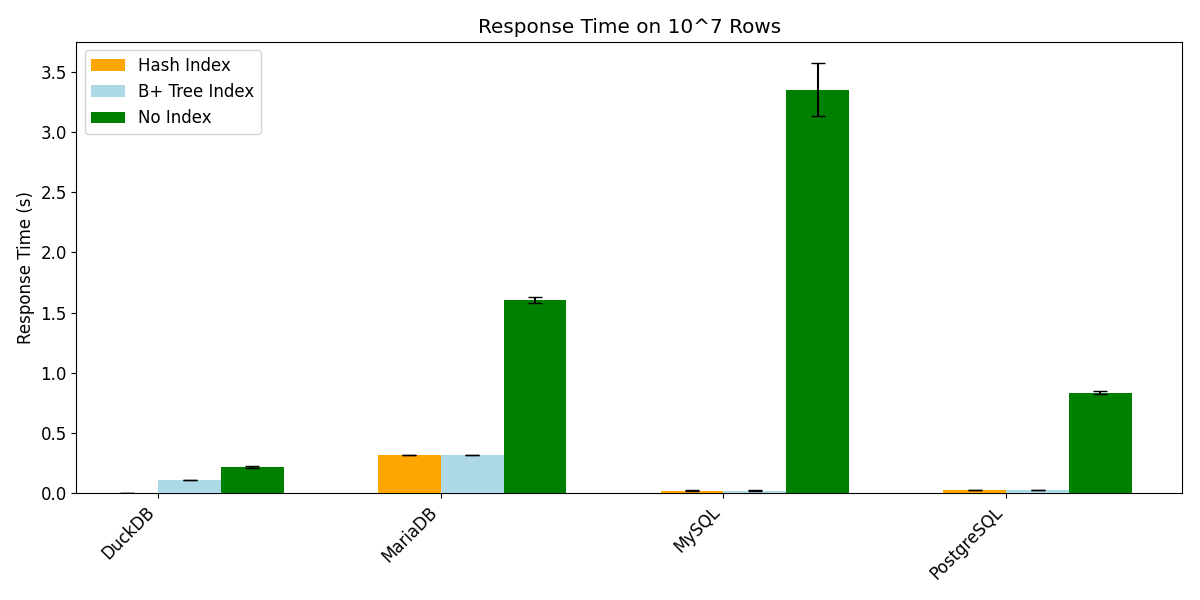}
  \caption{Impact of Hash Structure vs. B+ Tree  on Multipoint Query Performance on $10^7$ rows.  The performance of both kinds of data structures on MariaDB, MySQL and PostgreSQL are similar. No Index proves to be the worst strategy on all the dbms. }
  \label{fig: Hashvsbtree multipoint results for 10^7 dataset}
\end{figure}

\textbf{On the $10^7$ row dataset}(shown in Figure~\ref{fig: Hashvsbtree multipoint results for 10^7 dataset}), DuckDB on its B+ Tree-like structure, achieves a response time of 0.11 seconds with low variability, compared to 0.21 seconds when using no index.
MariaDB, MySQL and PostgreSQL again show no performance distinction between indexing strategies, with both hash and B+ Tree indices yielding identical average response times and much better performance compared to the no indexing strategy. Indexing is 5× faster for MariaDB, more than 100× faster for MySQL and 40× faster for PostgreSQL.

Refer to table (Table~\ref{tab:Hashvsbtree_multipoint7}) for the full set of average response times and their standard deviations.

\begin{table}[H]
  \centering
  \caption{Average Response Time (Standard Deviation) for Hash Structure and B+ Tree  in Multipoint Query Experiments on a $10^7$-Row Dataset (in seconds).}
  \label{tab:Hashvsbtree_multipoint7}

    \begin{tabular}{lccc}
        \toprule
        \textbf{System} & \textbf{Hash Index} & \textbf{B+ Tree Index} & \textbf{No Index} \\
        \midrule
        DuckDB      & N.A  & 0.11 (0.0020) & 0.21 (0.007) \\
        MariaDB     & 0.32 (0.0025) & 0.32 (0.0019) & 1.6 (0.026)\\
        MySQL       & 0.022 (0.0006) & 0.022 (0.0007) & 3.35 (0.21) \\
        PostgreSQL  & 0.025 (0.0005) & 0.025 (0.0004) & 0.83 (0.013)\\
        \bottomrule
    \end{tabular}
\end{table}

\section[\appendixname~\thesection]{Hash vs B+ Tree for Point Query: Quantitative Experimental Results}\label{hashvsbtree_point-appendix}
\begin{figure}[H]
  \centering
  \includegraphics[width=1\textwidth]{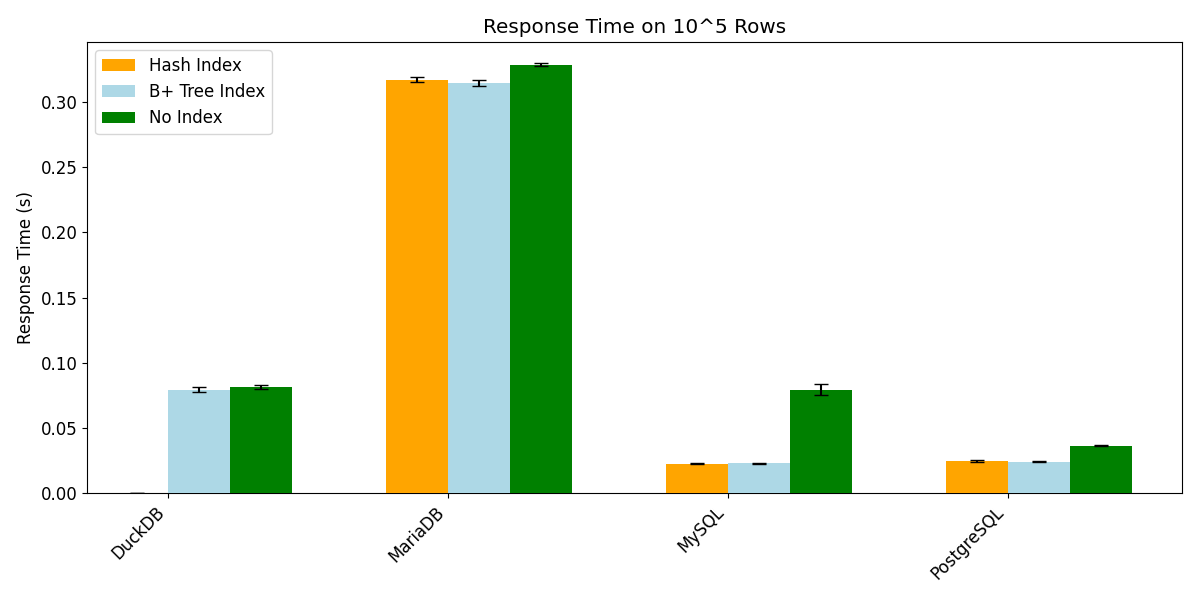}
  \caption{Impact of Hash Structure vs. B+ Tree  on Point Query Performance on $10^5$ rows. DuckDB does not support Hash index. MySQL, MariaDB and PostgreSQL perform roughly the same with Hash Indexes and B+ Trees. No index is the worst strategy in all the dbms except DuckDB where it is roughly the same as B+ Tree.}
  \label{fig: Hashvsbtree point results for 10^5 dataset}
\end{figure}

\textbf{On the $10^5$ row dataset}, (shown in Figure~\ref{fig: Hashvsbtree point results for 10^5 dataset}) MySQL yielded the lowest average response time of 0.022 seconds for B+ Tree indexes and 0.023 seconds for Hash indexes, with minimal variance but significant improvement over the no index query. PostgreSQL showed a small performance difference between indexing strategies: Hash indexing achieved 0.025 seconds, while B+ Tree indexing was at 0.024 seconds and no indexing was at 0.036 seconds. DuckDB, which supports only B+ Tree-like indexing, reported nearly the same  response time of about 0.080 seconds for both the B+ Tree and no index. MariaDB also demonstrated nearly identical performances, with average response times of 0.32 seconds for Hash and 0.31 seconds for B+ Trees and 0.33 seconds for no index.

Refer to Table~\ref{tab:Hashvsbtree_point5} for the full set of average response times and their standard deviations.

\begin{table}[H]
  \centering
  \caption{Average Response Time (Standard Deviation) for Hash and B+ Tree Indexes in Point Query Experiments on a $10^5$-Row Dataset (in seconds).}
  \label{tab:Hashvsbtree_point5}

    \begin{tabular}{lccc}
        \toprule
        \textbf{System} & \textbf{Hash Index} & \textbf{B+ Tree Index} & \textbf{No Index} \\
        \midrule
        DuckDB      & N.A & 0.079 (0.0021) & 0.081 (0.0017) \\
        MariaDB     & 0.32 (0.0019)  & 0.31 (0.0023) & 0.33 (0.001)\\
        MySQL       & 0.023 (0.00042) & 0.022 (0.00056) & 0.79 (0.0044) \\
        PostgreSQL  & 0.025 (0.00048) & 0.024 (0.00042) & 0.036 (0.00051) \\
        \bottomrule
    \end{tabular}
\end{table}

\begin{figure}[H]
  \centering
  \includegraphics[width=0.8\textwidth]{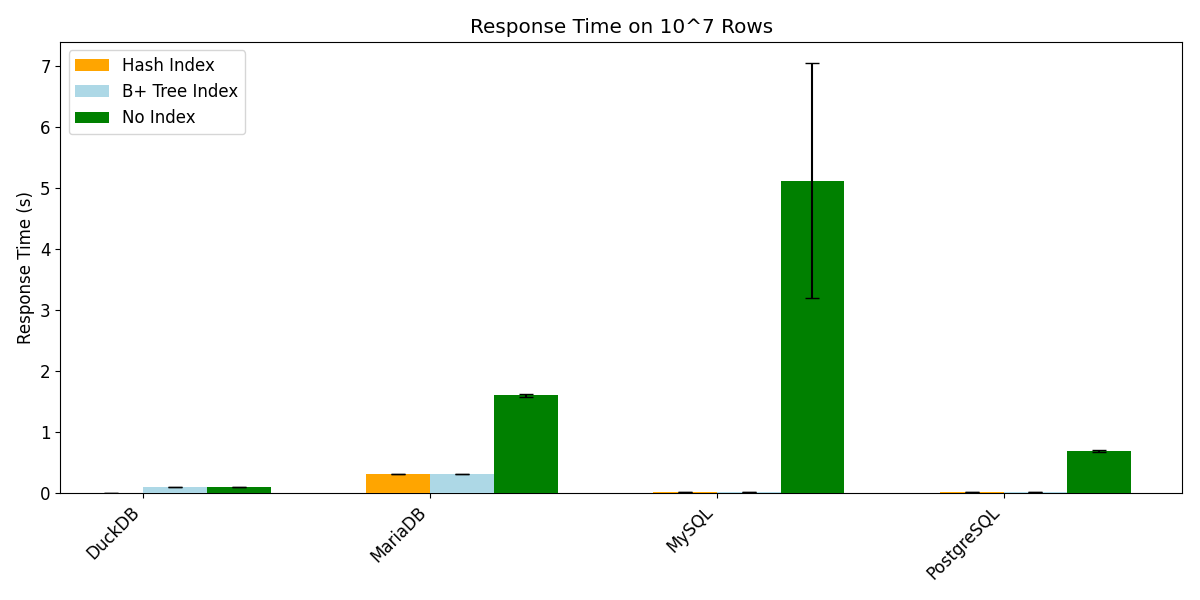}
  \caption{Impact of Hash Index vs. B+ Tree Index on Point Query Performance on $10^7$ rows. DuckDB does not support Hash indexes. The performance of the two kinds of indexes on MariaDB, MySQL and PostgreSQL is similar. No index is the worst strategy in all the dbms except DuckDB, where it performs as well  as the B+ Tree-like index.}
  \label{fig: Hashvsbtree point results for 10^7 dataset}
\end{figure}

\textbf{On the $10^7$ row dataset}, (shown in Figure~\ref{fig: Hashvsbtree point results for 10^7 dataset}) MySQL exhibited the best performance with a Hash and B+ Tree index, achieving an average response time of 0.022 seconds, outperforming its no index counterpart, which had a response time of 5.12 seconds. PostgreSQL also performed equally well  for Hash and B+ Tree indexing, with a response time of 0.025 seconds, nearly 20× faster than its no index performance. MariaDB showed no practical difference between index types, with both yielding an average response time of 0.31 seconds and negligible variance and both outperformed the no index query with 1.6 seconds.  DuckDB, which only supports B+ Tree indexes, recorded a response time of 0.1 seconds, with notably low variability for B+ Tree and no index query.

Refer to Table~\ref{tab:Hashvsbtree_multipoint7} for the full set of average response times and their standard deviations.

\begin{table}[H]
  \centering
  \caption{Average Response Time (Standard Deviation) for Hash and B+ Tree Indexes in Point Query Experiments  on a $10^7$-Row Dataset (in seconds).}
  \label{tab:Hashvsbtree_point5}

    \begin{tabular}{lccc}
        \toprule
        \textbf{System} & \textbf{Hash Index} & \textbf{B+ Tree Index} & \textbf{No Index} \\
        \midrule
        DuckDB      & N.A & 0.1 (0.0002) & 0.1 (0.0029) \\
        MariaDB     & 0.31 (0.0026)  & 0.31 (0.0019) & 1.6 (0.032)\\
        MySQL       & 0.022 (0.00042) & 0.022 (0.00073) & 5.12 (1.92)\\
        PostgreSQL  & 0.025 (0.00091) & 0.025 (0.00052) & 0.69 (0.014)\\
        \bottomrule
    \end{tabular}
\end{table}

\section[\appendixname~\thesection]{Clustered, Non-Clustered and No Index for Point Query: Quantitative Experimental Results}\label{clustered_point-appendix}
\textbf{On the $10^5$ row dataset} (Figure~\ref{fig:clustered-index-10e5}), both clustered and non-clustered indexes improve query performance on MySQL, PostgreSQL and MariaDB. DuckDB shows a minor performance gain with non-clustered indexing. 

MySQL sees the largest gain, with average response time dropping from 0.2s (no index) to 0.024s (non-clustered and clustered), yielding speedups of 10×. On MariaDB, both index types reduce response time from 0.385s to around 0.313s (non-clustered) and 0.314s (clustered), achieving a modest improvement of about 1.2×. For PostgreSQL, the clustered and non-clustered index delivers the best performance at 0.027s, compared to 0.042s (no index), indicating a slight benefit. DuckDB shows a small improvement with non-clustered indexing (1.13× speedup) vs. no indexing.

Refer to table (\ref{tab:clustered-index-10e5}) for a full list of response times and variance across all DBMS.


\begin{figure}[H]
  \centering
  \includegraphics[width=0.85\textwidth]{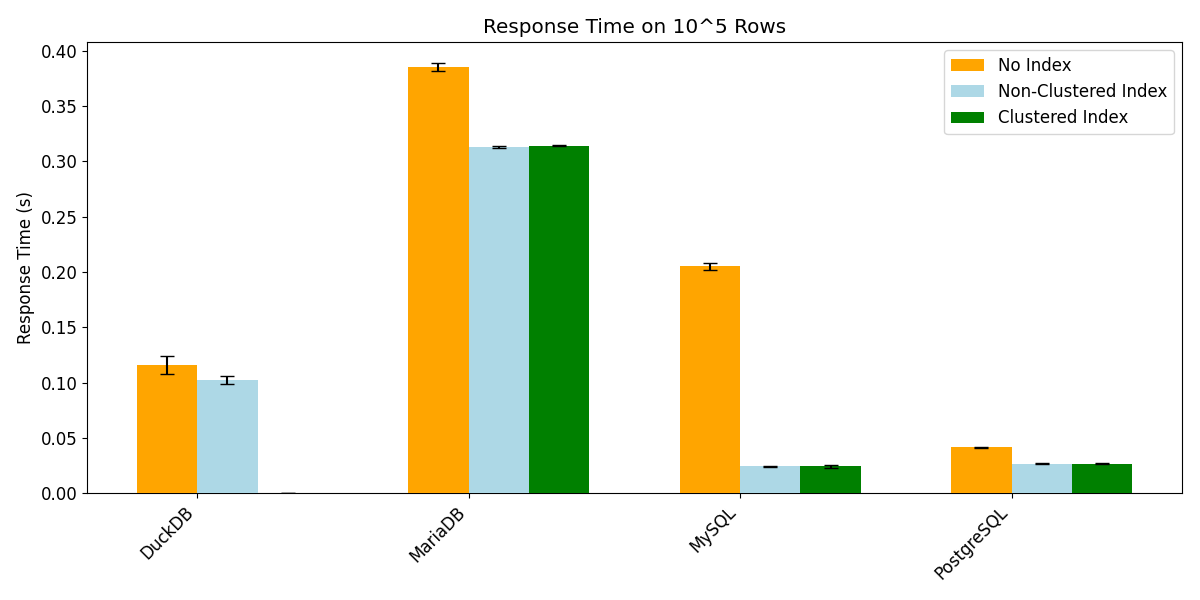}
  \caption{All the DBMS benefit from indexes, but show no difference in clustered and non-clustered index for $10^5$ rows. DuckDB does not support clustered indexes.}
  \label{fig:clustered-index-10e5}
\end{figure}

\begin{table}[H]
  \centering
  \caption{Average (Standard Deviation) of Response Time for Clustered, Non-clustered, and No Index Point Query Experiments on a $10^5$-Row Dataset (in seconds)}
  \label{tab:clustered-index-10e5}
  \begin{tabular}{lccc}
    \toprule
    \textbf{System} & \textbf{No Index} & \textbf{Non-clustered} & \textbf{Clustered} \\
    \midrule
    DuckDB      & 0.12 (0.0078) & 0.10 (0.0037) & --       \\
    MariaDB   & 0.39 (0.0036) & 0.31 (0.00074) & 0.31 (0.00047) \\
    MySQL       & 0.20 (0.0029)  & 0.024 (0.00078) & 0.024 (0.0014) \\
    PostgreSQL  & 0.042 (0.00052) & 0.027 (0.00063)  & 0.027 (0.00067)  \\
    \bottomrule
  \end{tabular}
\end{table}



\begin{figure}[H]
  \centering
  \includegraphics[width=0.8\textwidth]{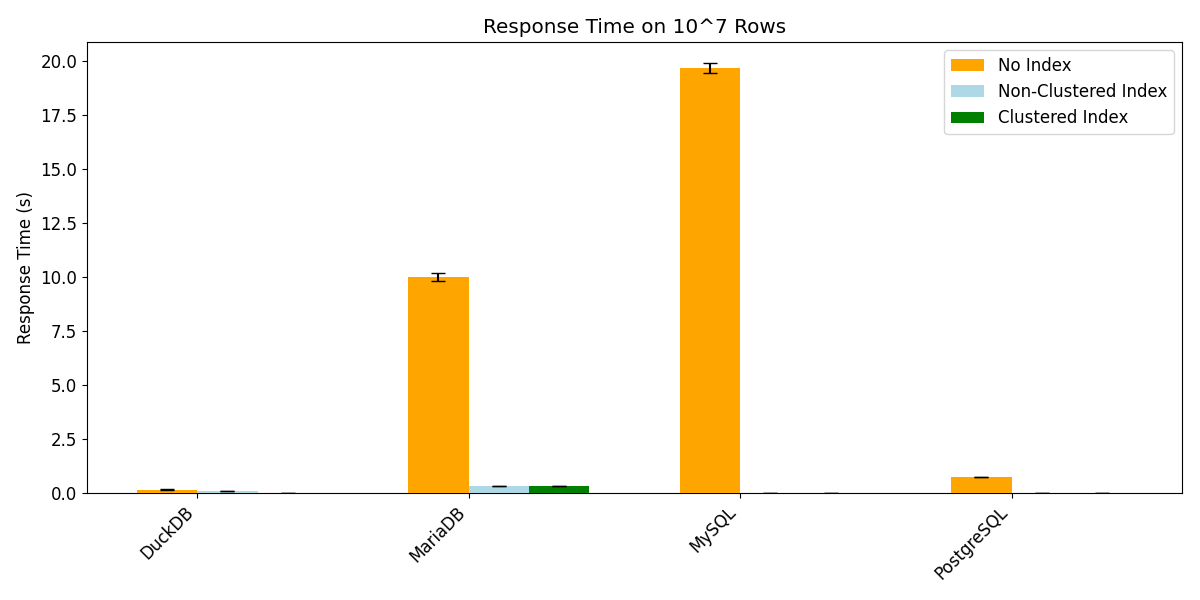}
  \caption{All the DBMS benefit from indexes, but show roughly the same performance for clustered and non-clustered indexes for $10^7$ rows. DuckDB does not support clustered indexes. } \label{fig:-clustered-index-results-for-10_7-dataset}
\end{figure}

\textbf{On the $10^7$ row dataset} (Figure~\ref{fig:-clustered-index-results-for-10_7-dataset}), indexing dramatically improves point query performance across all systems.

MySQL experiences the most substantial improvement: average response time drops from 19.7s (no index) to 0.024s (clustered) and 0.023s (non-clustered). MariaDB follows closely, with a reduction from 9.99s to 0.314s (clustered) and 0.3139s (non-clustered), achieving speedups of 31.8× for both index types. On PostgreSQL,  indexing reduces query time from 0.76s to 0.027s (clustered) and  0.026s (non-clustered). For DuckDB,  non-clustered indexing reduces response time compared to no indexing from 0.169s to 0.099s, improving performance by a factor of 1.71×.

As shown in Table~\ref{tab:clustered-index-10e7}, indexing also improves performance stability across most systems. MariaDB and MySQL see dramatic drops in standard deviation by using clustered indexes, indicating far more consistent performance. PostgreSQL and DuckDB also benefit from reduced response time variance under indexing.


\begin{table}[H]
  \centering
  \caption{Average (Standard Deviation) of Response Time for Clustered, Non-clustered, and No Index Point Query Experiments on a $10^7$-Row Dataset (in seconds)}
  \label{tab:clustered-index-10e7}
  \begin{tabular}{lccc}
    \toprule
    \textbf{System} & \textbf{No Index} & \textbf{Non-clustered} & \textbf{Clustered} \\
    \midrule
    DuckDB      & 0.17 (0.0084)  & 0.099 (0.0037) & -- \\
    MariaDB     & 10 (0.18) & 0.31 (0.00094) & 0.31 (0.00074)  \\
    MySQL       & 19.7 (0.23) & 0.023 (0.00057)  & 0.024 (0)   \\
    PostgreSQL  & 0.76 (0.0055) & 0.026 (0.0007) & 0.027 (0.00067)  \\
    \bottomrule
  \end{tabular}
\end{table}

\section[\appendixname~\thesection]{Clustered, Non-Clustered and No Index for Range Query: Quantitative Experimental Results}
\label{clustered_range-appendix}

\textbf{On the $10^5$ row dataset}
(Figure~\ref{fig:clustered-index-10e5}), indexing improves range query performance across all systems except DuckDB, though the degree of benefit varies.

MySQL sees a 6.5× speedup when using a non-clustered index and a 7.7× speedup with a clustered index, compared to no index. PostgreSQL sees a 1.37× speedup when using a non-clustered index and a 1.45× speedup with a clustered index. On MariaDB, both index types offer modest but consistent improvements of approximately 1.2×, with clustered index performing slightly better than the non-clustered one. DuckDB, by contrast, shows degraded performance when using a non-clustered index.


Regarding stability (Table~\ref{tab:clustered-index-10e5}), indexes generally reduce response time variance. Clustered indexing yields the most stable performance on MariaDB and PostgreSQL. On DuckDB, the non-clustered index offers the lowest variance. On MySQL, the differences in stability across the various indexes are negligible.


\begin{figure}[H]
  \centering
  \includegraphics[width=0.85\textwidth]{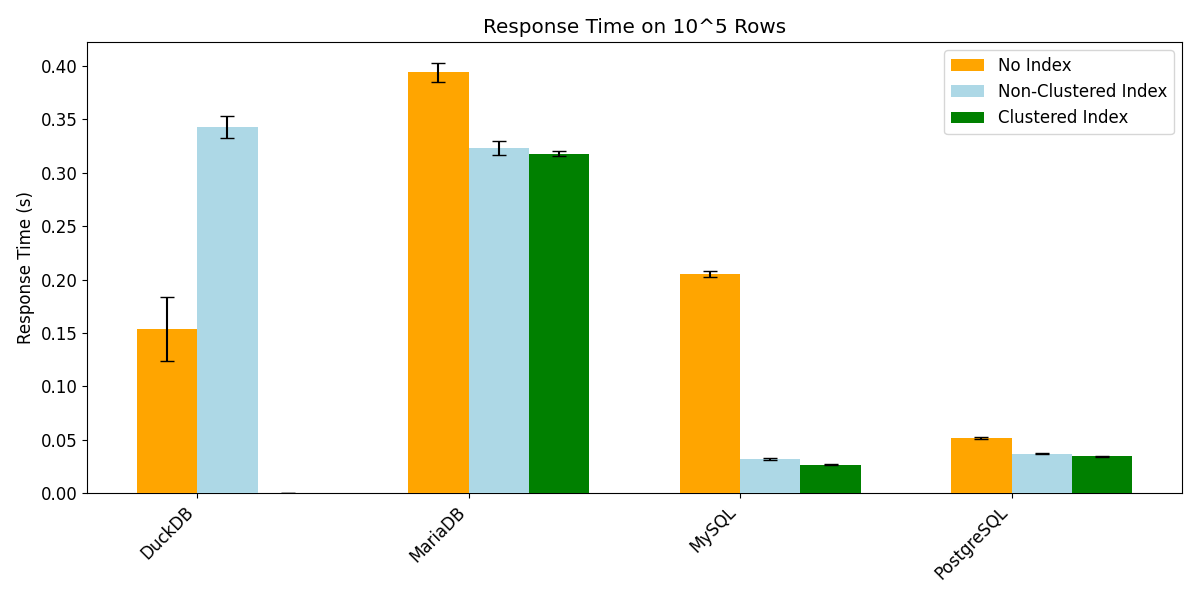}
  \caption{All the DBMS benefit from indexing except DuckDB on $10^5$ rows for range queries retrieving 1000 rows (1\% of the data). On all DBMS, clustered indexes perform slightly better than non-clustered indexes. DuckDB does not support clustered indexes.}
  \label{fig:clustered-index-10e5}
\end{figure}

\begin{table}[H]
  \centering
  \caption{Average (Standard Deviation) of Response Time for Clustered, Non-clustered, and No Index Range Query Experiments on a $10^5$-Row Dataset (in seconds)}
  \label{tab:clustered-index-10e5}
  \begin{tabular}{lccc}
    \toprule
    \textbf{System} & \textbf{No Index} & \textbf{Non-clustered} & \textbf{Clustered} \\
    \midrule
    DuckDB      & 0.15 (0.030) & 0.34 (0.011) & --            \\
    MariaDB     & 0.39 (0.0085) & 0.32 (0.0065) & 0.32 (0.0023) \\
    MySQL       & 0.21 (0.0028) & 0.032 (0.00063) & 0.027 (0.00063) \\
    PostgreSQL  & 0.051 (0.00096) & 0.037 (0.00057) & 0.035 (0.00053) \\
    \bottomrule
  \end{tabular}
\end{table}



\begin{figure}[H]
  \centering
  \includegraphics[width=0.8\textwidth]{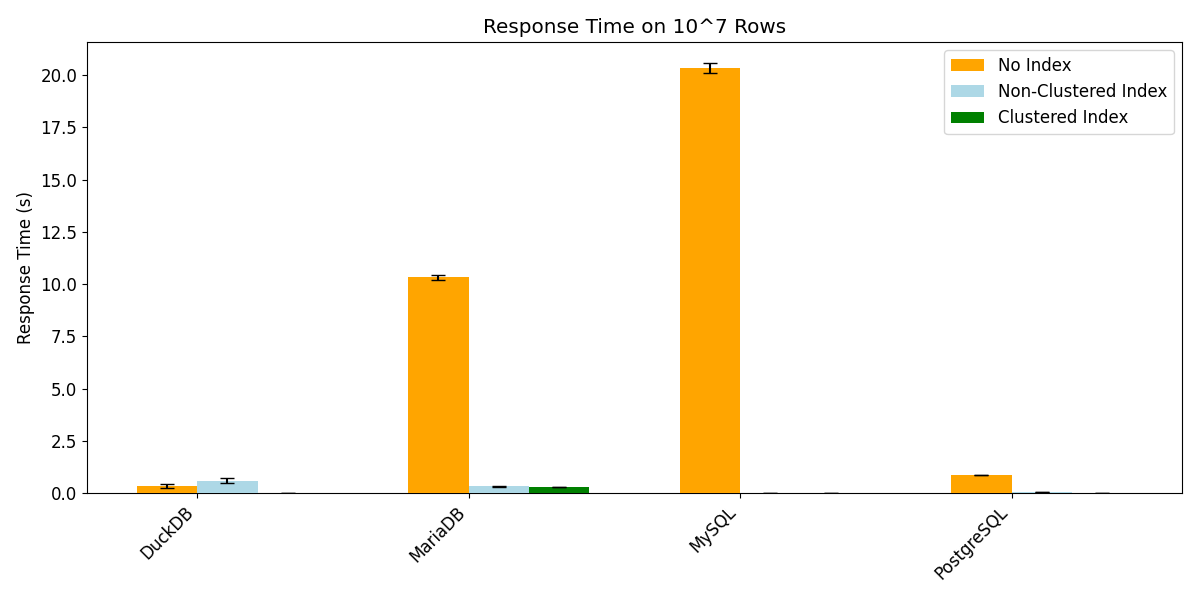}
  \caption{For range queries (retrieving 1000 rows), DuckDB does better with scanning than with a non-clustered index. (DuckDB does not support clustered indexes.)
 Clustered indexes out-performs non-clustered indexes on the rest of the DBMS on $10^7$ rows. }
  \label{fig: clustered index results for 10^7 dataset}
\end{figure}

\textbf{On the $10^7$ row dataset} (Figure~\ref{fig: clustered index results for 10^7 dataset}), indexing significantly improves range query performance for MariaDB, PostgreSQL and MySQL, with the reverse effect on DuckDB.

MySQL exhibits the most dramatic improvement: clustered and non-clustered indexes reduce response time by over 750× and 600×, respectively, compared to a full table scan. MariaDB also benefits substantially, with both index types offering speedups of over 30×, while the clustered index performs slightly better than the non-clustered index. PostgreSQL performs best with a clustered index, followed by a non-clustered index, while the no-index baseline is the slowest. DuckDB sees degraded performance with a non-clustered index, which is consistent with the fact that DuckDB is particularly strong at scans.

As shown in Table~\ref{tab:clustered-index-10e7}, indexing also enhances performance stability in most systems. MySQL, PostgreSQL and MariaDB show notably reduced standard deviations when using clustered indexes, indicating highly consistent execution. DuckDB maintains relatively stable performance across configurations, with minimal changes in variance.


\begin{table}[H]
  \centering
  \caption{Average (Standard Deviation) of Response Time for Clustered, Non-clustered, and No Index Range Query Experiments (retrieving 1000 rows) on a $10^7$-Row Dataset (in seconds)}
  \label{tab:clustered-index-10e7}
  \begin{tabular}{lccc}
    \toprule
    \textbf{System} & \textbf{No Index} & \textbf{Non-clustered} & \textbf{Clustered} \\
    \midrule
    DuckDB      & 0.28 (0.0059) & 0.51 (0.0057) & -- \\
    MariaDB     & 10 (0.13) & 0.33 (0.023)  & 0.32 (0.00070) \\
    MySQL       & 20.3 (0.23) & 0.034 (0.00074) & 0.027 (0.00052)  \\
    PostgreSQL  & 0.87 (0.012) & 0.037 (0.00074) & 0.035 (0.00079)  \\
    \bottomrule
  \end{tabular}
\end{table}

\section[\appendixname~\thesection]{Covering Index: Quantitative Experimental Results}\label{covering-appendix}
\begin{figure}[H]
  \centering
  \includegraphics[width=0.85\textwidth]{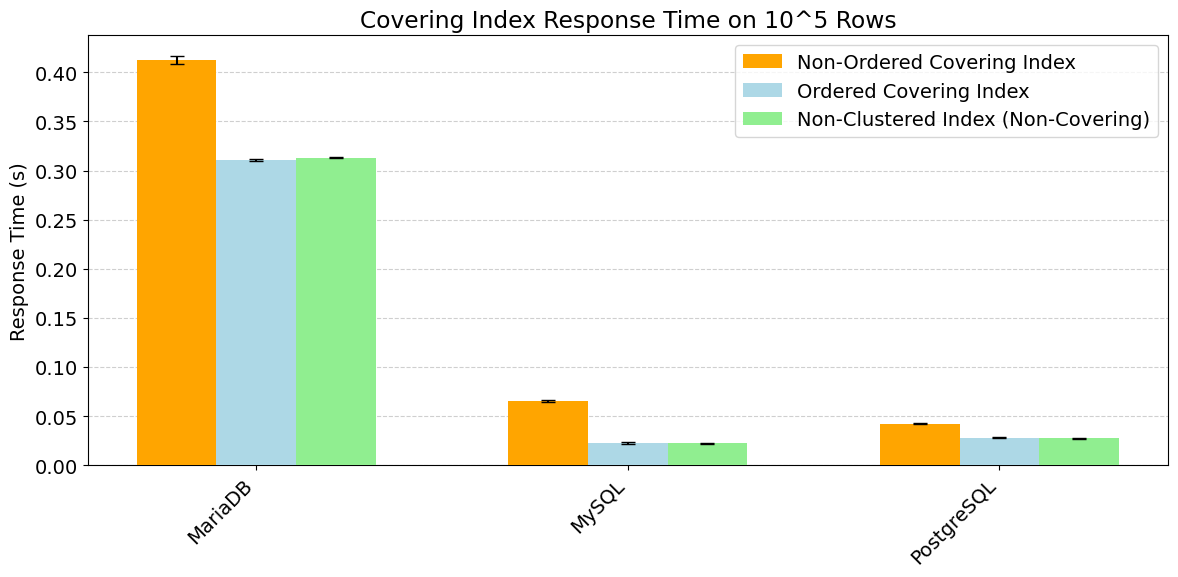}
  \caption{On the $10^5$-row dataset, the ordered covering index outperforms the unordered covering index on all systems, while the non-clustered non-covering index has nearly identical performance to the ordered covering index.}
  \label{fig:covering-index-10e5}
\end{figure}

\textbf{On the $10^5$ row dataset} (Figure~\ref{fig:covering-index-10e5} \& Table~\ref{tab:covering-clustered-10e5}), ordered covering indexes generally perform the best on MariaDB, MySQL, and PostgreSQL. However, non-clustered non-covering indexes achieve nearly the same performance as ordered covering indexes, while unordered covering indexes perform the worst overall. 

MySQL shows the most significant performance improvement (of 2.9x speedup) when using an ordered covering index compared to a non-clustered non-covering index. PostgreSQL and MariaDB also benefitx from the ordered covering index compared to non-clustered noncovering index, with response times reduced by a factor of 1.5 and 1.3, respectively. 

According to the query execution plans, DuckDB did not make use of multi-column indexes, even when we attempted to enforce their usage by setting the configuration parameters SET index\_scan\_percentage = 1 and SET index\_scan\_max\_count = 1000000. For this reason, we have omitted DuckDB from these experiments. The execution plans of MariaDB and MySQL show that, in the case of the ordered covering index, the \texttt{'key\_len = 5'} indicates that only the first column of the multi-column index \texttt{nc4(lat, ssnum, name)}, namely \texttt{lat}, is used for filtering. However, the note \texttt{Extra = 'Using index'} indicates that table access is avoided, confirming that the ordered covering index \texttt{nc4} is indeed utilized.

Across MySQL, MariaDB, and PostgreSQL, the performance of non-clustered non-covering indexes is nearly identical to that of ordered covering indexes, ranging from 0.99x (MariaDB) to 1.02x (PostgreSQL), and consistently better than that of unordered covering indexes.

In terms of stability, 
ordered covering indexes yield lower standard deviation on all systems, most notably on MySQL and MriaDB. 

\vspace{1em}

\begin{table}[H]
  \centering
  \caption{Average Response Time (Standard Deviation) for Unordered Covering, Ordered Covering and Non-clustered Index Experiments on a $10^5$-Row Dataset (in seconds).}
  \label{tab:covering-clustered-10e5}

  \begin{tabular}{lccc}
    \toprule
    \textbf{System} & \textbf{Unordered Covering} & \textbf{Ordered Covering} & \textbf{Non-clustered} \\
    \midrule
    MariaDB     & 0.41 (0.0041) & 0.31 (0.0011) & 0.31 (0.00067) \\
    MySQL       & 0.065 (0.0014) & 0.022 (0.00084) & 0.022 (0.00042) \\
    PostgreSQL  & 0.042 (0.00070) & 0.028 (0.00047) & 0.027 (0.00052) \\
    \bottomrule
  \end{tabular}
\end{table}

\begin{figure}[H]
  \centering
  \includegraphics[width=0.8\textwidth]{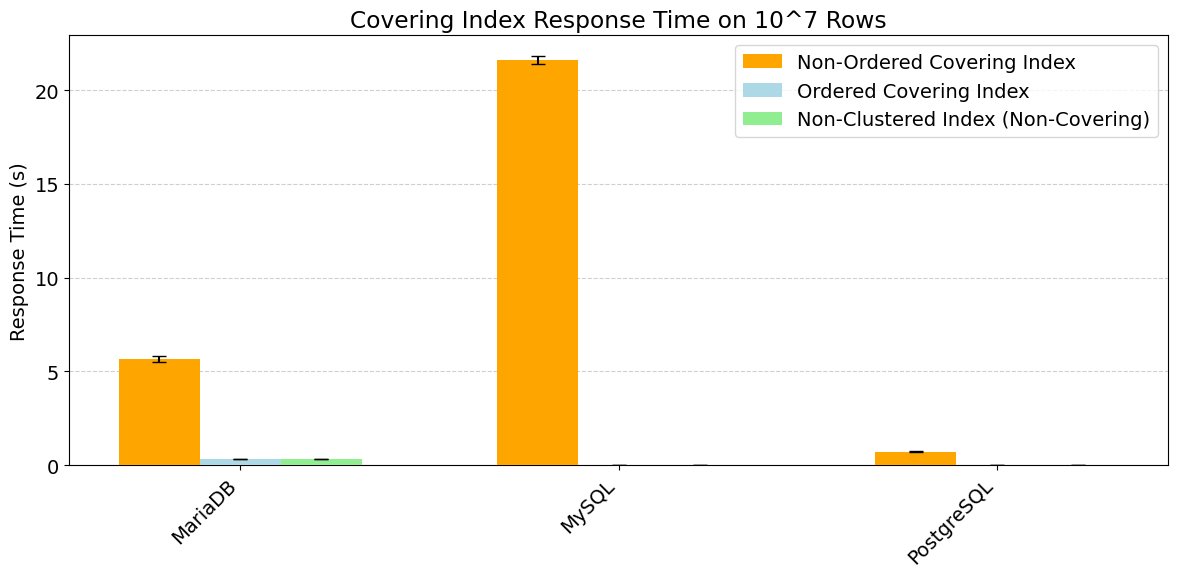}
  \caption{On the $10^7$-row dataset, the ordered covering index outperforms the unordered covering index on all systems, while the non-clustered non-covering index has nearly identical performance to the ordered covering index.}
  \label{fig: covering index results for 10^7 dataset}
\end{figure}

\textbf{On the $10^7$ row dataset} (Figure~\ref{fig: covering index results for 10^7 dataset} \& Table~\ref{tab:covering-clustered-10e7}), the ordered covering index consistently outperforms the unordered covering index across all DBMSs except DuckDB. On MySQL, the response time drops drastically from 21.63s (unordered covering index) to just 0.02s, yielding a 965.35× speedup. MariaDB and PostgreSQL also show benefits with ordered covering index, which is 18.17x and 26.13x faster than the unordered covering index. 


Ordered and non-clustered non-covering indexes perform similarly. The performance of the two cases is nearly identical, with the average response time ratio (ordered index / non-clustered non-covering index) ranging from 0.98$\times$ (MySQL) to 1.03$\times$ (PostgreSQL), and consistently outperforming the unordered covering index.

In terms of stability, ordered covering indexes and non-clustered non-covering indexes yield the lowest standard deviations on all systems. 

\begin{table}[H]
  \centering
  \caption{Average Response Time (Standard Deviation) for Unordered Covering, Ordered Covering, and Non-clustered Index Experiments on a $10^7$-Row Dataset (in seconds).}
  \label{tab:covering-clustered-10e7}

  \begin{tabular}{lccc}
    \toprule
    \textbf{System} & \textbf{Unordered Covering} & \textbf{Ordered Covering} & \textbf{Non-clustered Non-covering} \\
    \midrule
    MariaDB     & 5.7 (0.17)   & 0.31 (0.0015) & 0.31 (0.00048) \\
    MySQL       & 22 (0.21)    & 0.022 (0.00070) & 0.023 (0.00079) \\
    PostgreSQL  & 0.73 (0.014) & 0.028 (0.00032) & 0.027 (0.00057) \\
    \bottomrule
  \end{tabular}
\end{table}

\section[\appendixname~\thesection]{For which access fraction is scanning better than non-clustered index for Range query: Quantitative Experimental Results}\label{accessfraction_range-appendix}
\begin{figure}[H]
  \centering
  \includegraphics[width=0.8\textwidth]{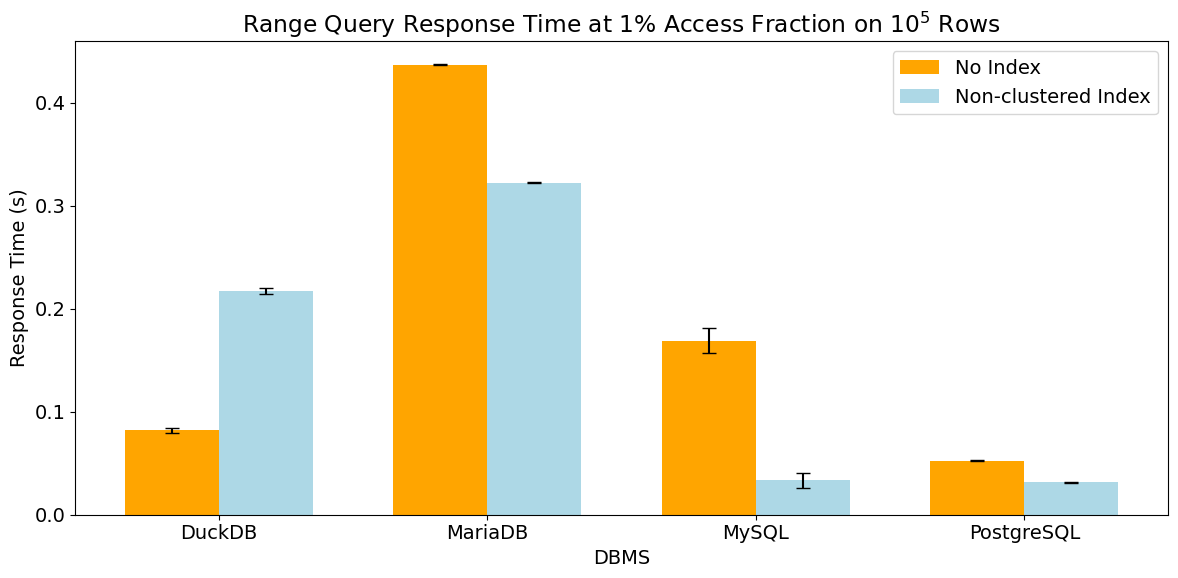}
  \caption{Impact of Non-clustered Indexes vs. Full Table Scan on Range Query Performance at 1\% access fraction on $10^5$ rows. For DuckDB, full table scan outperforms the non-clustered index. For MariaDB, MySQL and PostgreSQL, the non-clustered index yields better performance.}
  \label{fig: scanwin range perc1 10^5}
\end{figure}

\begin{figure}[H]
  \centering
  \includegraphics[width=0.8\textwidth]{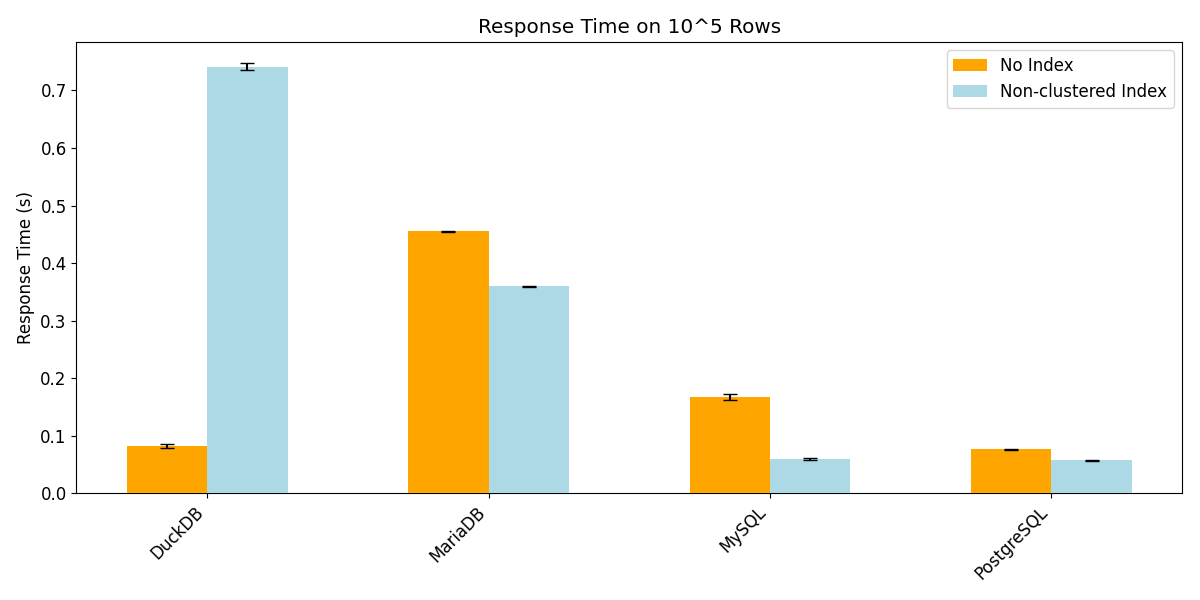}
  \caption{Impact of Non-clustered Indexes vs. Full Table Scan on Range Query Performance at 5\% access fraction on $10^5$ rows. For DuckDB, full table scan outperforms the non-clustered index. For MariaDB, MySQL and PostgreSQL, the non-clustered index yields better performance.}
  \label{fig: scanwin range perc5 10^5}
\end{figure}

\begin{figure}[H]
  \centering
  \includegraphics[width=0.8\textwidth]{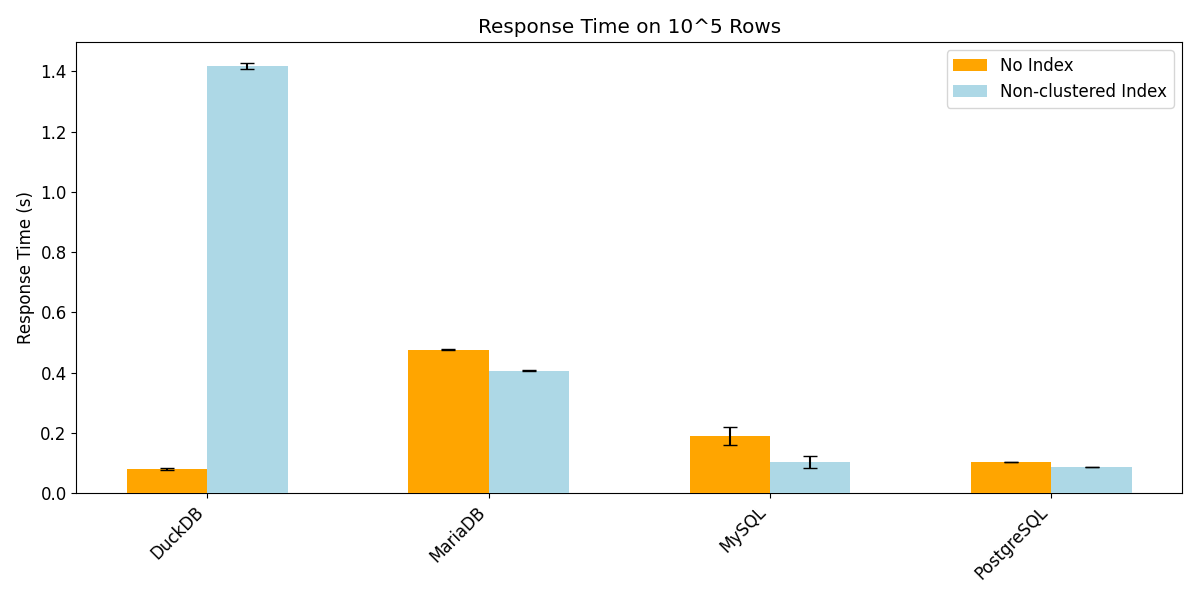}
  \caption{Impact of Non-clustered Indexes vs. Full Table Scan on Range Query Performance at 10\% access fraction on $10^5$ rows. For DuckDB, full table scan outperforms the non-clustered index. For MariaDB, MySQL and PostgreSQL, the non-clustered index yields better performance.}
  \label{fig: scanwin range perc10 10^5}
\end{figure}

\begin{figure}[H]
  \centering
  \includegraphics[width=0.8\textwidth]{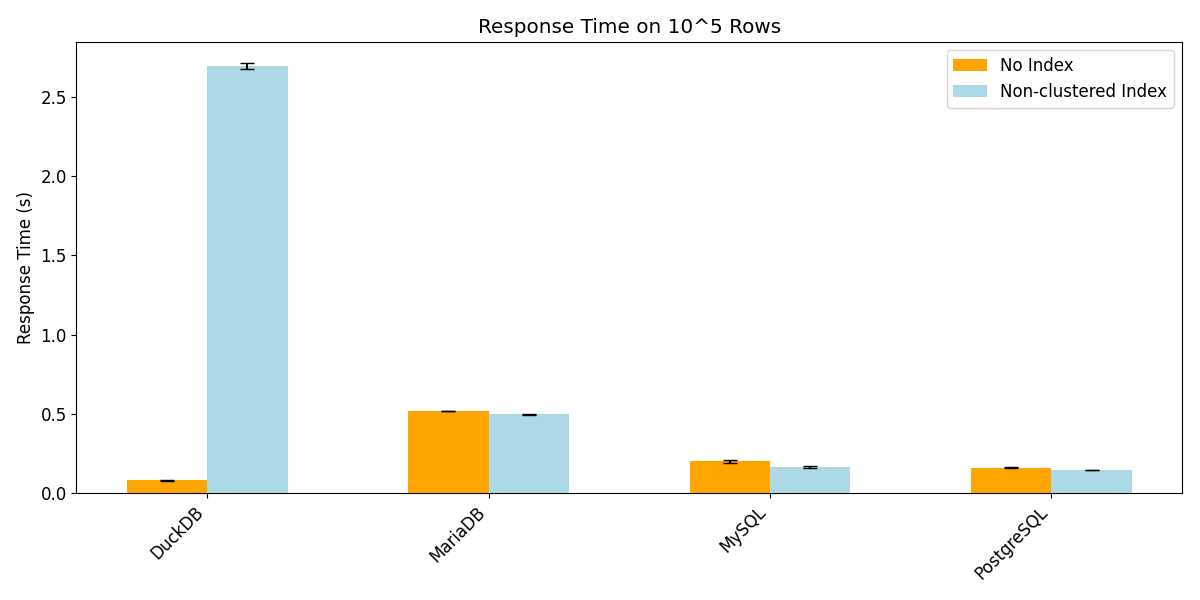}
  \caption{Impact of Non-clustered Indexes vs. Full Table Scan on Range Query Performance at 20\% access fraction on $10^5$ rows. For DuckDB, full table scan outperforms the non-clustered index. For MySQL, MariaDB and PostgreSQL, the non-clustered index yields better performance.}
  \label{fig: scanwin range perc20 10^5}
\end{figure}

\begin{figure}[H]
  \centering
  \includegraphics[width=0.8\textwidth]{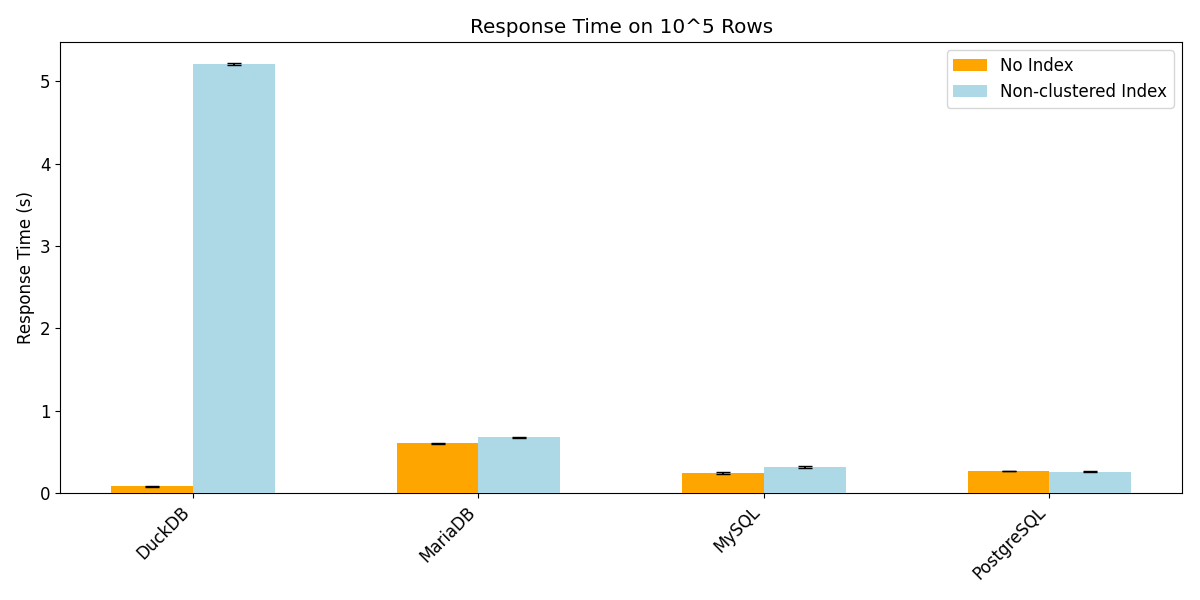}
  \caption{Impact of Non-clustered Indexes vs. Full Table Scan on Range Query Performance at 40\% access fraction on $10^5$ rows. For DuckDB, MariaDB and MySQL, full table scan outperforms the non-clustered index. For PostgreSQL, the non-clustered index and no index perform equally well.}
  \label{fig: scanwin range perc40 10^5}
\end{figure}

\textbf{On the \boldmath$10^5$ row dataset} (Figure~\ref{fig: scanwin range perc1 10^5}–\ref{fig: scanwin range perc40 10^5} and Table~\ref{tab:range-scan-vs-index-10e5}), performance varies across systems when comparing full table scan and non-clustered indexing for range queries, with system-specific behaviors shifting across access fractions.

At a low access fraction of 1\% (Figure~\ref{fig: scanwin range perc1 10^5}), indexing yields the largest gains on MySQL, where it outperforms scan by over 5×. PostgreSQL also benefits from indexing, with a  1.7× speedup. MariaDB shows a moderate 1.4× improvement with indexing. DuckDB, however, is an outlier: indexing is more than 2.6× slower than scanning at this level, indicating that full scan is significantly more efficient on small ranges.

At an access fraction of 5\% (Figure~\ref{fig: scanwin range perc5 10^5}), MySQL maintains a strong indexing advantage (approximately 2.8× faster), while MariaDB still favors indexing with a 1.3× speedup. PostgreSQL benefits from indexing by around 1.35×. On DuckDB, scan  enjoys an increasing edge (of 9×) over indexing.

As the access fraction increases to 10\% (Figure~\ref{fig: scanwin range perc10 10^5}), MySQL continues to favor indexing, albeit with a reduced margin (roughly 1.8× faster). MariaDB sees a consistent yet shrinking advantage from indexing, now around 1.2×. PostgreSQL still favors indexing, with an advantage of 1.15x. On DuckDB, scan  performs better than indexing (by a factor of 17.6×).

At 20\% access (Figure~\ref{fig: scanwin range perc20 10^5}), indexing remains beneficial for MySQL (about 1.2× faster), but the performance gap continues to shrink. On MariaDB, the difference between index and scan nearly disappears, with scan performing just slightly better. PostgreSQL still benefits modestly from indexing, while DuckDB continues to execute full table scans significantly faster than non-clustered index lookups.

At 40\% (Figure~\ref{fig: scanwin range perc40 10^5}), the trends shift. MySQL, PostgreSQL and MariaDB now perform slightly better with full scan, showing that the benefit of indexing diminishes for larger access ranges.  DuckDB consistently maintains its scan advantage..

In terms of response time variability, indexing generally leads to lower or comparable standard deviation in most systems. Refer to Table~\ref{tab:range-scan-vs-index-10e5} for more detailed data.

\begin{longtable}{llcc}
\caption{Response Time (Standard Deviation) of Range Queries Using Scan vs.\ Non-Clustered Index under Varying Access Fractions on a $10^5$-Row Dataset}
\label{tab:range-scan-vs-index-10e5} \\

\toprule
\textbf{Access Fraction} & \textbf{System} & \textbf{Non-Clustered Index} & \textbf{No Index} \\
\midrule
\endfirsthead

\toprule
\textbf{Access Fraction} & \textbf{System} & \textbf{Non-Clustered Index} & \textbf{No Index} \\
\midrule
\endhead

\endfoot
\bottomrule
\endlastfoot

\multirow{4}{*}{1\% ($10^3$ rows)} 
    & DuckDB     & 0.22 (0.0030) & 0.082 (0.0024) \\
    & MariaDB    & 0.32 (0.00082) & 0.44 (0.00063) \\
    & MySQL      & 0.034 (0.0074) & 0.17 (0.012) \\
    & PostgreSQL & 0.031 (0.00052) & 0.053 (0.00053) \\

\midrule
\multirow{4}{*}{5\% ($5 \times 10^3$ rows)} 
    & DuckDB     & 0.74 (0.0059) & 0.082 (0.0035) \\
    & MariaDB    & 0.36 (0.00084) & 0.46 (0.0013) \\
    & MySQL      & 0.060 (0.0012) & 0.17 (0.0051) \\
    & PostgreSQL & 0.057 (0.00082) & 0.076 (0.00099) \\

\midrule
\multirow{4}{*}{10\% ($10^4$ rows)} 
    & DuckDB     & 1.4 (0.0095) & 0.081 (0.0021) \\
    & MariaDB    & 0.41 (0.0013) & 0.48 (0.0013) \\
    & MySQL      & 0.10 (0.020) & 0.19 (0.029) \\
    & PostgreSQL & 0.087 (0.0010) & 0.10 (0.00063) \\

\midrule
\multirow{4}{*}{20\% ($2 \times 10^4$ rows)} 
    & DuckDB     & 2.7 (0.017) & 0.082 (0.0025) \\
    & MariaDB    & 0.50 (0.0024) & 0.52 (0.0021) \\
    & MySQL      & 0.17 (0.0047) & 0.20 (0.0076) \\
    & PostgreSQL & 0.15 (0.0013) & 0.16 (0.00094) \\

\midrule
\multirow{4}{*}{40\% ($4 \times 10^4$ rows)} 
    & DuckDB     & 5.2 (0.0094) & 0.084 (0.0022) \\
    & MariaDB    & 0.68 (0.0038) & 0.60 (0.0018) \\
    & MySQL      & 0.31 (0.010) & 0.25 (0.011) \\
    & PostgreSQL & 0.26 (0.0019) & 0.27 (0.0011) \\

\end{longtable}

\begin{figure}[H]
  \centering
  \includegraphics[width=0.8\textwidth]{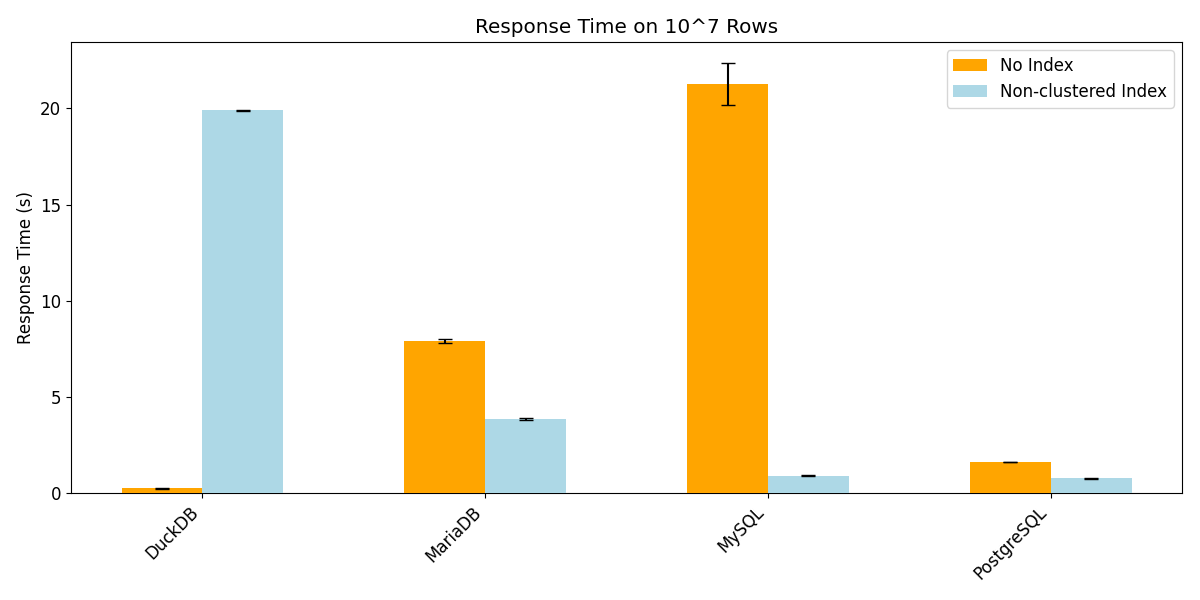}
   \caption{Impact of Non-clustered Indexes vs. Full Table Scan on Range Query Performance at 1\% access fraction on $10^7$ rows. For DuckDB, full table scan outperforms the non-clustered index. For MariaDB, MySQL and PostgreSQL, the non-clustered index yields better performance.}
  \label{fig: scanwin range perc1 10^7}
\end{figure}

\begin{figure}[H]
  \centering
  \includegraphics[width=0.8\textwidth]{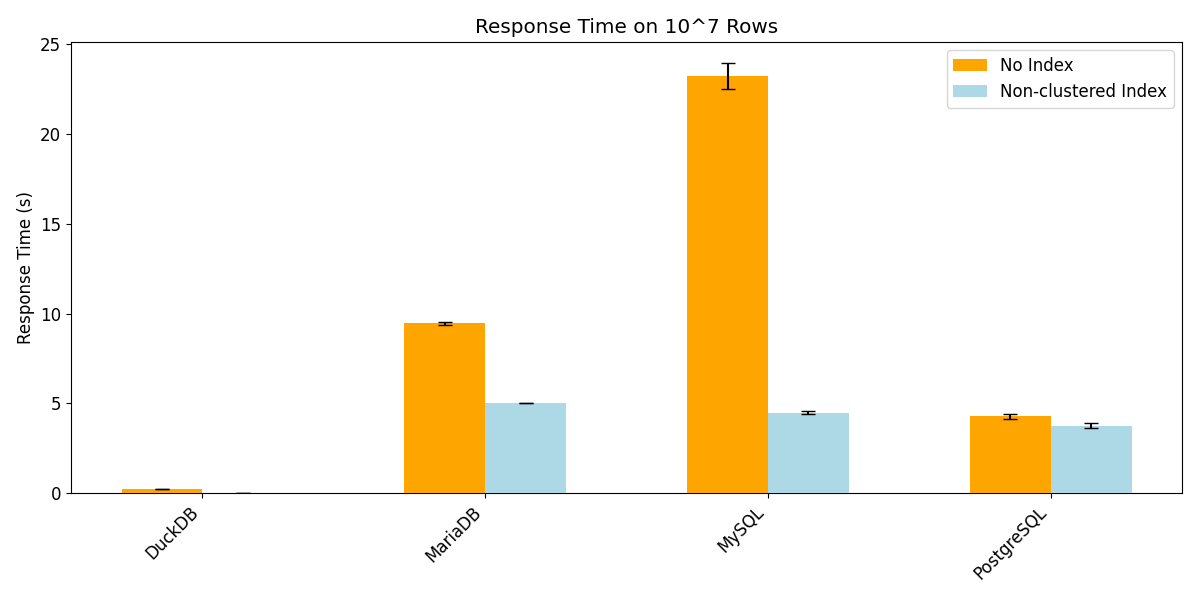}
   \caption{Impact of Non-clustered Indexes vs. Full Table Scan on Range Query Performance at 5\% access fraction on $10^7$ rows. For DuckDB, full table scan outperforms the non-clustered index. The non-clustered index exceeded the time budget and is not plotted. For MariaDB, MySQL and PostgreSQL, the non-clustered index yields marginally better performance.}
  \label{fig: scanwin range perc5 10^7}
\end{figure}

\begin{figure}[H]
  \centering
  \includegraphics[width=0.8\textwidth]{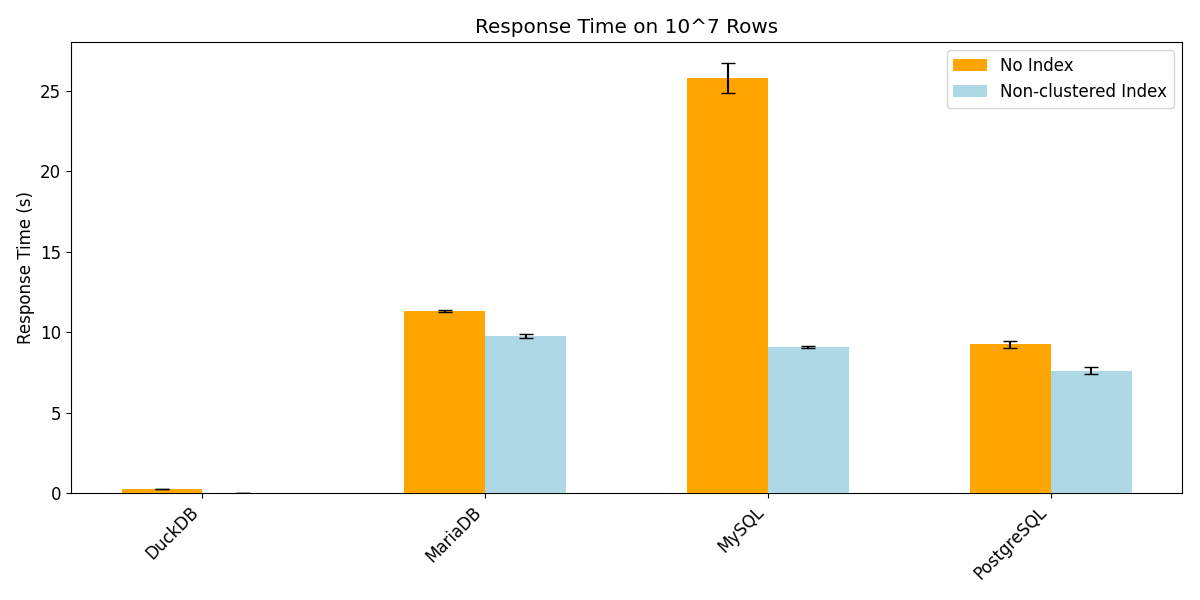}
   \caption{Impact of Non-clustered Indexes vs. Full Table Scan on Range Query Performance at 10\% access fraction on $10^7$ rows. For DuckDB, full table scan outperforms the non-clustered index. The non-clustered index exceeded the time budget and is not plotted. For MariaDB  and PostgreSQL, the non-clustered index yields marginally better performance. MySQL enjoys much better performance from non-clustered indexes.}
  \label{fig: scanwin range perc10 10^7}
\end{figure}

\begin{figure}[H]
  \centering
  \includegraphics[width=0.8\textwidth]{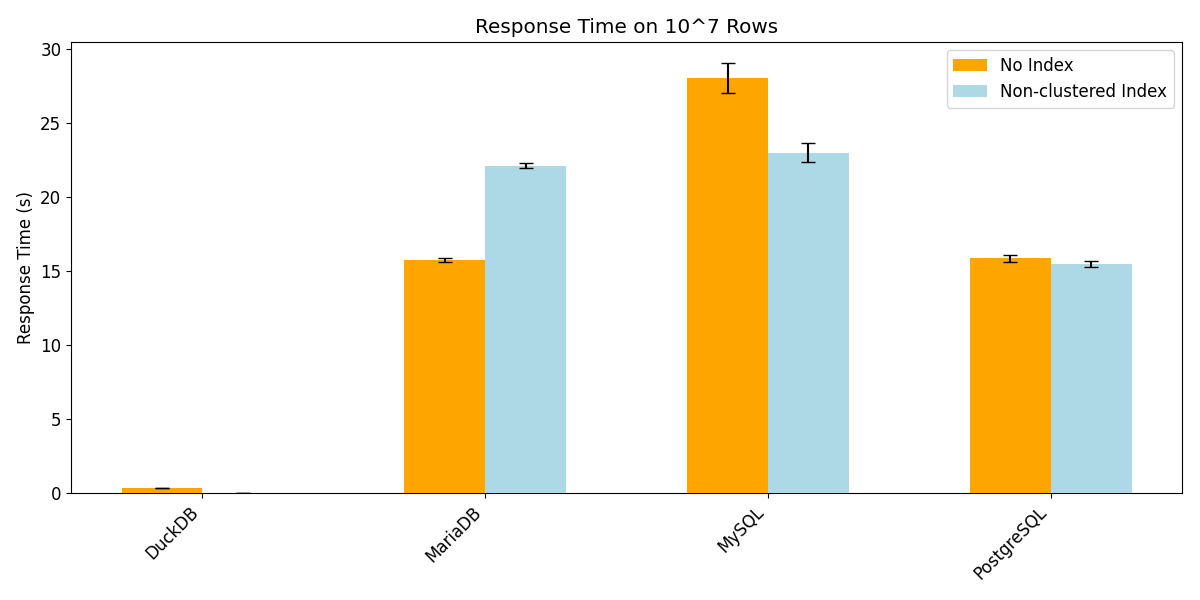}
   \caption{Impact of Non-clustered Indexes vs. Full Table Scan on Range Query Performance at 20\% access fraction on $10^7$ rows. For DuckDB and MariaDB full table scan outperforms the non-clustered index. The non-clustered index of DuckDB exceeded the time budget and is not plotted. For MySQL and PostgreSQL, the non-clustered index yields better performance.}
  \label{fig: scanwin range perc20 10^7}
\end{figure}

\begin{figure}[H]
  \centering
  \includegraphics[width=0.8\textwidth]{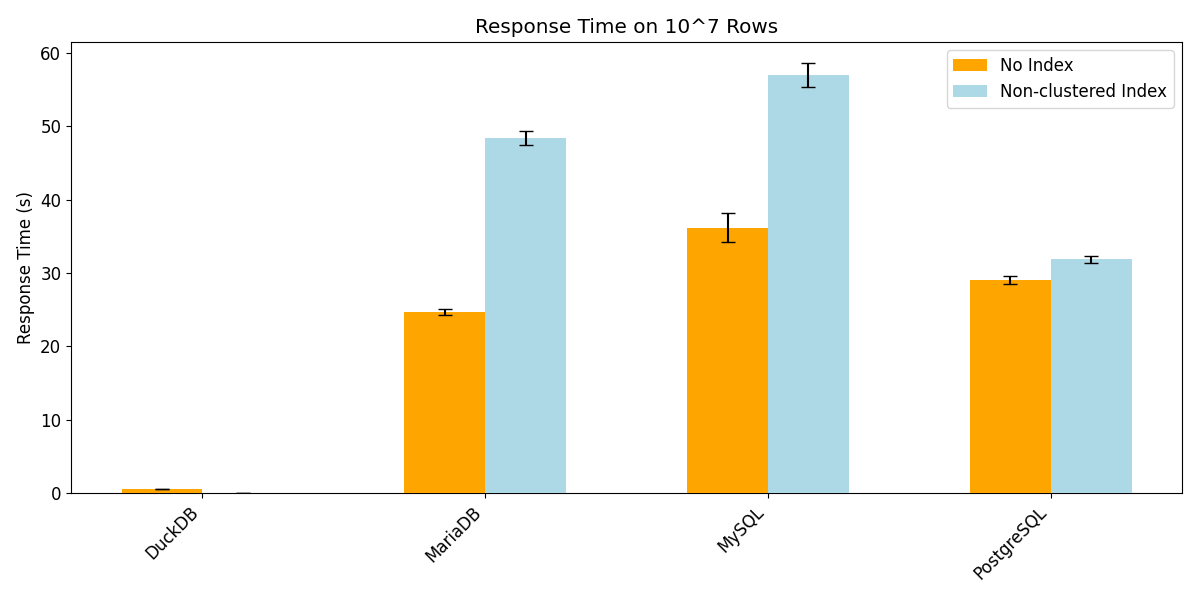}
   \caption{Impact of Non-clustered Indexes vs. Full Table Scan on Range Query Performance at 40\% access fraction on $10^7$ rows. For DuckDB, MariaDB, MySQL and PostgreSQL, full table scan outperforms the non-clustered index. The non-clustered index of DuckDB exceeded the time budget and is not plotted.}
  \label{fig: scanwin range perc40 10^7}
\end{figure}

\textbf{On the \boldmath$10^7$ row dataset} (Figure~\ref{fig: scanwin range perc1 10^7}–\ref{fig: scanwin range perc40 10^7}), scan performance of all the DBMSs generally surpasses that of non-clustered indexes as the access fraction increases.

At the 1\% access fraction (Figure~\ref{fig: scanwin range perc1 10^7}), indexing provides substantial benefits in most systems. MySQL shows the most dramatic improvement, where the index outperforms scan by over 20×. PostgreSQL and MariaDB also see significant gains, with indexing being approximately 2.1x and 2× faster than scan, respectively. On DuckDB, scanning performs better by a significant margin (around 77.4×).

At a 5\% access fraction (Figure~\ref{fig: scanwin range perc5 10^7}), the advantages of indexing remain consistent. MySQL continues to benefit with a roughly 5× speedup with indexing. PostgreSQL shows only a slight advantage from index-based access, with indexing performing about 1.16× faster than scanning. MariaDB maintains a 2× advantage for indexing. On DuckDB, indexing performs markedly better(0.24s), while the average response time of scanning exceeds 90s.

At a 10\% access fraction (Figure~\ref{fig: scanwin range perc10 10^7}), indexing continues to dominate on MySQL and PostgreSQL. MySQL's advantage narrows to around 3×, while PostgreSQL still shows a 1.2× speedup with indexing. On MariaDB, the performance gap between indexing and scanning shrinks significantly, with scan only slightly slower than index. On DuckDB, scan again performs significantly better than indexing.

At a 20\% access fraction (Figure~\ref{fig: scanwin range perc20 10^7}), on MariaDB, scan  performs better than index by a factor of 1.4. For MySQL, indexing remains faster but only by about 1.2×, showing diminishing returns. On PostgreSQL, indexing   outperforms scan by a very small margin, with response time remaining stable. DuckDB still noticeably favors scan.

At 40\% access fraction (Figure~\ref{fig: scanwin range perc40 10^7}), scan becomes preferable for PostgreSQL, MariaDB and MySQL. On MariaDB, scan is more than 1.9× faster than indexing. MySQL also experiences a reversal, with scan outperforming index by around 1.6× and PostgreSQL shows a slight degradation of 3 seconds with the index. DuckDB continues its trend of significantly better scan performance. 

Across all access fractions, the standard deviation of response time reveals distinct patterns of stability between index-based and scan-based access methods. DuckDB demonstrates slightly lower variability when using full table scan. MariaDB benefits from greater stability with indexing at lower access fractions (1\%–10\%), but this advantage diminishes at higher fractions, where scan becomes equally or slightly more stable. MySQL consistently exhibits significantly lower variability under index access, particularly at smaller access fractions where scan performance is highly unstable. Similarly, PostgreSQL shows much more stable response times with indexing across all fractions, with the gap in variability widening as access fraction increases.

\begin{longtable}{llcc}
\caption{Response Time (Standard Deviation) of Range Queries Using Scan vs.\ Non-Clustered Index under Varying Access Fractions on a $10^7$-Row Dataset}
\label{tab:range-scan-vs-index-10e7} \\

\toprule
\textbf{Access Fraction} & \textbf{System} & \textbf{Non-Clustered Index} & \textbf{No Index} \\
\midrule
\endfirsthead

\toprule
\textbf{Access Fraction} & \textbf{System} & \textbf{Non-Clustered Index} & \textbf{No Index} \\
\midrule
\endhead

\endfoot
\bottomrule
\endlastfoot

\multirow{4}{*}{1\% ($10^5$ rows)} 
    & DuckDB     & 20 (0.038)    & 0.26 (0.011) \\
    & MariaDB    & 3.9 (0.072)   & 7.9 (0.095)  \\
    & MySQL      & 0.92 (0.011)  & 21 (1.1)     \\
    & PostgreSQL & 0.77 (0.011)  & 1.6 (0.0067) \\

\midrule
\multirow{4}{*}{5\% ($5 \times 10^5$ rows)} 
    & DuckDB     & --            & 0.24 (0.0080) \\
    & MariaDB    & 5.0 (0.0078)  & 9.5 (0.081)   \\
    & MySQL      & 4.5 (0.085)   & 23 (0.72)     \\
    & PostgreSQL & 3.7 (0.13)    & 4.3 (0.12)    \\

\midrule
\multirow{4}{*}{10\% ($10^6$ rows)} 
    & DuckDB     & --            & 0.24 (0.0097) \\
    & MariaDB    & 9.8 (0.14)    & 11 (0.078)    \\
    & MySQL      & 9.1 (0.074)   & 26 (0.91)     \\
    & PostgreSQL & 7.6 (0.19)    & 9.2 (0.19)    \\

\midrule
\multirow{4}{*}{20\% ($2 \times 10^6$ rows)} 
    & DuckDB     & --            & 0.36 (0.012)  \\
    & MariaDB    & 22 (0.17)     & 16 (0.14)     \\
    & MySQL      & 23 (0.63)     & 28 (1.0)      \\
    & PostgreSQL & 15.4 (0.21)   & 15.8 (0.23)   \\

\midrule
\multirow{4}{*}{40\% ($4 \times 10^6$ rows)} 
    & DuckDB     & --            & 0.61 (0.023)  \\
    & MariaDB    & 48 (0.93)     & 25 (0.42)     \\
    & MySQL      & 57 (1.6)      & 36 (2.0)      \\
    & PostgreSQL & 32 (0.47)     & 29 (0.55)     \\

\end{longtable}

\section[\appendixname~\thesection]{For which access fraction is scanning better than non-clustered index for Multipoint Query: Quantitative Experimental Results}\label{accessfraction_multipoint-appendix}
\begin{figure}[H]
  \centering
  \includegraphics[width=0.8\textwidth]{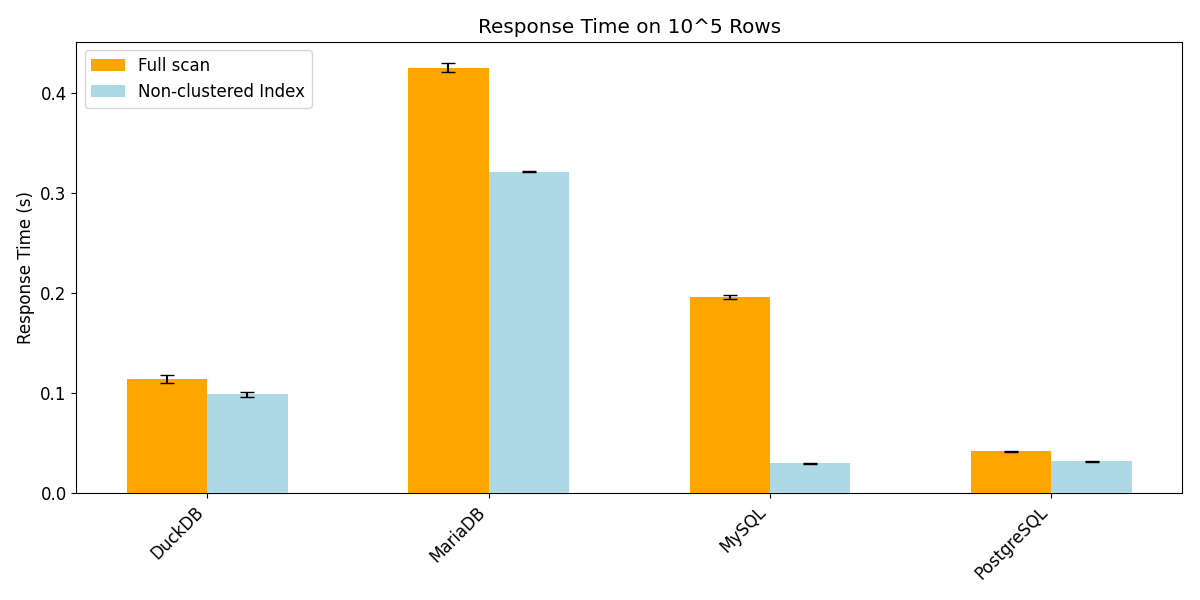}
  \caption{Impact of Non-clustered Indexes vs. Full Table Scan on Multipoint Query Performance of access fraction 1\% ($10^3$ rows) on $10^5$ rows. All engines benefit from non-clustered index.}
  \label{fig: scanwin_multip_1p_5}
\end{figure}

\begin{figure}[H]
  \centering
  \includegraphics[width=0.8\textwidth]{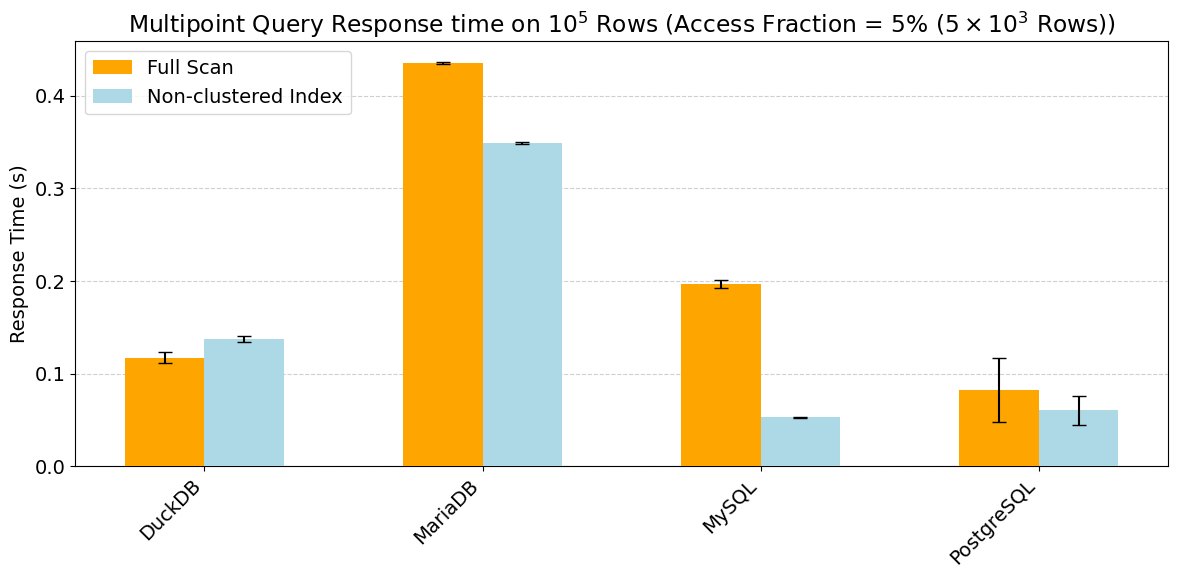}
  \caption{Impact of Non-clustered Indexes vs. Full Table Scan on Multipoint Query Performance of access fraction 5\% ($5 \times 10^3$ rows) on $10^5$ rows. All the database systems except DuckDB, benefit from non-clustered index.}
  \label{fig: scanwin_multip_5p_5}
\end{figure}

\begin{figure}[H]
  \centering
  \includegraphics[width=0.8\textwidth]{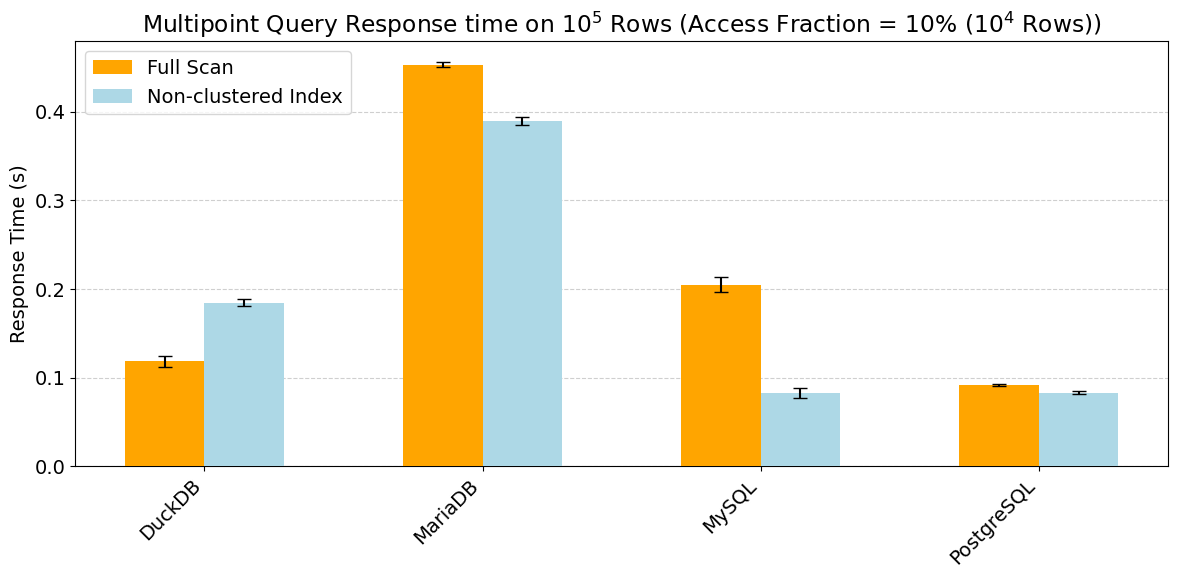}
  \caption{Impact of Non-clustered Indexes vs. Full Table Scan on Multipoint Query Performance of access fraction 10\% ($10^4$ rows) on $10^5$ rows. All systems except DuckDB show improved performance with non-clustered indexes, PostgreSQL exhibits only marginal differences between clustered indexing and full scan. }
  \label{fig: scanwin_multip_10p_5}
\end{figure}

\begin{figure}[H]
  \centering
  \includegraphics[width=0.8\textwidth]{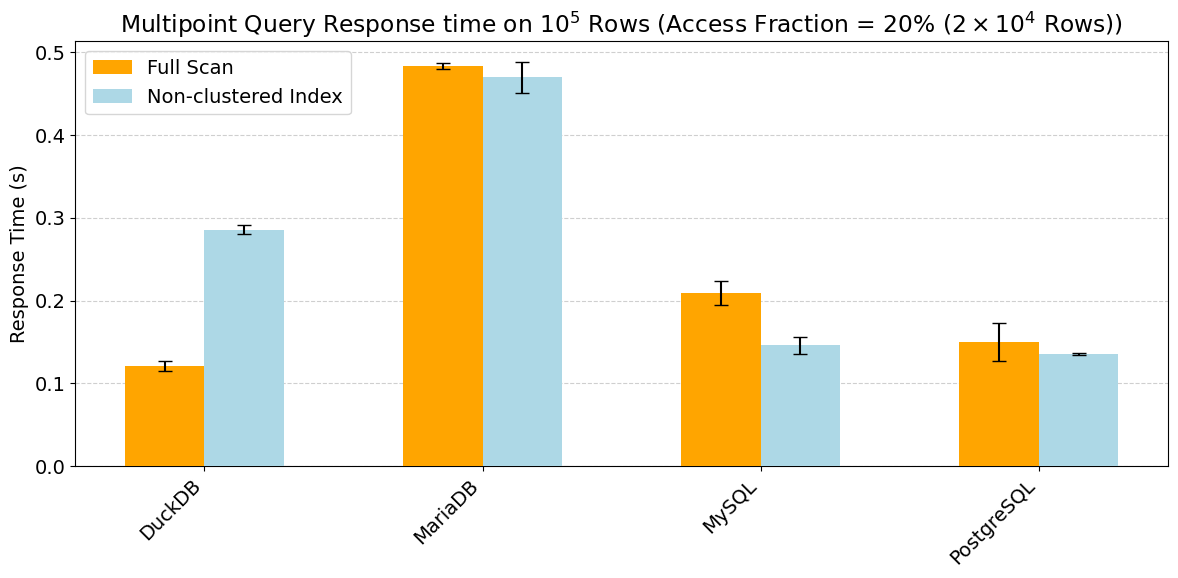}
  \caption{Impact of Non-clustered Indexes vs. Full Table Scan on Multipoint Query Performance of access fraction 20\% ($2 \times 10^4$ rows) on $10^5$ rows. MariaDB, MySQL and PostgreSQL benefit from non-clustered index. In contrast, scan performs better on DuckDB. MariaDB and PostgreSQL show little difference in performance between scan and indexing.}
  \label{fig: scanwin_multip_20p_5}
\end{figure}

\textbf{On the $10^5$ row dataset,}
Figure~\ref{fig: scanwin_multip_1p_5}–\ref{fig: scanwin_multip_20p_5} and Table~\ref{tab:scan-vs-index-10e5} illustrate how the performance of multipoint queries using full table scan and non-clustered indexing varies across DBMS as the access fraction increases.

At low access fractions (e.g., 1\%) (Figure~\ref{fig: scanwin_multip_1p_5}), indexing provides a clear advantage on most systems. MySQL exhibits the most significant improvement, with indexing over 7× faster than scan. MariaDB, DuckDB and PostgreSQL also benefit from indexing, showing speedups of approximately 1.3×, 1.2× and 1.3×, respectively.

As the access fraction increases to 5\% ($5 \times 10^3$ rows) (Figure~\ref{fig: scanwin_multip_5p_5}), indexing remains beneficial for MariaDB, MySQL, and PostgreSQL. MySQL retains a strong 3.7× improvement, while MariaDB and PostgreSQL achieve speedups of about 1.2× and 1.4×, respectively. However, on DuckDB, indexing underperforms, resulting in a 1.2× slowdown relative to scan. 

At an access fraction of 10\% ($10^4$ rows) (Figure~\ref{fig: scanwin_multip_10p_5}), MySQL continues to gain from indexing with a 2.5× speedup. MariaDB also maintains a slight advantage for indexing, outperforming scan by roughly 1.2×. PostgreSQL sees near parity, with indexing just slightly better than scan (1.1×). DuckDB's performance with indexing deteriorates further, being 1.6× slower than scan.

At 20\% (returns $2 \times 10^4$ rows) (Figure~\ref{fig: scanwin_multip_20p_5}), MySQL is the only system with meaningful gains from indexing (1.4× faster). MariaDB and PostgreSQL show minimal differences, but indexing still performs slightly better for both systems. DuckDB again favors scan, with indexing incurring a 2.4× slowdown.

In terms of response time variability, indexing generally leads to lower or comparable standard deviation in most systems. MySQL consistently shows reduced variance under indexing across all access fractions. DuckDB also maintains slightly better stability with indexing, especially at low access fractions. MariaDB’s variability under indexing remains low at small access fractions but increases notably at 20\%. On PostgreSQL, indexing incurs high variability at 1\% access, but becomes more stable than scan as the access fraction grows.

\begin{table}[H]
  \centering
  \caption{Response Time (Standard Deviation) for Scan vs. Non-Clustered Index for Multipoint queries under Varying Access Fractions on a $10^5$-Row Dataset}
  \label{tab:scan-vs-index-10e5}
  \begin{tabular}{llcc}
    \toprule
    \textbf{Access Fraction} & \textbf{System} & \textbf{Non-Clustered Index} & \textbf{No Index} \\
    \midrule
    \multirow{4}{*}{1\% ($10^3$ rows)} 
        & DuckDB       & 0.097 (0.0019) & 0.11 (0.0041) \\
        & MariaDB      & 0.32 (0.00048) & 0.43 (0.0043) \\
        & MySQL        & 0.028 (0.00042)& 0.20 (0.0028) \\
        & PostgreSQL   & 0.032 (0.00032)  & 0.042 (0.00056) \\
    \midrule
    \multirow{4}{*}{5\% ($5 \times 10^3$ rows)} 
        & DuckDB       & 0.14 (0.0030)  & 0.12 (0.0056) \\
        & MariaDB      & 0.35 (0.0010)  & 0.44 (0.0011) \\
        & MySQL        & 0.053 (0.00063)& 0.20 (0.0045) \\
        & PostgreSQL   & 0.061 (0.016)  & 0.083 (0.034) \\
    \midrule
    \multirow{4}{*}{10\% ($10^4$ rows)} 
        & DuckDB       & 0.18 (0.0038)  & 0.12 (0.0059) \\
        & MariaDB      & 0.39 (0.0045)  & 0.45 (0.0033) \\
        & MySQL        & 0.083 (0.0058) & 0.21 (0.0082) \\
        & PostgreSQL   & 0.083 (0.0016) & 0.092 (0.00079) \\
    \midrule
    \multirow{4}{*}{20\% ($2 \times 10^4$ rows)} 
        & DuckDB       & 0.29 (0.0055)  & 0.12 (0.0060) \\
        & MariaDB      & 0.47 (0.019)   & 0.48 (0.0037) \\
        & MySQL        & 0.15 (0.010)   & 0.21 (0.015) \\
        & PostgreSQL   & 0.13 (0.0016)  & 0.15 (0.023) \\
    \bottomrule
  \end{tabular}
\end{table}

\begin{figure}[H]
  \centering
  \includegraphics[width=0.8\textwidth]{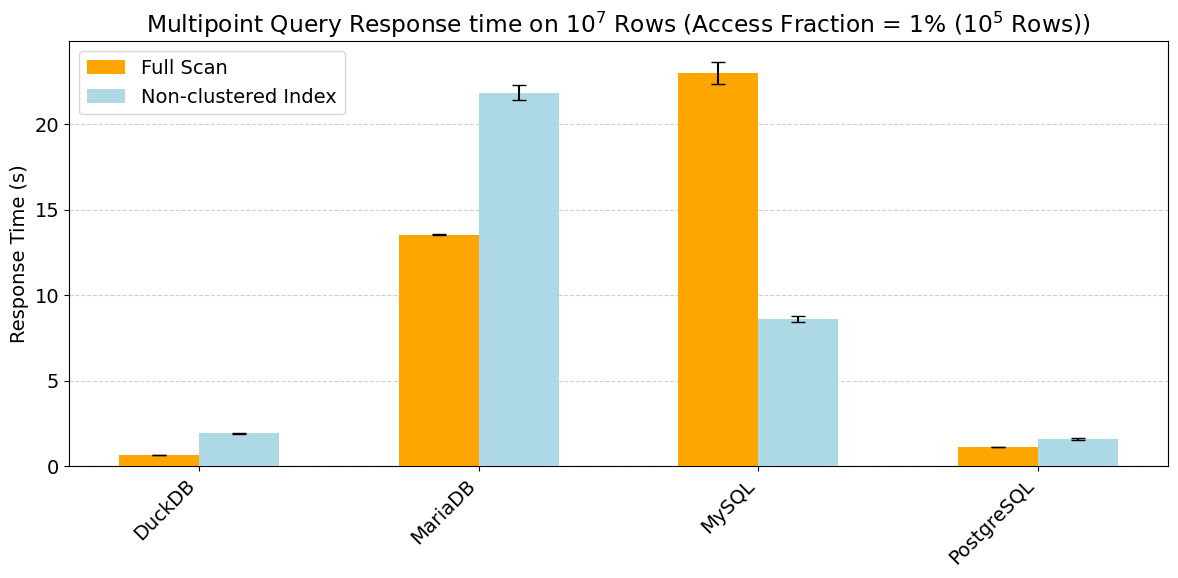}
  \caption{Impact of Non-clustered Indexes vs. Full Table Scan on Multipoint Query Performance of access fraction 1\% ($10^5$) on $10^7$ rows. MySQL benefits from a non-clustered index. In contrast, scan performs better on DuckDB, MariaDB and PostgreSQL.}
  \label{fig: scanwin_multip_1p_7}
\end{figure}

\begin{figure}[H]
  \centering
  \includegraphics[width=0.8\textwidth]{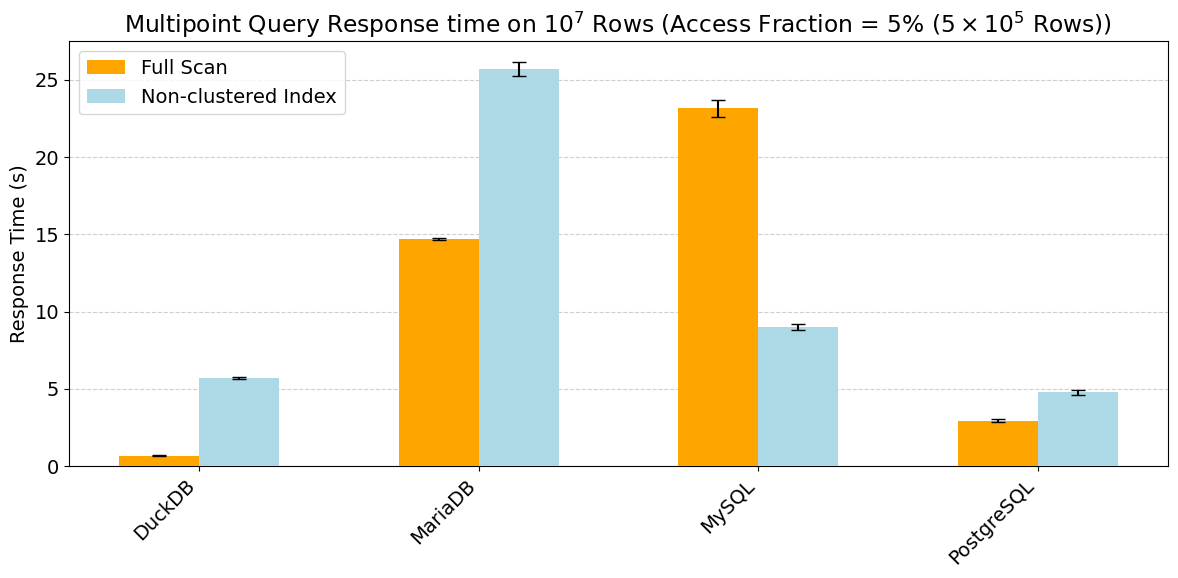}
  \caption{Impact of Non-clustered Indexes vs. Full Table Scan on Multipoint Query Performance of access fraction 5\% ($5 \times 10^5$) on $10^7$ rows. MySQL benefits from a non-clustered index. In contrast, scan performs better on DuckDB, MariaDB and PostgreSQL.}
  \label{fig: scanwin_multip_5p_7}
\end{figure}

\begin{figure}[H]
  \centering
  \includegraphics[width=0.8\textwidth]{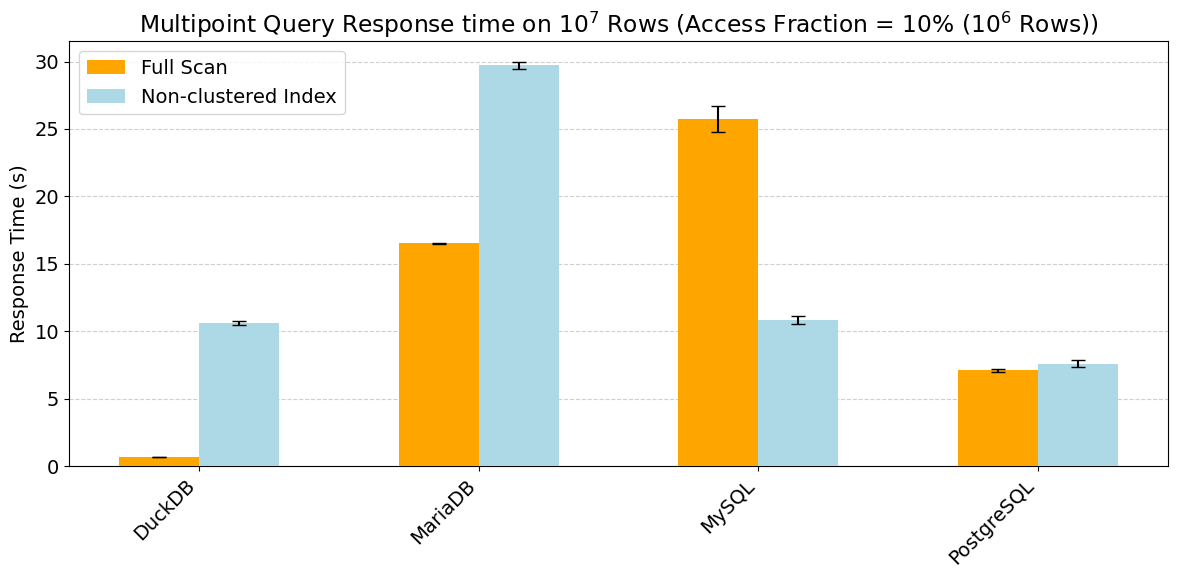}
  \caption{Impact of Non-clustered Indexes vs. Full Table Scan on Multipoint Query Performance of access fraction 10\% ($10^6$) on $10^7$ rows. MySQL benefit from non-clustered index. In contrast, scan performs better on MariaDB and DuckDB. PostgreSQL shows little difference in performance between scan and indexing.}
  \label{fig: scanwin_multip_10p_7}
\end{figure}

\begin{figure}[H]
  \centering
  \includegraphics[width=0.8\textwidth]{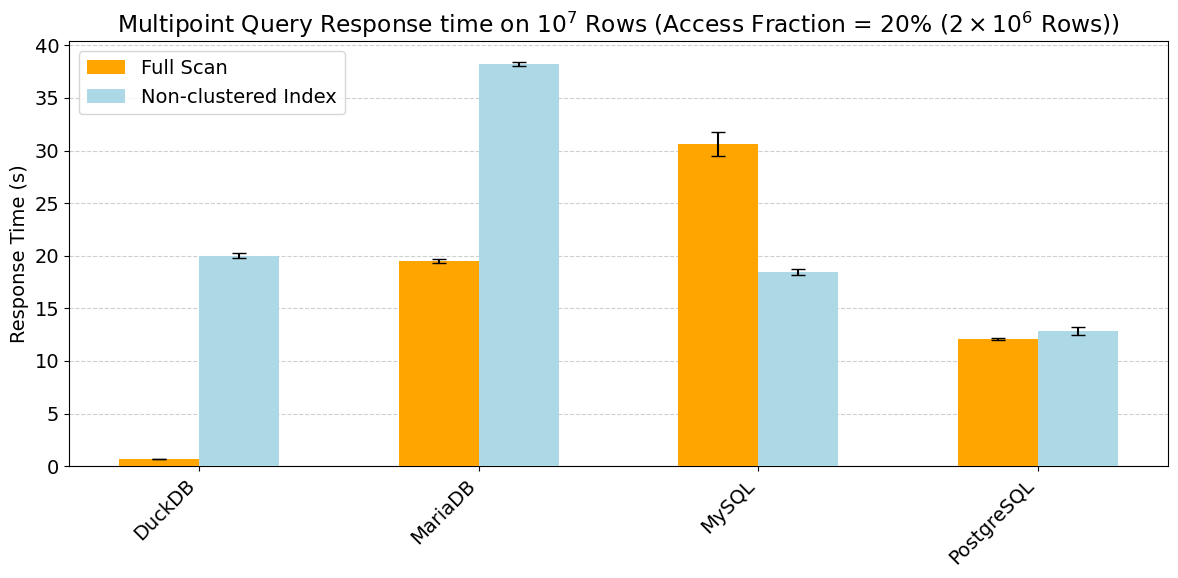}
  \caption{Impact of Non-clustered Indexes vs. Full Table Scan on Multipoint Query Performance of access fraction 20\% ($2 \times 10^6$) on $10^7$ rows.  MySQL benefits from a non-clustered index. In contrast, scan performs better on DuckDB and MariaDB. PostgreSQL shows little difference in performance between scan and indexing.}
  \label{fig: scanwin_multip_20p_7}
\end{figure}

\textbf{On the $10^7$ row dataset,} Figure~\ref{fig: scanwin_multip_1p_7}–\ref{fig: scanwin_multip_20p_7} and Table~\ref{tab:scan-vs-index-10e7} illustrate how the effectiveness of non-clustered indexing versus full table scan changes with access fraction for multipoint queries.

At an access fraction of 1\% ($10^5$ rows out of $10^7$) (Figure~\ref{fig: scanwin_multip_1p_7}), only MySQL shows a clear benefit from indexing, with a speedup of approximately 2.7×. In contrast, indexing performs worse on all other systems: MariaDB becomes over 1.6× slower with indexing, PostgreSQL suffers a 1.4× degradation, and DuckDB shows 2.9x slowdown.

At 5\% ($5 \times 10^5$ rows) (Figure~\ref{fig: scanwin_multip_5p_7}), the trend remains. MySQL continues to benefit from indexing (around 2.6× faster), while MariaDB and PostgreSQL see even greater slowdowns with indexing, around 1.7× and 1.6×, respectively. DuckDB still shows a meaningful difference, with scan outperforming index.

At 10\% ($10^6$ rows) (Figure~\ref{fig: scanwin_multip_10p_7}), MySQL maintains a strong advantage from indexing (about 2.4× speedup), while the other systems remain scan-favorable or nearly neutral. MariaDB and PostgreSQL still show degraded performance with indexing, although the difference shrinks slightly. The response time increased by a factor of 1.8 and 1.03 compared to scanning, respectively. DuckDB continuously show an advantage with scanning.

At 20\% ($2 \times 10^6$ rows) (Figure~\ref{fig: scanwin_multip_20p_7}), MySQL still benefits from indexing (roughly 1.7× faster), while MariaDB and PostgreSQL continue to favor scan. The response time increases by a factor of 2.0 and 1.03 compared to scanning, respectively. For DuckDB, scan performs 30x faster than indexing.

In terms of response time variability, MySQL consistently shows reduced standard deviation under indexing, especially at lower access fractions. For example, at 1\%, indexing improves stability by over 3×. DuckDB shows a general trend of slightly more stable performance with scanning. In contrast, MariaDB displays increased variance with indexing across all access levels, especially at lower access fractions, where standard deviation under index grows nearly 10×. PostgreSQL shows mixed behavior, scan is slightly more stable at 1\% and 10\%, but indexing catches up or slightly outperforms at other fractions. 

\begin{table}[H]
  \centering
  \caption{Response Time (Standard Deviation) for Scan vs. Non-Clustered Index under Varying Access Fractions on a $10^7$-Row Dataset}
  \label{tab:scan-vs-index-10e7}
  \begin{tabular}{llcc}
    \toprule
    \textbf{Access Fraction} & \textbf{System} & \textbf{Non-Clustered Index} & \textbf{No Index} \\
    \midrule
    \multirow{4}{*}{1\% ($10^5$ rows)} 
        & DuckDB       & 1.9 (0.015) & 0.66 (0.019) \\
        & MariaDB      & 22 (0.45)   & 14 (0.046)   \\
        & MySQL        & 8.6 (0.17)  & 23 (0.64)    \\
        & PostgreSQL   & 1.6 (0.083) & 1.1 (0.0092) \\
    \midrule
    \multirow{4}{*}{5\% ($5 \times 10^5$ rows)} 
        & DuckDB       & 5.7 (0.065) & 0.69 (0.011) \\
        & MariaDB      & 26 (0.47)   & 15 (0.051)   \\
        & MySQL        & 9.0 (0.18)  & 23 (0.53)    \\
        & PostgreSQL   & 4.8 (0.18)  & 3.0 (0.080)  \\
    \midrule
    \multirow{4}{*}{10\% ($10^6$ rows)} 
        & DuckDB       & 11 (0.14)   & 0.68 (0.012) \\
        & MariaDB      & 30 (0.26)   & 17 (0.065)   \\
        & MySQL        & 11 (0.30)   & 26 (0.97)    \\
        & PostgreSQL   & 7.6 (0.28)  & 7.1 (0.12)   \\
    \midrule
    \multirow{4}{*}{20\% ($2 \times 10^6$ rows)} 
        & DuckDB       & 20 (0.23)   & 0.66 (0.0097) \\
        & MariaDB      & 38 (0.23)   & 20 (0.18)     \\
        & MySQL        & 18 (0.28)   & 31 (1.2)      \\
        & PostgreSQL   & 13 (0.39)   & 12 (0.091)    \\
    \bottomrule
  \end{tabular}
\end{table}

\section[\appendixname~\thesection]{Index on Small Table: Quantitative Experimental Results for Update Query}\label{smalltable-appendix}
\begin{figure}[H]
  \centering
  \includegraphics[width=0.8\textwidth]{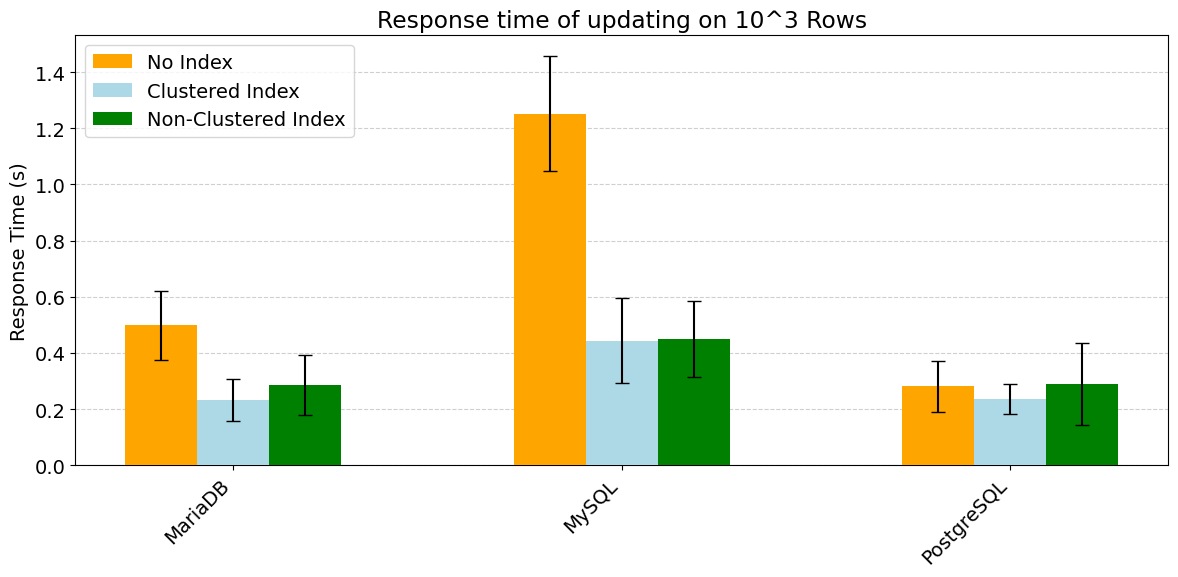}
  \caption{Clustered indexes and non-clustered indexes improve concurrent updating performance on $10^3$ rows for MySQL, MariaDB, and PostgreSQL compared to having no index. Clustered indexes outperform non-clustered ones on all these systems. DuckDB does not support concurrent processing.}
  \label{fig:smalltable_update3}
\end{figure}

\textbf{On the $10^3$ row dataset} (Figure~\ref{fig:smalltable_update3} and Table~\ref{tab:smalltable_update3}), the experiment shows that enabling indexes significantly improves concurrent \texttt{UPDATE} performance across all tested systems. Compared to the no-index case, MariaDB achieved a 2.14× speedup with clustered indexing, while MySQL and PostgreSQL improved by 2.82× and 1.19×, respectively.

When using non-clustered indexes, performance gains remain notable. MySQL’s response time improved by a factor of 2.79 over the no-index baseline, which is almost equivalent to its clustered index performance. PostgreSQL’s non-clustered index performed slightly worse than no index (0.97×), indicating that the overhead of traversing the index  outweighed the selectivity benefit for this workload. MariaDB saw a 1.75× speedup with non-clustered indexing, slightly below its 2.14× gain with clustered indexing.

In the comparison of index types, non-clustered indexing was slightly slower than clustered indexing on all systems. Specifically, on MariaDB, the non-clustered index took 1.22× longer than the clustered index. On MySQL, the performance difference was minimal, with the non-clustered index being 1.01× slower. PostgreSQL also showed a modest gap, where the non-clustered index was 1.23× slower than the clustered index.

Regarding performance variability (Table~\ref{tab:smalltable_update3}), indexes reduced the standard deviation of response times compared to the no-index case for both MariaDB and PostgreSQL. MariaDB’s clustered index achieved the lowest standard deviation (0.0749s), while MySQL and PostgreSQL also saw reduced variance under clustered indexing. 

\begin{table}[H]
  \centering
  \caption{Average Response Time (Standard Deviation) for Concurrent \texttt{UPDATE} With Clustered Index, With Non-clustered Index or Without Indexes on a $10^3$-Row Table.}
  \label{tab:smalltable_update3}
  \begin{tabular}{lccc}
    \toprule
    \textbf{System} & \textbf{No Index} & \textbf{Clustered Index} & \textbf{Non-Clustered Index} \\
    \midrule
    DuckDB      & --              & --              & --              \\
    MariaDB     & 0.50 (0.12)     & 0.23 (0.075)    & 0.28 (0.11)     \\
    MySQL       & 1.3 (0.21)      & 0.44 (0.15)     & 0.45 (0.13)     \\
    PostgreSQL  & 0.28 (0.090)    & 0.24 (0.054)    & 0.29 (0.15)     \\
    \bottomrule
  \end{tabular}
\end{table}

\begin{figure}[H]
  \centering
  \includegraphics[width=0.8\textwidth]{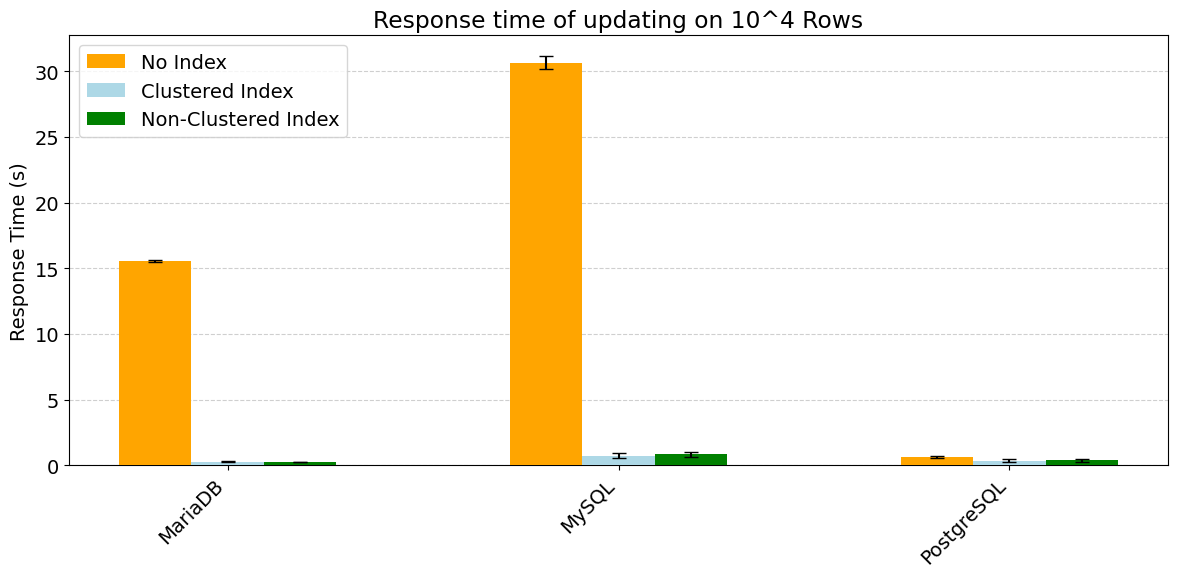}
  \caption{Clustered indexes and non-clustered indexes improve concurrent updating performance on $10^4$ rows for MySQL, MariaDB, and PostgreSQL compared to having no index. Clustered index performs the best on MySQL and PostgreSQL. For MariaDB, non-clustered index performs slightly better. DuckDB does not support concurrent processing.}
  \label{fig:smalltable_update4}
\end{figure}

\textbf{On the $10^4$ row dataset} (Figure~\ref{fig:smalltable_update4} and Table~\ref{tab:smalltable_update4}), enabling clustered indexes significantly improved concurrent \texttt{UPDATE} performance across all tested systems. On MariaDB, clustered indexing reduced the average response time by a factor of 61.6 compared to the no-index case. MySQL showed a 42.3× improvement, and PostgreSQL also benefited with a 1.80× speedup.

Non-clustered indexing also yielded strong performance improvements in certain systems. On MariaDB, the non-clustered index achieved a 64.7× speedup over the no-index configuration—slightly outperforming the clustered index, with a performance ratio of 0.95×. In contrast, on PostgreSQL, the non-clustered index improved performance by 1.62× over the no-index case and was only 1.11× slower than the clustered index. MySQL’s non-clustered index, while still offering a 37.6× speedup over no index, performed slightly worse than its clustered counterpart (1.13× slower).

These results suggest that both clustered and non-clustered indexes can dramatically improve update performance, particularly in systems like MariaDB and MySQL. However, the relative advantage of non-clustered indexing varies across DBMSs and does not always outperform clustered indexing.

In terms of performance stability (Table~\ref{tab:smalltable_update4}), response time variability was substantially higher in the absence of indexes. Enabling either type of index led to improved stability. Notably, on MariaDB and MySQL, the non-clustered index achieved the lowest standard deviation of response time, indicating the most consistent performance. On PostgreSQL, clustered indexing provided slightly better stability than non-clustered indexing, but both significantly outperformed the no-index baseline.

\begin{table}[H]
  \centering
  \caption{Average Response Time (Standard Deviation) for concurrent \texttt{UPDATE} With or Without Indexes on $10^4$ rows.}
  \label{tab:smalltable_update4}
  \begin{tabular}{lccc}
    \toprule
    \textbf{System} & \textbf{No Index} & \textbf{Clustered Index} & \textbf{Non-Clustered Index} \\
    \midrule
    DuckDB      & --            & --            & --            \\
    MariaDB     & 16 (0.085)    & 0.25 (0.034)  & 0.24 (0.033)  \\
    MySQL       & 31 (0.51)     & 0.73 (0.18)   & 0.82 (0.17)   \\
    PostgreSQL  & 0.61 (0.083)  & 0.34 (0.099)  & 0.38 (0.12)   \\
    \bottomrule
  \end{tabular}
\end{table}

\section[\appendixname~\thesection]{Index on Small Table: Quantitative Experimental Results for Search Query}\label{smalltable-appendix}
\begin{figure}[H]
  \centering
  \includegraphics[width=0.8\textwidth]{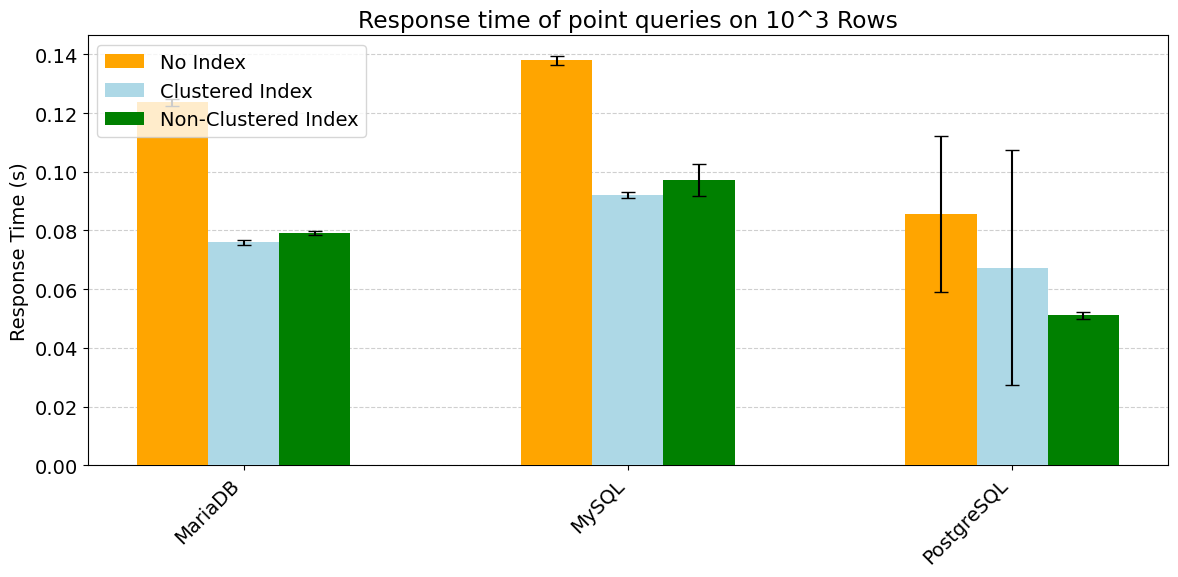}
  \caption{Enabling indexes improve concurrent point queries performance on $10^3$ rows for MySQL, MariaDB, and PostgreSQL compared to having no index. The non-clustered index performs better than clustered index on PostgreSQL, slightly worse than clustered index on PostgreSQL and MariaDB. DuckDB does not support concurrent processing.}
  \label{fig:smalltable_search3}
\end{figure}

\textbf{On the $10^3$ row dataset}  (Figure~\ref{fig:smalltable_search3} and Table~\ref{tab:smallsearch3}), enabling clustered indexes significantly improved concurrent point query performance compared to no index across all systems. MariaDB and MySQL saw speedups of 1.63× and 1.50×, respectively, while PostgreSQL improved by 1.27×.

Non-clustered indexes also brought notable gains over the no-index baseline: PostgreSQL achieved a 1.67× speedup, the highest among the three systems, followed by MariaDB (1.56×) and MySQL (1.42×). Overall, all systems benefited from indexing, with varying degrees of effectiveness depending on the DBMS and index type.

In terms of index comparison, performance was nearly identical between clustered and non-clustered indexes on MariaDB and MySQL, with differences within 5\%. On PostgreSQL, the non-clustered index was faster than the clustered index by a factor of 1.32×.

From a stability standpoint (Table~\ref{tab:smallsearch3}), non-clustered indexes consistently showed the lowest response time variance across all systems. For example, standard deviation dropped to 0.00070s on MariaDB and 0.00119s on PostgreSQL, highlighting their advantage for predictable, low-latency point queries under concurrent workloads.

\begin{table}[H]
  \centering
  \caption{Average Response Time (Standard Deviation) for concurrent \texttt{SELECT} With or Without Indexes on $10^3$ rows.}
  \label{tab:smallsearch3}
  \begin{tabular}{lccc}
    \toprule
    \textbf{System} & \textbf{No Index} & \textbf{Clustered} & \textbf{Non-Clustered} \\
    \midrule
    DuckDB      & --              & --              & --              \\
    MariaDB     & 0.12 (0.0012)   & 0.076 (0.00075) & 0.079 (0.00070) \\
    MySQL       & 0.14 (0.0016)   & 0.092 (0.00098) & 0.097 (0.0055)  \\
    PostgreSQL  & 0.086 (0.027)   & 0.067 (0.040)   & 0.051 (0.0012)  \\
    \bottomrule
  \end{tabular}
\end{table}

\begin{figure}[H]
  \centering
  \includegraphics[width=0.8\textwidth]{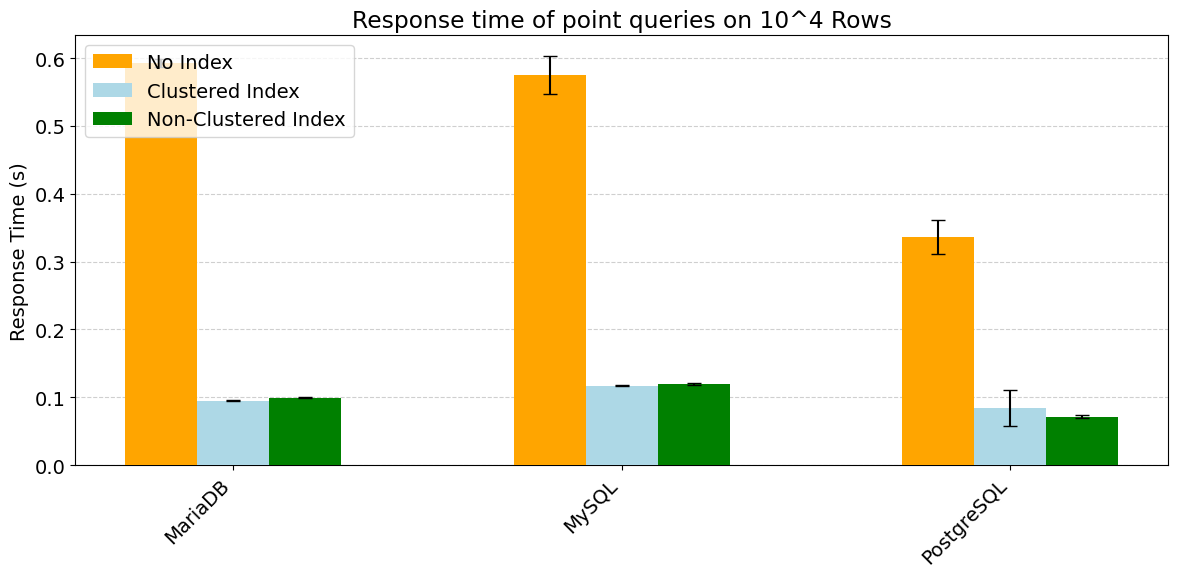}
  \caption{Clustered indexes and non-clustered indexes improve concurrent point queries performance on $10^4$ rows for MySQL, MariaDB, and PostgreSQL compared to having no index. Nonclustered indexes performs nearly as well as clustered ones on MariaDB and MySQL, and worse than clustered ones on PostgreSQL. DuckDB does not support concurrent processing.}
  \label{fig:smalltable_search4}
\end{figure}

\textbf{On the $10^4$ row dataset} (Figure~\ref{fig:smalltable_search4} and Table~\ref{tab:smallsearch4}), enabling indexes significantly improved concurrent \texttt{SELECT} performance across all tested systems. MariaDB showed the largest improvement, with response time reduced by a factor of 6.2× using clustered indexing and 6.0× with non-clustered indexing. MySQL followed with speedups of 4.9× and 4.8×, respectively. PostgreSQL also benefited, achieving a 4.0× speedup with clustered indexing and 4.7× with non-clustered indexing.

Comparing index types, performance was similar between clustered and non-clustered indexes on MariaDB and MySQL, with differences within 5\%. On PostgreSQL, the non-clustered index was slightly faster than the clustered index, completing queries 1.2× faster.

In terms of performance stability (Table~\ref{tab:smallsearch4}), both index types greatly reduced response time variance compared to the no-index baseline. Non-clustered indexes yielded the most consistent performance on MariaDB and PostgreSQL, while on MySQL, clustered indexing produced slightly more stable results.

\begin{table}[H]
  \centering
  \caption{Average Response Time (Standard Deviation) for concurrent \texttt{SELECT} With or Without Indexes on $10^4$ rows.}
  \label{tab:smallsearch4}
  \begin{tabular}{lccc}
    \toprule
    \textbf{System} & \textbf{No Index} & \textbf{Clustered} & \textbf{Non-Clustered} \\
    \midrule
    DuckDB      & --              & --              & --              \\
    MariaDB     & 0.59 (0.0060)   & 0.095 (0.00065) & 0.099 (0.00060) \\
    MySQL       & 0.58 (0.028)    & 0.12 (0.00071)  & 0.12 (0.0011)   \\
    PostgreSQL  & 0.34 (0.025)    & 0.084 (0.027)   & 0.072 (0.0018)  \\
    \bottomrule
  \end{tabular}
\end{table}

\section[\appendixname~\thesection]{Looping in Stored Procedures: Quantitative Experimental Results}\label{looping-appendix}
\begin{figure}[H]
  \centering
  \includegraphics[width=0.8\textwidth]{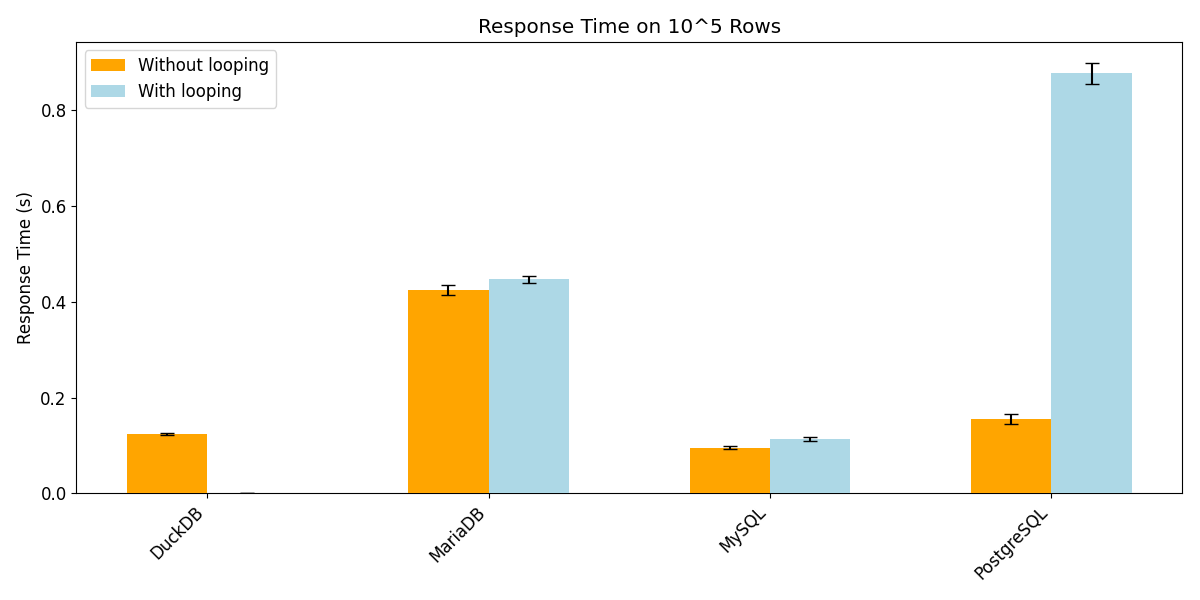}
  \caption{All engines with range queries (without looping) perform better than their looping counterparts. DuckDB  does not support loops.}
  \label{fig: Looping results for 10^5 dataset}
\end{figure}

\textbf{On the $10^5$ row dataset} (Figure \ref{fig: Looping results for 10^5 dataset}), replacing the range scan with a point-query loop caused a slowdown on every engine.
\emph{PostgreSQL} experienced the most significant slowdown due to loops, running $5.6\times$ slower.  \emph{MySQL} incurred a moderate penalty, executing $1.19\times$ slower than the range query. \emph{MariaDB} was the least impacted, with only a mild performance drop of $1.05\times$. \emph{DuckDB} was excluded from the looping comparison because it does not support loop-based point queries; however, its range scan performance is included for reference, allowing comparison with the other engines' non-looping performance. 
See Table~\ref{tab:looping-10e5} for the full set of average response times and standard deviations.

\begin{table}[H]
  \centering
  \caption{Average Response Time along with their Standard Deiations for Looping Experiment on a $10^5$ Row Dataset (in seconds)}
  \label{tab:looping-10e5}

  \label{tab:looping-response-10e5}
  \begin{tabular}{lccc}
    \toprule
    \textbf{System} & \textbf{No Looping} & \textbf{Looping} \\
    \midrule
    MariaDB     & 0.43 (0.0099) & 0.45 (0.0072) \\
    MySQL       & 0.095 (0.011) & 0.11 (0.0212) \\
    PostgreSQL  & 0.16 (0.0036) & 0.88 (0.0037) \\
    DuckDB      & 0.12 (0.0018) & N.A \\
    \bottomrule
  \end{tabular}
\end{table}

\begin{figure}[H]
  \centering
  \includegraphics[width=0.8\textwidth]{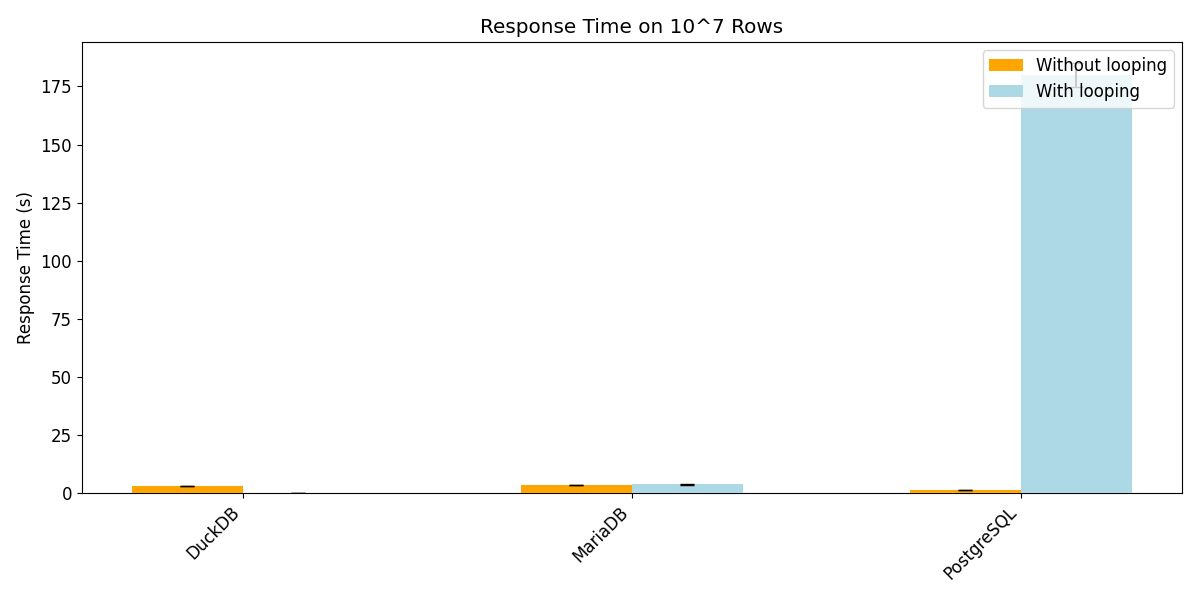}
  \caption{On $10^7$ rows, PostgreSQL performs significantly better without the loop and MariaDB maintains a modest edge without the loop. MySQL could not complete the experiment.}
  \label{fig: Looping results for 10^7 dataset}
\end{figure}

\textbf{On the $10^7$ row dataset} (Figure \ref{fig: Looping results for 10^7 dataset}), the performance penalty from looping became even more pronounced compared to on the $10^5$ dataset. \emph{PostgreSQL} experienced severe degradation, with mean response times increasing by a factor of $146$, rendering looping impractical for large tables. In contrast, \emph{MariaDB} incurred only a modest but consistent slowdown ($1.03\times$). \emph{MySQL} did not create the table at this size within our 10 minute time limit. \emph{DuckDB} was excluded from the looping comparison because it does not support loop-based point queries within stored procedures. DuckDB's range scan performance is included for reference, allowing comparison with the other engines' non-looping performance.

\begin{table}[H]
  \centering
  \caption{Average Response Times along with their Standard Deviations for Looping Experiment on a $10^7$ Row Dataset (in seconds)}
  \label{tab:looping-10e7}

  \label{tab:looping-response-10e7}
  \begin{tabular}{lccc}
    \toprule
    \textbf{System} & \textbf{No Looping} & \textbf{Looping} \\
    \midrule
    MariaDB     & 3.66 (0.066) & 3.77 (0.041) \\
    PostgreSQL  & 1.23 (0.029) & 179.85 (5.08) \\
    DuckDB      & 3.15 (0.007) & N.A \\
    \bottomrule
  \end{tabular}
\end{table}

\section[\appendixname~\thesection]{Looping with Cursors: Quantitative Experimental Results}\label{cursor-appendix}
\begin{enumerate}

\item
\textbf{Experiment Goal:} \\[3pt]
The goal of the experiment is to quantify the performance impact of replacing a single set-based query with a cursor-driven loop that fetches each row individually. While cursors allow precise control by iteratively performing parameterized lookups, this method introduces per-iteration statement execution overhead, potentially offsetting any advantages from avoiding unnecessary scans.  \\
\item 
\textbf{Experiment Setup:}
We conduct experiments on two data scales: $10^5$ and $10^7$ rows. The same dataset schema, \texttt{employee(ssnum, name, lat, longitude, hundreds1, hundreds2)} (described in Section~\ref{subsec:benchmark employee}), is used throughout the experiments, with database settings adjusted accordingly.

The queries are listed in Listings \ref{lst:cursor} and \ref{lst:no-cursor}. Both return the rows with \texttt{l\_partkey} $<$ 200.

\begin{lstlisting}[language=SQL,caption={Query to fetch and process `employee` rows without cursor.},label={lst:no-cursor}] 
SELECT * FROM employee; 
\end{lstlisting} 

\begin{lstlisting}[language=SQL,caption={Cursor-based queries that fetch and process `employee` rows one at a time.},label={lst:cursor}] 
DELIMITER //
CREATE PROCEDURE fetch_employee()
BEGIN
    DECLARE done INT DEFAULT FALSE;
    DECLARE emp_ssnum INT;
    DECLARE emp_name VARCHAR(255);
    DECLARE emp_lat DECIMAL(10,2);
    DECLARE emp_longitude DECIMAL(10,2);
    DECLARE emp_hundreds1 INT;
    DECLARE emp_hundreds2 INT;

    -- DECLARE cursor
    DECLARE emp_cursor CURSOR FOR 
        SELECT ssnum, name, lat, longitude, hundreds1, hundreds2 FROM employee;

    -- finish dealing with cursor
    DECLARE CONTINUE HANDLER FOR NOT FOUND SET done = TRUE;

    -- open cursor
    OPEN emp_cursor;

    -- read cursor
    read_loop: LOOP
        FETCH emp_cursor INTO emp_ssnum, emp_name, emp_lat, emp_longitude, emp_hundreds1, emp_hundreds2;
        IF done THEN
            LEAVE read_loop;
        END IF;
        
        -- print line by line
        SELECT emp_ssnum AS SSN, emp_name AS Name, emp_lat AS Latitude, emp_longitude AS Longitude, emp_hundreds1, emp_hundreds2;
    END LOOP;

    -- close cursor
    CLOSE emp_cursor;
END//
DELIMITER ;

DECLARE d_cursor CURSOR FOR select * from employee;
OPEN d_cursor while (@@FETCH_STATUS = 0)
BEGIN
    FETCH NEXT from d_cursor 
END
CLOSE d_cursor
go
\end{lstlisting} 


\item
\textbf{Quantitative Experimental Results}

\textbf{On the $10^5$ row dataset} (Figure \ref{fig: Cursor results for 10^5 dataset}), using the cursor degraded performance across all engines.
\emph{MariaDB} suffered the steepest decline, running approximately $9.4\times$ slower with a cursor. \emph{MySQL} followed closely, incurring a $9.1\times$ penalty.  \emph{PostgreSQL} showed a lesser slowdown at $5.1\times$ compared to the no-cursor version.
These results highlight the substantial overhead introduced by cursor-based iteration, even on moderate-sized datasets. \emph{DuckDB} was excluded from the cursor comparison because it does not support cursor-based point queries; however, its range scan performance is included for reference, allowing comparison with the other engines' non-cursor performance. See Table~\ref{tab:cursor-10e5} for the full set of average response times and standard deviations.

\begin{figure}[H]
  \centering
  \includegraphics[width=0.8\textwidth]{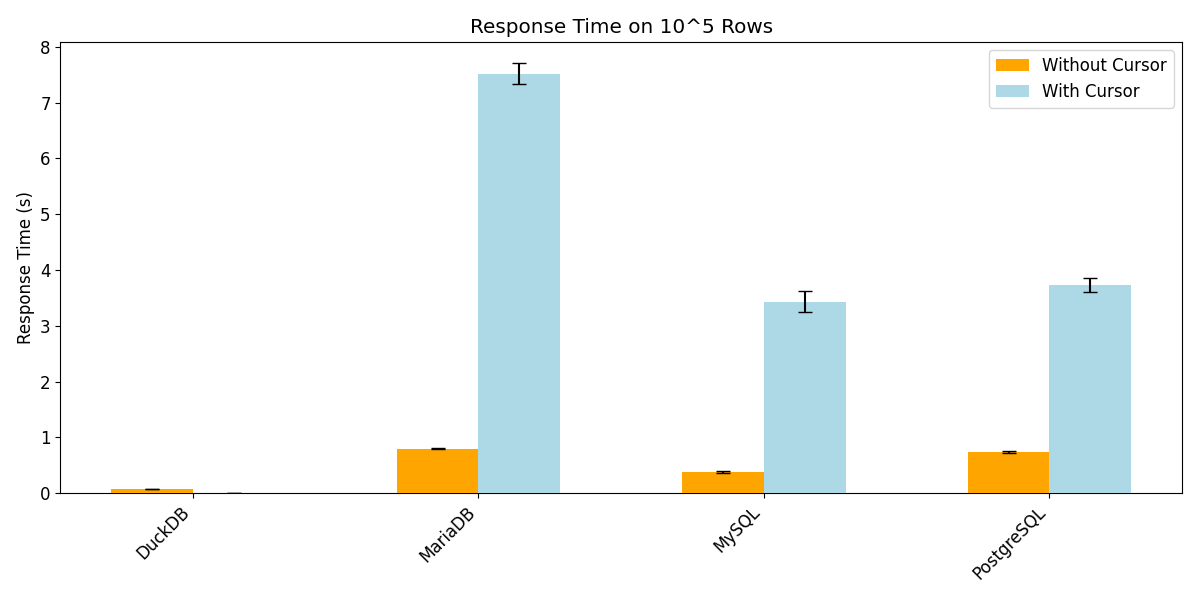}
  \caption{All engines perform better without using cursors to process the rows.}
  \label{fig: Cursor results for 10^5 dataset}
\end{figure}

\begin{table}[H]
  \centering
  \caption{Average Response Times along with their Standard Deviations for Cursor Experiment on a $10^5$ Row Dataset (in seconds).}
  \label{tab:cursor-10e5}

  \label{tab:cursor-response-10e5}
  \begin{tabular}{lccc}
    \toprule
    \textbf{System} & \textbf{No Cursor} & \textbf{Cursor} \\
    \midrule
    MariaDB     & 0.80 (0.0069) & 7.52 (0.18)\\
    MySQL       & 0.38 (0.015) & 3.43 (0.19)\\
    PostgreSQL  & 0.73 (0.016) & 3.74 (0.12)\\
    DuckDB      & 0.07 (0.0012) & N.A \\
    \bottomrule
  \end{tabular}
\end{table}

\textbf{On the $10^7$ row dataset} (Figure \ref{fig: Cursor results for 10^7 dataset}), the performance cost of using cursors became dramatically more pronounced.
\emph{MariaDB} slowed down by nearly $11.8\times$, \emph{MySQL} by $9.2\times$, and \emph{PostgreSQL} by $4.6\times$ compared to their respective set-based baselines.
In addition to longer runtimes, all three systems saw substantial increases in response time variability: standard deviation rose over $30\times$ in \emph{PostgreSQL}, more than $65\times$ in \emph{MariaDB}, and nearly $40\times$ in \emph{MySQL}.
This suggests that cursor-based iteration not only degrades performance but also introduces greater unpredictability, especially on large datasets. \emph{DuckDB} was excluded from the cursor comparison because it does not support cursor-based point queries; however, its range scan performance is included for reference, allowing comparison with the other engines' non-cursor performance. See Table~\ref{tab:cursor-10e7} for the full set of average response times and standard deviations.

\begin{figure}[H]
  \centering
  \includegraphics[width=0.8\textwidth]{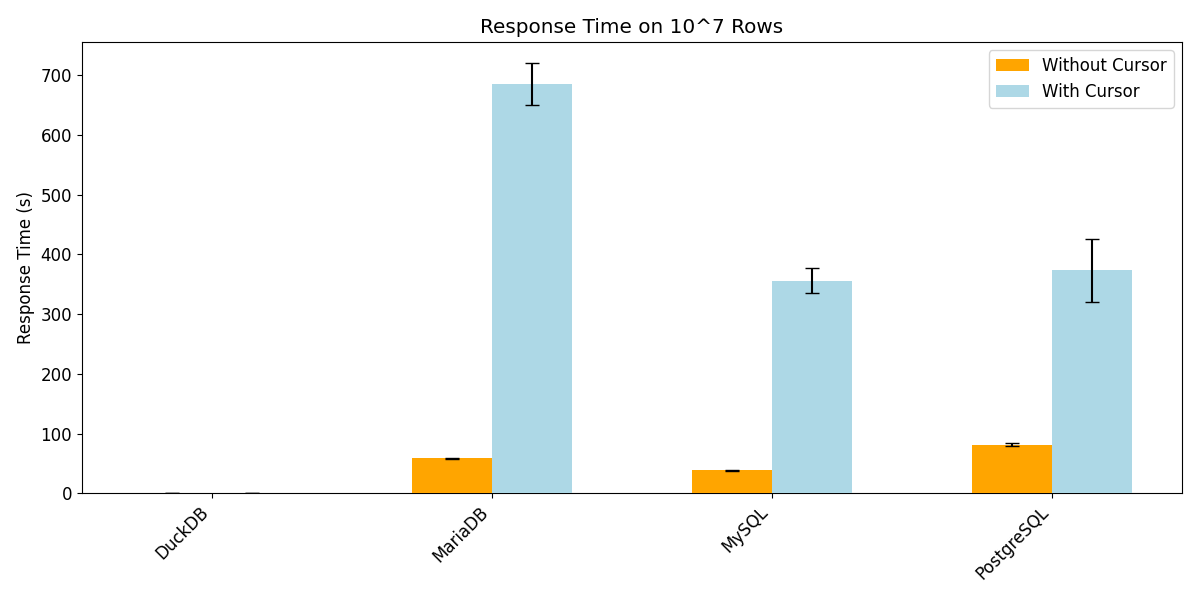}
  \caption{All engines perform better without using cursors to process the rows.}
  \label{fig: Cursor results for 10^7 dataset}
\end{figure}

\begin{table}[H]
  \centering
  \caption{Average Response Times along with their Standard Deviations for Cursor Experiment on a $10^7$ Row Dataset (in seconds)}
  \label{tab:cursor-10e7}
  
  \label{tab:cursor-response-10e7}
  \begin{tabular}{lccc}
    \toprule
    \textbf{System} & \textbf{No Cursor} & \textbf{Cursor} \\
    \midrule
    MariaDB     & 58.30 (1.098) & 685.58 (34.59)\\
    MySQL       & 38.58 (0.53) & 356.38 (20.89)\\
    PostgreSQL  & 81.32 (2.62) & 373.52 (52.40)\\
    DuckDB      & 1.22 (0.0038) & N.A \\
    \bottomrule
  \end{tabular}
\end{table}

\item
\textbf{Qualitative Conclusions}
\\
Cursor-based iteration is consistently detrimental to performance in our experiments, which is consistent with the findings of Shasha and Bonnet~\cite{10.1145/1024694.1024720}. While cursor-based iteration does offer the advantage of providing  a more familiar programming interface to developers familiar with scalar programming languages, organizations should generally discourage their use.

Replacing a single set-based scan with hundreds of cursor-driven fetches forces the engine to repeatedly enter and exit execution context, amplifying statement dispatch overhead.
At $10^5$ rows, this penalty was already severe.
At $10^7$ rows, the impact became extreme.
\end{enumerate}

\section[\appendixname~\thesection]{Retrieve Needed Columns: Quantitative Experimental Results}\label{retrieve-appendix}
\begin{figure}[H]
  \centering
  \includegraphics[width=0.8\textwidth]{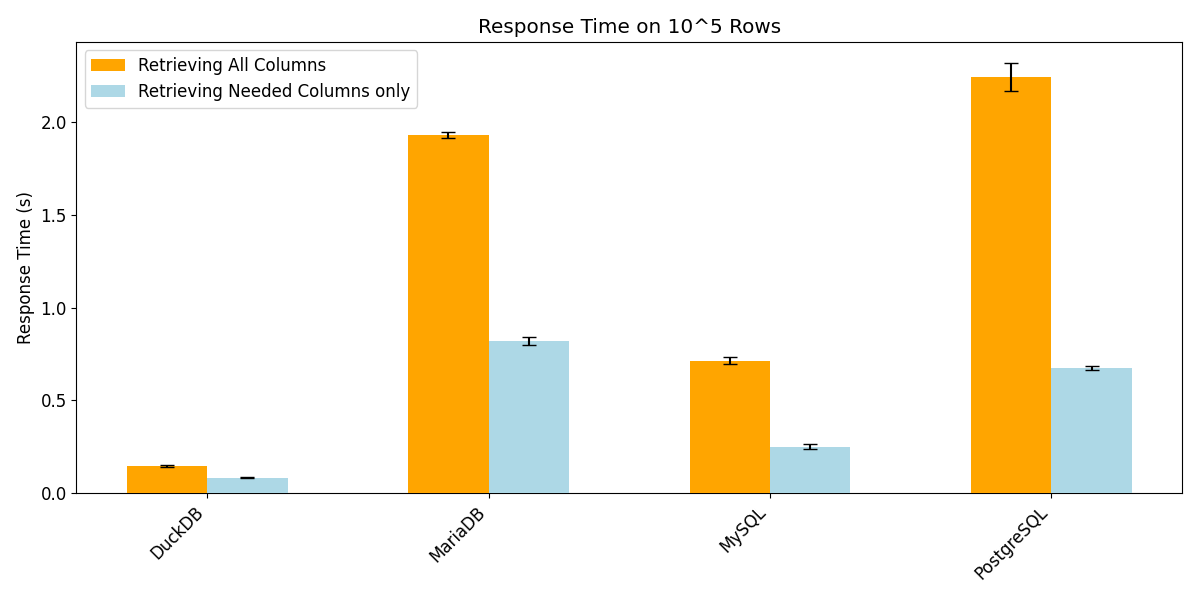}
  \caption{All engines perform better when retrieving  needed columns only. }
  \label{fig: Retrieval results for 10^5 dataset}
\end{figure}

\textbf{On the $10^5$ row dataset} (Figure \ref{fig: Retrieval results for 10^5 dataset}), retrieving only the required columns significantly improved performance across all database engines.
\emph{PostgreSQL} exhibited the most dramatic improvement, running approximately $3.3\times$ faster when selecting only the needed columns. \emph{MySQL} and \emph{MariaDB} followed with $2.8\times$ and $2.4\times$ speedups, respectively. \emph{DuckDB}, benefiting from its columnar architecture, showed a $1.7\times$ improvement.
These results confirm that explicitly requesting columns instead of writing  'SELECT *' reduces unnecessary data retrieval and enhances execution efficiency even on moderately sized datasets. See Table \ref{tab:retrieval-10e5} for the full set of average response times and standard deviations.

\begin{table}[H]
  \centering
  \caption{Average Response Times along with their Standard Deviations for Retrieval Experiment on a $10^5$ Row Dataset (in seconds)}
  \label{tab:retrieval-10e5}

  \label{tab:retrieval-response-10e5}
  \begin{tabular}{lccc}
    \toprule
    \textbf{System} & \textbf{All columns} & \textbf{Selected Columns} \\
    \midrule
    DuckDB     & 0.15 (0.0037) & 0.085 (0.0019) \\
    MariaDB     & 1.93 (0.016) & 0.82 (0.022) \\
    MySQL       & 0.71 (0.020) & 0.25 (0.014) \\
    PostgreSQL  & 2.24 (0.074) & 0.68 (0.010)\\
    \bottomrule
  \end{tabular}

\end{table}

\begin{figure}[H]
  \centering
  \includegraphics[width=0.8\textwidth]{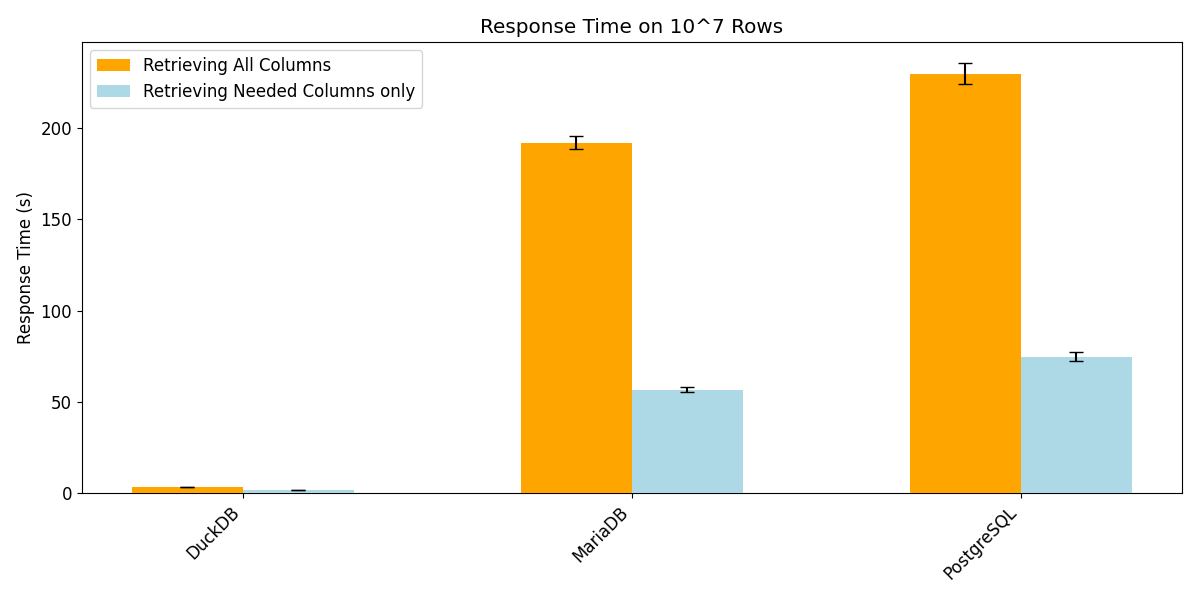}
  \caption{All engines perform better when retrieving  needed columns only. }
  \label{fig: Retrieval results for 10^7 dataset}
\end{figure}

\textbf{On the $10^7$ row dataset} (Figure \ref{fig: Retrieval results for 10^7 dataset}), selecting only the necessary columns led to substantial performance gains across all tested engines.
\emph{MariaDB} exhibited the most pronounced improvement, executing $3.4\times$ faster with column pruning. \emph{PostgreSQL} followed closely with a $3.1\times$ speedup, while \emph{DuckDB} achieved a $2.1\times$ reduction in execution time.
In addition to faster runtimes, selecting fewer columns consistently reduced the variability in performance. Standard deviations dropped across all engines, indicating more stable and predictable query behavior when excess columns were excluded.
\emph{MySQL} is excluded from this comparison because it was unable to create the $10^7$ row \texttt{lineitem} table within a reasonable time limit, at which point the process was manually terminated.
See Table \ref{tab:retrieval-10e7} for the full set of average response times and standard deviations.

\begin{table}[H]
  \centering
  \caption{Average Response Times along with their Standard Deviations for Retrieval Experiment on a $10^7$ Row Dataset (in seconds)}
  \label{tab:retrieval-10e7}

  \label{tab:retrieval-response-10e7}
  \begin{tabular}{lccc}
    \toprule
    \textbf{System} & \textbf{All columns} & \textbf{Needed Columns Only} \\
    \midrule
    DuckDB     & 3.35 (0.012) & 1.58 (0.0041) \\
    MariaDB     & 192.13 (3.53) & 56.80 (1.31) \\
    PostgreSQL  & 229.99 (5.69) & 74.85 (2.52)\\
    \bottomrule
  \end{tabular}

\end{table}

\section[\appendixname~\thesection]{Eliminate Unneeded DISTINCT clause: Quantitative Experimental Results}\label{distinct-appendix}
\begin{figure}[H]
  \centering
  \includegraphics[width=0.8\textwidth]{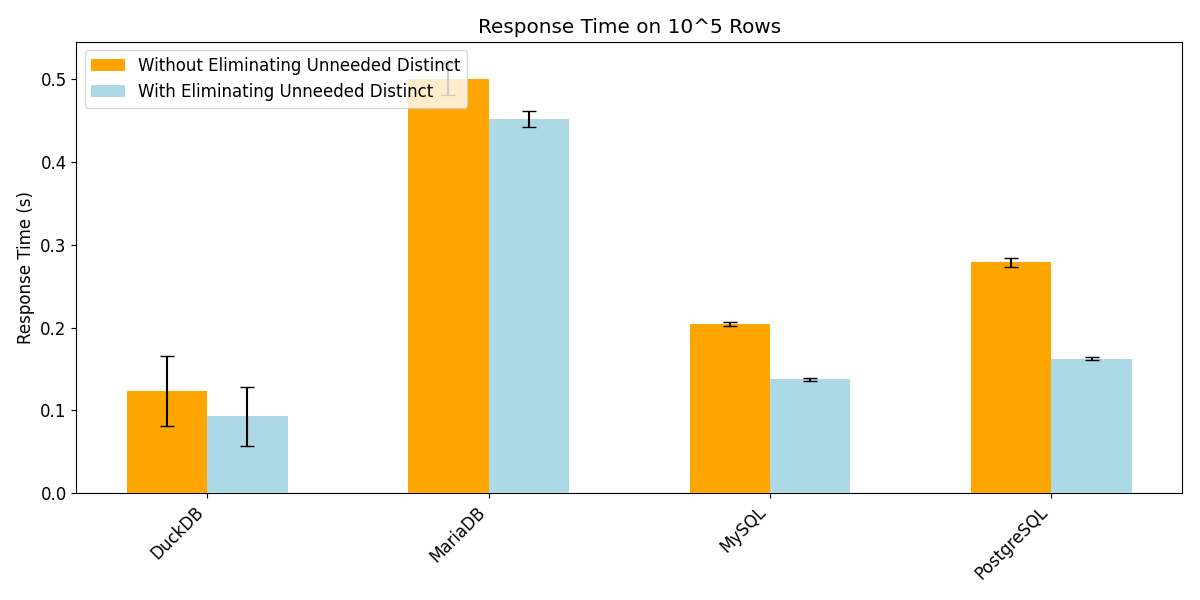}
  \caption{All engines perform faster after eliminating the 'DISTINCT' clause on uniformly-distributed data.}
  \label{fig: Eliminate Unneeded Uniform results for 10^5 dataset}
\end{figure}

\begin{figure}[H]
  \centering
  \includegraphics[width=0.8\textwidth]{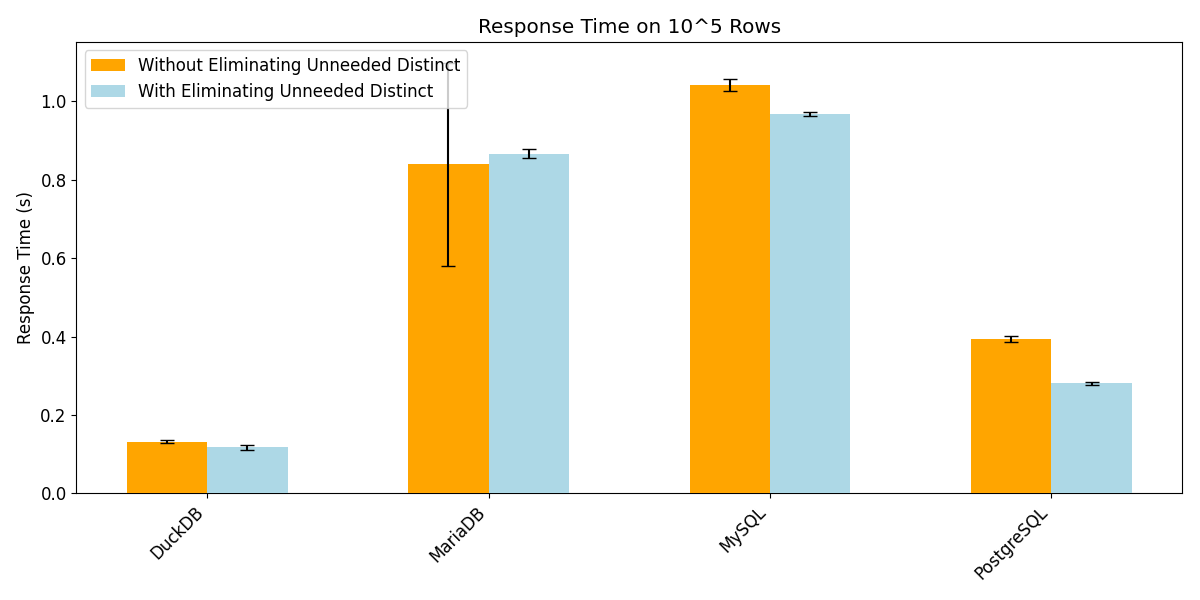}
  \caption{Eliminating the 'DISTINCT' clause helps the performance on fractally-distributed data except for MariaDB, where the effect is unclear because DISTINCT caused very high variability.}
  \label{fig: Eliminate Unneeded Fractal results for 10^5 dataset}
\end{figure}

\textbf{On the $10^5$ row dataset} (Tables \ref{fig: Eliminate Unneeded Uniform results for 10^5 dataset} and~\ref{fig: Eliminate Unneeded Fractal results for 10^5 dataset}), eliminating the unnecessary \texttt{DISTINCT} clause consistently improved query performance across most database engines and data distributions.

\emph{PostgreSQL} benefited the most, showing a $1.75\times$ speedup on uniformly distributed data and a $1.39\times$ speedup on fractally distributed data. \emph{MySQL} also improved markedly, with execution time reduced by $1.43\times$ on uniform data and $1.08\times$ on fractal data. \emph{DuckDB} showed a $1.29\times$ improvement on uniform data, though only a marginal change on fractal data, where performance remained nearly constant.

\emph{MariaDB} exhibited the smallest improvement on uniform data, with a modest $1.11\times$ speedup, and was the only engine to show a slight slowdown on the fractal dataset, where average response time increased slightly. However, the high standard deviation in the original run suggests this regression may be due to statistical noise rather than a consistent performance trend.
These results demonstrate that eliminating unnecessary \texttt{DISTINCT} clauses is a generally effective optimization. 

See Table \ref{tab:distinct-uniform-10e5} and \ref{tab:distinct-fractal-10e5} for the full set of average response times and standard deviations.

\begin{table}[H]
  \centering
  \caption{Average Response Times along with their Standard Deviations when testing the effect of  Eliminating Unneeded DISTINCT on a uniformly-distributed $10^5$ Row Dataset (in seconds)}
  \label{tab:distinct-uniform-10e5}

  \label{tab:distinct-uniform-response-10e5}
  \begin{tabular}{lccc}
    \toprule
    \textbf{System} & \textbf{Without eliminating distinct} & \textbf{With eliminating distinct} \\
    \midrule
    DuckDB     & 0.12 (0.043) & 0.093 (0.036)\\
    MariaDB     & 0.50 (0.019) & 0.45 (0.0097)\\
    MySQL       & 0.20 (0.002) & 0.14 (0.0016)\\
    PostgreSQL  & 0.28 (0.0053) & 0.16 (0.0017)\\
    \bottomrule
  \end{tabular}
\end{table}

\begin{table}[H]
  \centering
  \caption{Average Response Times along with their Standard Deviations when testing the effect of  Eliminating Unneeded DISTINCT on a fractally-distributed $10^5$ Row Dataset (in seconds)}
  \label{tab:distinct-fractal-10e5}

  \label{tab:distinct-fractal-response-10e5}
  \begin{tabular}{lccc}
    \toprule
    \textbf{System} & \textbf{Without eliminating distinct} & \textbf{With eliminating distinct} \\
    \midrule
    DuckDB     & 0.13 (0.0045) & 0.12 (0.0076)\\
    MariaDB     & 0.84 (0.26) & 0.87 (0.011)\\
    MySQL       & 1.04 (0.016) & 0.96 (0.0058)\\
    PostgreSQL  & 0.39 (0.0076) & 0.28 (0.0040)\\
    \bottomrule
  \end{tabular}
\end{table}

\begin{figure}[H]
  \centering
  \includegraphics[width=0.8\textwidth]{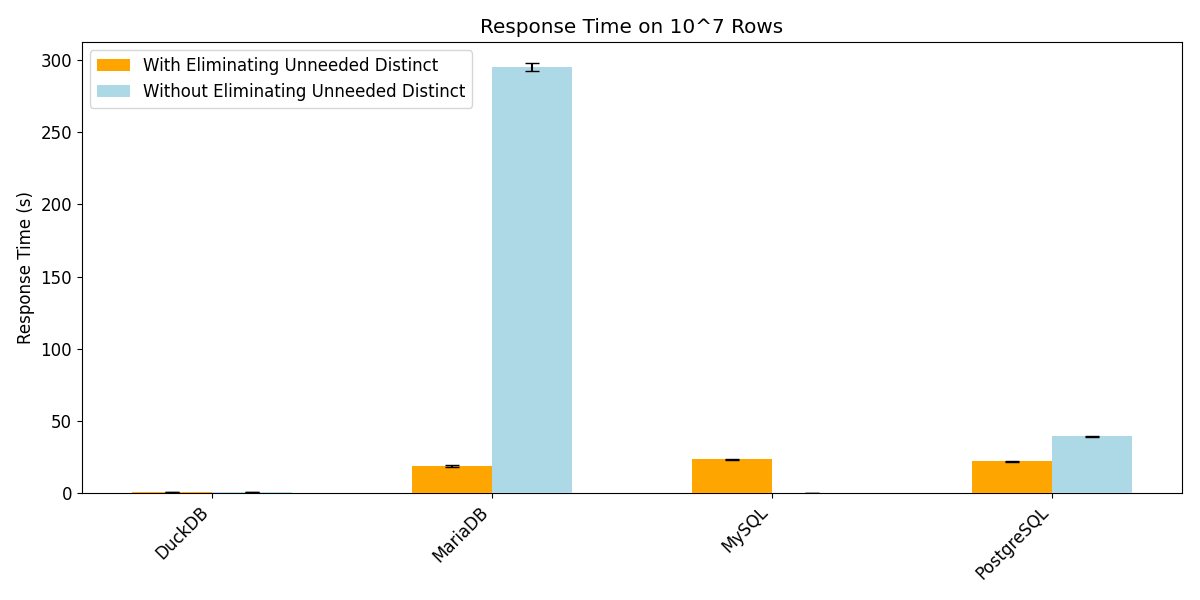}
  \caption{All engines perform better after eliminating the 'DISTINCT' clause on uniformly-distributed data of $10^7$ rows. MySQL did not complete with the DISTINCT.}
  \label{fig: Eliminate Unneeded Uniform results for 10^7 dataset}
\end{figure}

\begin{figure}[H]
  \centering
  \includegraphics[width=0.8\textwidth]{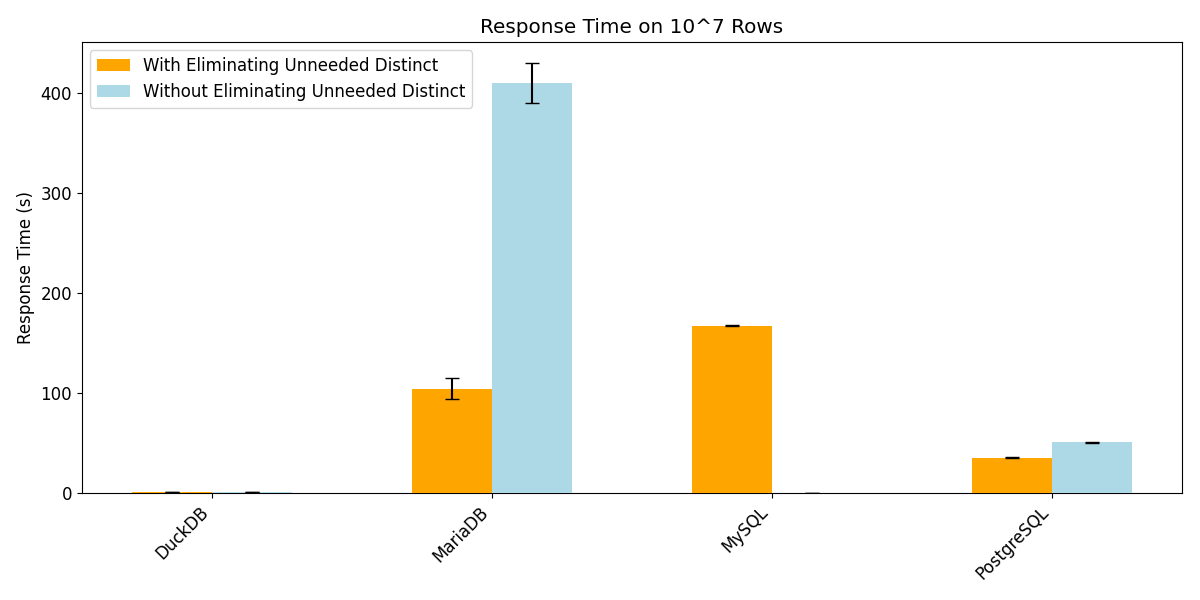}
  \caption{As for uniformly distributed data, all engines perform better after eliminating the 'DISTINCT' clause on fractally-distributed data of $10^7$ rows. MySQL did not complete with the DISTINCT.}
  \label{fig: Eliminate Unneeded Fractal results for 10^7 dataset}
\end{figure}

\textbf{On the $10^7$ row dataset} (Tables \ref{fig: Eliminate Unneeded Uniform results for 10^7 dataset} and \ref{fig: Eliminate Unneeded Fractal results for 10^7 dataset}), eliminating the unnecessary \texttt{DISTINCT} clause yielded dramatic improvements in query performance across all systems that completed both runs. Due to prohibitively long runtimes, \emph{MySQL} was unable to complete the query using the \texttt{DISTINCT} clause for either data distribution, so it has been excluded from this comparison. However, MySQL results without the \texttt{DISTINCT} clause are included for comparison with other databases. 

\emph{MariaDB} demonstrated the most extreme gains. On uniformly-distributed data, query time showed a $15.4\times$ improvement. On fractally-distributed data, it achieved a $3.9\times$ speedup. \emph{PostgreSQL} also benefited significantly, with execution time reduced by $1.79\times$ on uniform data and $1.47\times$ on fractal data. \emph{DuckDB}, although already much faster than the other engines, still showed consistent improvement: a $1.28\times$ gain on uniform data and a $1.15\times$ gain on fractal data.

These results highlight the substantial overhead that unnecessary \texttt{DISTINCT} operations impose at scale, particularly on engines like \emph{MariaDB}, where the cost of sorting or deduplication grows non-linearly with input size. The performance gap widened significantly from the $10^5$ to the $10^7$ row dataset, reinforcing that query optimizations such as eliminating redundant \texttt{DISTINCT} clauses become increasingly critical as data volumes grow.

See Table \ref{tab:distinct-uniform-10e7} and \ref{tab:distinct-fractal-10e7} for the full set of average response times and standard deviations.

\begin{table}[H]
  \centering
  \caption{Average Response Times along with their Standard Deviations when testing the effect of  Eliminating Unneeded DISTINCT  on a uniformly-distributed $10^7$ Row Dataset (in seconds)}
  \label{tab:distinct-uniform-10e7}

  \label{tab:distinct-uniform-response-10e7}
  \begin{tabular}{lccc}
    \toprule
    \textbf{System} & \textbf{With distinct} & \textbf{Without  distinct}\\
    \midrule
    DuckDB     & 0.92 (0.017) & 0.72 (0.013)\\
    MariaDB     & 294.83 (2.81) & 19.13 (0.65)\\
    MySQL       & N.A           & 23.68 (0.37)\\
    PostgreSQL  & 39.37 (0.47) & 21.99 (0.16)\\
    \bottomrule
  \end{tabular}
\end{table}

\begin{table}[H]
  \centering
  \caption{Average Response Times along with their Standard Deviations when testing the effect of  Eliminating Unneeded DISTINCT on a fractally-distributed $10^7$ Row Dataset (in seconds)}
  \label{tab:distinct-fractal-10e7}

  \label{tab:distinct-fractal-response-10e7}
  \begin{tabular}{lccc}
    \toprule
    \textbf{System} & \textbf{With  distinct} & \textbf{Without  distinct}\\
    \midrule
    DuckDB     & 1.60 (0.089) & 1.39 (0.042)\\
    MariaDB     & 410.45 (19.72) & 104.77 (10.053)\\
    MySQL       & N.A           & 167.65 (0.455)\\
    PostgreSQL  & 51.08 (0.40) & 34.75 (0.35)\\
    \bottomrule
  \end{tabular}
\end{table}

\section[\appendixname~\thesection]{Correlated Subqueries vs Temporary Tables: Quantitative Experimental Results}\label{correlated-appendix}
\begin{figure}[H]
  \centering
  \includegraphics[width=0.8\textwidth]{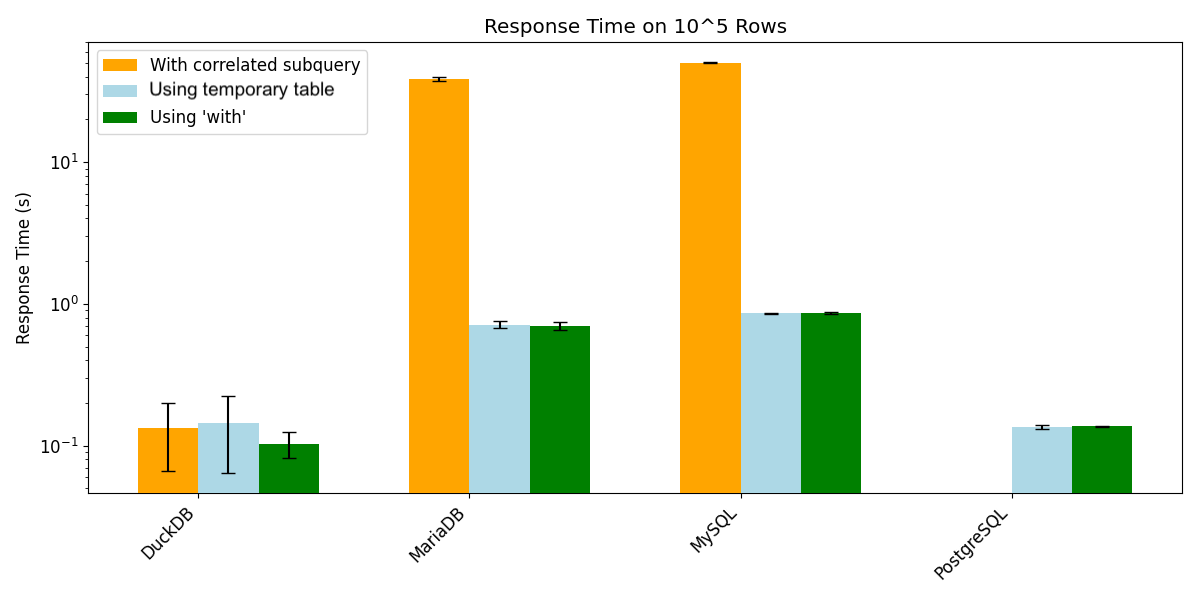}
  \caption{For uniformly distributed data, DUCKDB performed best when using the WITH formulation. MariaDB and MySQL showed much higher response times when using correlated subqueries than using the temporary table method or the WITH clause.  The correlated subquery on PostgreSQL exceeded the timeout period.}
  \label{fig: Co-related Subquery Uniform results for 10^5 dataset}
\end{figure}

\begin{figure}[H]
  \centering
  \includegraphics[width=0.8\textwidth]{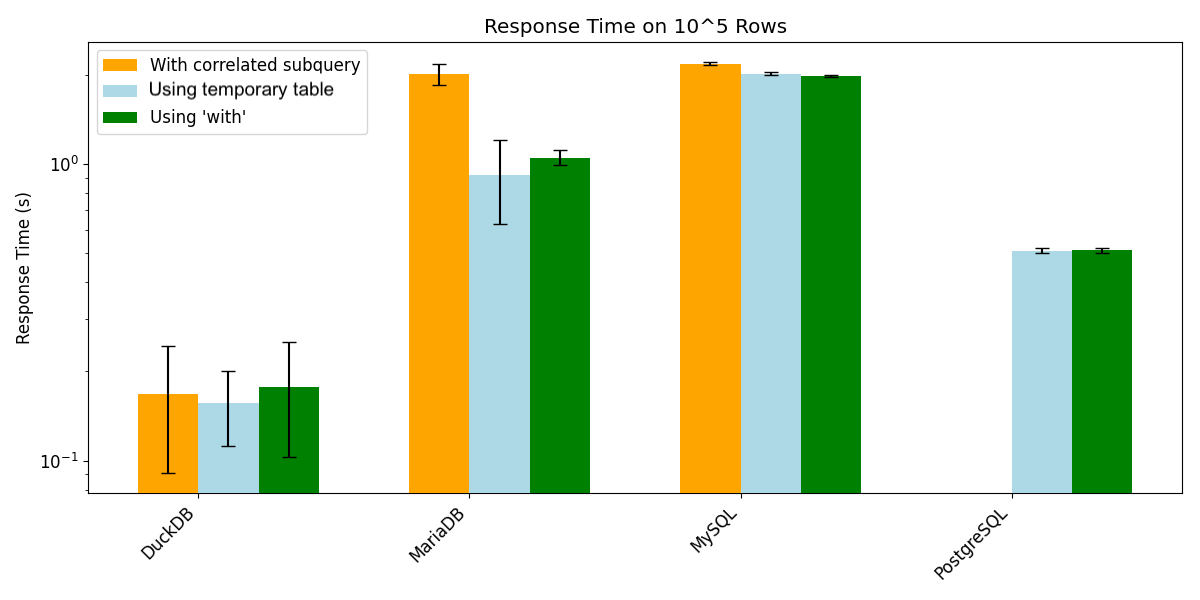}
  \caption{On fractally distributed data, the results on DuckDB and PostgreSQL are qualitatively similar to the results on uniformly distributed data. Correlated Subqueries run poorly on MariaDB and MySQL for this kind of data, but not as badly relative to the temporary table or WITH clause formulations as for uniformly distributed data.}
  \label{fig: Co-related Subquery Fractal results for 10^5 dataset}
\end{figure}

\textbf{On the $10^{5}$ row dataset} (Tables \ref{fig: Co-related Subquery Uniform results for 10^5 dataset} and \ref{fig: Co-related Subquery Fractal results for 10^5 dataset}), the three query styles diverged sharply in cost.
For uniformly distributed data, the correlated form behaved much like the join in DuckDB, but the with-clause shows an improvement of 30\%. By contrast, the correlated subquery performed very poorly in both MariaDB and MySQL, while the temporary table/with-clause formulations slashed runtime by roughly fifty-to-sixty-fold. The correlated variant could not complete on PostgreSQL, yet the join and with-clause version finished in well under a second.  The coefficients of variation were below 10\%.

Repeating the test on fractally distributed keys flattened the picture.  In DuckDB all three forms converged, differing by under 10\%.  MariaDB still punished the correlated form -- temporary table and with-clause formulations were about twice as fast -- but the gap shrank from orders of magnitude to a single-digit factor.  Using the temporary table or WITH clause, MySQL improved by  8–10\% compared with the correlated subquery.  As before, PostgreSQL’s correlated version stalled, while its join and with-clause plans remained steady at about half a second with negligible variation.

See Table \ref{tab:uniform-subquery-10e5} and \ref{tab:fractal-subquery-10e5} for the full set of average response times and standard deviations.

\begin{table}[H]
  \centering
  \caption{Average Response Times along with their Standard Deviations for correlated subqueries on a $10^5$ uniformly-distributed row dataset (in seconds)}
  \label{tab:uniform-subquery-10e5}

  \label{tab:uniform-subquery-response-10e5}
  \begin{tabular}{lccc}
    \toprule
    \textbf{System} & \textbf{With correlated subquery} & \textbf{Join} & \textbf{With-clause}\\
    \midrule
    DuckDB     & 0.13 (0.067) & 0.14 (0.079) & 0.103 (0.022)\\
    MariaDB     & 38.69 (1.34) & 0.71 (0.042) & 0.69 (0.045)\\
    MySQL       & 50.12 (0.38) & 0.86 (0.0070) & 0.87 (0.013)\\
    PostgreSQL  & N.A & 0.14 (0.0044) & 0.14 (0.0018)\\
    \bottomrule
  \end{tabular}
\end{table}

\begin{table}[H]
  \centering
  \caption{Average Response Times along with their Standard Deviations for correlated subqueries on a $10^5$ fractally-distributed row dataset (in seconds)}
  \label{tab:fractal-subquery-10e5}

  \label{tab:fractal-subquery-response-10e5}
  \begin{tabular}{lccc}
    \toprule
    \textbf{System} & \textbf{With correlated subquery} & \textbf{Join} & \textbf{With-clause}\\
    \midrule
    DuckDB     & 0.17 (0.076) & 0.16 (0.043) & 0.17 (0.073)\\
    MariaDB     & 2.01 (0.16) & 0.92 (0.28) & 1.05 (0.061)\\
    MySQL       & 2.18 (0.022) & 2.01 (0.023) & 1.98 (0.012)\\
    PostgreSQL  & N.A & 0.51 (0.0089) & 0.51 (0.0101)\\
    \bottomrule
  \end{tabular}
\end{table}

\begin{figure}[H]
  \centering
  \includegraphics[width=0.8\textwidth]{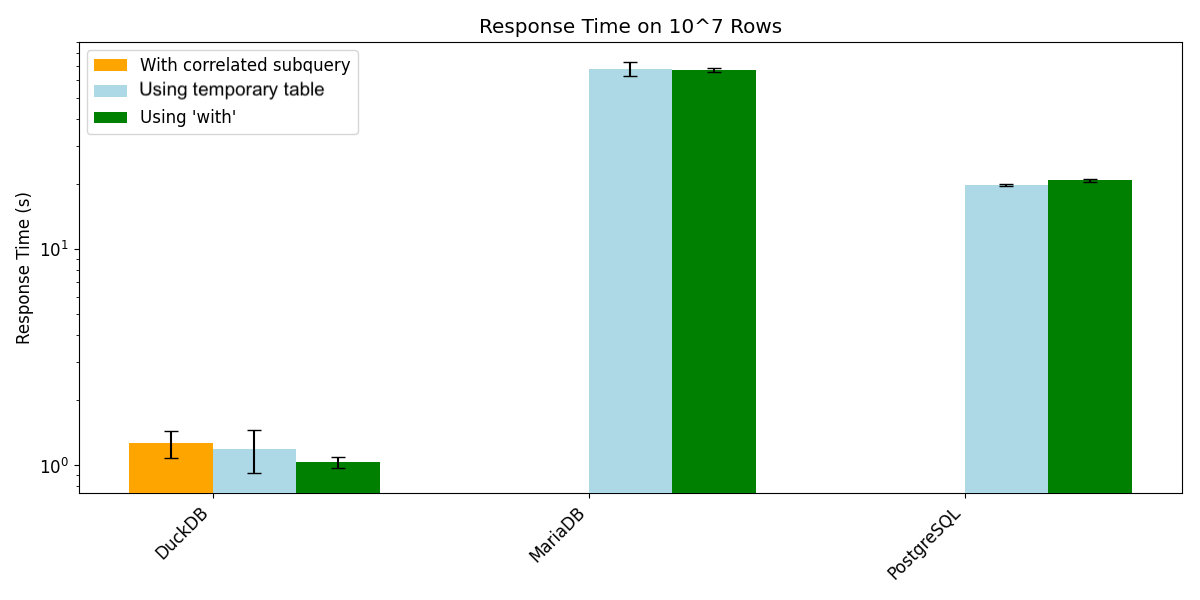}
  \caption{On uniformly distributed data with $10^7$ rows, MySQL could not complete any formulation. PostgreSQL and MariaDB could not run correlated subqueries at this size, but could run the temporary table/With-clause formulations. DuckDB continued to show outstanding performance regardless of query formulation. }
  \label{fig: correlated subquery uniform results for 10^7 dataset}
\end{figure}

\begin{figure}[H]
  \centering
  \includegraphics[width=0.8\textwidth]{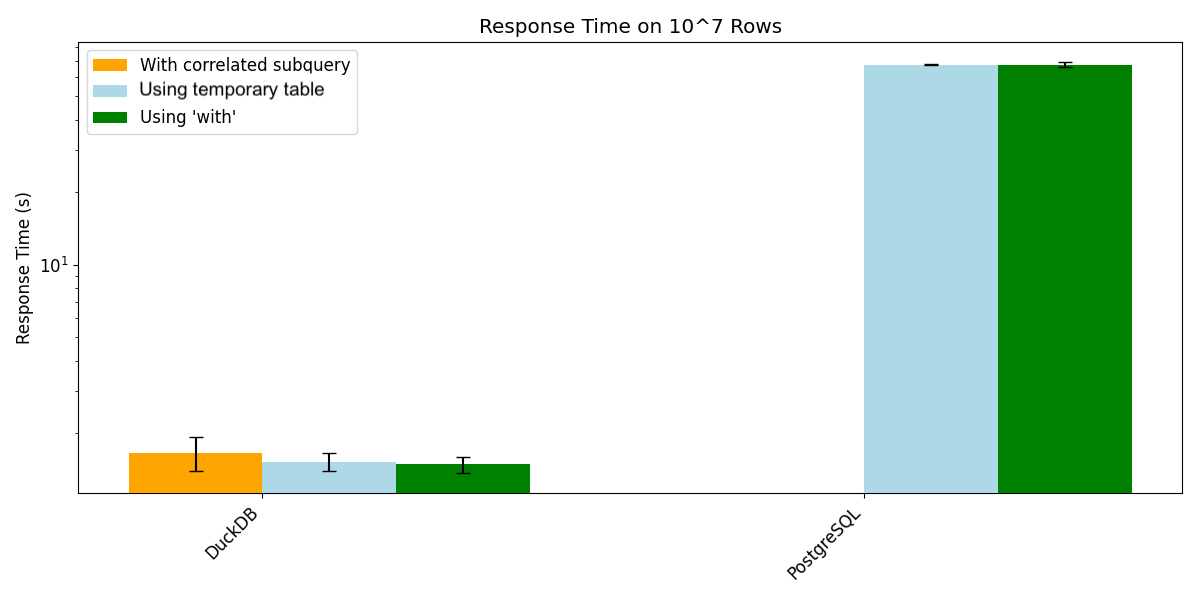}
  \caption{DuckDB shows excellent performance at scale, regardless of the data distribution and PostgreSQL stays consistent with its previous results. MariaDB and MySQL could not be tested at such a large scale for fractally-distributed data. }
  \label{fig: correlated subquery fractal results for 10^7 dataset}
\end{figure}

\textbf{On the $10^{7}$ row dataset} (Tables \ref{fig: correlated subquery uniform results for 10^7 dataset} and \ref{fig: correlated subquery fractal results for 10^7 dataset}), engine behaviour split into two clear camps.

DuckDB handled every formulation briskly: rewriting the correlated query into a join or an in-lined \texttt{WITH} clause trimmed latency by a modest 15–20,\%, leaving all three plans clustered between 1–1.7s.
In stark contrast, the correlated version did not finish on MariaDB or PostgreSQL. Only after recasting the logic as a join or with-clause did those engines return a result. The latter two engines performed roughly the same, when using the join/with-clause.

Switching to a fractal distribution changed little: DuckDB again hovered near 1.5s regardless of form, while PostgreSQL’s join/with-clause plateaued around 68s and the correlated plan remained impractically slow.

MySQL proved too slow to finish any of the three query styles on either distribution, and MariaDB likewise did not complete on the fractal dataset, so results for those cases are omitted.

\begin{table}[H]
  \centering
  \caption{Average Response Times along with their Standard Deviations for correlated subqueries on a $10^7$ uniformly-distributed row dataset (in seconds)}
  \label{tab:uniform-subquery-10e7}

  \label{tab:uniform-subquery-response-10e7}
  \begin{tabular}{lccc}
    \toprule
    \textbf{System} & \textbf{With correlated subquery} & \textbf{Join} & \textbf{With-clause}\\
    \midrule
    DuckDB     & 1.26 (0.18) & 1.19 (0.26) & 1.03 (0.055)\\
    MariaDB     & N.A & 67.71 (5.03) & 67.05 (1.45)\\
    PostgreSQL  & N.A & 19.76 (0.22) & 20.79 (0.34)\\
    \bottomrule
  \end{tabular}
\end{table}

\begin{table}[H]
  \centering
  \caption{Average Response Times along with their Standard Deviations for correlated subqueries on a $10^7$ fractally-distributed row dataset (in seconds). Neither MySQL nor MariaDB could execute.}
  \label{tab:fractal-subquery-10e7}

  \label{tab:fractal-subquery-response-10e7}
  \begin{tabular}{lccc}
    \toprule
    \textbf{System} & \textbf{With correlated subquery} & \textbf{Join} & \textbf{With-clause}\\
    \midrule
    DuckDB     & 1.66 (0.26) & 1.52 (0.13) & 1.48 (0.11)\\
    PostgreSQL  & N.A & 67.52 (0.45) & 67.51 (1.56)\\
    \bottomrule
  \end{tabular}
\end{table}

\section[\appendixname~\thesection]{Aggregate Maintenance -- triggers: Quantitative Experimental Results for Data Insertion}\label{triggers_insert-appendix}
\begin{figure}[H]
  \centering
  \includegraphics[width=0.8\textwidth]{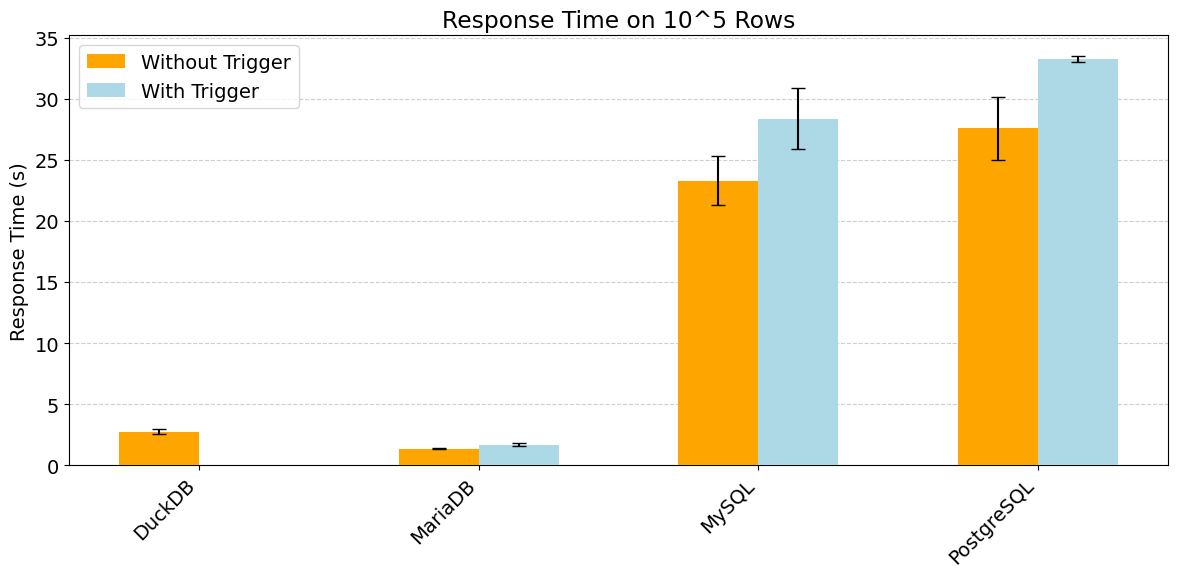}
  \caption{Impact of aggregate maintenance using \texttt{AFTER INSERT} triggers on Insertion (1000 rows) performance on $10^5$ rows. DuckDB does not support triggers. The performance of INSERT without triggers on MySQL, PostgreSQL and MariaDB is (unsurprisingly) better than the performance of using triggers.}
  \label{fig: trigger insertion results for 10^5 dataset}
\end{figure}

\textbf{On the $10^5$ row dataset}, Figure~\ref{fig: trigger insertion results for 10^5 dataset} and Table~\ref{tab:trigger_insert5} present the average response times and standard deviations for 1000 single-row insertions into the \texttt{orders} table. These insertions trigger automatic updates to related aggregate tables.

Across MariaDB, MySQL, and PostgreSQL, enabling triggers generally led to an increase in insertion latency. On MariaDB, the average response time slightly increased from 1.62s to 1.64s, reflecting a marginal slowdown of 1.01×. MySQL experienced a much larger increase, representing a 2.44× slowdown. PostgreSQL also saw a moderate increase in insertion time, from 1.30s to 1.50s (a 1.15× slowdown). DuckDB does not support triggers and achieved an average insertion time of 2.82s without any trigger mechanism.

As shown in Table~\ref{tab:trigger_insert5}, the impact of triggers on performance variability differs across DBMSs. On MariaDB, triggers slightly improved consistency, while PostgreSQL exhibited increased variability. MySQL's performance remained relatively stable despite the added overhead.

\begin{table}[H]
  \centering
  \caption{Average Response Time (Standard Deviation) with or without aggregate maintenance using \texttt{AFTER INSERT} trigger in Insertion (1000 rows) Experiments on a $10^5$-Row Dataset (in seconds) .}
  \label{tab:trigger_insert5}

  \begin{tabular}{lcc}
    \toprule
    \textbf{System} & \textbf{Without Trigger} & \textbf{With Trigger} \\
    \midrule
    DuckDB      & 2.8 (0.15)   & --            \\
    MariaDB     & 1.6 (0.026)  & 1.6 (0.019)    \\
    MySQL       & 2.5 (0.11)   & 6.1 (0.10)     \\
    PostgreSQL  & 1.3 (0.028)  & 1.5 (0.070)    \\
    \bottomrule
  \end{tabular}
\end{table}

\begin{figure}[H]
  \centering
  \includegraphics[width=0.8\textwidth]{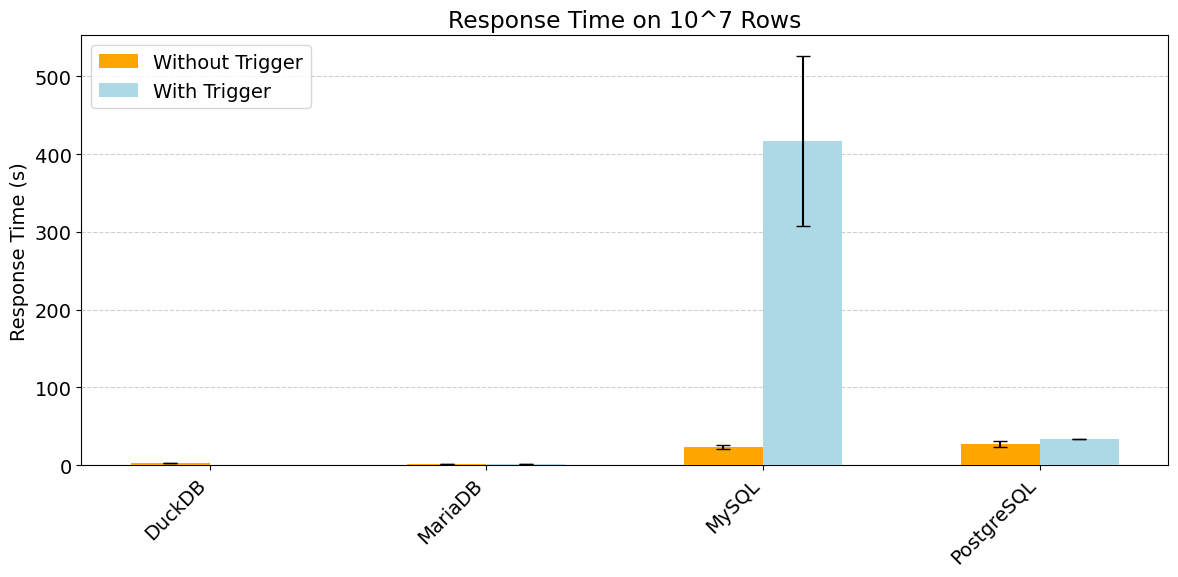}
  \caption{Impact of aggregate maintenance using \texttt{AFTER INSERT} triggers on Insertion (1000 rows) performance on $10^7$ rows. DuckDB does not support triggers. The performance  without triggers on MySQL, PostgreSQL and MariaDB is (again, unsurprisingly) better than the performance of using triggers.}
  \label{fig: trigger insertion results for 10^7 dataset}
\end{figure}

\textbf{On the $10^7$ row dataset} (Figure~\ref{fig: trigger insertion results for 10^7 dataset} and Table~\ref{tab:trigger_insert7}), DuckDB, which does not support triggers, achieved a baseline insertion time of 3.02 seconds. Across MariaDB, MySQL, and PostgreSQL, enabling triggers led to varying degrees of performance degradation. The most dramatic impact was observed on MySQL, where the average insertion time increased more than 120× with triggers enabled. PostgreSQL also saw a performance drop, with insertion latency rising by a factor of 1.21. On MariaDB, the overhead was minimal, with only a slight increase in insertion time.

As shown in Table~\ref{tab:trigger_insert7}, the effects on performance consistency also varied. MySQL exhibited a sharp increase in variability under triggers, with standard deviation growing substantially. PostgreSQL experienced a similar trend, though to a lesser extent. In contrast, MariaDB showed slightly improved consistency, as its standard deviation decreased when triggers were enabled.

\begin{table}[H]
  \centering
  \caption{Average Response Time (Standard Deviation) for Insertions (1000 rows) With or Without Aggregate Maintenance Using \texttt{AFTER INSERT} Triggers on a $10^7$-Row Dataset (in seconds).}
  \label{tab:trigger_insert7}

  \begin{tabular}{lcc}
    \toprule
    \textbf{System} & \textbf{Without Trigger} & \textbf{With Trigger} \\
    \midrule20
    DuckDB      & 3.0 (0.081)   & --              \\
    MariaDB     & 1.6 (0.013)   & 1.6 (0.011)      \\
    MySQL       & 2.6 (0.12)    & 320 (5.9)        \\
    PostgreSQL  & 1.3 (0.036)   & 1.6 (0.28)       \\
    \bottomrule
  \end{tabular}
\end{table}

\section[\appendixname~\thesection]{Aggregate Maintenance -- triggers: Quantitative Experimental Results for Data Retrieval}\label{triggers_retrieve-appendix}
\begin{figure}[H]
  \centering
  \includegraphics[width=1\textwidth]{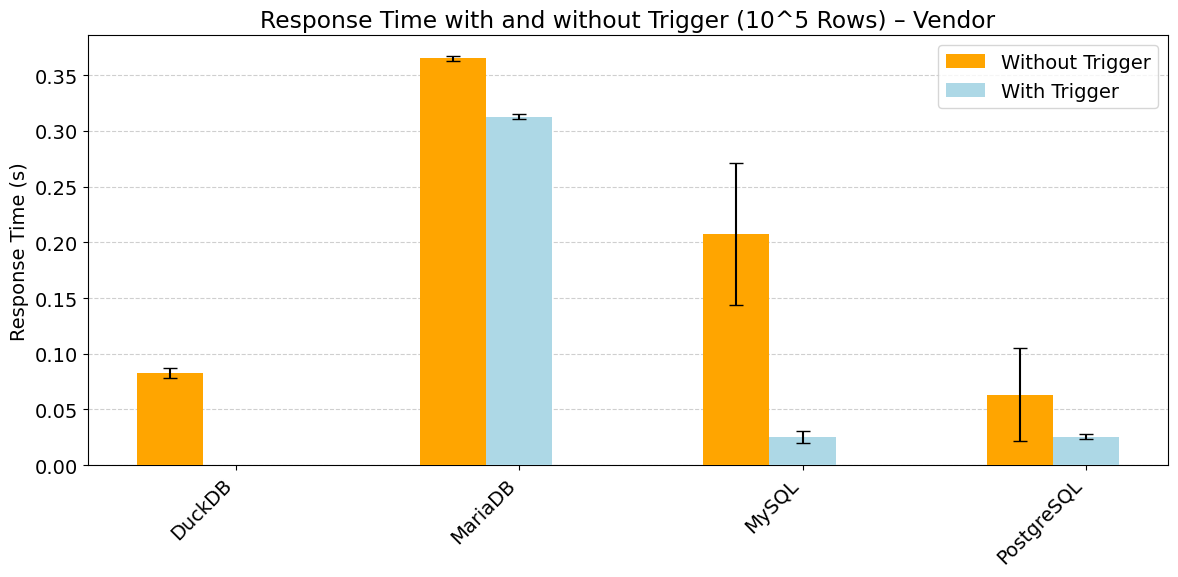}
  \caption{Impact of aggregate maintenance using \texttt{AFTER INSERT} triggers on data retrieval performance on \texttt{vendor} on $10^5$ rows. DuckDB does not support triggers. The performance  with triggers on MySQL, PostgreSQL and MariaDB is better than their performance  without triggers.}
  \label{fig: trigger vendor 5}
\end{figure}

\textbf{On the $10^5$ row dataset}, Figure~\ref{fig: trigger vendor 5} and Table~\ref{tab:triggerretrieve5} show that enabling \texttt{AFTER INSERT} triggers significantly improved query performance for vendor-related queries on MySQL, MariaDB, and PostgreSQL. MySQL exhibited the largest speedup, with query time reduced by a factor of 8.2. PostgreSQL achieved a 2.5× speedup, while MariaDB improved modestly by 1.2×. DuckDB, which does not support triggers, achieved a response time of 0.083 seconds.

Standard deviation was also reduced across all trigger-supporting systems, indicating more consistent performance. MySQL’s variability dropped notably, and PostgreSQL's standard deviation decreased by over 18×. MariaDB's variability remained low and nearly unchanged.

\begin{figure}[H]
  \centering
  \includegraphics[width=1\textwidth]{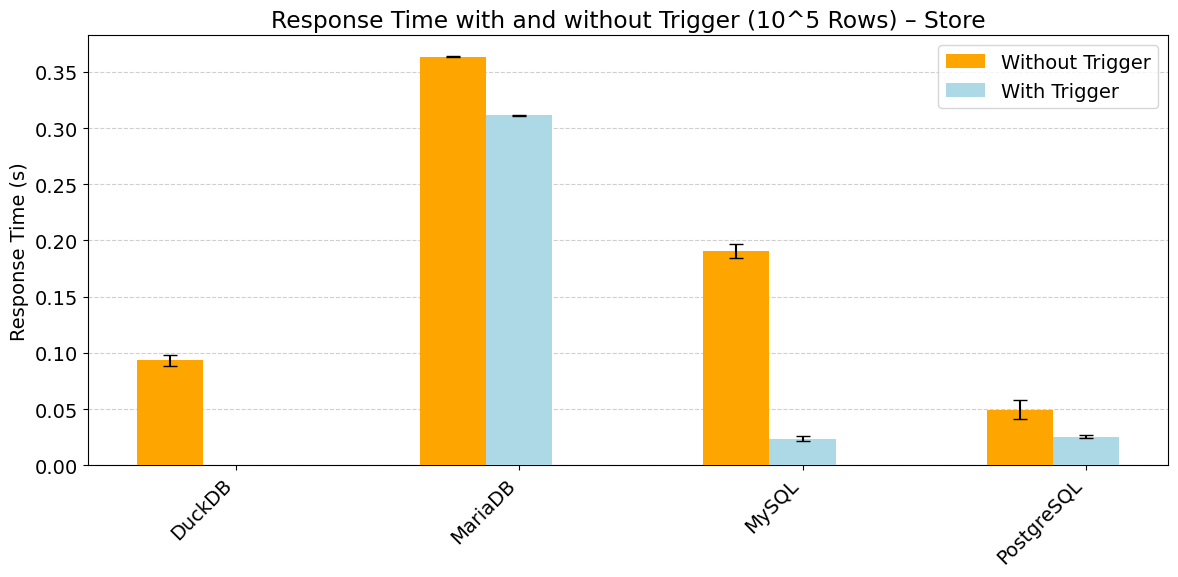}
  \caption{Impact of aggregate maintenance using \texttt{AFTER INSERT} triggers on data retrieval performance on \texttt{store} on $10^5$ rows. DuckDB does not support triggers. The performance  with triggers on MySQL, PostgreSQL and MariaDB is better than the performance  without triggers.}
  \label{fig: trigger store 5}
\end{figure}

As shown in Figure~\ref{fig: trigger store 5} and Table~\ref{tab:triggerretrieve5}, triggers also improved store query performance. MySQL achieved an 8.1× speedup, followed by PostgreSQL with a 1.9× improvement. MariaDB again saw a modest 1.2× gain. DuckDB, operating without triggers, reported a store query response time of 0.093 seconds.

Trigger-based maintenance also improved consistency: all three systems saw reduced standard deviation. MySQL and PostgreSQL in particular became noticeably more stable, while MariaDB’s already low variance decreased slightly.

\begin{table}[H]
  \centering
  \caption{Average Response Time (Standard Deviation) of Data Retrieval for Vendor and Store Queries With or Without Aggregate Maintenance Using \texttt{AFTER INSERT} Triggers on $10^5$ dataset (in seconds).}
  \label{tab:triggerretrieve5}
  \begin{tabular}{llcc}
    \toprule
    \textbf{Query Type} & \textbf{System} & \textbf{Without Trigger} & \textbf{With Trigger} \\
    \midrule
    \multirow{4}{*}{Vendor}
        & DuckDB      & 0.083 (0.0047)     & --                  \\
        & MariaDB     & 0.37 (0.0024)      & 0.31 (0.0023)       \\
        & MySQL       & 0.21 (0.064)       & 0.025 (0.0056)      \\
        & PostgreSQL  & 0.063 (0.042)      & 0.026 (0.0023)      \\
    \midrule
    \multirow{4}{*}{Store}
        & DuckDB      & 0.093 (0.0051)     & --                  \\
        & MariaDB     & 0.36 (0.00067)     & 0.31 (0.00052)      \\
        & MySQL       & 0.19 (0.0063)      & 0.024 (0.0023)      \\
        & PostgreSQL  & 0.049 (0.0083)     & 0.025 (0.0014)      \\
    \bottomrule
  \end{tabular}
\end{table}

\textbf{On the $10^7$ row dataset}, Figure~\ref{fig: trigger insertion results for 10^7 dataset} and Table~\ref{tab:triggerretrieve7} show that enabling \texttt{AFTER INSERT} triggers greatly improved vendor query performance on MySQL, MariaDB, and PostgreSQL. On MySQL, the query time dropped from 19.9s to 0.16s (a 127× speedup); MariaDB improved from by a factor of 22, and PostgreSQL from 1.08s to 0.026s (45×). DuckDB, which does not support triggers, achieved a response time of 0.77s.

As for consistency, all trigger-supporting systems showed substantial reductions in variability. MySQL and PostgreSQL achieved much more stable performance with triggers—standard deviation dropped to 0.0036s and 0.0013s, respectively. MariaDB also remained stable.

\begin{figure}[H]
  \centering
  \includegraphics[width=1\textwidth]{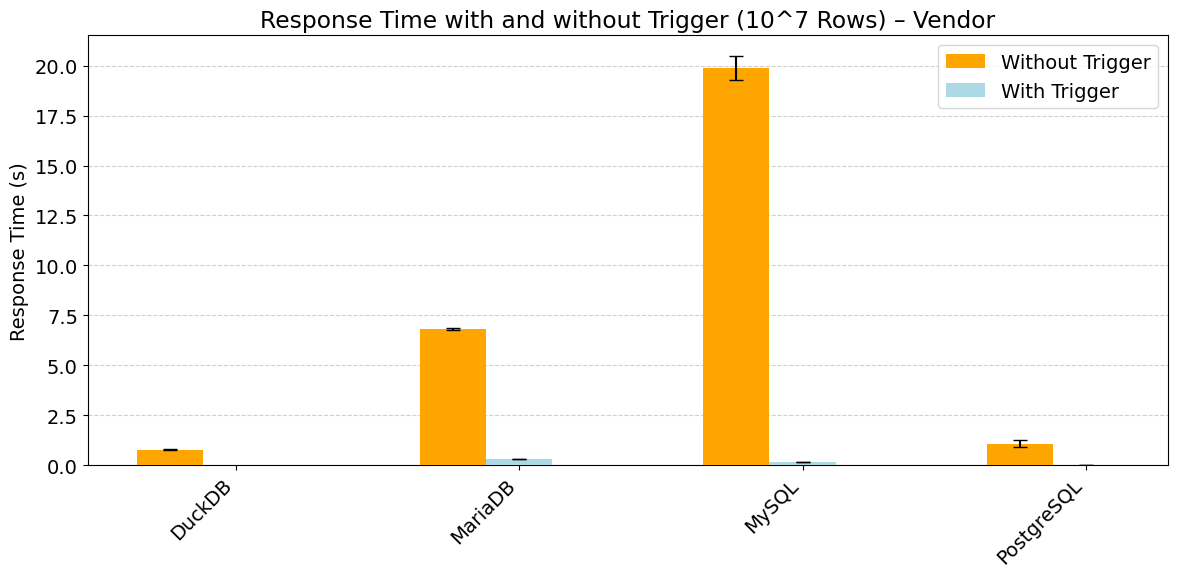}
  \caption{Impact of aggregate maintenance using \texttt{AFTER INSERT} triggers on Insertion performance on \texttt{vendor} on $10^7$ rows. DuckDB does not support triggers. The performance with triggers on MySQL, PostgreSQL and MariaDB is better than the performance without triggers.}
  \label{fig: trigger vendor 7}
\end{figure}

\begin{figure}[H]
  \centering
  \includegraphics[width=1\textwidth]{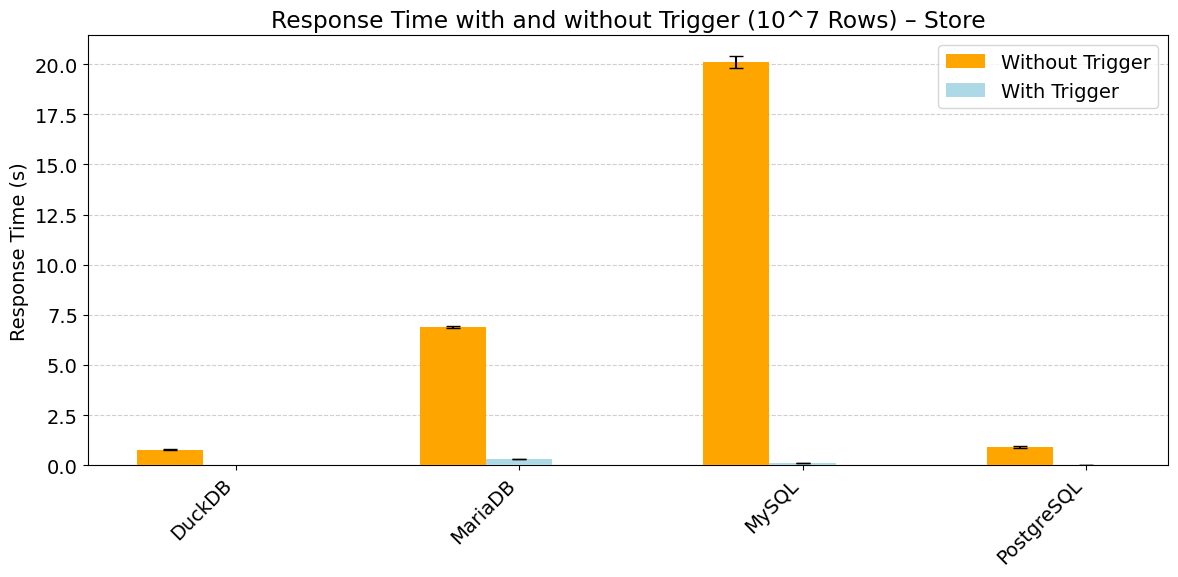}
  \caption{Impact of aggregate maintenance using \texttt{AFTER INSERT} triggers on Insertion performance on \texttt{store} on $10^7$ rows. DuckDB does not support triggers. The performance with triggers on MySQL, PostgreSQL and MariaDB is better than the performance without triggers.}
  \label{fig: trigger store 7}
\end{figure}

As shown in Figure~\ref{fig: trigger store 7} and Table~\ref{tab:triggerretrieve7}, store query performance also benefited greatly from precomputed aggregates. MySQL achieved the largest gain, with response time dropping from 20.1s to 0.092s (a 219× improvement). MariaDB followed, improving from 6.88s to 0.31s (151×), and PostgreSQL from 0.90s to 0.025s (36×). DuckDB, without trigger support, completed the store query in 0.78s.

All three systems with triggers showed improved consistency. Standard deviation decreased significantly: PostgreSQL reached 0.0010s, MySQL 0.0014s, and MariaDB 0.0019s, indicating highly stable performance.

\begin{table}[H]
  \centering
  \caption{Average Response Time (Standard Deviation) of Data Retrieval for Vendor and Store Queries With or Without Aggregate Maintenance Using \texttt{AFTER INSERT} Triggers on $10^7$ dataset (in seconds).}
  \label{tab:triggerretrieve7}
  \begin{tabular}{llcc}
    \toprule
    \textbf{Query Type} & \textbf{System} & \textbf{Without Trigger} & \textbf{With Trigger} \\
    \midrule
    \multirow{4}{*}{Vendor}
        & DuckDB      & 0.77 (0.019)     & --               \\
        & MariaDB     & 6.8 (0.032)      & 0.31 (0.0037)    \\
        & MySQL       & 20 (0.60)        & 0.16 (0.0036)    \\
        & PostgreSQL  & 1.1 (0.17)       & 0.026 (0.0013)   \\
    \midrule
    \multirow{4}{*}{Store}
        & DuckDB      & 0.78 (0.019)     & --               \\
        & MariaDB     & 6.9 (0.055)      & 0.31 (0.0019)    \\
        & MySQL       & 20 (0.30)        & 0.092 (0.0014)   \\
        & PostgreSQL  & 0.90 (0.042)     & 0.025 (0.0010)   \\
    \bottomrule
  \end{tabular}
\end{table}

\section[\appendixname~\thesection]{Vertical Partitioning: Quantitative Experimental Results for Scan Query}\label{vertical_partition-appendix}
\begin{figure}[H]
  \centering
  \includegraphics[width=0.8\textwidth]{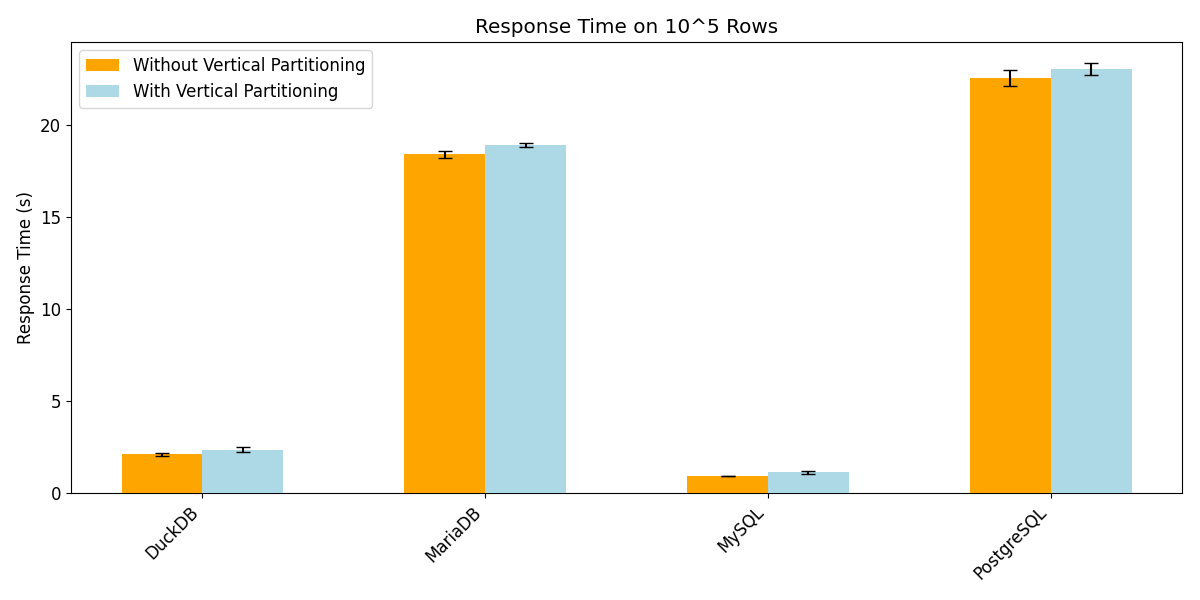}
  \caption{All engines perform better without vertical partitioning when accessing all fields.  }
  \label{fig: All fields vertical partioning results for 10^5 dataset}
\end{figure}

\begin{figure}[H]
  \centering
  \includegraphics[width=0.8\textwidth]{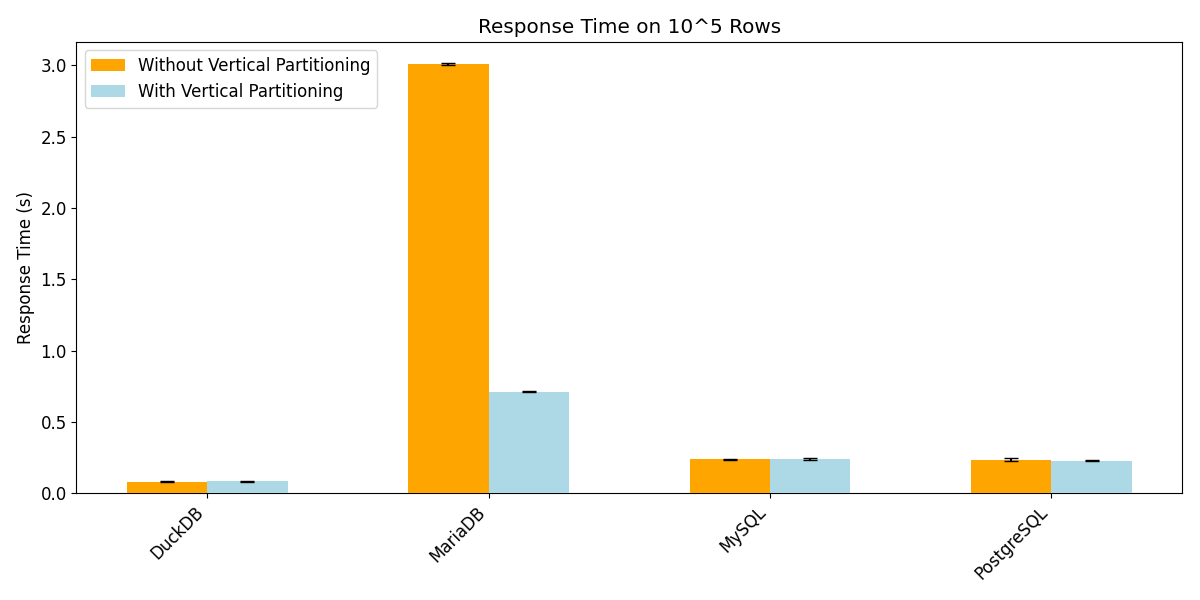}
  \caption{When accessing the subset of columns of one partition (account1), vertical partitioning improves performance for MariaDB. While, for DuckDB, MySQL, and PostgreSQL, vertical partitioning shows no benefit.  }
  \label{fig: Part fields vertical partioning results for 10^5 dataset}
\end{figure}

\textbf{On the $10^5$ row dataset} (Tables \ref{fig: All fields vertical partioning results for 10^5 dataset} and \ref{fig: Part fields vertical partioning results for 10^5 dataset}), vertical partitioning exhibited a clear query-dependent performance.  

When all columns were retrieved, vertical partitioning consistently degraded performance across most engines. Specifically,  DuckDB slowed from 2.12s to 2.35s, MariaDB from 18.42s to 18.94s, PostgreSQL from 22.55s to 23.05s, and MySQL from 0.93s to 1.13s. Standard deviations remained low, confirming the reliability of these results.

When queries accessed only the fields in \texttt{account1}, the impact of vertical partitioning varied by system. MariaDB, showed a large improvement (3.01s to 0.71s), suggesting it benefits substantially when fewer columns are scanned and data is vertically partitioned. DuckDB, PostgreSQL and MySQL remained mostly unaffected.

These results indicate that vertical partitioning can benefit partial-field access in some row-based systems like MariaDB but may not help systems like DuckDB and PostgreSQL that already  optimize columnar access internally.
\\
See Table \ref{tab:all-partition-10e5} and \ref{tab:part-partition-10e5} for the full set of average response times and standard deviations.

\begin{table}[H]
  \centering
  \caption{Average Response Times along with their Standard Deviations for accessing all fields on a $10^5$ Row Dataset (in seconds)}
  \label{tab:all-partition-10e5}

  \label{tab:all-partition-response-10e5}
  \begin{tabular}{lccc}
    \toprule
    \textbf{System} & \textbf{Without Vertical Partitioning} & \textbf{With Vertical Partitioning}\\
    \midrule
    DuckDB     & 2.12 (0.081) & 2.35 (0.14)\\
    MariaDB     & 18.42 (0.21) & 18.94 (0.11)\\
    MySQL       & 0.93 (0.0033) & 1.13 (0.096)\\
    PostgreSQL  & 22.55 (0.43) & 23.05 (0.32)\\
    \bottomrule
  \end{tabular}
\end{table}

\begin{table}[H]
  \centering
  \caption{Average Response Times along with their Standard Deviations for accessing partial fields on a $10^5$ Row Dataset (in seconds). In no case is the advantage of vertical partitioning statistically significant.}
  \label{tab:part-partition-10e5}

  \label{tab:part-partition-response-10e5}
  \begin{tabular}{lccc}
    \toprule
    \textbf{System} & \textbf{Without Vertical Partitioning} & \textbf{With Vertical Partitioning}\\
    \midrule
    DuckDB     & 0.082 (0.002) & 0.084 (0.002)\\
    MariaDB     & 3.01 (0.005) & 0.71 (0.003)\\
    MySQL       & 0.23 (0.0022) & 0.23 (0.0044)\\
    PostgreSQL  & 0.23 (0.010) & 0.22 (0.0041)\\
    \bottomrule
  \end{tabular}
\end{table}

\begin{figure}[H]
  \centering
  \includegraphics[width=0.8\textwidth]{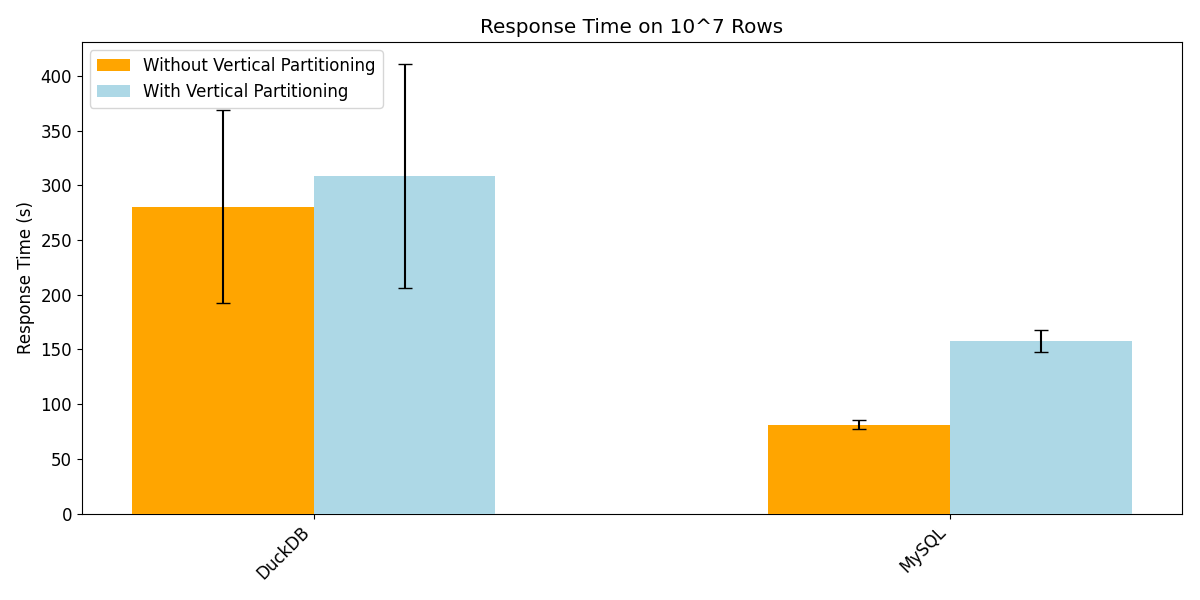}
  \caption{MySQL and DuckDB perform better without vertical partitioning when accessing all fields. }
  \label{fig: All fields vertical partioning results for 10^7 dataset}
\end{figure}

\begin{figure}[H]
  \centering
  \includegraphics[width=0.8\textwidth]{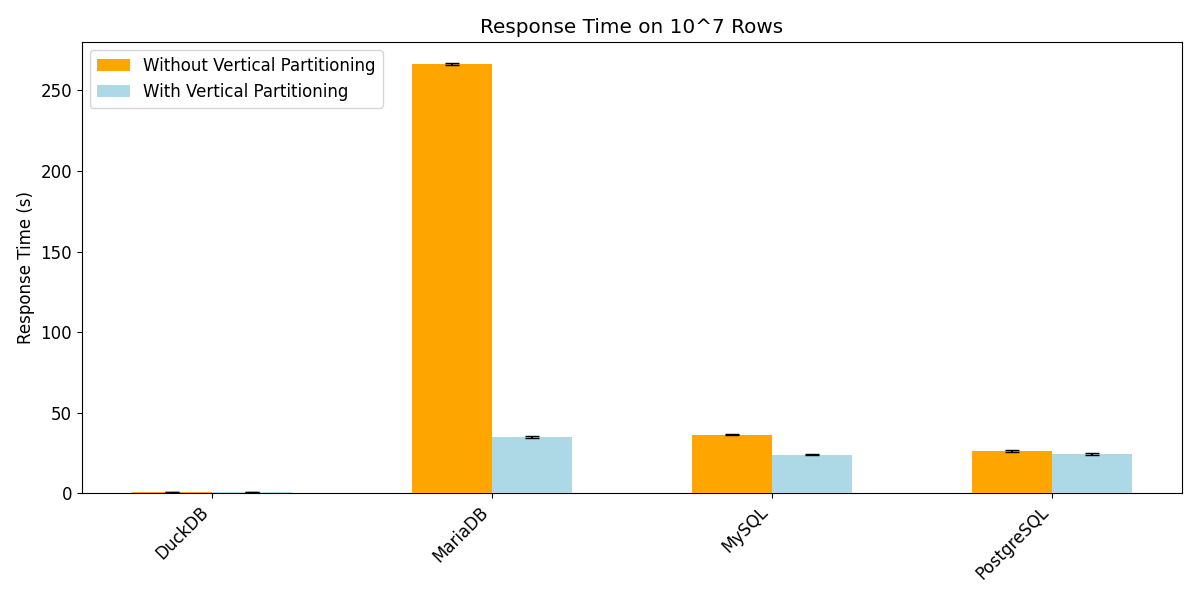}
  \caption{When accessing only the \texttt{account1} columns, the performance benefit of vertical partitioning is significant for MariaDB and MySQL, while DuckDB and PostgreSQL show modest changes.}
  \label{fig: Part fields vertical partioning results for 10^7 dataset}
\end{figure}

\textbf{On the $10^7$ row dataset} (Tables \ref{fig: All fields vertical partioning results for 10^7 dataset} and \ref{fig: Part fields vertical partioning results for 10^7 dataset}), vertical partitioning showed very engine-specific results.

When accessing all fields, DuckDB slowed from 280.43s to 308.21s, when using vertical partitioning, while the results for MySQL, reporting 81s without partitioning and 157.36s with partitioning.  MariaDB and PostgreSQL were not able to execute the query within the time constraint of 10 minutes  of this experiment. Standard deviations for DuckDB were relatively large, but the general trend remains consistent.

When queries accessed only the fields in \texttt{account1} (e.g., \texttt{id}, \texttt{balance}), vertical partitioning yielded mixed results. MariaDB showed a massive speedup (266.29s to 35.02s), suggesting that avoiding wide row scans has a significant benefit for its execution engine. PostgreSQL improved slightly (26.04s to 24.45s), DuckDB too saw a modest gain (0.74s to 0.62s), and MySQL showed significant improvement as well from 36.42s to 24.02s. 

See Table \ref{tab:all-partition-10e7} and \ref{tab:part-partition-10e7} for the full set of average response times and standard deviations.

\begin{table}[H]
  \centering
  \caption{Average Response Times along with their Standard Deviations when accessing all \texttt{account} fields on a $10^7$ Row Dataset (in seconds)}
  \label{tab:all-partition-10e7}

  \label{tab:all-partition-response-10e7}
  \begin{tabular}{lccc}
    \toprule
    \textbf{System} & \textbf{Without Vertical Partitioning} & \textbf{With Vertical Partitioning}\\
    \midrule
    DuckDB     & 280.43 (88.39) & 308.21 (102.39)\\
    MySQL       & 81.00 (4.17) & 157.36 (10.04)\\
    \bottomrule
  \end{tabular}
\end{table}

\begin{table}[H]
  \centering
  \caption{Average Response Times along with their Standard Deviations when accessing partial fields on a $10^7$ Row Dataset (in seconds)}
  \label{tab:part-partition-10e7}

  \label{tab:part-partition-response-10e7}
  \begin{tabular}{lccc}
    \toprule
    \textbf{System} & \textbf{Without Vertical Partitioning} & \textbf{With Vertical Partitioning}\\
    \midrule
    DuckDB     & 0.74 (0.17) & 0.62 (0.13)\\
    MariaDB     & 266.29 (0.54) & 35.02 (0.53)\\
    MySQL       & 36.42 (0.54) & 24.02 (0.56)\\
    PostgreSQL  & 26.04 (0.54) & 24.45 (0.80)\\
    \bottomrule
  \end{tabular}
\end{table}

\section[\appendixname~\thesection]{Vertical Partitioning: Quantitative Experimental Results for Point Query}\label{vertical_partition-appendix}
\begin{figure}[H]
  \centering
  \includegraphics[width=0.8\textwidth]{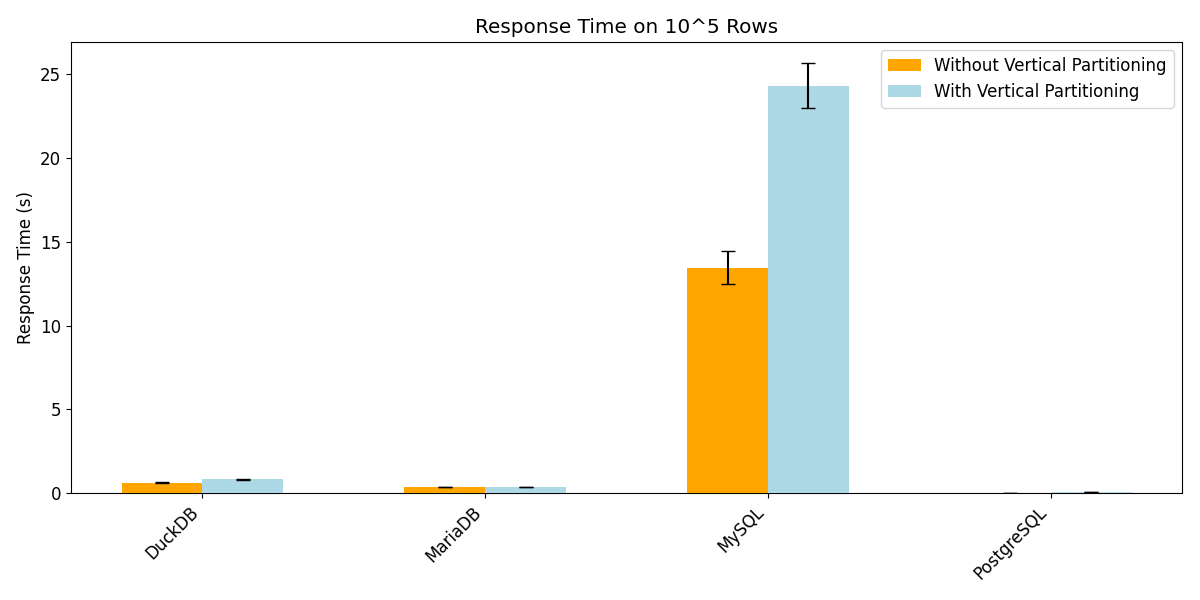}
  \caption{For the $10^5$ row dataset, all engines performed better in the case where none of the queries (0\%) used vertical partitioning.}
  \label{fig: Vertical Partitioning for n=0 results for 10^5 dataset}
\end{figure}

\begin{figure}[H]
  \centering
  \includegraphics[width=0.8\textwidth]{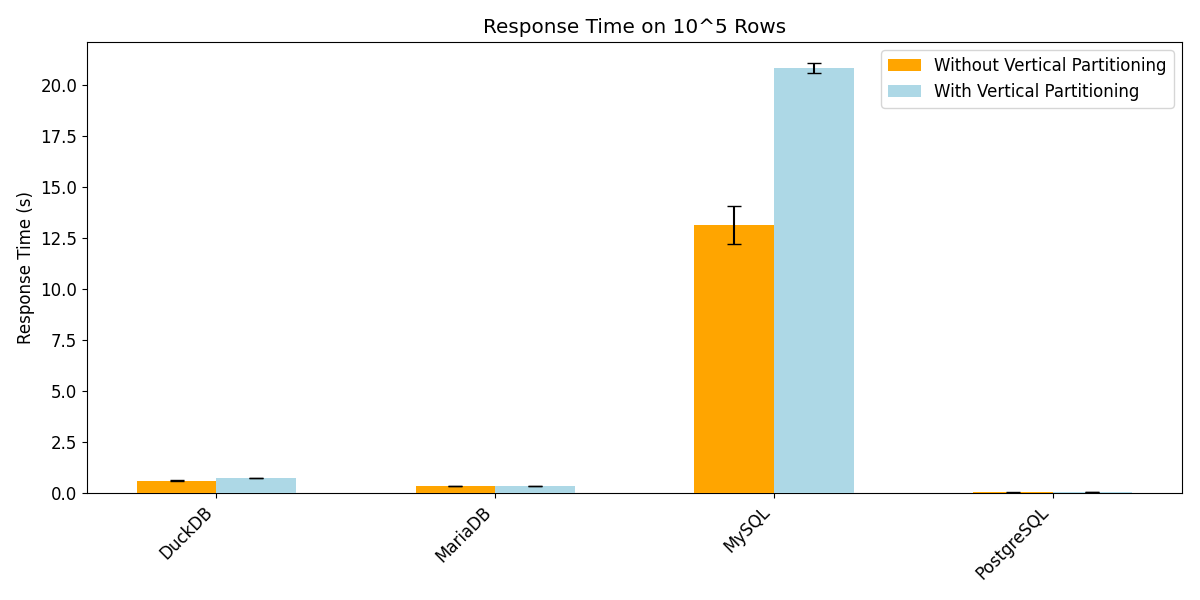}
  \caption{In the $10^5$ row dataset, queries executed without vertical partitioning performed better across all engines except MariaDB, where 20\% of the queries used vertical partitioning.}
  \label{fig: Vertical Partitioning for n=20 results for 10^5 dataset}
\end{figure}

\begin{figure}[H]
  \centering
  \includegraphics[width=0.8\textwidth]{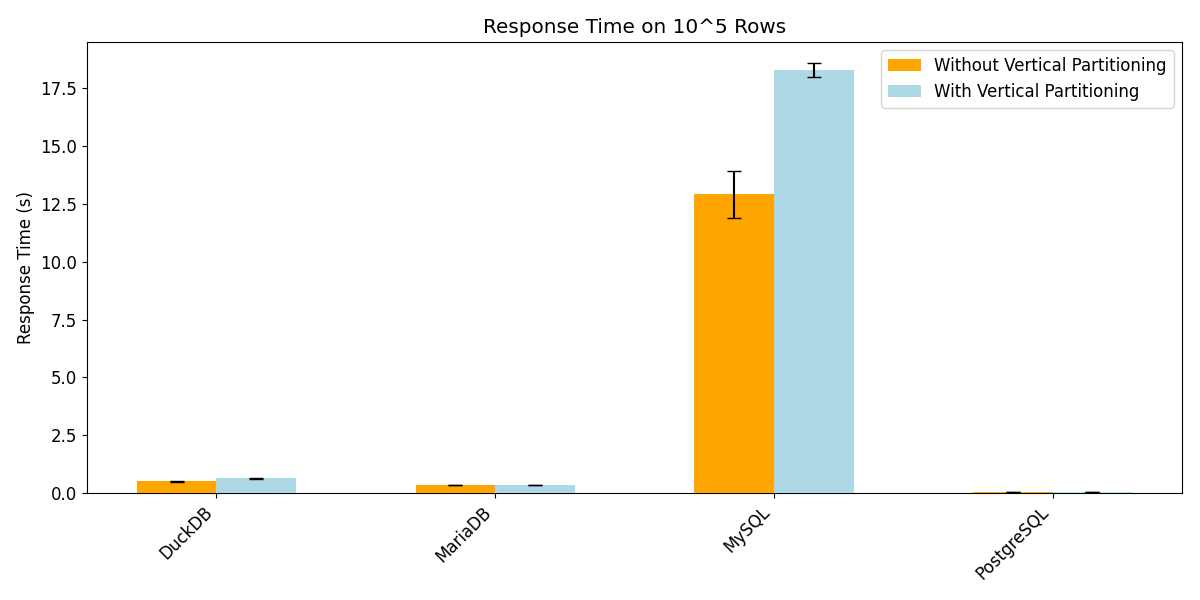}
  \caption{In the $10^5$  row dataset, queries executed without vertical partitioning performed better across all engines, under a workload where 40\% of the queries used vertical partitioning.}
  \label{fig: Vertical Partitioning for n=40 results for 10^5 dataset}
\end{figure}

\begin{figure}[H]
  \centering
  \includegraphics[width=0.8\textwidth]{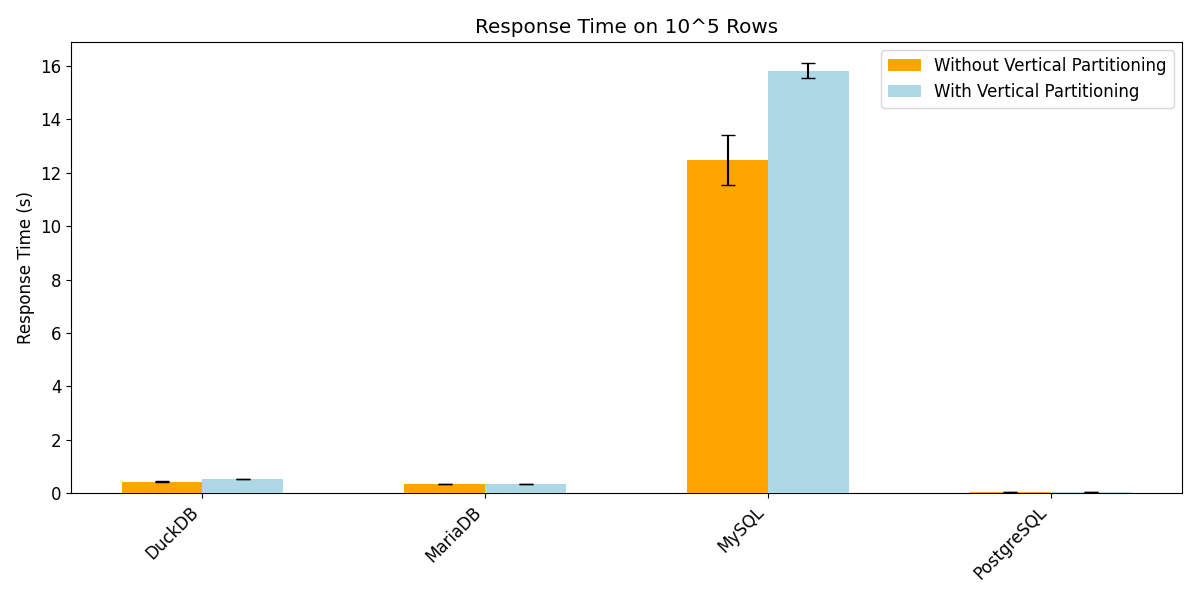}
  \caption{In the $10^5$ row dataset, queries executed without vertical partitioning performed better across all engines except MariaDB, under a workload where 60\% of the queries used vertical partitioning.}
  \label{fig: Vertical Partitioning for n=60 results for 10^5 dataset}
\end{figure}

\begin{figure}[H]
  \centering
  \includegraphics[width=0.8\textwidth]{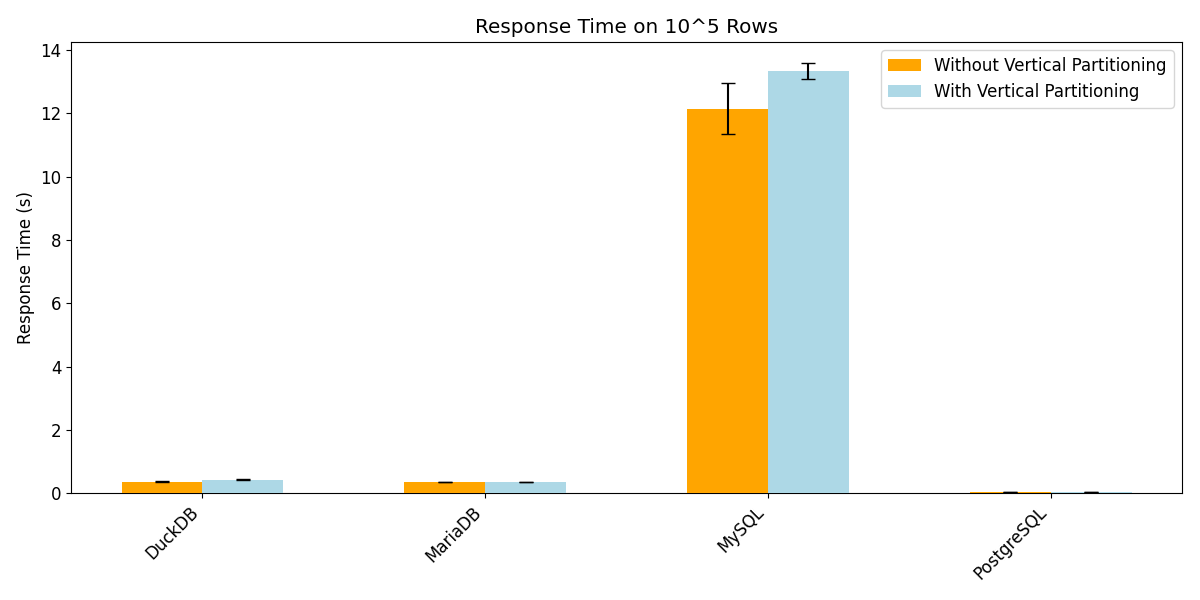}
  \caption{In the $10^5$ row dataset, queries executed without vertical partitioning performed better across all engines except MariaDB, under a workload where 80\% of the queries used vertical partitioning.}
  \label{fig: Vertical Partitioning for n=80 results for 10^5 dataset}
\end{figure}

\begin{figure}[H]
  \centering
  \includegraphics[width=0.8\textwidth]{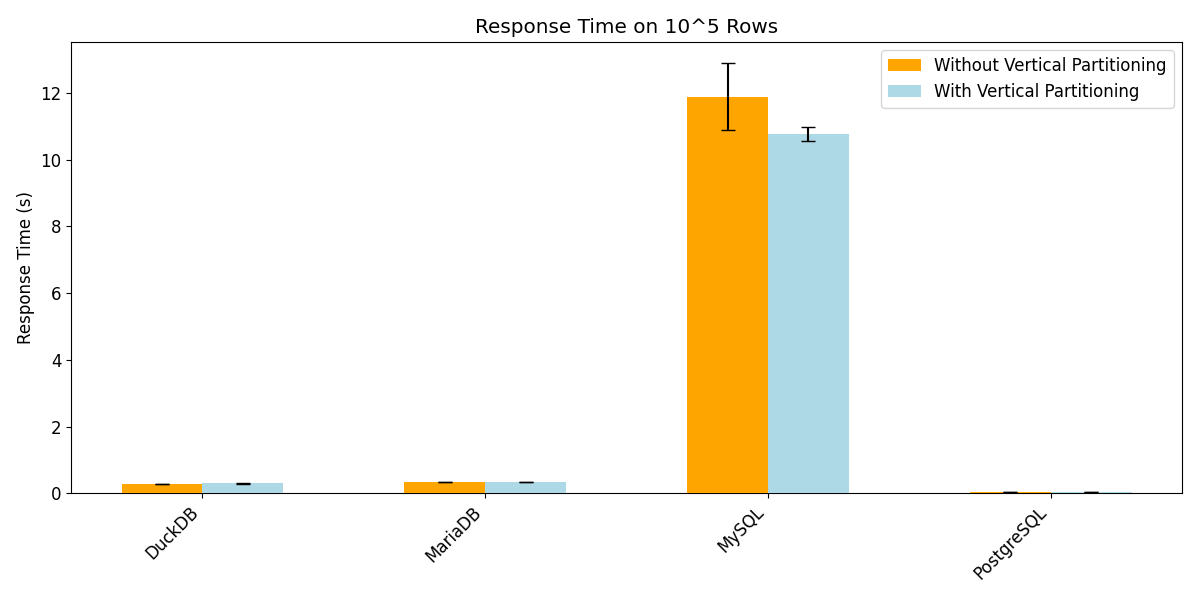}
  \caption{In the $10^5$ row dataset, queries executed with vertical partitioning performed better across all engines except DuckDB, under a workload where 100\% of the queries  accessed only the attributes of account1.}
  \label{fig: Vertical Partitioning for n=100 results for 10^5 dataset}
\end{figure}

\textbf{On the $10^{5}$ row dataset} (Table \ref{fig: Vertical Partitioning for n=0 results for 10^5 dataset}, \ref{fig: Vertical Partitioning for n=20 results for 10^5 dataset}, \ref{fig: Vertical Partitioning for n=40 results for 10^5 dataset}, \ref{fig: Vertical Partitioning for n=60 results for 10^5 dataset}, \ref{fig: Vertical Partitioning for n=80 results for 10^5 dataset} and \ref{fig: Vertical Partitioning for n=100 results for 10^5 dataset}), vertical partitioning exhibited varying effects across different database engines, with consistency observed across multiple query fractions (denoted  by f). Notably, MySQL demonstrates a pronounced effect as vertical partitioning query fractions increase.

\begin{itemize}
    \item  For DuckDB, vertical partitioning consistently increased response times across all query fractions, with the penalty decreasing as the query fraction increases. At f = 0, response time increased by approximately 29.7\% (from 0.64s to 0.83s). This performance gap narrowed as f increased, becoming negligible at f = 1, where response times were nearly identical (0.28s vs. 0.29s), suggesting vertical partitioning is detrimental when vertical partition queries dominates.
    \item For MariaDB, vertical partitioning had minimal impact across all query fractions, with response times remaining nearly identical (e.g., 0.35s vs. 0.36s at f = 0, and 0.34s for both at f = 100 ).
    \item For MySQL, vertical partitioning introduced a significant penalty at lower f, with the penalty decreasing as query fraction increases and proved to be slightly helpful only at f = 1.
    \item For PostgreSQL, vertical partitioning caused a mild penalty at f = 0 (0.038s vs. 0.047s), but this overhead diminished as f increased. By f = 1, both layouts performed equivalently (0.03s for both), indicating no significant overall impact.
\end{itemize}
See Table \ref{tab:vp-10e5} for the full set of average response times and standard deviations.


\begin{longtable}{lcc}
\caption{Average Response Times along with their Standard Deviations for vertical partitioning on multiple fractions of queries (denoted by $f$) that access only the attributes of account1 on a $10^5$ row dataset (in seconds).}
\label{tab:vp-10e5} \\

\toprule
\textbf{System} & \textbf{Without Vertical Partitioning} & \textbf{With Vertical Partitioning} \\
\midrule
\endfirsthead

\toprule
\textbf{System} & \textbf{Without Vertical Partitioning} & \textbf{With Vertical Partitioning} \\
\midrule
\endhead

\midrule
\multicolumn{3}{r}{\textit{Continued on next page}} \\
\endfoot

\bottomrule
\endlastfoot

\multicolumn{3}{l}{\textbf{f=0}} \\
\midrule
DuckDB      & 0.64 (0.01) & 0.83 (0.017) \\
MariaDB     & 0.35 (0.00085) & 0.36 (0.00067) \\
MySQL       & 13.5 (0.99) & 24.3 (1.3) \\
PostgreSQL  & 0.038 (0.00073) & 0.047 (0.00092) \\

\midrule
\multicolumn{3}{l}{\textbf{f=0.2}} \\
\midrule
DuckDB      & 0.61 (0.024) & 0.75 (0.011) \\
MariaDB     & 0.35 (0.00042) & 0.35 (0.00069) \\
MySQL       & 13.1 (0.92) & 20.8 (0.22) \\
PostgreSQL  & 0.037 (0.00048) & 0.044 (0.00063) \\

\midrule
\multicolumn{3}{l}{\textbf{f=0.4}} \\
\midrule
DuckDB      & 0.51 (0.01) & 0.64 (0.01) \\
MariaDB     & 0.34 (0.00091) & 0.35 (0.0019) \\
MySQL       & 12.9 (1.03) & 18.2 (0.029) \\
PostgreSQL  & 0.036 (0.0010) & 0.042 (0.0005) \\

\midrule
\multicolumn{3}{l}{\textbf{f=0.6}} \\
\midrule
DuckDB      & 0.44 (0.0099) & 0.53 (0.011) \\
MariaDB     & 0.34 (0.0014) & 0.34 (0.00063) \\
MySQL       & 12.5 (0.94) & 15.8 (0.27) \\
PostgreSQL  & 0.035 (0.00073) & 0.039 (0.00051) \\

\midrule
\multicolumn{3}{l}{\textbf{f=0.8}} \\
\midrule
DuckDB      & 0.36 (0.0076) & 0.43 (0.018) \\
MariaDB     & 0.34 (0.00081) & 0.34 (0.00056) \\
MySQL       & 12.1 (0.81) & 13.3 (0.24) \\
PostgreSQL  & 0.035 (0.0078) & 0.037 (0.00071) \\

\midrule
\multicolumn{3}{l}{\textbf{f=1}} \\
\midrule
DuckDB      & 0.28 (0.0083) & 0.29 (0.012) \\
MariaDB     & 0.34 (0.0016) & 0.34 (0.0013) \\
MySQL       & 11.8 (1.00) & 10.7 (0.21) \\
PostgreSQL  & 0.034 (0.00048) & 0.034 (0.00052) \\

\end{longtable}

\begin{figure}[H]
  \centering
  \includegraphics[width=0.8\textwidth]{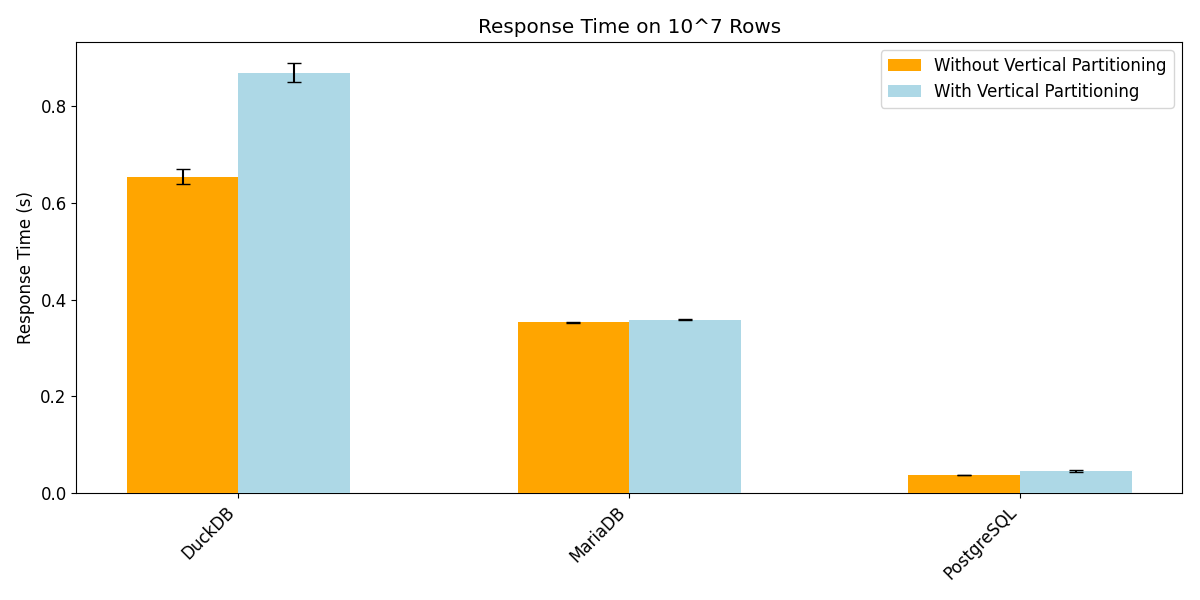}
  \caption{In the $10^7$ row dataset, queries executed without vertical partitioning performed better across all engines, under a workload where 0\% of the queries  accessed only the attributes of account1.}
  \label{fig: Vertical Partitioning for n=0 results for 10^7 dataset}
\end{figure}

\begin{figure}[H]
  \centering
  \includegraphics[width=0.8\textwidth]{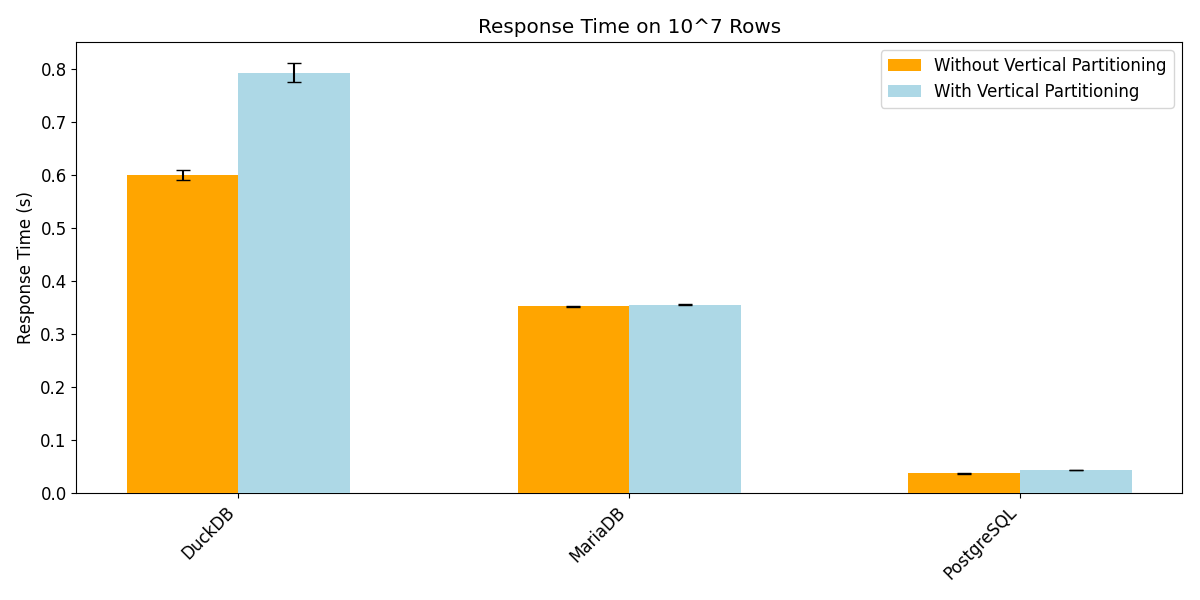}
  \caption{In the $10^7$ row dataset, queries executed without vertical partitioning performed better across all engines, under a workload where 20\% of the queries  accessed only the attributes of account1.}
  \label{fig: Vertical Partitioning for n=20 results for 10^7 dataset}
\end{figure}

\begin{figure}[H]
  \centering
  \includegraphics[width=0.8\textwidth]{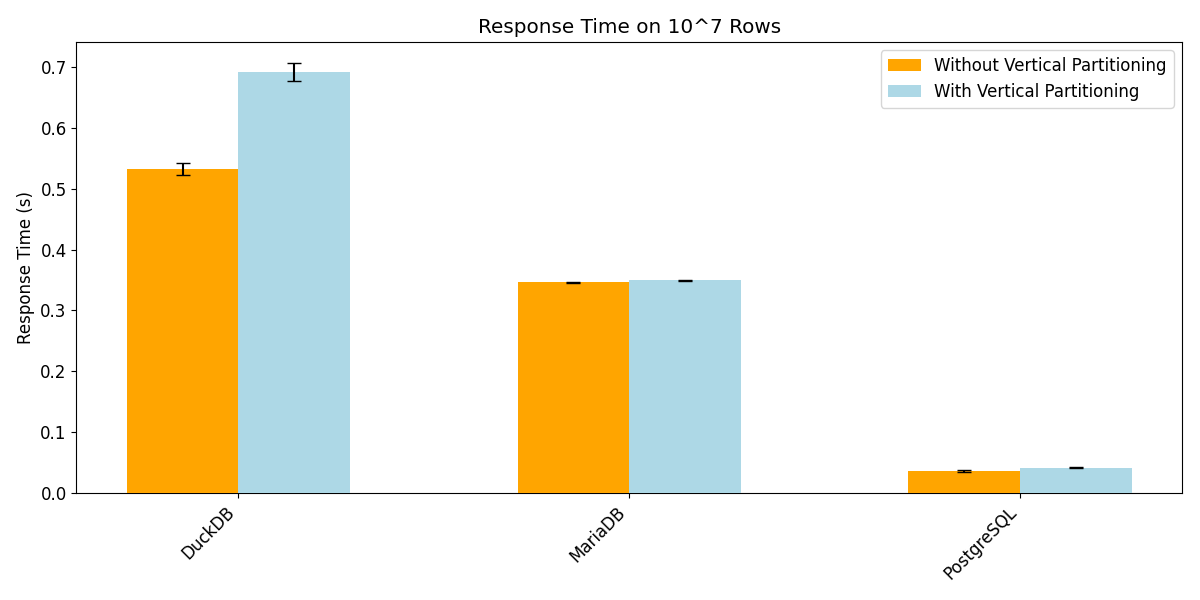}
  \caption{In the $10^7$ row dataset, queries executed without vertical partitioning performed better across all engines except MariaDB, under a workload where 40\% of the queries  accessed only the attributes of account1.}
  \label{fig: Vertical Partitioning for n=40 results for 10^7 dataset}
\end{figure}

\begin{figure}[H]
  \centering
  \includegraphics[width=0.8\textwidth]{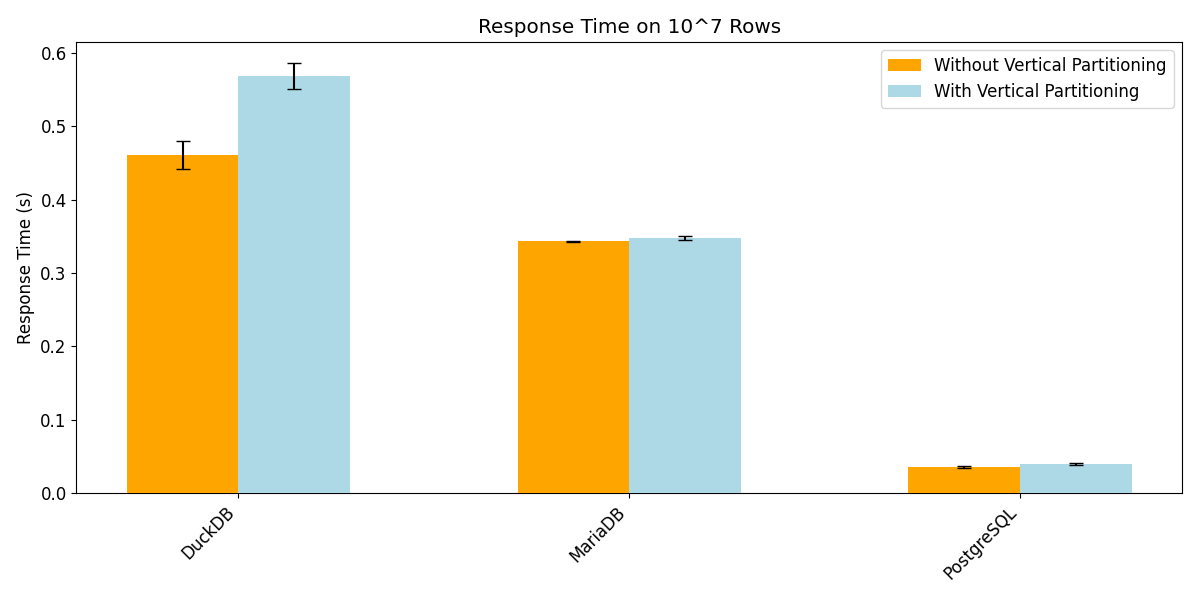}
  \caption{In the $10^7$ row dataset, queries executed without vertical partitioning performed better across all engines, under a workload where 60\% of the queries  accessed only the attributes of account1.}
  \label{fig: Vertical Partitioning for n=60 results for 10^7 dataset}
\end{figure}

\begin{figure}[H]
  \centering
  \includegraphics[width=0.8\textwidth]{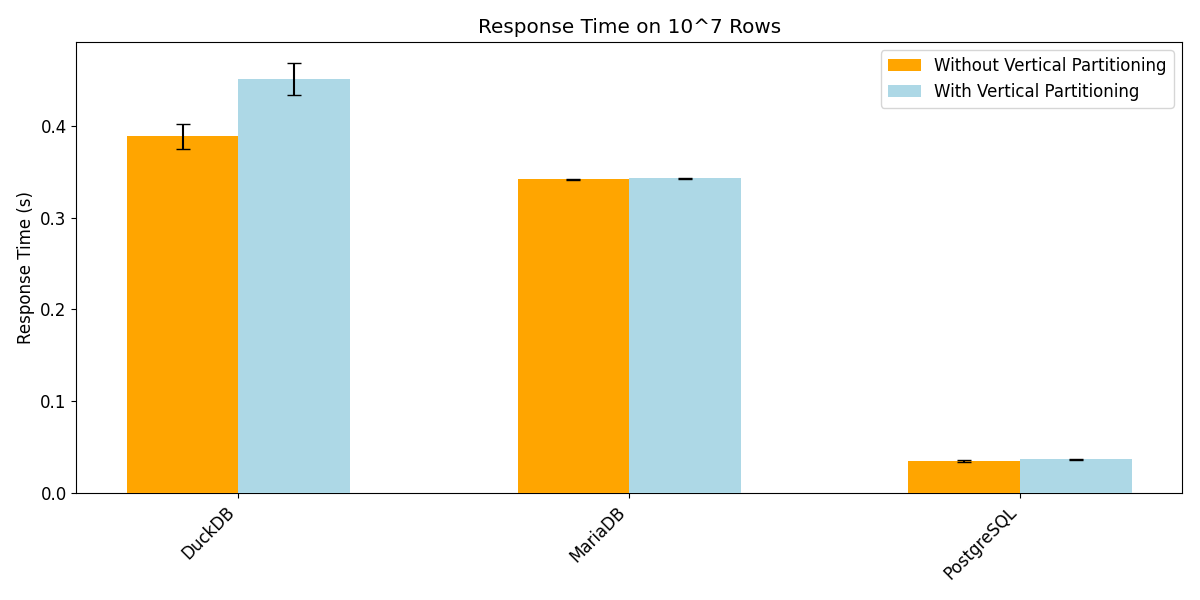}
  \caption{In the $10^7$ row dataset, queries executed without vertical partitioning performed better across all engines except MariaDB, under a workload where 80\% of the queries  accessed only the attributes of account1.}
  \label{fig: Vertical Partitioning for n=80 results for 10^7 dataset}
\end{figure}

\begin{figure}[H]
  \centering
  \includegraphics[width=0.8\textwidth]{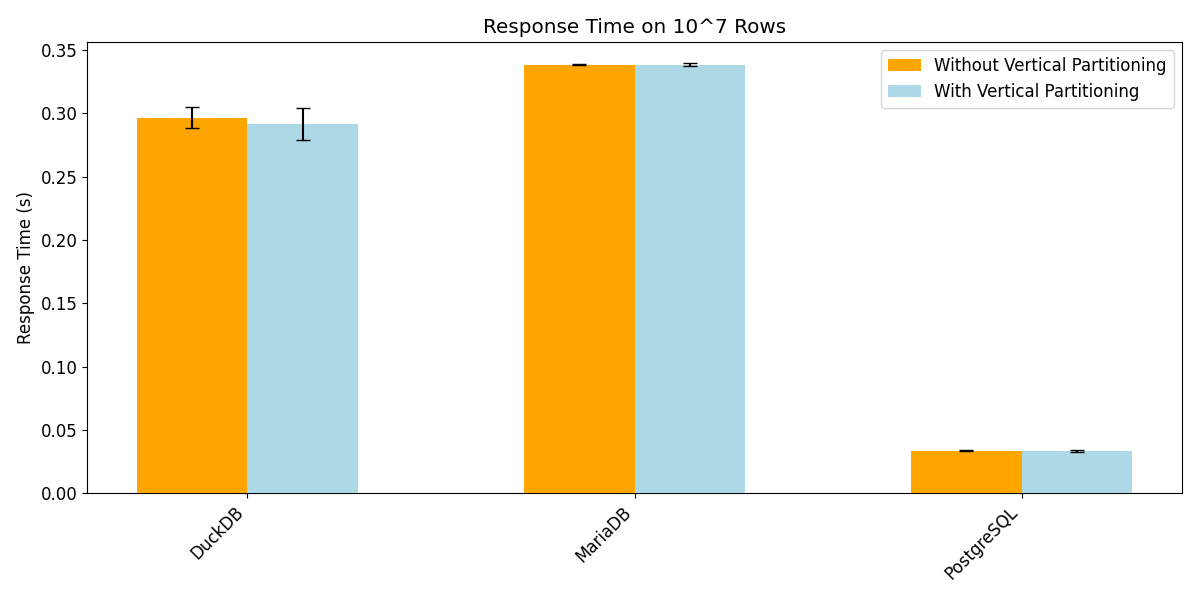}
  \caption{In the $10^7$ row dataset, queries executed with vertical partitioning performed better for DuckDB and showed the same performance for MariaDB, under a workload where 100\% of the queries  accessed only the attributes of account1.}
  \label{fig: Vertical Partitioning for n=100 results for 10^7 dataset}
\end{figure}

\textbf{On the $10^{7}$ row dataset} (Table \ref{fig: Vertical Partitioning for n=0 results for 10^7 dataset}, \ref{fig: Vertical Partitioning for n=20 results for 10^7 dataset}, \ref{fig: Vertical Partitioning for n=40 results for 10^7 dataset}, \ref{fig: Vertical Partitioning for n=60 results for 10^7 dataset}, \ref{fig: Vertical Partitioning for n=80 results for 10^7 dataset} and \ref{fig: Vertical Partitioning for n=100 results for 10^7 dataset}), vertical partitioning showed varied impacts across different database engines, with effects generally consistent across multiple query fractions. DuckDB exhibited the most noticeable impact as query fractions increased, while other systems showed minimal or no significant changes. MySQL was excluded from the comparison as it failed to execute the query within the time constraint of 10 minutes.

\begin{itemize}
    \item For DuckDB, vertical partitioning increased response times across most values of f, with the penalty most pronounced at lower query fractions (e.g., f = 0 to f = 0.4). At f = 0, response time increased by approximately 33.8\% (from 0.65s to 0.87s). The performance gap narrowed as f increased, becoming negligible at f = 1, where response times were nearly identical (0.30s vs. 0.29s), indicating vertical partitioning is never helpful. This is unsurprising because DuckDB has a columnar architecture.
    \item For MariaDB, vertical partitioning had a negligible impact across all query fractions, with response times remaining nearly identical (e.g., 0.35s vs. 0.36s at f = 0, and 0.34s for both at f = 1). 
    \item For PostgreSQL, vertical partitioning consistently introduced a penalty, with response times increasing from 0.038s to 0.047s at f = 0. This overhead diminished as f increased, with both layouts performing equivalently at f = 1 (0.034s for both), suggesting no significant helpful effect.
\end{itemize}


\begin{longtable}{lcc}
\caption{Average Response Times (in seconds) along with Standard Deviations for vertical partitioning on a $10^7$ row dataset. The table is organized based on a parameter $f$, where $f$ indicates the fraction of queries that access only the attributes of account1.}
\label{tab:vp-10e7} \\
\toprule
\textbf{System} & \textbf{Without Vertical Partitioning} & \textbf{With Vertical Partitioning} \\
\midrule
\endfirsthead

\toprule
\textbf{System} & \textbf{Without Vertical Partitioning} & \textbf{With Vertical Partitioning} \\
\midrule
\endhead

\midrule
\multicolumn{3}{r}{\textit{Continued on next page}} \\
\endfoot

\bottomrule
\endlastfoot

\multicolumn{3}{l}{\textbf{f=0}} \\
\midrule
DuckDB      & 0.65 (0.015) & 0.87 (0.019) \\
MariaDB     & 0.35 (0.0011) & 0.36 (0.00052) \\
PostgreSQL  & 0.038 (0.00063) & 0.047 (0.0023) \\

\midrule
\multicolumn{3}{l}{\textbf{f=0.2}} \\
\midrule
DuckDB      & 0.60 (0.010) & 0.79 (0.018) \\
MariaDB     & 0.35 (0.00067) & 0.36 (0.00069) \\
PostgreSQL  & 0.037 (0.0011) & 0.044 (0.00063) \\

\midrule
\multicolumn{3}{l}{\textbf{f=0.4}} \\
\midrule
DuckDB      & 0.53 (0.0097) & 0.69 (0.015) \\
MariaDB     & 0.35 (0.00082) & 0.35 (0.00052) \\
PostgreSQL  & 0.037 (0.0012) & 0.042 (0.00069) \\

\midrule
\multicolumn{3}{l}{\textbf{f=0.6}} \\
\midrule
DuckDB      & 0.46 (0.019) & 0.57 (0.017) \\
MariaDB     & 0.34 (0.00032) & 0.35 (0.0027) \\
PostgreSQL  & 0.036 (0.00096) & 0.04 (0.0012) \\

\midrule
\multicolumn{3}{l}{\textbf{f=0.8}} \\
\midrule
DuckDB      & 0.39 (0.013) & 0.45 (0.017) \\
MariaDB     & 0.34 (0.00063) & 0.34 (0.00069) \\
PostgreSQL  & 0.035 (0.00084) & 0.037 (0.00047) \\

\midrule
\multicolumn{3}{l}{\textbf{f=1}} \\
\midrule
DuckDB      & 0.30 (0.0082) & 0.29 (0.012) \\
MariaDB     & 0.34 (0.00071) & 0.34 (0.0011) \\
PostgreSQL  & 0.034 (0.00067) & 0.034 (0.000069) \\

\end{longtable}

\section[\appendixname~\thesection]{Denormalization: Quantitative Experimental Results}\label{denormalization-appendix}
\begin{figure}[H]
  \centering
  \includegraphics[width=0.8\textwidth]{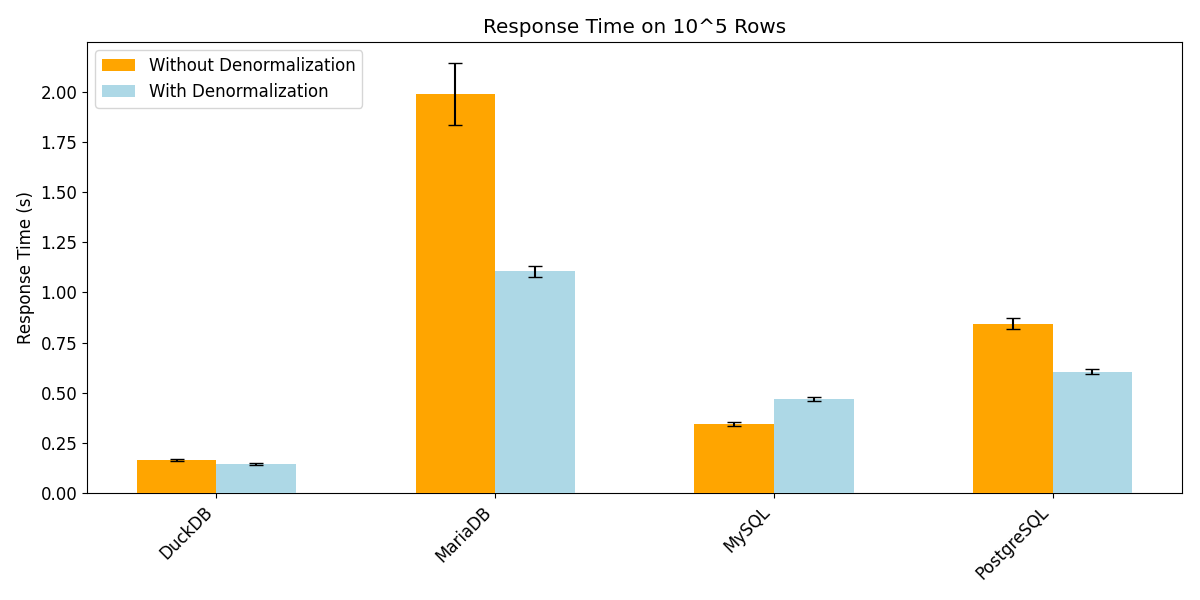}
  \caption{MariaDB shows a most pronounced advantage for denormalization, while MySQL shows a disadvantage for denormalization. }
  \label{fig: Denormalization results for 10^5 dataset}
\end{figure}

\textbf{On the $10^5$ row dataset} (Figure \ref{fig: Denormalization results for 10^5 dataset}), denormalization enjoyed an advantage  on DuckDB, MariaDB and PostgreSQL.
\emph{DuckDB} executed the denormalized query \(1.13\times\) faster than the normalized one.  
\emph{MariaDB} posted the largest gain, achieving a \(1.80\times\) speed-up, while \emph{PostgreSQL} improved by \(1.39\times\) and showed a small reduction in variability.  
By contrast, \emph{MySQL} slowed down by \(1.37\times\); however, its run-to-run variance fell by more than an order of magnitude, indicating that the denormalized plan was consistently, but uniformly slower.  
See Table~\ref{tab:denormalized-10e5} for the full set of average response times and standard deviations.

\begin{table}[H]
  \centering
  \caption{Average Response Times along with their standard deviations for Normalized and Denormalized Experiments on a $10^5$ Row Dataset (in seconds)}
  \label{tab:denormalized-10e5}

  \label{tab:denormalized-response-10e5}
  \begin{tabular}{lccc}
    \toprule
    \textbf{System} & \textbf{Normalized} & \textbf{Denormalized} \\
    \midrule
    DuckDB      & 0.17 (0.0050) & 0.15 (0.0042)\\
    MariaDB     & 1.99 (0.15) & 1.11 (0.027)\\
    MySQL       & 0.34 (0.0092) & 0.47 (0.0089)\\
    PostgreSQL  & 0.84 (0.027) & 0.61 (0.013)\\
    \bottomrule
  \end{tabular}
\end{table}

\begin{figure}[H]
  \centering
  \includegraphics[width=0.8\textwidth]{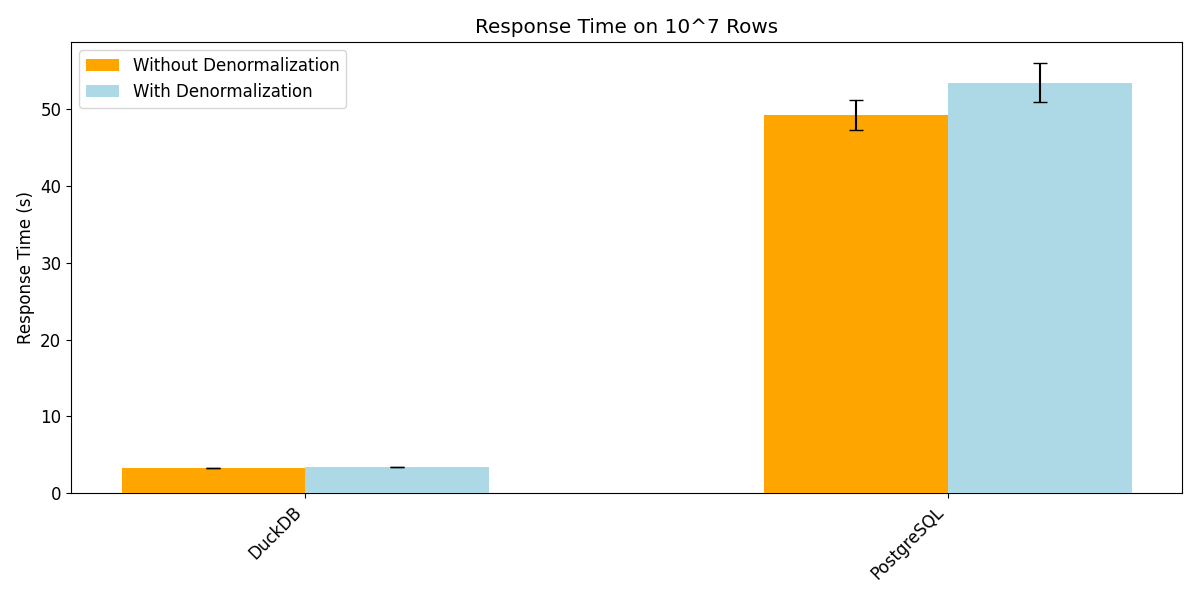}
  \caption{PostgreSQL and DuckDB shows a disadvantage of using denormalization. MariaDB and MySQL could not support the creation of foreign keys on such a big dataset.}
  \label{fig: Denormalization results for 10^7 dataset}
\end{figure}

\textbf{On the $10^7$ row dataset} (Figure \ref{fig: Denormalization results for 10^7 dataset}), denormalization offers no performance gain. 
\emph{DuckDB} shows a slight slowdown with a small reduction in run-to-run spread, whereas \emph{PostgreSQL} incurs a more noticeable latency penalty accompanied by greater variability.  
\emph{MySQL} and \emph{MariaDB} are absent from this comparison because they could not create the foreign-keys needed for the normalized schema at this size. See Table~\ref{tab:denormalized-10e7} for the full set of average response times and standard deviations.

\begin{table}[H]
  \centering
  \caption{Average Response Times along with their Standard Deviations for Normalized and Denormalized Experiments on a $10^7$ Row Dataset (in seconds)}
  \label{tab:denormalized-10e7}

  \label{tab:denormalized-response-10e7}
  \begin{tabular}{lccc}
    \toprule
    \textbf{System} & \textbf{Normalized} & \textbf{Denormalized} \\
    \midrule
    DuckDB      & 3.30 (0.019) & 3.40 (0.017) \\
    PostgreSQL  & 49.27 (1.90) & 53.49 (2.50) \\
    \bottomrule
  \end{tabular}

\end{table}

\section[\appendixname~\thesection]{Connection Pooling: Quantitative Experimental Results}\label{connection_pooling-appendix}
\begin{figure}[H]
  \centering
  \includegraphics[width=0.8\textwidth]{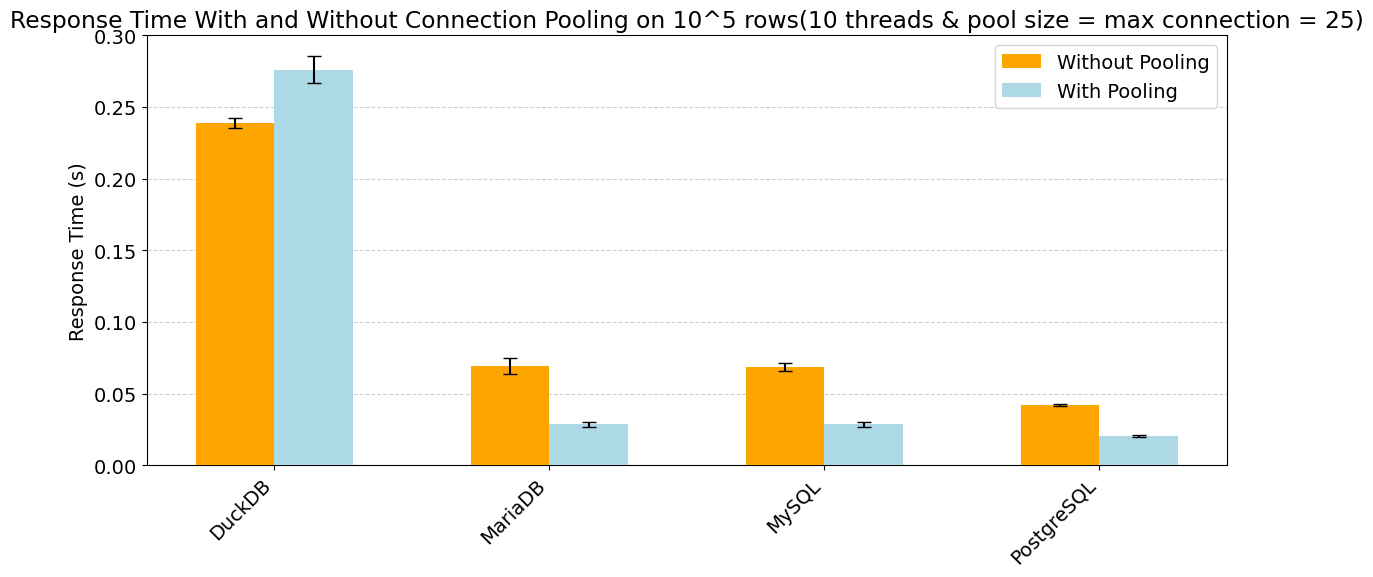}
  \caption{Under 10-thread workloads, the average response time with connection pooling (pool size = max connections = 25) outperforms simple connections across all tested systems except DuckDB on $10^5$ rows.}
  \label{fig: connection pooling t10p25 results for 10^5 dataset}
\end{figure}

\begin{figure}[H]
  \centering
  \includegraphics[width=0.8\textwidth]{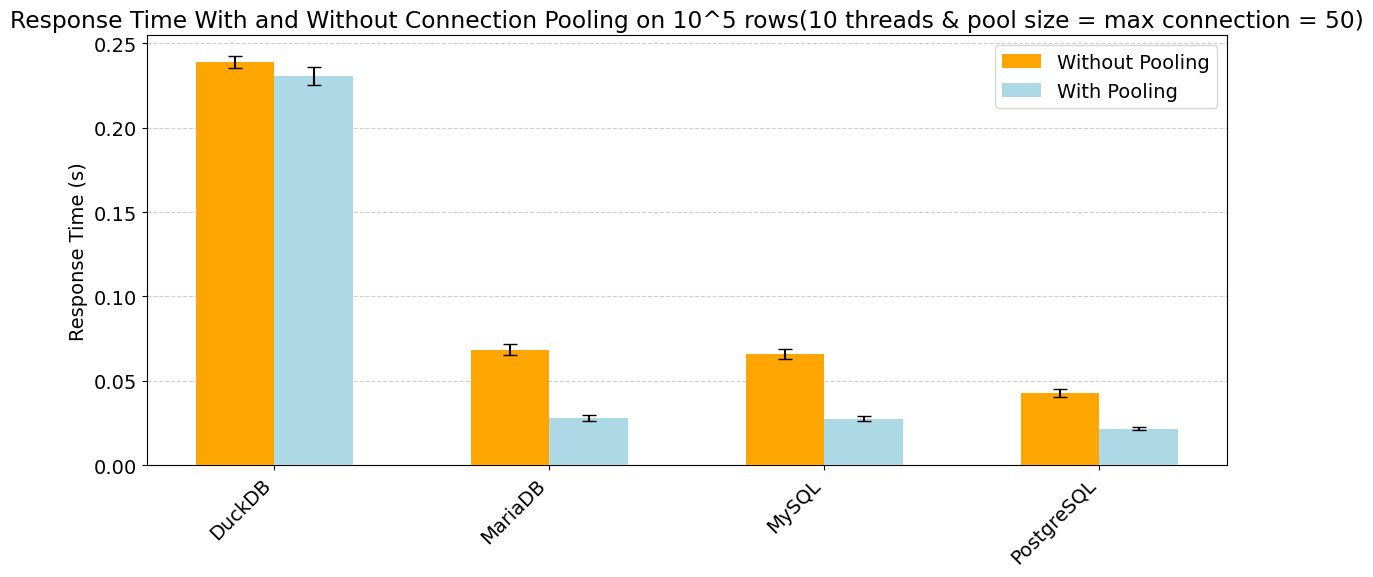}
  \caption{Under 10-thread workloads, the average response time with connection pooling (pool size = max connections = 50) outperforms simple connections across all tested systems: DuckDB, MariaDB, MySQL and PostgreSQL on $10^{5}$ rows.}
  \label{fig: connection pooling t10p50 results for 105 dataset}
\end{figure}

\begin{figure}[H]
  \centering
  \includegraphics[width=0.8\textwidth]{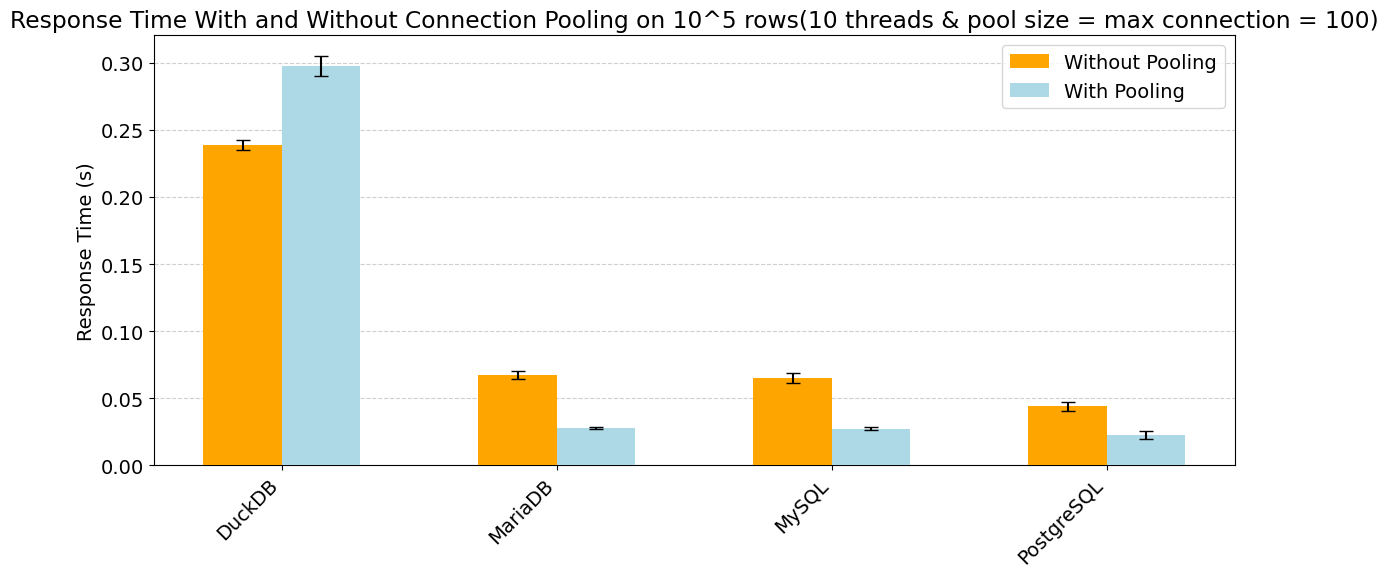}
  \caption{Under 10-thread workloads, the average response time with connection pooling (pool size = max connections = 100) outperforms simple connections across all tested systems except DuckDB on $10^5$ rows.}
  \label{fig: connection pooling t10p100 results for 10^5 dataset}
\end{figure}

\textbf{On the $10^5$ row dataset}(Table~\ref{tab:conn-pooling5}), when there are 10 client threads (Figure~\ref{fig: connection pooling t10p25 results for 10^5 dataset}, Figure~\ref{fig: connection pooling t10p50 results for 105 dataset} and Figure~\ref{fig: connection pooling t10p100 results for 10^5 dataset}), almost all of the systems enjoy improved performance under pooling. MariaDB achieves the largest improvement when the pool size and max\_connections are set equal to 50, reducing the response time by a factor of 2.41–2.44. MySQL improves by 2.38–2.41×, and PostgreSQL by up to 2.06×. While DuckDB shows a modest 1.04× improvement with pooling when the pool size is set to 50, its performance degrades when the pool size is either increased or decreased. The improvements are relatively modest at this scale due to the low concurrency level and the fact that the thread count does not yet challenge the system's connection limit.

\begin{figure}[H]
  \centering
  \includegraphics[width=0.8\textwidth]{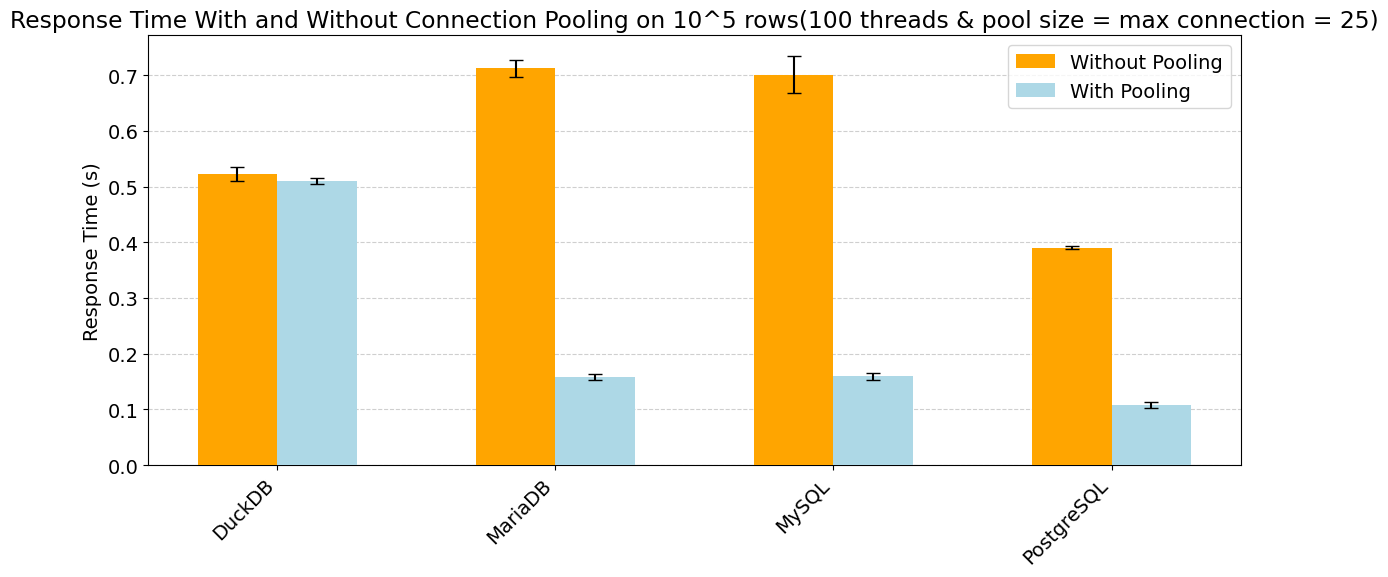}
  \caption{Under 100-thread workloads, the average response time with connection pooling (pool size = max connections = 25) outperforms simple connections across all tested systems: DuckDB, MariaDB, MySQL, and PostgreSQL on $10^5$ rows.}
  \label{fig: connection pooling t100p25 results for 10^5 dataset}
\end{figure}

\begin{figure}[H]
  \centering
  \includegraphics[width=0.8\textwidth]{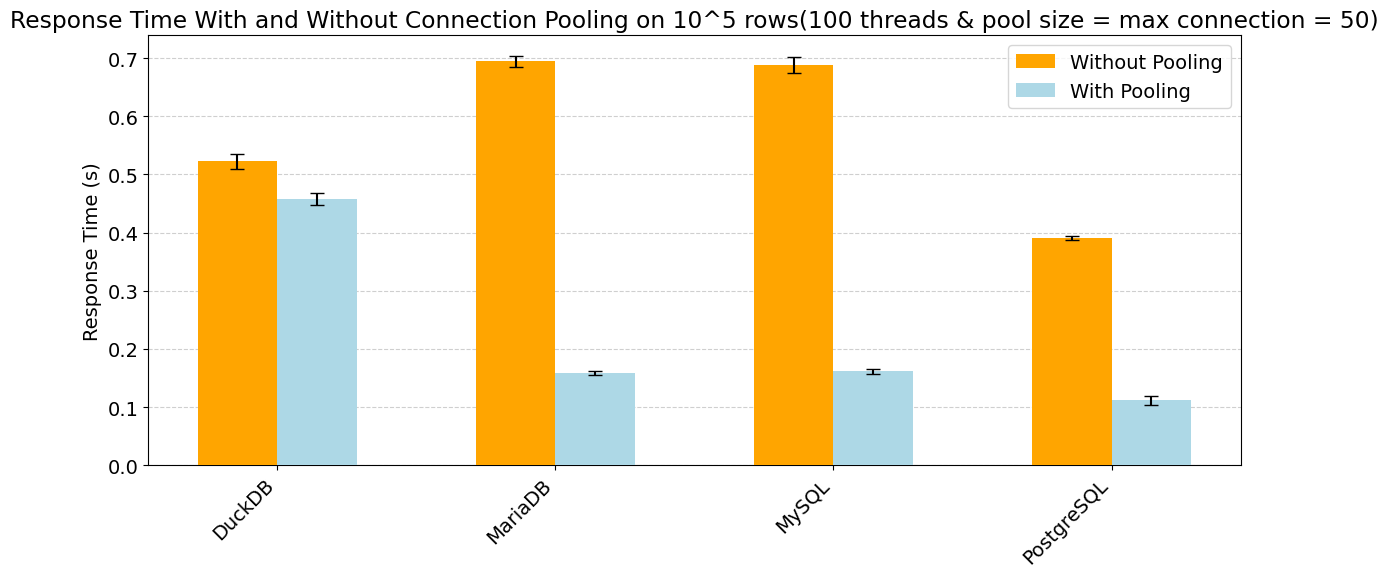}
  \caption{Under 100-thread workloads, the average response time with connection pooling (pool size = max connections = 50) outperforms simple connections across all tested systems: DuckDB, MariaDB, MySQL, and PostgreSQL on $10^5$ rows.}
  \label{fig: connection pooling t100p50 results for 10^5 dataset}
\end{figure}

\begin{figure}[H]
  \centering
  \includegraphics[width=0.8\textwidth]{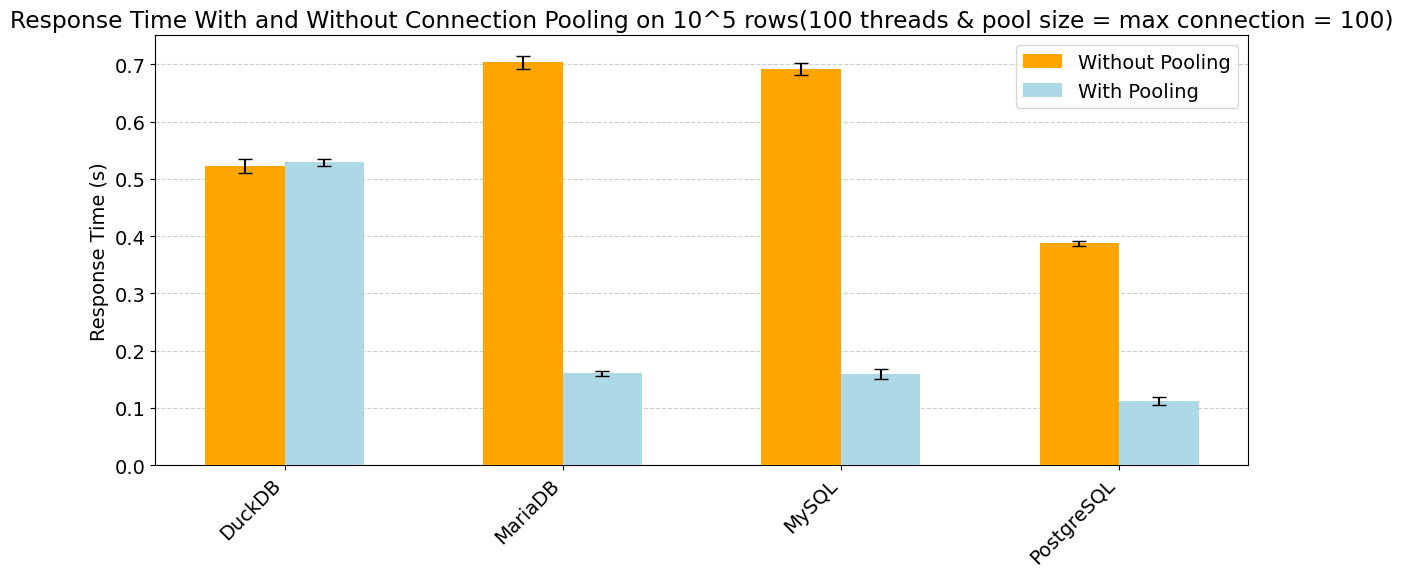}
  \caption{Under 100-thread workloads, the average response time with connection pooling (pool size = max connections = 100) outperforms simple connections across all tested systems except DuckDB on $10^5$ rows.}
  \label{fig: connection pooling t100p100 results for 10^5 dataset}
\end{figure}

As the concurrency increases to 100 threads (Figure~\ref{fig: connection pooling t100p25 results for 10^5 dataset}, Figure~\ref{fig: connection pooling t100p50 results for 10^5 dataset} and Figure~\ref{fig: connection pooling t100p100 results for 10^5 dataset}), the advantages of pooling become significantly more pronounced. MariaDB, in particular, shows a dramatic performance gain of up to 4.48× with a pool size of 25. MySQL and PostgreSQL see their highest improvements of 4.39× and 3.6× with a pool size and max connection of 25, respectively. DuckDB remains relatively stable, with pooling yielding only modest improvements (up to 1.14× at a pool size of 50). At a pool size of 25, the gain is about 1.02×, while performance essentially breaks even ($\approx 0.99\times$) when the pool size reaches 100. Interestingly, increasing the pool size from 25 to 50 or 100 for PostgreSQL and MariaDB continues to yield substantial gains, while DuckDB shows little to no benefit from larger pool sizes.

\begin{figure}[H]
  \centering
  \includegraphics[width=0.8\textwidth]{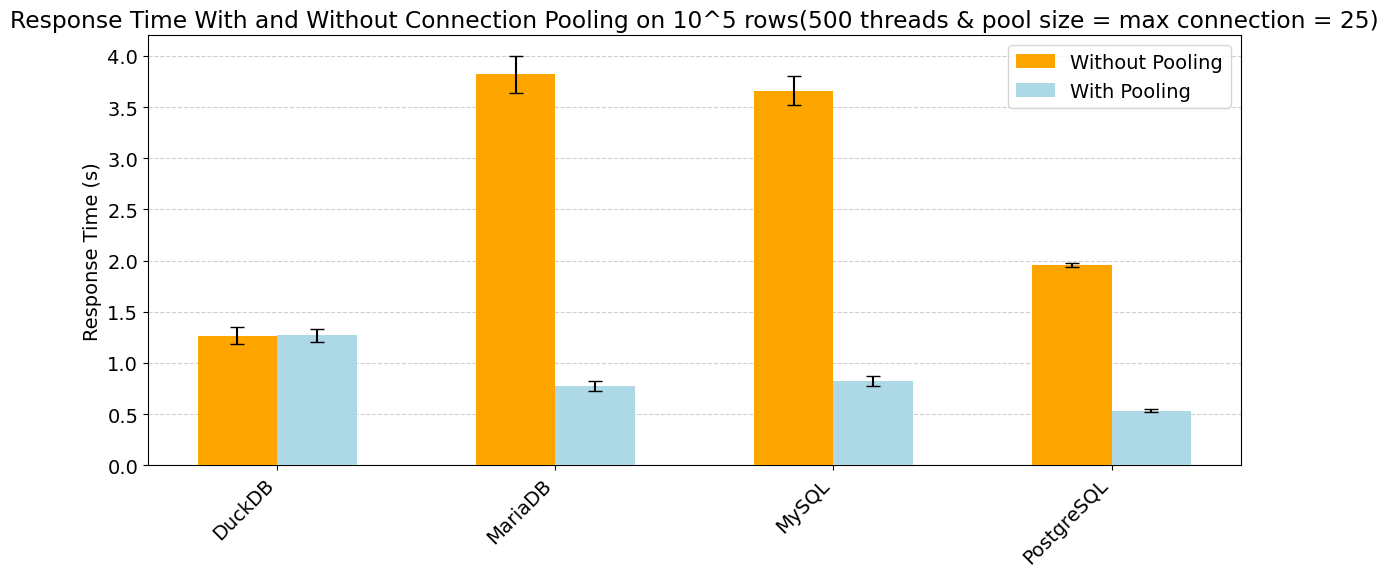}
  \caption{Under 500-thread workloads, the average response time with connection pooling (pool size = max connections = 25) outperforms simple connections across all tested systems except DuckDB on $10^5$ rows.}
  \label{fig: connection pooling t500p25 results for 10^5 dataset}
\end{figure}

\begin{figure}[H]
  \centering
  \includegraphics[width=0.8\textwidth]{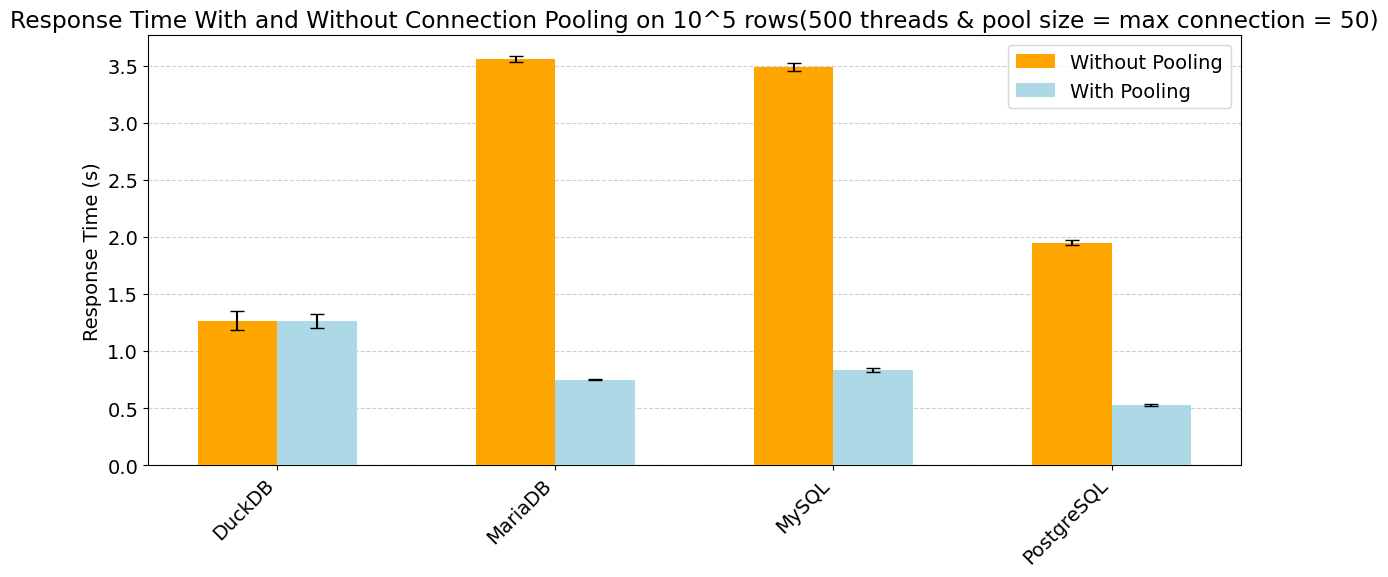}
  \caption{Under 500-thread workloads, the average response time with connection pooling (pool size = max connections = 50) outperforms simple connections across all tested systems except DuckDB on $10^5$ rows.}
  \label{fig: connection pooling t500p50 results for 10^5 dataset}
\end{figure}

\begin{figure}[H]
  \centering
  \includegraphics[width=0.8\textwidth]{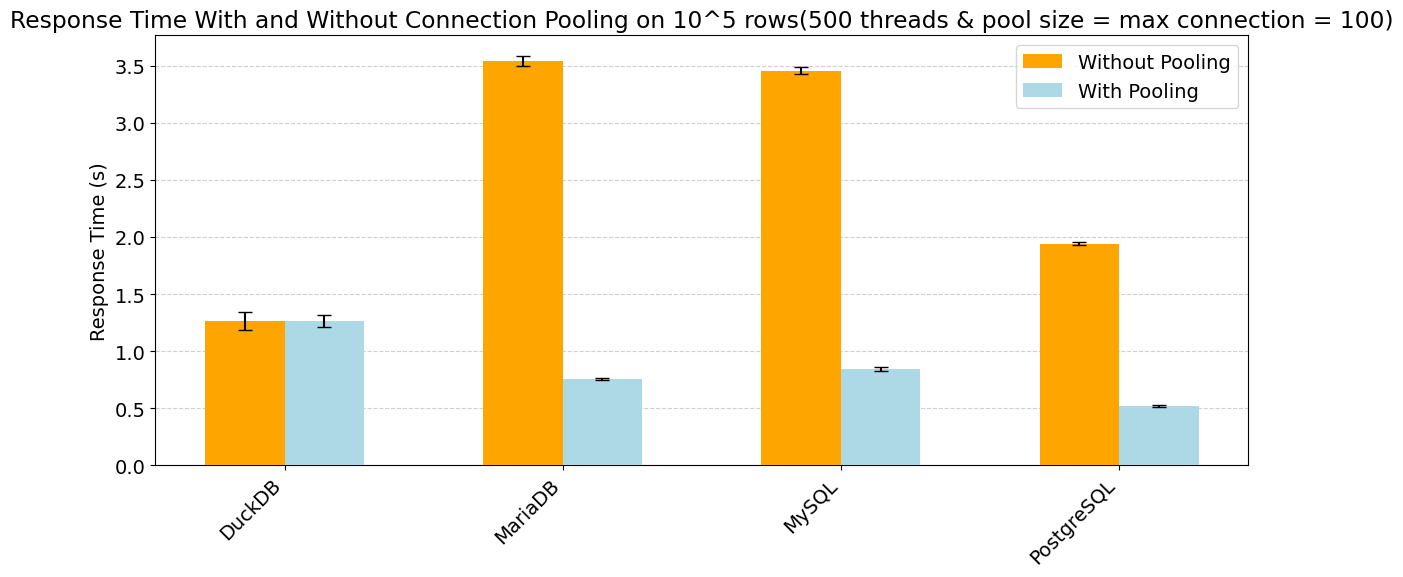}
  \caption{Under 500-thread workloads, the average response time with connection pooling (pool size = max connections = 100) outperforms simple connections across all tested systems except DuckDB on $10^5$ rows.}
  \label{fig: connection pooling t500p100 results for 10^5 dataset}
\end{figure}

At the highest concurrency of 500 threads(Figure~\ref{fig: connection pooling t500p25 results for 10^5 dataset}, Figure~\ref{fig: connection pooling t500p50 results for 10^5 dataset} and Figure~\ref{fig: connection pooling t500p100 results for 10^5 dataset}), the benefits of pooling are still obvious in all the systems except DuckDB. MariaDB’s response time improves by up to 4.96× when switching from simple connections to a pooled setup at pool size and max\_connection equal to 25. PostgreSQL and MySQL also show major gains of 3.73× and 4.44×, respectively. DuckDB shows virtually no difference between the two cases, with both improvements and degradations remaining minimal at around 1.0×. For all systems except DuckDB, pooling with a larger pool size (50 or 100) seems to generally provide better or comparable performance, though marginal returns may diminish beyond 50 connections, especially for MySQL and MariaDB.

In terms of stability(Table~\ref{tab:conn-pooling5}), the standard deviation of response time tends to be lower under simple connections in MySQL, PostgreSQL and MariaDB. In contrast, simple connection is more stable in DuckDB.

\begin{longtable}{lllcc}
\caption{Average response time (Standard deviation) of using connection pooling and using simple connections under various connection settings on $10^5$ rows (in seconds).}
\label{tab:conn-pooling5} \\

\toprule
\textbf{DBMS} &  &  & \textbf{Simple Connection} & \textbf{Connection Pooling} \\
\midrule
\endfirsthead

\toprule
\textbf{DBMS} &  &  & \textbf{Simple Connection} & \textbf{Connection Pooling} \\
\midrule
\endhead
\endfoot

\bottomrule
\endlastfoot

\multicolumn{5}{l}{\textbf{Threads = 10, Max Connection = Pool Size = 25}} \\
\midrule
DuckDB       & & & 0.24 (0.0037) & 0.28 (0.0096) \\
MariaDB      & & & 0.069 (0.0058) & 0.029 (0.0016) \\
MySQL        & & & 0.069 (0.0028) & 0.028 (0.0014) \\
PostgreSQL   & & & 0.042 (0.00070) & 0.020 (0.00056) \\

\midrule
\multicolumn{5}{l}{\textbf{Threads = 10, Max Connection = Pool Size = 50}} \\
\midrule
DuckDB       & & & 0.24 (0.0037) & 0.23 (0.0052) \\
MariaDB      & & & 0.068 (0.0033) & 0.028 (0.0019) \\
MySQL        & & & 0.066 (0.0031) & 0.027 (0.0016) \\
PostgreSQL   & & & 0.043 (0.0026) & 0.021 (0.00090) \\

\midrule
\multicolumn{5}{l}{\textbf{Threads = 10, Max Connection = Pool Size = 100}} \\
\midrule
DuckDB       & & & 0.24 (0.0037) & 0.30 (0.0076) \\
MariaDB      & & & 0.067 (0.0030) & 0.028 (0.0010) \\
MySQL        & & & 0.065 (0.0036) & 0.027 (0.0010) \\
PostgreSQL   & & & 0.044 (0.0035) & 0.022 (0.0031) \\

\midrule
\multicolumn{5}{l}{\textbf{Threads = 100, Max Connection = Pool Size = 25}} \\
\midrule
DuckDB       & & & 0.52 (0.013) & 0.51 (0.0063) \\
MariaDB      & & & 0.71 (0.014) & 0.16 (0.0052) \\
MySQL        & & & 0.70 (0.034) & 0.16 (0.0064) \\
PostgreSQL   & & & 0.39 (0.0031) & 0.11 (0.0053) \\

\midrule
\multicolumn{5}{l}{\textbf{Threads = 100, Max Connection = Pool Size = 50}} \\
\midrule
DuckDB       & & & 0.52 (0.013) & 0.46 (0.010) \\
MariaDB      & & & 0.69 (0.0092) & 0.16 (0.0036) \\
MySQL        & & & 0.69 (0.014) & 0.16 (0.0040) \\
PostgreSQL   & & & 0.39 (0.0031) & 0.11 (0.0080) \\

\midrule
\multicolumn{5}{l}{\textbf{Threads = 100, Max Connection = Pool Size = 100}} \\
\midrule
DuckDB       & & & 0.52 (0.013) & 0.53 (0.0059) \\
MariaDB      & & & 0.70 (0.012) & 0.16 (0.0047) \\
MySQL        & & & 0.69 (0.011) & 0.16 (0.0089) \\
PostgreSQL   & & & 0.39 (0.0039) & 0.11 (0.0074) \\

\midrule
\multicolumn{5}{l}{\textbf{Threads = 500, Max Connection = Pool Size = 25}} \\
\midrule
DuckDB       & & & 1.27 (0.080) & 1.27 (0.064) \\
MariaDB      & & & 3.8 (0.18) & 0.77 (0.049) \\
MySQL        & & & 3.7 (0.14) & 0.82 (0.048) \\
PostgreSQL   & & & 2.0 (0.018) & 0.53 (0.018) \\

\midrule
\multicolumn{5}{l}{\textbf{Threads = 500, Max Connection = Pool Size = 50}} \\
\midrule
DuckDB       & & & 1.27 (0.080) & 1.26 (0.062) \\
MariaDB      & & & 3.6 (0.028) & 0.75 (0.0072) \\
MySQL        & & & 3.5 (0.037) & 0.83 (0.016) \\
PostgreSQL   & & & 1.95 (0.021) & 0.53 (0.011) \\

\midrule
\multicolumn{5}{l}{\textbf{Threads = 500, Max Connection = Pool Size = 100}} \\
\midrule
DuckDB       & & & 1.27 (0.080) & 1.27 (0.053) \\
MariaDB      & & & 3.5 (0.043) & 0.75 (0.0090) \\
MySQL        & & & 3.5 (0.029) & 0.84 (0.021) \\
PostgreSQL   & & & 1.94 (0.011) & 0.52 (0.0087) \\

\end{longtable}

\begin{figure}[H]
  \centering
  \includegraphics[width=0.8\textwidth]{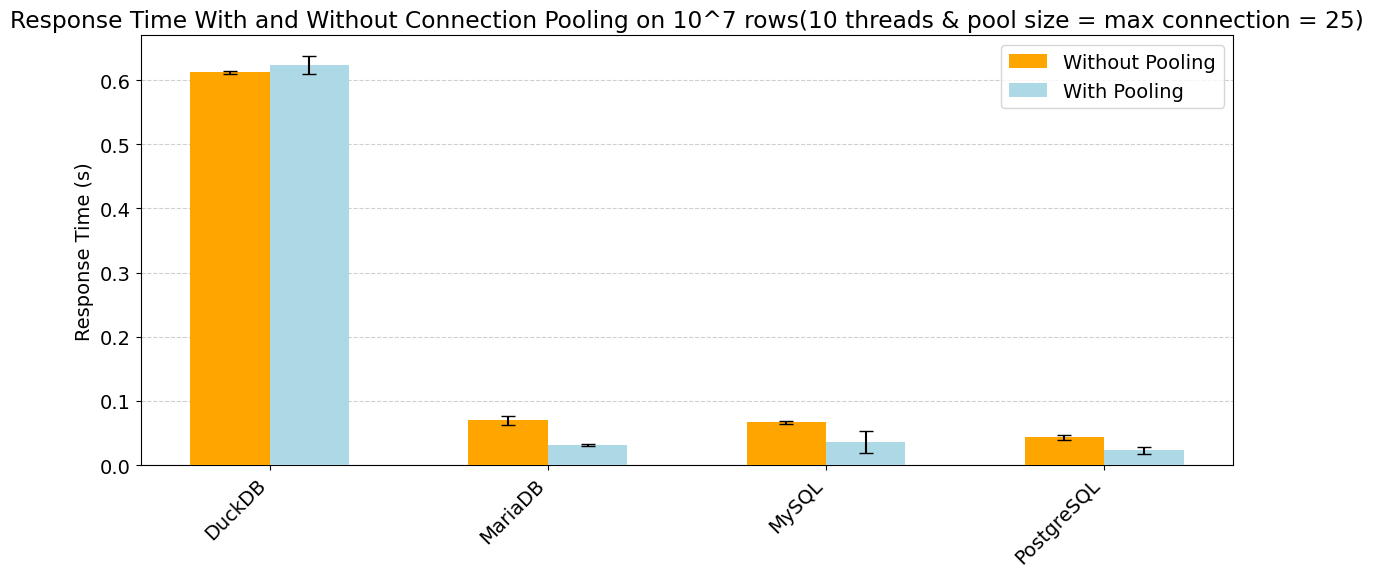}
  \caption{Under 10-thread workloads, the average response time with connection pooling (pool size = max connections = 25) outperforms simple connections except DuckDB on $10^7$ rows.}
  \label{fig: connection pooling t10p25 results for 10^7 dataset}
\end{figure}

\begin{figure}[H]
  \centering
  \includegraphics[width=0.8\textwidth]{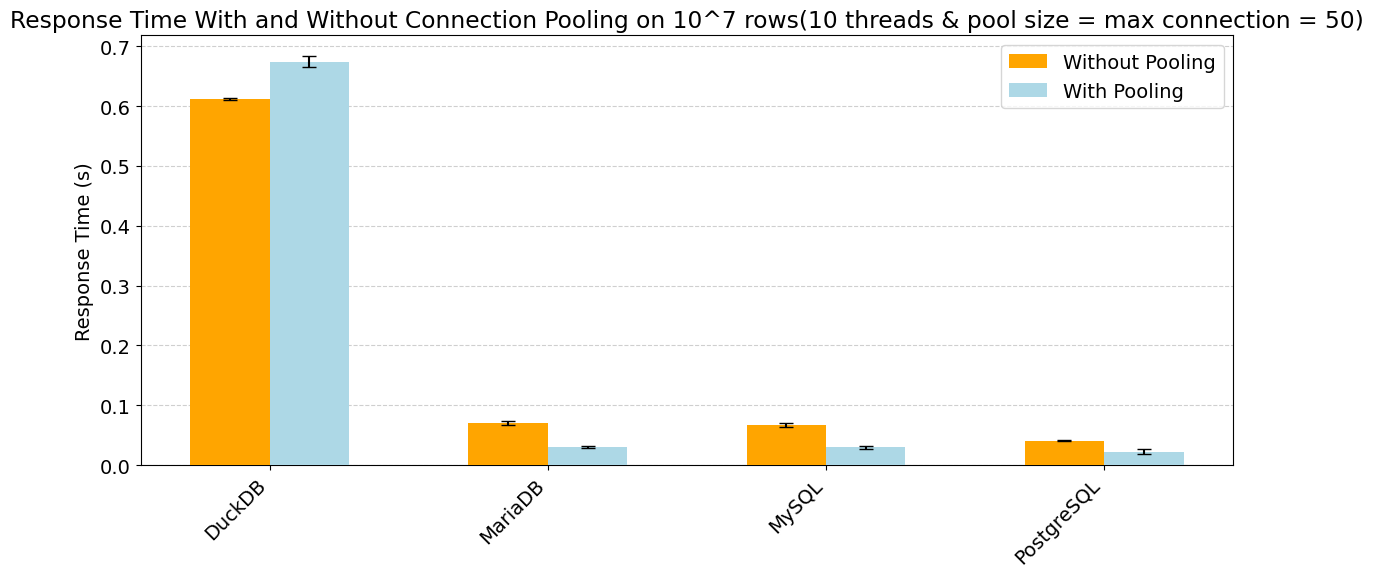}
  \caption{Under 10-thread workloads, the average response time with connection pooling (pool size = max connections = 50) outperforms simple connections across all tested systems except DuckDB on $10^7$ rows.}
  \label{fig: connection pooling t10p50 results for 107 dataset}
\end{figure}

\begin{figure}[H]
  \centering
  \includegraphics[width=0.8\textwidth]{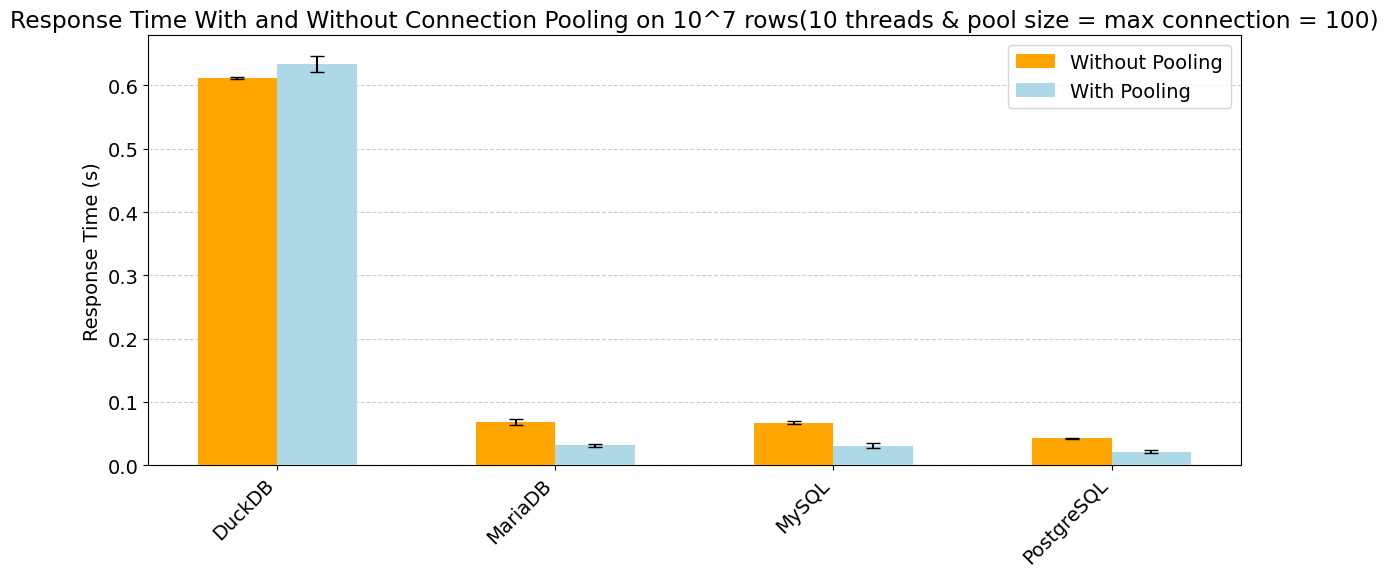}
  \caption{Under 10-thread workloads, the average response time with connection pooling (pool size = max connections = 100) outperforms simple connections across MariaDB, MySQL, and PostgreSQL on $10^7$ rows. For DuckDB, simple connections perform better.}
  \label{fig: connection pooling t10p100 results for 10^7 dataset}
\end{figure}

\textbf{On the $10^7$ row dataset}(Table~\ref{tab:conn-pooling7}), in the low-concurrency scenario (10 threads) (Figure~\ref{fig: connection pooling t10p25 results for 10^7 dataset}, Figure~\ref{fig: connection pooling t10p50 results for 107 dataset} and Figure~\ref{fig: connection pooling t10p100 results for 10^7 dataset}), connection pooling showed clear advantages for all systems except DuckDB. Increasing the pool size to 50 and 100 generally amplified the benefits. For example, at 10 threads with a pool size of 50, MariaDB showed nearly a 2.3× reduction in average response time with pooling. PostgreSQL and MySQL also exhibited effective improvements (around 2× and 2.25×, respectively). In contrast, DuckDB sees slight degradations in performance with connection pooling.

\begin{figure}[H]
  \centering
  \includegraphics[width=0.8\textwidth]{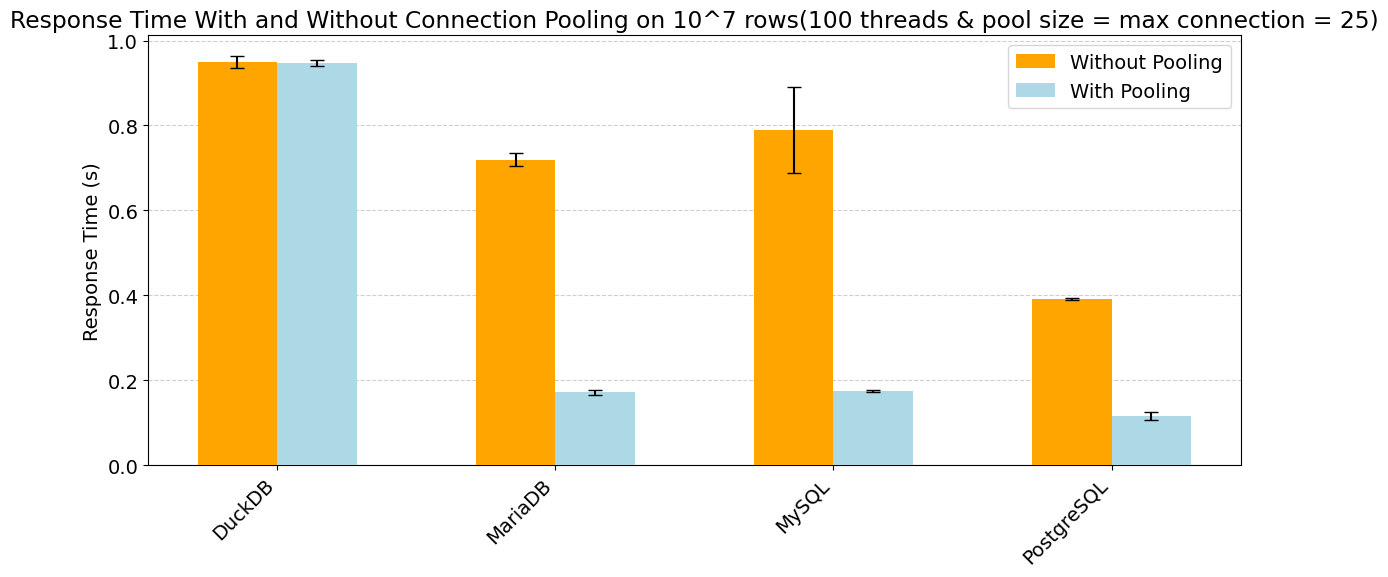}
  \caption{Under 100-thread workloads, the average response time with connection pooling (pool size = max connections = 25) outperforms simple connections across all tested systems except DuckDB on $10^7$ rows. }
  \label{fig: connection pooling t100p25 results for10^7 dataset}
\end{figure}

\begin{figure}[H]
  \centering
  \includegraphics[width=0.8\textwidth]{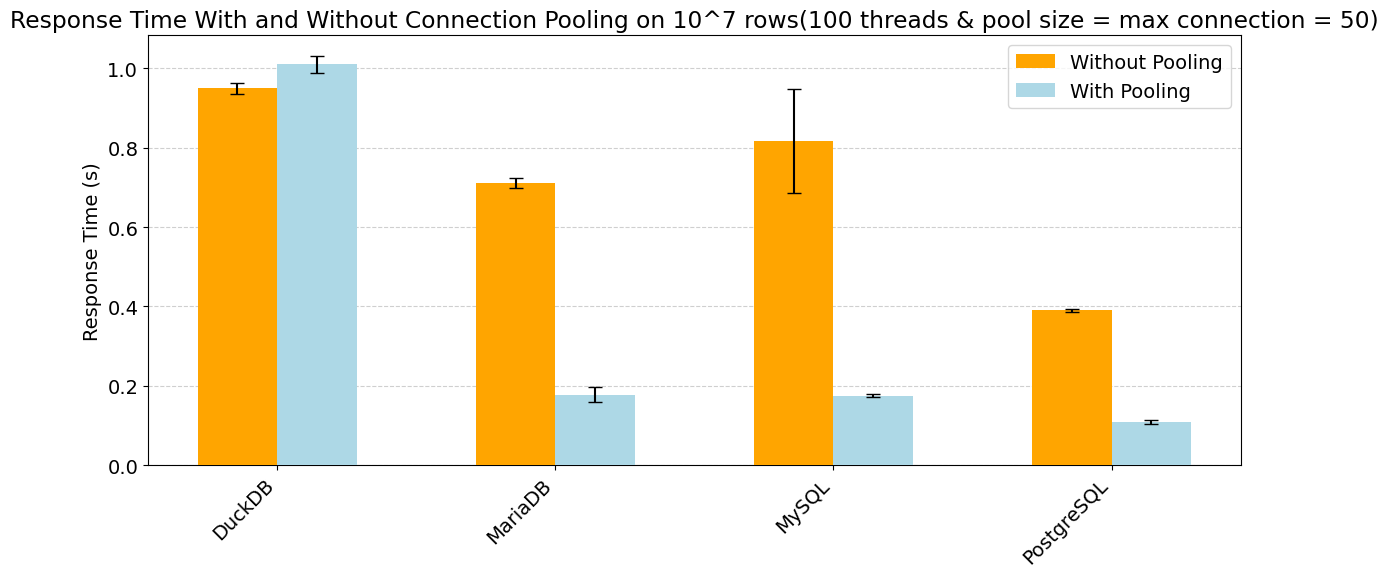}
  \caption{Under 100-thread workloads, the average response time with connection pooling (pool size = max connections = 50) outperforms simple connections across MariaDB, MySQL, and PostgreSQL on $10^7$ rows. For DuckDB, simple connections perform better.}
  \label{fig: connection pooling t100p50 results for 10^7 dataset}
\end{figure}

\begin{figure}[H]
  \centering
  \includegraphics[width=0.8\textwidth]{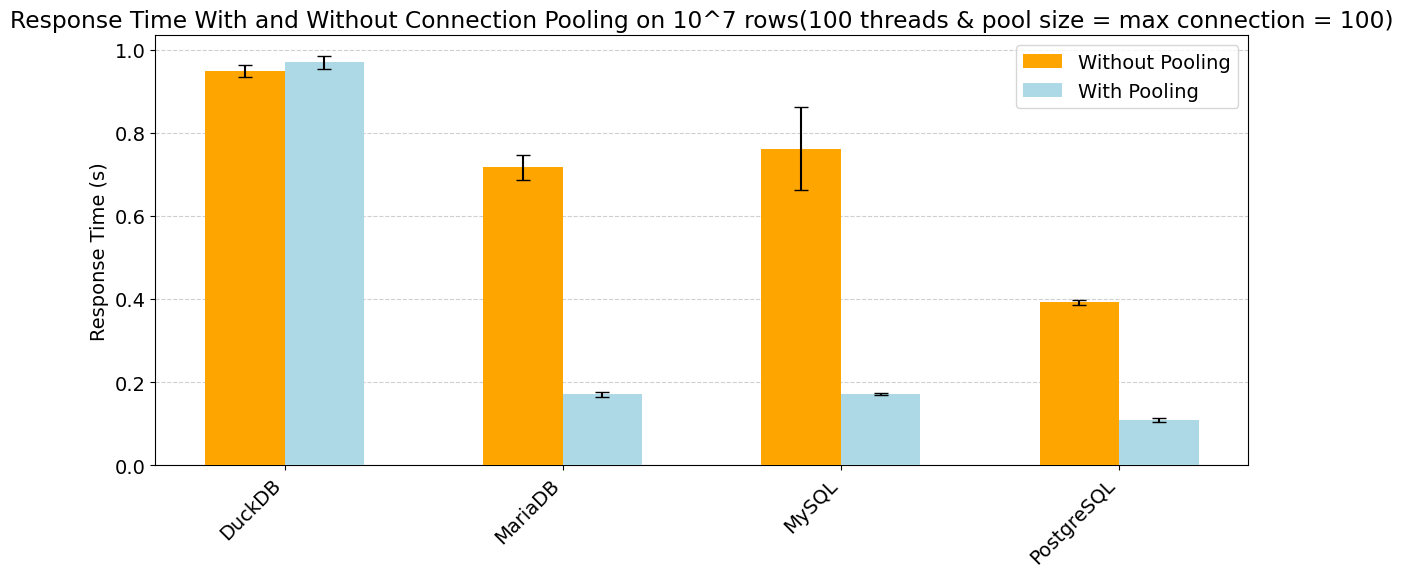}
  \caption{Under 100-thread workloads, the average response time with connection pooling (pool size = max connections = 100) outperforms simple connections across MariaDB, MySQL, and PostgreSQL on $10^7$ rows. For DuckDB, simple connections perform better.}
  \label{fig: connection pooling t100p100 results for 10^7 dataset}
\end{figure}

As the concurrency increased to 100 threads (Figure~\ref{fig: connection pooling t100p25 results for10^7 dataset},  Figure~\ref{fig: connection pooling t100p50 results for 10^7 dataset} and Figure~\ref{fig: connection pooling t100p100 results for 10^7 dataset}), pooling began to deliver more substantial performance gains in MySQL, MariaDB and PostgreSQL. MySQL experienced significant improvements, with response times improving by up to 4.7× when the pool size was set to 50, and remained significantly better (by 4.5× at pool size and max\_connection equal to 25 and 100). PostgreSQL also benefitted, especially at larger pool sizes (up to 3.6× improvement). MariaDB saw consistent benefits, with pooling reducing latency by 4.2×, 4.0×, and 4.2× as pool size increased. For DuckDB, connection pooling shows slightly worse performance compared to using simple connections across when pool sizes increase to larger than 50.

\begin{figure}[H]
  \centering
  \includegraphics[width=0.8\textwidth]{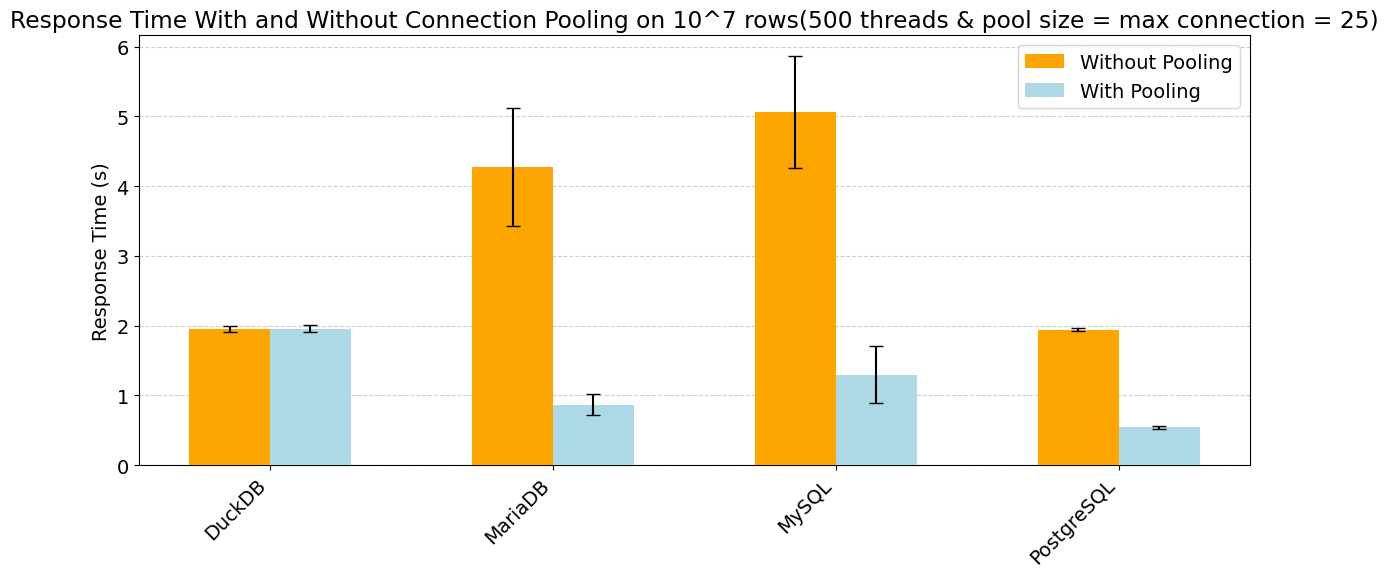}
  \caption{Under 500-thread workloads, the average response time with connection pooling (pool size = max connections = 25) outperforms simple connections across all tested systems except DuckDB on $10^7$ rows.}
  \label{fig: connection pooling t500p25 results for 10^7 dataset}
\end{figure}

\begin{figure}[H]
  \centering
  \includegraphics[width=0.8\textwidth]{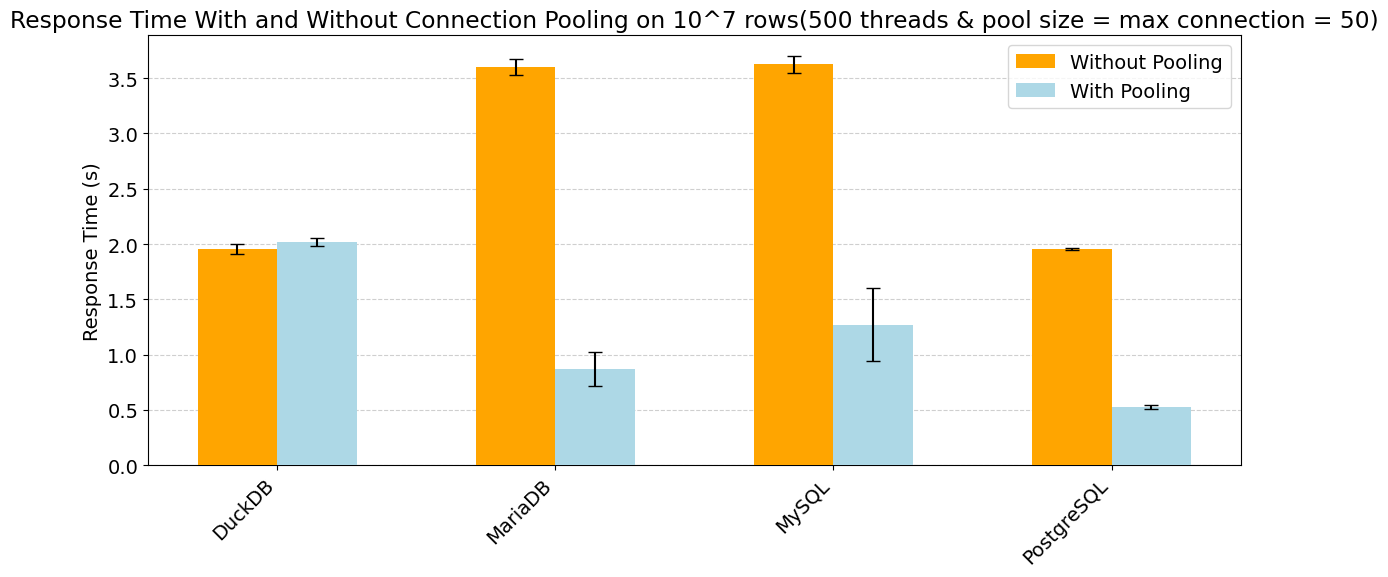}
  \caption{Under 500-thread workloads, the average response time with connection pooling (pool size = max connections = 50) outperforms simple connections across MariaDB, MySQL, and PostgreSQL on $10^7$ rows. For DuckDB, simple connections perform better.}
  \label{fig: connection pooling t500p50 results for 10^7 dataset}
\end{figure}

\begin{figure}[H]
  \centering
  \includegraphics[width=0.8\textwidth]{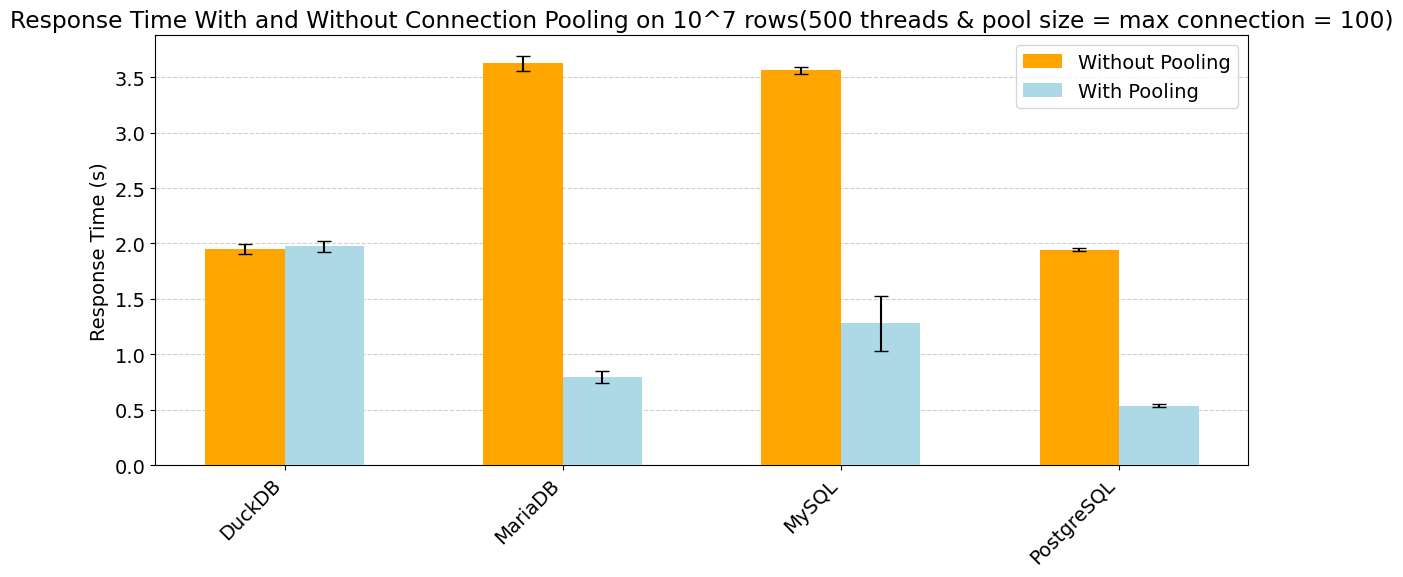}
  \caption{Under 500-thread workloads, the average response time with connection pooling (pool size = max connections = 100) outperforms simple connections across MariaDB, MySQL, and PostgreSQL on $10^7$ rows. For DuckDB, simple connections perform better.}
  \label{fig: connection pooling t500p100 results for 10^7 dataset}
\end{figure}

Under very high concurrency (500 threads) (Figure~\ref{fig: connection pooling t500p25 results for 10^7 dataset}, Figure~\ref{fig: connection pooling t500p50 results for 10^7 dataset} and Figure~\ref{fig: connection pooling t500p100 results for 10^7 dataset}), connection pooling appeared to be critical to maintain performance. For MariaDB and MySQL, response times dropped by 4.9× and 3.9×, respectively, at a small pool size of 25. Even at larger pool sizes (50 and 100), the benefits remained substantial (4.1× and 4.6x for MariaDB, 2.9× and 2.8x for MySQL). PostgreSQL followed a similar trend, with pooling improving performance by 3.69× at pool size 50 and remaining over at least 3× better at all pool sizes. DuckDB still works better with simple connection across all pool sizes.

In terms of response time variability (Table~\ref{tab:conn-pooling7}), pooling consistently reduced standard deviation across MySQL, MariaDB and PostgreSQL under lower load. When the number of threads increased to 500, simple connection shows better variability. In contrast, DuckDB displayed increased variability under pooling in many cases. 

\begin{longtable}{lllcc}
\caption{Average response time (Standard deviation) of using connection pooling and using simple connections under various connection settings on $10^7$ rows (in seconds).}
\label{tab:conn-pooling7} \\

\toprule
\textbf{DBMS} &  &  & \textbf{Simple Connection} & \textbf{Connection Pooling} \\
\midrule
\endfirsthead

\toprule
\textbf{DBMS} &  &  & \textbf{Simple Connection} & \textbf{Connection Pooling} \\
\midrule
\endhead

\endfoot
\bottomrule
\endlastfoot

\multicolumn{5}{l}{\textbf{Threads = 10, Max Connection = Pool Size = 25}} \\
\midrule
DuckDB        & & & 0.61 (0.0021) & 0.62 (0.014) \\
MariaDB       & & & 0.070 (0.0068) & 0.031 (0.0014) \\
MySQL         & & & 0.067 (0.0024) & 0.036 (0.018) \\
PostgreSQL    & & & 0.043 (0.0039) & 0.023 (0.0050) \\

\midrule
\multicolumn{5}{l}{\textbf{Threads = 10, Max Connection = Pool Size = 50}} \\
\midrule
DuckDB        & & & 0.61 (0.0021) & 0.67 (0.0096) \\
MariaDB       & & & 0.070 (0.0030) & 0.031 (0.0013) \\
MySQL         & & & 0.067 (0.0031) & 0.030 (0.0025) \\
PostgreSQL    & & & 0.041 (0.00055) & 0.023 (0.0046) \\

\midrule
\multicolumn{5}{l}{\textbf{Threads = 10, Max Connection = Pool Size = 100}} \\
\midrule
DuckDB        & & & 0.61 (0.0021) & 0.63 (0.013) \\
MariaDB       & & & 0.068 (0.0042) & 0.031 (0.0020) \\
MySQL         & & & 0.067 (0.0027) & 0.031 (0.0038) \\
PostgreSQL    & & & 0.042 (0.00088) & 0.021 (0.0026) \\

\midrule
\multicolumn{5}{l}{\textbf{Threads = 100, Max Connection = Pool Size = 25}} \\
\midrule
DuckDB        & & & 0.95 (0.015) & 0.95 (0.0076) \\
MariaDB       & & & 0.72 (0.015) & 0.17 (0.0063) \\
MySQL         & & & 0.79 (0.10) & 0.17 (0.0021) \\
PostgreSQL    & & & 0.39 (0.0033) & 0.12 (0.0086) \\

\midrule
\multicolumn{5}{l}{\textbf{Threads = 100, Max Connection = Pool Size = 50}} \\
\midrule
DuckDB        & & & 0.95 (0.015) & 1.0 (0.021) \\
MariaDB       & & & 0.71 (0.012) & 0.18 (0.019) \\
MySQL         & & & 0.82 (0.13) & 0.18 (0.0042) \\
PostgreSQL    & & & 0.39 (0.0035) & 0.11 (0.0048) \\

\midrule
\multicolumn{5}{l}{\textbf{Threads = 100, Max Connection = Pool Size = 100}} \\
\midrule
DuckDB        & & & 0.95 (0.015) & 0.97 (0.016) \\
MariaDB       & & & 0.72 (0.030) & 0.17 (0.0059) \\
MySQL         & & & 0.76 (0.10) & 0.17 (0.0027) \\
PostgreSQL    & & & 0.39 (0.0049) & 0.11 (0.0049) \\

\midrule
\multicolumn{5}{l}{\textbf{Threads = 500, Max Connection = Pool Size = 25}} \\
\midrule
DuckDB        & & & 2.0 (0.046) & 2.0 (0.048) \\
MariaDB       & & & 4.3 (0.85) & 0.86 (0.15) \\
MySQL         & & & 5.1 (0.80) & 1.3 (0.41) \\
PostgreSQL    & & & 1.9 (0.021) & 0.54 (0.021) \\

\midrule
\multicolumn{5}{l}{\textbf{Threads = 500, Max Connection = Pool Size = 50}} \\
\midrule
DuckDB        & & & 2.0 (0.046) & 2.0 (0.036) \\
MariaDB       & & & 3.6 (0.073) & 0.87 (0.16) \\
MySQL         & & & 3.6 (0.077) & 1.3 (0.33) \\
PostgreSQL    & & & 2.0 (0.010) & 0.53 (0.017) \\

\midrule
\multicolumn{5}{l}{\textbf{Threads = 500, Max Connection = Pool Size = 100}} \\
\midrule
DuckDB        & & & 2.0 (0.046) & 2.0 (0.047) \\
MariaDB       & & & 3.6 (0.067) & 0.80 (0.054) \\
MySQL         & & & 3.6 (0.029) & 1.3 (0.25) \\
PostgreSQL    & & & 1.9 (0.0098) & 0.54 (0.018) \\

\end{longtable}






\end{document}